\documentclass[a4paper,titlepage,11pt]{book}
\pdfoutput=1
\usepackage{fancyhdr}
\usepackage[Lenny]{fncychap}

\usepackage{emptypage}
\usepackage{slashed}

\makeatletter
\newcommand\HUGE{\@setfontsize\Huge{35}{50}}	
\makeatother

\renewcommand{\headrulewidth}{0pt}
\fancypagestyle{FirstPage}{
	\lfoot{\begin{wrapfigure}{l}{0.15\textwidth}
			\centering
			\includegraphics[width=0.15\textwidth]{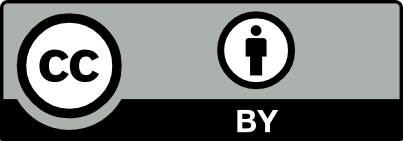}
		\end{wrapfigure}
	\smallskip\newline\tiny{This document is licensed under the Creative Commons Attribution 4.0 International License (CC BY 4.0):
			http://creativecommons.org/licenses/by/4.0 This CC license does not apply to third party material (attributed to another source) in this publication.}}
}

	{
	\end{minipage}
	\vspace*{\stretch{3}}
	\clearpage
}

\usepackage[T1]{fontenc}                    
\usepackage[utf8]{inputenc}                 
\usepackage[italian,english]{babel}         
\setactivedoublequote                       
\usepackage{microtype}                      
\usepackage{lmodern}
\usepackage{epigraph}						
\usepackage{appendix}               
\usepackage{comment}

\usepackage[autostyle,italian=guillemets]{csquotes}
\usepackage[backend=biber,bibstyle=numeric,citestyle=numeric-comp,minnames=3,maxnames=4,hyperref=auto,sorting=none]{biblatex}
\usepackage{amssymb}                
\usepackage{amsmath,esint}          
\usepackage{mathrsfs}               
\usepackage{mathtools}              
\usepackage{caption}                
\usepackage{graphicx}               
\usepackage{subcaption}
\usepackage{wrapfig}
\usepackage{siunitx}                
\DeclareSIUnit{\angstrom}{\textup{\AA}}
\usepackage{dsfont}					
\usepackage{booktabs}
\usepackage{array}
\usepackage{afterpage}

\newcommand*\chem[1]{\ensuremath{\mathrm{#1}}}

\DeclarePairedDelimiter{\abs}{\lvert}{\rvert}       

\newcommand*\dif{\mathop{}\!\mathrm{d}}             
\newcommand{\vect}[1]{\boldsymbol{\mathbf{#1}}}     

\let\temp\varepsilon                                
\let\varepsilon\epsilon
\let\epsilon\temp

\let\tmp\oddsidemargin
\let\oddsidemargin\evensidemargin
\let\evensidemargin\tmp
\reversemarginpar

\usepackage[pdfauthor={Luca Martinoia},bookmarks=false,colorlinks,pdftitle={PhD-Thesis},pdfpagelabels=true,linkcolor=blue,plainpages=false]{hyperref}      
\newcommand{\mail}[1]{\href{mailto:#1}{\texttt{#1}}}            

\usepackage{doi}

\DeclareFieldFormat{doilink}{%
	\iffieldundef{doi}{%
		\iffieldundef{url}{%
		}{%
			\href{\thefield{url}}{#1}%
		}%
	}{%
		\href{http://dx.doi.org/\thefield{doi}}{#1}%
	}%
}

\renewbibmacro{in:}{
	\ifentrytype{article}
	{}
	{\bibstring{in}%
		\printunit{\intitlepunct}}}
\renewbibmacro{in:}{
	\ifentrytype{inproceedings}
	{}
	{\bibstring{in}%
		\printunit{\intitlepunct}}}
\renewbibmacro{in:}{
	\ifentrytype{incollection}
	{}
	{\bibstring{in}%
		\printunit{\intitlepunct}}}

\renewbibmacro*{eprint}{%
	\iffieldundef{eprinttype}
	{\printfield{eprint}}
	{\printfield[eprint:\strfield{eprinttype}]{eprint}}}

\DeclareBibliographyDriver{article}{%
	\usebibmacro{bibindex}%
	\usebibmacro{begentry}%
	\usebibmacro{author/translator+others}%
	\setunit{\labelnamepunct}\newblock
	\usebibmacro{title}\addcomma\space%
	\newunit\newblock
	\usebibmacro{byauthor}%
	\newunit\newblock
	\usebibmacro{bytranslator+others}%
	\newunit\newblock
	\printfield{version}%
	\newunit\newblock
	\printtext[doilink]{%
		\usebibmacro{journal+issuetitle}%
		\newunit
		\usebibmacro{byeditor+others}%
		\newunit
		\usebibmacro{note+pages}%
	}%
	\newunit\newblock
	\iftoggle{bbx:isbn}
	{\printfield{issn}}
	{}%
	\newunit\newblock
	\usebibmacro{doi+eprint+url}%
	\newunit\newblock
	\usebibmacro{addendum+pubstate}%
	\setunit{\bibpagerefpunct}\newblock
	\usebibmacro{pageref}%
	\usebibmacro{finentry}}
	
\DeclareBibliographyDriver{book}{%
	\usebibmacro{bibindex}%
	\usebibmacro{begentry}%
	\usebibmacro{author/translator+others}%
	\setunit{\labelnamepunct}\newblock
	\usebibmacro{title}\addcomma\space%
	\newunit\newblock
	\usebibmacro{byauthor}%
	\newunit\newblock
	\usebibmacro{bytranslator+others}%
	\newunit\newblock
	\printfield{version}%
	\newunit\newblock
	\printtext[doilink]{%
		\printlist{publisher}%
		\newunit
		\usebibmacro{date}%
		\newunit
		\usebibmacro{note+pages}%
	}%
	\newunit\newblock
	\iftoggle{bbx:isbn}
	{\printfield{issn}}
	{}%
	\newunit\newblock
	\usebibmacro{eprint}%
	\newunit\newblock
	\usebibmacro{addendum+pubstate}%
	\setunit{\bibpagerefpunct}\newblock
	\usebibmacro{pageref}%
	\usebibmacro{finentry}}
	
\DeclareBibliographyDriver{incollection}{%
	\usebibmacro{bibindex}%
	\usebibmacro{begentry}%
	\usebibmacro{author/translator+others}%
	\setunit{\labelnamepunct}\newblock
	\usebibmacro{title}\addcomma\space%
	\newunit\newblock
	\usebibmacro{byauthor}%
	\newunit\newblock
	\usebibmacro{bytranslator+others}%
	\newunit\newblock
	\printfield{version}%
	\newunit\newblock
	\printtext[doilink]{%
		\printlist{publisher}%
		\newunit
		\usebibmacro{date}%
		\newunit
		\usebibmacro{note+pages}%
	}%
	\newunit\newblock
	\iftoggle{bbx:isbn}
	{\printfield{issn}}
	{}%
	\newunit\newblock
	\usebibmacro{doi+eprint+url}%
	\newunit\newblock
	\usebibmacro{addendum+pubstate}%
	\setunit{\bibpagerefpunct}\newblock
	\usebibmacro{pageref}%
	\usebibmacro{finentry}}

\DeclareCiteCommand{\fullcite}
{\usebibmacro{prenote}}
{\usedriver
	{\defcounter{minnames}{6}%
		\defcounter{maxnames}{6}}
	{\thefield{entrytype}}}
{\multicitedelim}
{\usebibmacro{postnote}}

\begin{document}

\hypersetup{pageanchor=false}
\pagenumbering{alph}
\thispagestyle{empty}
\begin{center}
	\textcolor{white}{.}
	\vspace{6cm}
	
	{\HUGE\textbf{Developments in}}
	\vspace{0.7cm}
	
	{\HUGE\textbf{quasihydrodynamics}}
\end{center}

\begin{titlepage}
	\begin{center}
		\HUGE
		\textbf{Developments in quasihydrodynamics}\\

		\vspace{0.5cm}
		\LARGE
		Theory and applications

		\vspace{1.2cm}
		\renewcommand{\thefootnote}{\fnsymbol{footnote}}
		
		\textbf{Luca Martinoia}\footnote{\mail{luca.martinoia@ge.infn.it}}
		\renewcommand{\thefootnote}{\arabic{footnote}}
		
		\vspace{1.5cm}
		A thesis presented for the degree of\\
		Doctor of Philosophy in Physics and Nanoscience\\
		\vspace{0.2cm}
		\Large{Theoretical Physics curriculum}
		\vspace{1cm}

		\includegraphics[width=0.7\textwidth]{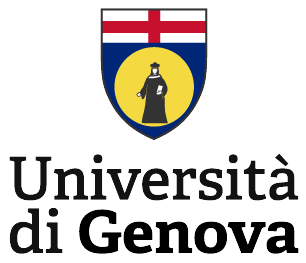}
		
	\end{center}
\end{titlepage}

\makeatletter		
\@openrightfalse
\makeatother

\frontmatter
	\hypersetup{pageanchor=true}	
	\pagestyle{plain}
	\tableofcontents
\chapter*{Referees and defense details}
\addcontentsline{toc}{chapter}{Examining Committee}
\thispagestyle{FirstPage}
Referees for this thesis:
\begin{itemize}
	\item Prof.~Koenraad Schalm -- Leiden University
	\item Prof.~Piotr Surówka -- Wrocław University of Science and Technology
\end{itemize}

\noindent The defense of the doctoral degree in Physics was held at the University of Genoa in the Department of Physics on the 1$^\text{st}$ of March 2024.

\medskip
\noindent Examination Committee for the defense:
\begin{itemize}
	\item Prof.~Paolo Solinas -- University of Genoa
	\item Prof.~Piotr Surówka -- Wrocław University of Science and Technology
	\item Prof.~Daniele Passerone -- ETH Zürich
	\item Prof.~Fernando Dominguez -- Technische Universität Braunschweig
\end{itemize}

\medskip
\noindent Supervisor: Prof.~Andrea Amoretti

\medskip
\noindent Candidate: Luca Martinoia\newline
University ID: 3630563\newline
Italian Ph.D. cycle:	XXXVI

\chapter*{Summary}
\addcontentsline{toc}{chapter}{Summary}
At its core, hydrodynamics is a many-body low-energy effective theory for the long-wavelength, long-timescale dynamics of conserved charges in systems close to thermodynamic equilibrium. It has a wide range of applications, that span from nuclear physics, astrophysics, cosmology, and more recently strongly-interacting electronic phases of matter. In solid state systems, however, symmetries are often only approximate, and softly broken by the presence of the lattice, impurities and defects, or because the symmetry is accidental. Therefore, the hydrodynamic regime must be expanded to include weak non-conservation effects, which lead to a theory known as quasihydrodynamics.

In this thesis we make progress in understanding the theory of (quasi) hydrodynamics, with a specific focus on applications to condensed matter systems and their holographic description. First, we consider an electron fluid in a strong magnetic field for which translations are broken by the presence of Charge Density Waves. Therefore, the low-energy theory contains Goldstone modes associated with the broken symmetries, which modify the spectrum and transport properties. We focus on a new regime at non-zero lattice pressure and compare with holographic models, finding perfect agreement between the two descriptions.

Next we consider a simple system that mimics the weakly-coupled Drude model from a hydrodynamic perspective. Specifically, a charged fluid in an external electric field in the presence of impurities that relax momentum and energy. We look for steady states, and we find that stationarity constraints should be modified to include relaxations, which consequently give novel predictions for the thermoelectric transport.

Finally, we study the effect of the axial anomaly on the transport properties of Weyl semimetals in the hydrodynamic regime. We suggest a better approach to deal with the derivative counting of the magnetic field, correcting mistakes in the literature. Subsequently, we discuss the DC values of the conductivities and look for models that obey fundamental and phenomenological considerations. We find that generalized relaxations, which we study in depth using variational methods and kinetic-theory approaches, are a necessary ingredient to have finite DC conductivity, conserve electric charge, and have the correlators obey Onsager relations. Moreover, our model provides qualitatively new predictions for the thermoelectric transport, which could be used to probe the hydrodynamic regime in such materials.

\chapter*{Publications}
\addcontentsline{toc}{chapter}{Publications}

This thesis is based on the following original publications:
\begin{itemize}
	\item Chapter~\ref{chapter:charge_density_waves} is based on \cite{Amoretti:HydrodynamicMagnetotransportHolographic}, \fullcite{Amoretti:HydrodynamicMagnetotransportHolographic}
	\item Chapter~\ref{chapter:electrically_driven_fluids} is based on \cite{Amoretti:NondissipativeElectricallyDriven}, \fullcite{Amoretti:NondissipativeElectricallyDriven}
	\item Chapter~\ref{chapter:onsager} is based on \cite{Amoretti:RestoringTimereversalCovariance}, \fullcite{Amoretti:RestoringTimereversalCovariance}
	\item Chapter~\ref{chapter:anomalous_hydrodynamics} is based on \cite{Amoretti:LeadingOrderMagnetic}, \fullcite{Amoretti:LeadingOrderMagnetic}\\and \cite{Amoretti:RelaxationTermsAnomalous}, \fullcite{Amoretti:RelaxationTermsAnomalous}
\end{itemize}
During the Ph.D. I also worked on the following paper, which will not be discussed here:
\begin{itemize}
	\item \cite{Amoretti:DestroyingSuperconductivityThin} \fullcite{Amoretti:DestroyingSuperconductivityThin}.
\end{itemize}

\makeatletter		
\@openrighttrue
\makeatother

\chapter*{Aknowledgment}
\addcontentsline{toc}{chapter}{Acknowledgement}
First, I express my sincere gratitude to my advisor, Andrea Amoretti, for his thoughtful guidance throughout the Ph.D. I am especially thankful for his support and trust in my research, but also for encouraging me to step out of my comfort zone.

I would also like to acknowledge my close collaborators: Danny Brattan, Ioannis Matthaiakakis, and more recently, Jonas Rongen. I had many interesting and insightful discussions about physics with them over the years, which have taught me a lot and helped me better understand many topics.

On a more personal note, I am extremely grateful to my wonderful wife, Valeria, who has patiently supported me throughout the doctorate and always manages to help in challenging times. I am a better person because of her. These years together have been amazing, and I am truly excited to see what awaits us in the future.

Finally, I would really thank my parents, Sergio and Laura, my family, Simone, Anna, Giovanni, Anouck, Nhoack, Annina, and all my friends, for still keeping me around after all these years.

\mainmatter
	\newpage

	\pagestyle{fancy}
\chapter{Introduction and context}\label{chapter:introduction}
\epigraph{``I don’t see how he can ever finish if he doesn’t begin.''}{Lewis Carroll, \emph{Alice in Wonderland}}

\section{Hydrodynamic regime and applications}\label{sec:ch1:hydrodynamic_regime}
Hydrodynamics and hydrostatics are very old subjects, which date back more than two thousands years ago to Archimedes' work, however it is only in the 18th century that physicists started a systematic study of hydrodynamics to understand the flow of water and liquids in general.

To this day hydrodynamics is fundamentally well understood \cite{Landau:FluidMechanicsVolume}, but it remains one of the topics at the forefront of research. This is due to the numerical and theoretical challenges on one side, but also to its universality, which makes it a great tool to describe the phenomenology of many different effects in all branches of physics: astrophysics (like neutron stars \cite{Shibata:GeneralRelativisticViscous}, mergers \cite{Faber:HydrodynamicsNeutronStar}, accretion disks \cite{Balbus:InstabilityTurbulenceEnhanced}), cosmology \cite{Weinberg:GravitationCosmologyPrinciples}, condensed matter (graphene \cite{Lucas:HydrodynamicsElectronsGraphene,Narozhny:ElectronicHydrodynamicsGraphene}, high-$T_c$ superconductors \cite{Hartnoll:TheoryNernstEffect,Baggioli:ColloquiumHydrodynamicsHolography}, Dirac and Weyl semimetals \cite{Lucas:HydrodynamicTheoryThermoelectric,Gorbar:NonlocalTransportWeyl}, \dots), active and soft matter \cite{Toner:FlocksHerdsSchools}, and high-energy physics (quark-gluon-plasma produced at RHIC and LHC \cite{Arslandok:HotQCDWhite,Becattini:PolarizationVorticityQuark,Heinz:CollectiveFlowViscosity}), \dots

From a theoretical perspective, hydrodynamics has gained a lot of attentions in recent years, thanks to a renewed interest following the discovery of the fluid/gravity duality in holography \cite{Ammon:GaugeGravityDuality,Rangamani:GravityHydrodynamicsLectures,Hartnoll:HolographicQuantumMatter}. This has opened the paths to many new research avenues in theoretical hydrodynamics, of which some examples are: a classification of dissipative superfluid terms \cite{Bhattacharya:TheoryFirstOrder}, and of second- \cite{Baier:RelativisticViscousHydrodynamics,Bhattacharyya:NonlinearFluidDynamics,Romatschke:RelativisticViscousFluid} and third-order hydrodynamics \cite{El:ThirdorderRelativisticDissipative,Grozdanov:ConstructingHigherorderHydrodynamics}, parity-odd fluids \cite{Jensen:ParityViolatingHydrodynamicsDimensions,Lier:PassiveOddViscoelasticity,Lucas:PhenomenologyNonrelativisticParityviolating}, quantum anomalies \cite{Son:HydrodynamicsTriangleAnomalies,Jensen:TriangleAnomaliesThermodynamics}, spin hydrodynamics \cite{Montenegro:LinearResponseTheory,Gallegos:HydrodynamicsSpinCurrents,Becattini:SpinTensorIts}, viscoelastic and viscoplastic fluids with broken translation symmetry (both spontaneous and explicit) \cite{Armas:ApproximateSymmetriesPseudoGoldstones,Armas:HydrodynamicsPlasticDeformations,Armas:HydrodynamicsChargeDensity,Amoretti:HydrodynamicMagnetotransportHolographic,Baggioli:ColloquiumHydrodynamicsHolography}, developments in fluctuating hydrodynamics as an EFT from an action principle \cite{Kovtun:EffectiveActionRelativistic,Grozdanov:ViscosityDissipativeHydrodynamics,Glorioso:LecturesNonequilibriumEffective,Haehl:FluidManifestoEmergent,Haehl:EffectiveActionRelativistic}, a full classification of the possible classes of hydrodynamic transport \cite{Haehl:AdiabaticHydrodynamicsEightfold,Haehl:EightfoldWayDissipation}, quasihydrodynamics \cite{Baggioli:QuasihydrodynamicsSchwingerKeldyshEffective,Grozdanov:HolographyHydrodynamicsWeakly}, Hydro+ to extend hydrodynamics to include slow modes near critical points \cite{Stephanov:HydroHydrodynamicsParametric}, fracton hydrodynamics \cite{Glodkowski:HydrodynamicsDipoleconservingFluids,Grosvenor:HydrodynamicsIdealFracton,Guo:FractonHydrodynamicsTimereversal}, hydrodynamics with higher-form symmetries \cite{Das:HigherformSymmetriesAnomalous,Grozdanov:GeneralizedGlobalSymmetries,Armas:ApproximateHigherformSymmetries}.

Before dealing with some of these research paths, however, we should discuss what hydrodynamics is first. It can be understood as a universal low-energy theory for many-body thermal systems that describes the collective macroscopic dynamics of conserved charges (or other low-energy modes) in the long-wavelength and small-frequency regime $\omega,\vect{k}\ll T$, where $T$ is the temperature of the system. To understand the emergence of hydrodynamics we will follow the heuristic argument given in \cite{Glorioso:LecturesNonequilibriumEffective}.

Consider some quantum many-body system. At zero temperature, the low-energy theory is characterized by massless quasiparticle excitations above the ground state. Generically, these modes are long-lived and give rise to e.g. Fermi-liquid behaviour. If the system is thermal, however, then there is a large bath of gapless quasiparticles and any new excitation becomes quickly incoherent in the bath, on a time (and length) scale fixed by the microscopic dynamics $\tau$, $l$. This effectively means that the quasiparticles become gapped due to the presence of the thermal bath, and decay on very short scale, leading to a system that thermalizes quickly and can thus be assumed in local thermodynamic equilibrium on macroscopic scales. This should be understood from a coarse-grained perspective by dividing the system in small volume elements, such that each element is small enough to be effectively point-like in the effective-theory description, but microscopically large enough to have a well-defined thermodynamic limit.

\begin{figure}
	\centering
	\includegraphics[width=0.7\textwidth]{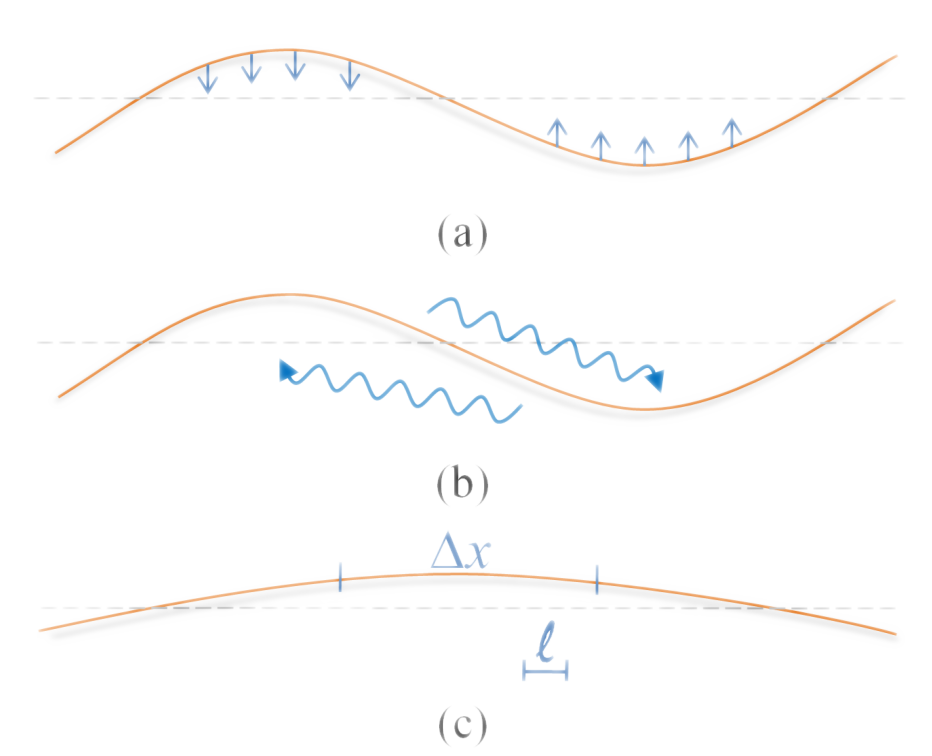}
	\caption{Figure taken from \cite{Glorioso:LecturesNonequilibriumEffective}. (a) Microscopic degrees of freedom thermalize quickly to equilibrium. On the other hand, (b) conserved charges cannot be destroyed locally and diffuse slowly. (c) The typical microscopic length scale $l$ must be small compared to the length over which conserved charges vary.}
	\label{fig:ch1:hydrodynmic_regime}
\end{figure}

Thus, in this regime, the low-energy dynamics is governed by the conserved charges, because they cannot be destroyed locally, but can only diffuse from regions with an excess of charge to regions with a deficit of charge, and this process is parametrically slower than the local microscopic thermalization, see Figure~\ref{fig:ch1:hydrodynmic_regime}. Therefore, the hydrodynamic limit is reached, and we end up with a low-energy theory for long-wavelength/long-timescale dynamics that is governed by the local conservation of charges (and eventually other low-energy collective degrees of freedom which might survive, like in a superfluid phase).

\begin{figure}
	\centering
	\includegraphics[width=0.45\textwidth]{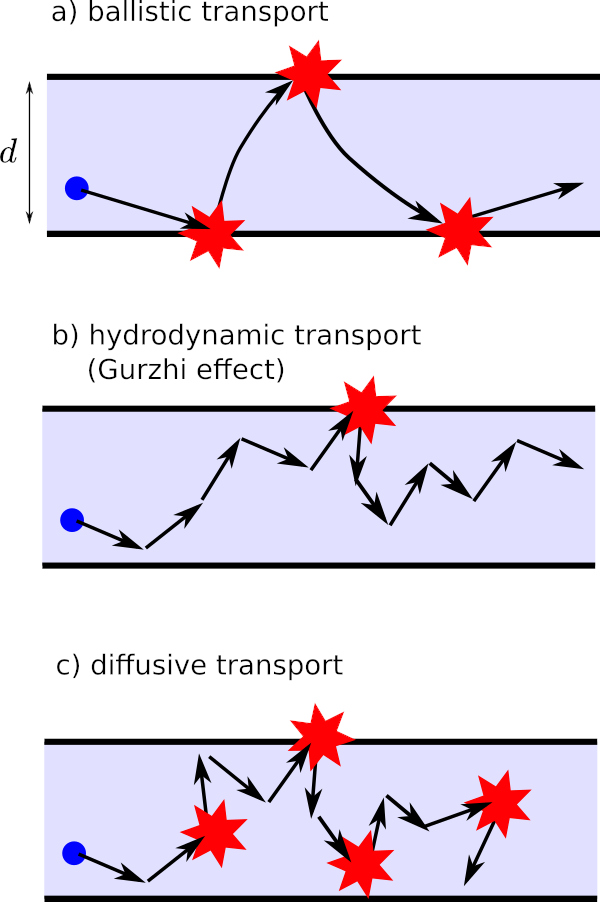}
	\caption{If the system is strongly coupled and clean, scatterings with impurities are rare, while electron-electron interactions happen on a microscopic timescale. Then the system thermalizes locally very quickly, developing hydrodynamic behaviour.}
	\label{fig:ch1:hydrodynamic_regime_condensed_matter}
\end{figure}

Clearly, the separation of scales between microscopic thermalization and macroscopic diffusion needed for the emergence of hydrodynamics happens more sharply in phases which are strongly coupled and clean of impurities. In these systems the scattering time between quasiparticles becomes extremely short, see Figure~\ref{fig:ch1:hydrodynamic_regime_condensed_matter}. In particular, we will be interested in describing the low-energy dynamics of strongly-coupled condensed matter systems. While the intention of this thesis is not to describe the detailed phenomenology of specific condensed matter phases, the low-energy description of solid-state physics is indeed one of the main motivation to expand on the theory of hydrodynamics, specifically in its relaxed version, i.e. quasihydrodynamics.

\section{Canonical approach to hydrodynamics}\label{sec:ch1:canonical_approach}
In the canonical approach, (relativistic) hydrodynamics is understood as a perturbative expansion in a small parameter, the Knudsen number $\text{Kn}=l/L\ll1$, where $L$ is the system size or the length over which thermodynamic quantities vary, while $l$ is the microscopic length scale. Specifically, we can construct the derivative operator $l\partial\sim\text{Kn}$ which act as a perturbative parameter expansion. From this point of view, hydrodynamics is a theory based on the following constraints:
\begin{itemize}
	\item a derivative expansion,
	\item the symmetries (continuous, discrete and broken),
	\item the local form of the second law of thermodynamics.
\end{itemize}
This leads to the formulation of hydrodynamics as a formal power series in which the order zero (the ideal fluid) is non-dissipative, while the higher-order corrections produce entropy, thus being dissipative. Like in any effective theory, ideally adding higher-order terms should lead to a more accurate description, at least within some convergence radius\footnote{There is strong evidence that the hydrodynamic series does not converge \cite{Heller:HydrodynamicsGradientExpansion,Santos:DivergenceChapmanEnskogExpansion}, nonetheless it can be Borel resummed, leading to all-order hydrodynamics.}. From the above discussion, hydrodynamics stops being a sensible effective theory at length scales shorter than the microscopic mean free path $l$.

Each term in the derivative expansion at order one or higher comes with its own \emph{transport coefficient}, which parametrizes the non-universal part of fluid dynamics. These are functions of the thermodynamic quantities whose values cannot be obtained directly from hydrodynamics alone, but depends on the underlying microscopic theory, and can be computed from Kubo formulae using various methods depending on the situation (lattice calculations, kinetic theory, holography, QFT, \dots).

Although, as we claimed, hydrodynamics has been very successful in describing the physics of many different systems, as a classical theory it suffers from fundamental problems. The first one is related to the fact that the standard equations of hydrodynamics are not causal, and contain solutions with super-luminal velocity, which ruin the stability of the theory (super-luminal modes in one frame correspond to modes that travel back in time in a boosted frame). Specifically, the group velocity diverges $v_g=\frac{\dif\omega}{\dif k}=2\gamma_\eta k$ for the shear mode dispersion relation, and becomes super-luminal, which renders the theory acausal and unstable \cite{Hiscock:GenericInstabilitiesFirstorder,Hiscock:LinearPlaneWaves,Speranza:ChallengesSolvingChiral}. This prevents numerically solving the equations of relativistic hydrodynamics from a set of initial condition on a time-slice, ruining the predictability of the theory.

The classical solution to this problem comes from Maxwell and Cattaneo for diffusion first \cite{Maxwell:IVDynamicalTheory,Cattaneo:SullaConduzioneCalore}, and M\"uller, Israel, Stewart (MIS) for hydrodynamics later \cite{Mueller:ParadoxonWaermeleitungstheorie,Israel:TransientRelativisticThermodynamics,Israel:NonstationaryIrreversibleThermodynamics}. Formally, the idea amounts to adding soft UV cut-offs $\tau_i$ that tame the high-momentum behaviour of the hydrodynamic fluctuations. Specifically, defining as $\pi^{\mu\nu}$ the dissipative shear tensor, $\Pi$ the dissipative shear bulk, and $\nu^\mu$ the charge flux, the MIS theory paradigm prescribes to modify the constitutive relations in dynamical equations as
\begin{subequations}
	\begin{align}
		\left(\tau_\pi u^\mu\partial_\mu+1\right)\pi^{\mu\nu}&=\dots\\
		\left(\tau_\Pi u^\mu\partial_\mu+1\right)\Pi&=\dots\\
		\left(\tau_\nu u^\mu\partial_\mu+1\right)\nu^\mu&=\dots
	\end{align}
\end{subequations}
where the dots represent the standard terms appearing in the constitutive relations. The corrections provided by this approach enter only at very large frequency, however now the equations of hydrodynamics are strongly-hyperbolic, can be solved numerically given a set of initial conditions, and have a well-behaved thermal equilibrium state that is stable and causal. The new parameters $\tau_i$ are associated with UV degrees of freedom: indeed in MIS theory one finds a new non-hydrodynamic mode $\omega(\vect{k}=0)\sim-\frac{i}{\tau_\pi}$ which decays quickly. To this day MIS theory, which can be formally obtained from kinetic-theory arguments, is the most used formalisms in hydrodynamic simulations \cite{Romatschke:RelativisticFluidDynamics}.

In recent years there has been a new proposal to solve the problem of causality and stability of relativistic hydrodynamics \cite{Bemfica:NonlinearCausalityGeneral,Bemfica:CausalityExistenceSolutions,Kovtun:FirstorderRelativisticHydrodynamics,Hoult:StableCausalRelativistic,Abboud:CausalStableFirstorder}. This approach is based on taking advantage of an ambiguity that is present in any hydrodynamic description. What happens is that the fluid variables are not well-defined out of equilibrium, hence they are ambiguous up to derivative corrections, the so-called frame redefinitions. Following this approach it is possible to find a hydrodynamic frame (known as BDNK frame) in which first-order relativistic hydrodynamics is strongly-hyperbolic, causal and stable, at the cost of having more complicated constitutive relations.

The solution by M\"uller, Israel and Stewart requires working with second-order hydrodynamics to be fully consistent \cite{Baier:RelativisticViscousHydrodynamics}. This, however, leads to new fundamental issues related to hydrodynamics. In particular, hydrodynamic modes $\omega(\vect{k})$ are long-lived solutions to the conservation equations which can cause instabilities in the system, since they are against the assumption of local thermal equilibrium \cite{Kovtun:LecturesHydrodynamicFluctuations}.

Classical hydrodynamics is only concerned with the dissipative part of the system, without systematically taking into account the effects of thermal fluctuations. However, fluctuations should always appear on par with dissipative effects, according to the fluctuation-dissipation theorem \cite{Landau:StatisticalPhysicsVolume}. This can be amended via a bottom-up construction by including fluctuations as noise in the equations, thus turning hydrodynamics into a proper EFT, following the Martin-Siggia-Rose formalism \cite{Martin:StatisticalDynamicsClassical,Landau:HydrodynamicFluctuations,Hohenberg:TheoryDynamicCritical,Kovtun:EffectiveActionRelativistic,Kovtun:LecturesHydrodynamicFluctuations}. More recently, also a top-down view on dissipative fluctuating hydrodynamics has been developed \cite{Glorioso:LecturesNonequilibriumEffective,Grozdanov:ViscosityDissipativeHydrodynamics,Harder:ThermalFluctuationsGenerating,Haehl:TwoRoadsHydrodynamic}, based on the closed time-path formalism of Schwinger and Keldysh \cite{Kamenev:FieldTheoryNonEquilibrium,Chou:EquilibriumNonequilibriumFormalisms}, which has the advantage of working at the full non-linear level, providing access to the $n$-point correlators.

Specifically, beyond linear order, modes can interact with themselves via thermal fluctuations, and contribute to the renormalization of the transport coefficients compared to their bare, linear-response, values. The renormalization of transport coefficients depends on the UV scale $\Lambda_\text{UV}$ at which hydrodynamics is expected to fail and is due to interaction between modes close to the cut-off. These interactions can appear even in equilibrium due to thermal fluctuations, and can cause the breakdown of the hydrodynamic derivative expansion \cite{Kovtun:LecturesHydrodynamicFluctuations}. From a technical perspective, this is because the interactions between modes modify the retarded correlators by non-analytical terms which scale like $\mathcal{O}(\omega^{3/2})$ in $3+1d$\footnote{In $2+1$-dimensions this is a logarithmic correction \cite{Forster:LargedistanceLongtimeProperties}.} \cite{DeSchepper:NonexistenceLinearDiffusion,Kovtun:StickinessSoundAbsolute}, hence these corrections are effectively larger than the first-order hydrodynamics, which are linear in $\mathcal{O}(\omega)$, but smaller than second order corrections, which scale as $\mathcal{O}(\omega^2)$, and cannot be captured by local derivative corrections. This non-analytic behaviour is related to the well-known effect of long-time tails: namely, the non-analytic $\omega^{3/2}$ dependence of e.g. $3+1d$ hydrodynamics corresponds to a real-time evolution as $t^{3/2}$ of the response function \cite{Pomeau:TimeDependentCorrelation}, which is slower than exponential and decays slowly.

Finally, as we explained above, hydrodynamics stops being a sensible EFT at length scales shorter than the microscopic scale $l$. This assumption, while being at the core of the canonical approach, is challenged by the observation that hydrodynamics seems to work well as a model for systems in which the separation of scales does not happen sharply. This has sparkled a lot of research in this direction, trying to understand this \emph{unreasonable effectiveness of hydrodynamics}. Research in this direction focused on resummation methods for all-order hydrodynamics \cite{Heller:HydrodynamicsGradientExpansion,Romatschke:RelativisticFluidDynamics,Santos:DivergenceChapmanEnskogExpansion,Heller:HydrodynamizationKineticTheory} and the hydrodynamic attractor \cite{Romatschke:FluidDynamicsFar,Bu:LinearizedFluidGravity}, but also an action-like description of hydrodynamics via the Schwinger-Keldysh formalism \cite{Glorioso:LecturesNonequilibriumEffective,Haehl:TwoRoadsHydrodynamic,Grozdanov:ViscosityDissipativeHydrodynamics}.

We remark that the problems associated with causality, stability and fluctuations appear in interpreting the equations of hydrodynamics as a set of classical PDE, when solving for the real-time evolution of the fluid, and are due to high-frequency, high-wavevector modes, far from the hydrodynamic regime (hydrodynamics is not expected to hold at microscopic length scale anyway). We decided to include this section for completeness and to give a better context on the state of hydrodynamics, however in this thesis we will be mostly concerned with the linear response of fluids in the hydrodynamic regime, focusing on the $\omega,\vect{k}\rightarrow0$ properties of small fluctuations about some global thermodynamic equilibrium state. Hence, we will simply assume that suitable corrections to standard hydrodynamics exist, which render the theory sensible, and that can be expanded to include the various extensions of hydrodynamics discussed in this thesis.

\section{Other approaches}
In the previous section we discussed the canonical approach, some of its pitfalls and recent research avenues. There are however other possible points of view to formulate hydrodynamics, that might give different insights into what it is, how it relates to other theories, and how to expand upon the framework of hydrodynamics.

\subsubsection{Kinetic theory}
One of the first methods, also from an historical perspective, comes from kinetic theory \cite{Huang:StatisticalMechanics,Denicol:MicroscopicFoundationsRelativistic}. This might seem strange at first sight, since hydrodynamics is mostly understood as describing strongly-coupled phases of matter that thermalize quickly, while kinetic theory sits on the other side of the spectrum and is usually understood as describing weakly-coupled theories. This apparent contradiction disappears once we appreciate that hydrodynamics is extremely universal: in the presence of a single species of quasiparticles (as it is often the case in kinetic theory) and no impurities there is nothing that can stops the onset of hydrodynamics, which appears naturally at long wavelength even at small couplings.

MIS hydrodynamics comes quite naturally from kinetic theory \cite{Israel:TransientRelativisticThermodynamics,Denicol:MicroscopicFoundationsRelativistic,Denicol:DerivationTransientRelativistic}, providing a justification for the timescales $\tau_i$ which were introduced phenomenologically in the previous section. More recently, the kinetic theory approach to hydrodynamics has been generalized to naturally include different hydrodynamic frames \cite{Rocha:NovelRelaxationTime,Hoult:CausalFirstorderHydrodynamics}, in particular the already-cited BDNK frame. Furthermore, since kinetic theory is a microscopic description, this approach also gives direct access to expressions for the transport coefficients in terms of microscopic and thermodynamic parameters.

\subsubsection{Schwinger-Keldysh formalism}
Hydrodynamics is a very powerful tool, but from a microscopic point of view the arguments used in the first Section~\ref{sec:ch1:hydrodynamic_regime} to explain the emergence of hydrodynamics are purely heuristic and not satisfactory. First, the canonical formulation does not systematically deal with thermal and quantum fluctuations, which must be included separately in bottom-up approaches, even if they are important in many situations such as turbulence, chaotic systems, and the renormalization of transport coefficients (long-time tails). Another aspect is related to the regime of validity of the theory, which in classical hydrodynamics is subject to the scale separation $l\ll L$, but experiments in heavy-ion collisions suggest this bound is too strict. Indeed, recent action-principle formulations of fluid dynamics and resummed all-order hydrodynamics manage to go beyond this paradigm. Finally, the symmetries and the thermodynamic constraints, such as first and second law of thermodynamics, are imposed phenomenologically, while in the effective-action approach these come directly from fundamental properties of the dissipative effective action, which provides the theory with a strong theoretical ground.

From a theoretical perspective, there has always been a strong interest in trying to develop field theory action-like methods (based e.g. on path integrals) to describe dissipative systems like hydrodynamics \cite{Lazo:ActionPrincipleActiondependent,Dubovsky:EffectiveFieldTheory,Dubovsky:EffectiveFieldTheorya}. While perfect fluids and certain other non-dissipative aspects of hydrodynamics are already captured by action-like principles \cite{Jensen:HydrodynamicsEntropyCurrent,Banerjee:ConstraintsFluidDynamics}, a full discussion of the dissipative corrections from path integral methods is still a topic of research \cite{Jain:SchwingerKeldyshEffectiveField,Armas:EffectiveFieldTheory,Jain:EffectiveFieldTheory,Baggioli:QuasihydrodynamicsSchwingerKeldyshEffective}.

In the past decade however the Schwinger-Keldysh formalism has been successfully applied to formalize hydrodynamics (and more generally thermal dissipative systems) from an action-like principle \cite{Haehl:EffectiveActionRelativistic,Haehl:ThermalOutoftimeorderCorrelators,Glorioso:LecturesNonequilibriumEffective,Grozdanov:ViscosityDissipativeHydrodynamics}. This framework is based on realizing that in order to describe thermal dissipative systems it is necessary to recast the standard problem of statistical field theory on a closed time path, which leads to a doubling of the hydrodynamic degrees of freedom (advanced and retarded fields depending on which branch of the time contour they live on). After integrating out the UV degrees of freedom one finds a Wilsonian effective action that must obey certain constraints, such as unitarity and the novel $\mathbb{Z}_2$ dynamical KMS symmetry. This construction provides an effective action that encodes all the information related to the constitutive relations and the conservation equations of hydrodynamics. Furthermore, contrary to the canonical approach, this method automatically incorporates the effects of thermal (and eventually quantum) fluctuations at the full non-linear level, in agreement with the fluctuation-dissipation theorem.

\subsubsection{Fluid/gravity duality}
This approach is based on the gauge/gravity duality, grounded in string theory \cite{Maldacena:LargeLimitSuperconformal,Gubser:GaugeTheoryCorrelators,Witten:SitterSpaceHolography}. It comes in various different form, but the simplest and most practical one, which does not make reference to quantum gravity and strings, is that it is a duality between a strongly-coupled (large 't~Hooft coupling) large-$N$ quantum field theory in $d$ dimensions living on the boundary, and a theory of classical gravity in $d+1$ dimensions that lives in bulk Anti de Sitter space. In particular, a black hole in the interior of AdS space gives rise to thermodynamics in the dual boundary theory, whose temperature and entropy are fixed by the black-hole Gibbons-Hawking temperature and Bekenstein–Hawking entropy.

Even before the discovery of holography, after the realization that black holes are thermal systems, it was quickly argued that the analogy is not limited to thermal equilibrium, and works for generic hydrodynamic perturbations of the event horizon, implying that black holes are dissipative systems, which is summarized by the membrane paradigm \cite{Thorne:BlackHolesMembrane}. This correspondence is clearer in AdS space \cite{Iqbal:UniversalityHydrodynamicLimit,Policastro:AdSCFTCorrespondence,Policastro:AdSCFTCorrespondencea}, in which it acts as a basis for the fluid/gravity duality.

What holography managed to show is that the membrane paradigm analogy actually goes beyond linear perturbations, therefore the fluid/gravity correspondence works at the full non-linear level \cite{Bhattacharyya:NonlinearFluidDynamics}, allowing one to obtain not only the full set of hydrodynamic correlators, but also the equations of hydrodynamics from General Relativity, see \cite{Hartnoll:HolographicQuantumMatter,Ramallo:IntroductionAdSCFT,Rangamani:GravityHydrodynamicsLectures,Ammon:GaugeGravityDuality,Natsuume:AdSCFTDuality} for some books and reviews.

Holography is a powerful tool, which can be used in conjunction with hydrodynamics to study strongly-coupled systems from two different approaches, and as a cross-check on the correctness of the hydrodynamic effective theory. Historically, many interesting features of hydrodynamics (such as the effect of anomalies on fluids, to name one) were first discovered using holographic methods, and only subsequently obtained in hydrodynamics. Furthermore, like kinetic theory, holographic models also provide us with the non-universal part of the fluid description, namely the equation of state and the transport coefficients, which can then be used to set bounds on transport coefficients, e.g. the famous $\eta/s\geq\frac{1}{4\pi}$ KSS bound \cite{Kovtun:HolographyHydrodynamicsDiffusion}.

\section{Hydrodynamics in condensed matter}

\subsection{Overview}
In conventional metals the behaviour of electrons at low temperature is observed to be diffusive \cite{Lifshitz:PhysicalKineticsVolume,Ashcroft:SolidStatePhysics}. They are assumed to be (almost) non-interacting, and relax only via electron-impurity scatterings on some characteristic time $\tau_\text{ei}$, so that a Fermi-liquid picture in terms of long-lived quasiparticles applies. Indeed, electrons in metals are $\sim\qty{2}{\angstrom}$ apart, however their mean free path at room temperature is of order $\qty{e4}{\angstrom}$, implying a weakly-coupled phase. Furthermore, contrary to molecules in standard fluids, electrons move in a lattice background and interact with impurities, defects and phonons, thus losing momentum. Hence, if the sample is large $L\gg l=v_F\tau_\text{ei}$ with $v_F$ the Fermi velocity, and the temperature is low $T\tau_\text{ei}\ll1$, the system reaches a diffusive regime \cite{Chaikin:PrinciplesCondensedMatter}. In terms of collective excitations, diffusion is characterized by a single decaying mode, while fluids also have propagating sound modes with linear dispersion relation associated with energy and momentum conservation\footnote{The modern point of view on hydrodynamics is that it also includes diffusion, which is based on the same fundamental principles, only with a different set of conserved charges. In the context of condensed matter, however, we keep the distinction.}. One exception, that emerges in very clean samples for which the system is similar to the mean free path in size, is the \emph{ballistic} motion of electrons, see Figure~\ref{fig:ch1:hydrodynamic_regime_condensed_matter}.

Scattering mechanisms in solids are characterized by mean free paths that depend on the temperature. In conventional metals the low-temperature regime is dominated by electron-impurity elastic scatterings, leading to the residual resistance. On the other hand, at high temperatures usually electron-phonon scatterings dominate, leading to a characteristic $T$-dependence behaviour of the conductivity \cite{Lifshitz:PhysicalKineticsVolume}. For most materials, at any temperature either electron-impurity or electron-phonon scatterings are dominant compared to electron-electron interactions $l_\text{ee}\gg l_\text{ei},l_\text{ep}$, leaving no room for hydrodynamic electronic transport.

Despite the above discussion, historically, hydrodynamics has been successfully applied in condensed matter to study the dynamics of phonons \cite{Gurzhi:HydrodynamicEffectsSolids} (see \cite{Machida:PhononHydrodynamicsUltrahighroomtemperature} for experiments and references on the topic), spin waves (magnons) \cite{Halperin:HydrodynamicTheorySpin}, and more recently, based on the work by Gurzhi \cite{Gurzhi:HydrodynamicEffectsSolids}, also to electronic systems \cite{Narozhny:HydrodynamicApproachElectronica,Lucas:HydrodynamicsElectronsGraphene,Lucas:HydrodynamicTheoryThermoelectric,Andreev:HydrodynamicDescriptionTransport,Scopelliti:HydrodynamicChargeHeat,DiSante:TurbulentHydrodynamicsStrongly,Erdmenger:StronglyCoupledElectron,Xian:HallViscosityHydrodynamic}, see also the reviews \cite{Narozhny:HydrodynamicApproachTwodimensional,Fritz:HydrodynamicElectronicTransport}.

From an experimental perspective, only recently very clean materials in which $l_\text{ee}\ll l_\text{ei},l_\text{ep}$ have been realized, paving the road to electron hydrodynamics \cite{Bandurin:NegativeLocalResistance,Crossno:ObservationDiracFluid,Moll:EvidenceHydrodynamicElectron,Moll:EvidenceHydrodynamicElectron,Braem:ScanningGateMicroscopy,Jaoui:DepartureWiedemannFranzLaw,Gusev:StokesFlowObstacle,Varnavides:ElectronHydrodynamicsAnisotropic,Vool:ImagingPhononmediatedHydrodynamic,Jaoui:ThermalResistivityHydrodynamics,Gupta:HydrodynamicBallisticTransport}. In particular, in $2d$ materials $l_\text{ee}\sim T^{-2}$, while acoustic phonons interactions scale as $l_\text{ep}\sim T^{-1}$ and in ultra-pure samples the impurity scattering are rare $l_\text{ei}\gg l_\text{ee}$, allowing for a parameter range with hydrodynamic behaviour. In graphene \cite{Bandurin:NegativeLocalResistance,Crossno:ObservationDiracFluid,Kumar:SuperballisticFlowViscous,Bandurin:FluidityOnsetGraphene,Berdyugin:MeasuringHallViscosity,Gallagher:QuantumcriticalConductivityDirac}, for example, the mean free path with impurities and phonons remains very large $l_\text{ei},l_\text{ep}\geq\qty{1}{\micro\meter}$ up to room temperature, while at $T\geq\qty{150}{\kelvin}$ the electron-electron scattering has a characteristic length of $l_\text{ee}\sim\qty{0.1}{\micro\meter}$, which opens a window to hydrodynamic transport.

More recently, nonlocal resistance \cite{Bandurin:NegativeLocalResistance} and violation of the Wiedemann-Franz law \cite{Crossno:ObservationDiracFluid} have been used as signatures of hydrodynamic behaviour \cite{Bandurin:FluidityOnsetGraphene,Berdyugin:MeasuringHallViscosity,Gallagher:QuantumcriticalConductivityDirac}. In particular, hydrodynamic flow in electronic solid state systems should exhibit an enhanced conductivity, higher than the ballistic limit \cite{Kumar:SuperballisticFlowViscous,Jenkins:ImagingBreakdownOhmic}. This can happen because the Sharvin limit is obtained by assuming no electron-electron interactions in the ballistic regime, however collective hydrodynamic motion driven by electron-electron scatterings can overcome the ballistic bound. Recently, experiments \cite{Braem:ScanningGateMicroscopy,Vool:ImagingPhononmediatedHydrodynamic} employed various imaging methods to directly observe the viscous flow of electrons in graphene.

A different class of materials which also show hydrodynamic behaviour are ultra-pure heterostructure semiconductors \cite{Braem:ScanningGateMicroscopy,Gusev:StokesFlowObstacle,Gupta:HydrodynamicBallisticTransport}, since the first experimental observation of the Gurzhi effect \cite{deJong:HydrodynamicElectronFlow}. Furthermore, other materials such as Weyl semimetals \cite{Lucas:HydrodynamicTheoryThermoelectric,Gorbar:NonlocalTransportWeyl} might exhibit fluid dynamic behaviour \cite{Jaoui:DepartureWiedemannFranzLaw,Gooth:ElectricalThermalTransport}. Finally, hydrodynamic transport is often associated with the unusual linear-in-$T$ dependence of the resistivity in the strange metal phase of high-temperature superconductors \cite{Davison:HolographicDualityResistivity,Hartnoll:TheoryUniversalIncoherent,Amoretti:HydrodynamicalDescriptionMagnetotransport}.

\subsection{Gurzhi effect}
The Gurzhi effect \cite{Gurzhi:HydrodynamicEffectsSolids} is often considered a defining signature of hydrodynamic transport in solid state systems. Consider an electric current flowing in a thin, clean wire of metal. In this situation electrons can undergo two different scattering processes: scatterings with the walls of the wire (the boundaries), and electron-electron scatterings (eventually mediated by phonons). Assume that at very low temperature the electron-electron scattering length is larger than the width of the wire $l_\text{ee}\gg W$. In this regime, scatterings off the walls of the wire dominate, leading to a characteristic temperature-independence of the resistivity $\rho_\text{lowT}\sim1/W$.

\captionsetup[subfigure]{labelformat=empty}
\begin{figure}[t]
	\begin{subfigure}{.57\textwidth}
		\centering
		\includegraphics[width=\linewidth]{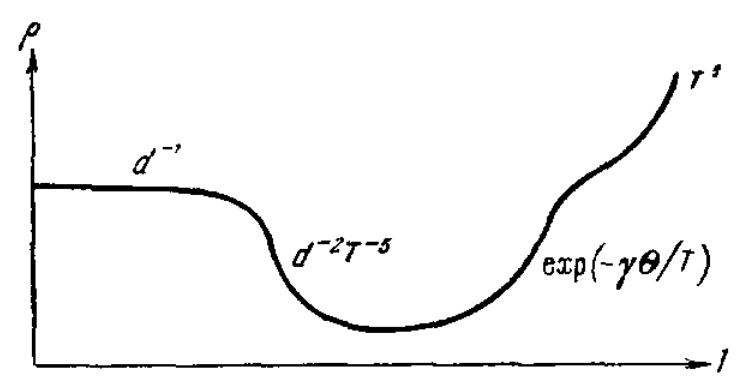}
	\end{subfigure} \hfill %
	\begin{subfigure}{.35\textwidth}
		\centering
		\includegraphics[width=\linewidth]{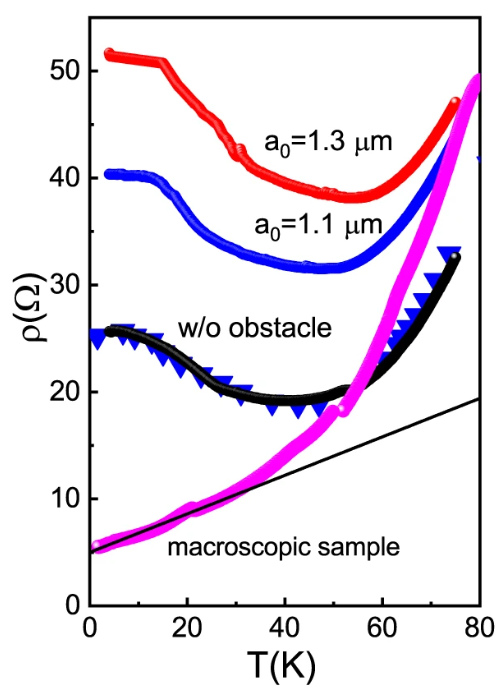}
	\end{subfigure}
	\caption{\textbf{Left:} Figure taken from \cite{Gurzhi:HydrodynamicEffectsSolids}. Theoretical prediction of the Gurzhi effect, with its characteristic $\dif\rho/\dif T\leq0$. \textbf{Right:} Figure taken from \cite{Gusev:StokesFlowObstacle}. In red and blue samples with an obstacle, in black without it, and in purple a macroscopic sample. The blue triangles are a theoretical prediction and the solid thin line represent the resistivity due to scatterings with acoustic phonons. Notice the negative slope for $T\leq\qty{40}{\kelvin}$.}
	\label{fig:ch1:gurzhi_effect}
\end{figure}

On the other hand, the electron-electron interaction is instead temperature dependent, e.g. $l_\text{ee}\sim T^{-2}$ for direct scatterings, and $l_\text{ee}\sim T^{-5}$ for phonon-mediated ones. Hence, as $T$ increases, the electron-electron scattering length decreases, until $l_\text{ee}\ll W$ and the resistivity becomes dominated by electron-electron interactions, leading to a dependence $\rho_\text{highT}\sim l_\text{ee}/W^2<\rho_\text{lowT}$ which is smaller than the low-temperature resistivity\footnote{In terms of standard fluids, the Gurzhi effect can be understood as the electronic realization of the crossover between Knudsen flow (particle based) and Poiseuille (viscous) flow.}. This result clearly assumes that the effective mean free path $W^2/l_\text{ee}$ in the wire, is much smaller than bulk momentum-relaxing processes, such as electron-impurities and electron-phonon scatterings, implying that we are in the hydrodynamic regime. As the temperature increases $W^2/l_\text{ee}\gg l_\text{ei}$ and a crossover between hydrodynamics and diffusive regime happens, leading to the usual $\rho(T)$ dependence. Thus, the Gurzhi effect identifies a minimum in the resistivity, which decreases $\rho\sim T^{-2}$ as a function of $T$, and is understood as a signature of hydrodynamic transport.

From an experimental perspective, several factors hinder the observation of the Gurzhi effect in real metals, such as impurities and Umklapp scatterings, or the non-spherical shape of the Fermi surface, which contribute to corrections to the $T$ dependence of the resistivity, eventually washing away its characteristic signature. Nonetheless, it was observed in \cite{deJong:HydrodynamicElectronFlow} in heterostructures and more recently in \cite{Gusev:StokesFlowObstacle}.

\subsection{Nonlocal transport experiment}
Some experiments manage to measure negative vicinity resistance \cite{Bandurin:NegativeLocalResistance} and Wiedemann-Franz law violation \cite{Crossno:ObservationDiracFluid} in graphene, both of which are usually interpreted as smoking guns of hydrodynamic electronic transport in \cite{Moll:EvidenceHydrodynamicElectron}.

\begin{figure}[t]
	\centering
	\includegraphics[width=0.7\linewidth]{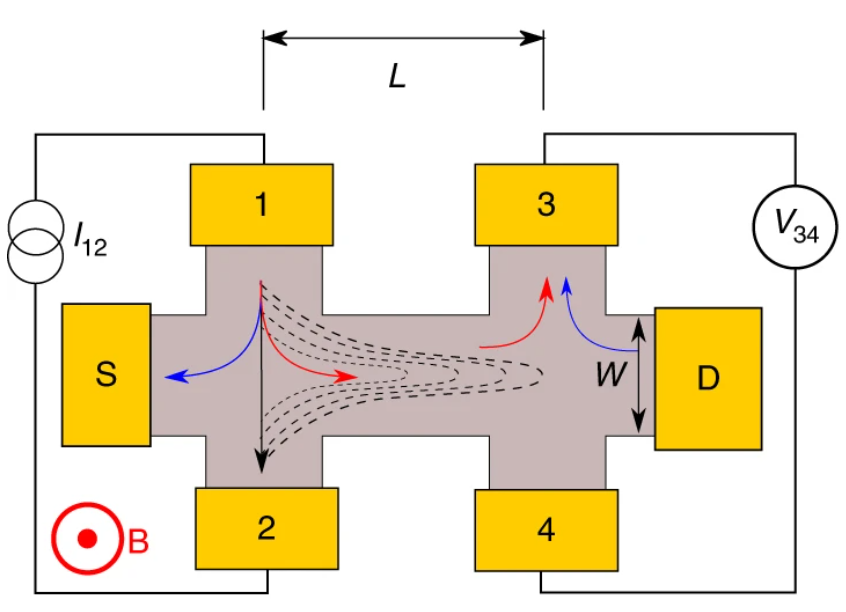}
	\caption{Figure taken from \cite{Ribeiro:ScaleinvariantLargeNonlocality}. Conventional four-terminals measurements consist in having a current flow between Source $S$ and Drain $D$, while recording the voltage drop between 1 and 3 (or 2 and 4). The resistance is then $R_{SD,13}=V_{13}/I_{SD}$ and is related to the longitudinal resistivity $\rho_{xx}=R_{SD,13}W/L$, with $W$ and $L$ the width and the length of the sample, as in Figure. In nonlocal measurements, instead, a current flows between 1 and 2, while the voltage is measured from 3 to 4. In the case of diffusive transport the voltage drop $V_{34}$ should be exponentially suppressed $R_{12,35}\sim\rho_{xx}e^{-L/W}$, but it is not in hydrodynamic transport.}
	\label{fig:ch1:nonlocal_transport}
\end{figure}

Standard transport measurements at two or four terminals are concerned with extracting the current-voltage characteristic, thus they focus only on the total current flowing inside the sample to obtain the resistance $R$. On the contrary, nonlocal transport experiments consist in measuring the spatial dependence of the current density in the system, by studying the voltage drops measured from multiple gates, see Figure~\ref{fig:ch1:nonlocal_transport}.

Nonlocal measurements were first intended to identify ballistic transport in mesoscopic systems, however they have been recently argued to be possible indicators of hydrodynamic transport \cite{Narozhny:HydrodynamicApproachElectronica,Lucas:HydrodynamicsElectronsGraphene,Bandurin:NegativeLocalResistance,Bandurin:FluidityOnsetGraphene,Berdyugin:MeasuringHallViscosity}. In particular, hydrodynamic flows can give rise to vortices in confined geometries, which lead to a characteristic negative nonlocal resistance at the opposite side of the vortex called \emph{vicinity resistance}. Although this effect can be a sign of hydrodynamic behaviour \cite{Bandurin:NegativeLocalResistance,Levitov:ElectronViscosityCurrent}, it can also appear in ballistic transport \cite{Mihajlovic:NegativeNonlocalResistance}, thus to differentiate between the two regimes one needs either to check the temperature dependence \cite{Bandurin:FluidityOnsetGraphene} or to employ spatial imaging techniques to directly observe the electron flow \cite{Ku:ImagingViscousFlow,Sulpizio:VisualizingPoiseuilleFlow,Ella:SimultaneousVoltageCurrent}.

\subsection{Wiedemann-Franz law violation}
Unconventional charge and heat transport in hydrodynamic electronic systems leads to a violation of the Wiedemann-Franz law \cite{Lifshitz:PhysicalKineticsVolume,Ashcroft:SolidStatePhysics}. If both charge and heat are due to the transport of the same quasiparticles, and are influenced by the same scattering processes (as is the case in most microscopic models of non-interacting electrons), then the ratio between charge and heat conductivity must be a constant
\begin{equation}
	\frac{\kappa}{\sigma}=\mathcal{L}T\qquad\qquad\mathcal{L}=\mathcal{L}_0=\frac{\pi^2}{3e^2}
\end{equation}
where $\sigma$ and $\kappa$ are the electric and thermal conductivities, while $\mathcal{L}$ is called Lorenz ratio and $\mathcal{L}_0$ is valid for free carriers. In most conventional metals electrons are indeed weakly interacting and departures from the WF law are usually weak. On the contrary, a strong violation of the WF law should be regarded as due to new effects, like strongly-coupled hydrodynamic transport \cite{Lucas:HydrodynamicTheoryThermoelectric,Lucas:HydrodynamicsElectronsGraphene,Hartnoll:TheoryNernstEffect}.

\captionsetup[subfigure]{labelformat=empty}
\begin{figure}[t]
	\begin{subfigure}{.32\textwidth}
		\centering
		\includegraphics[width=\linewidth]{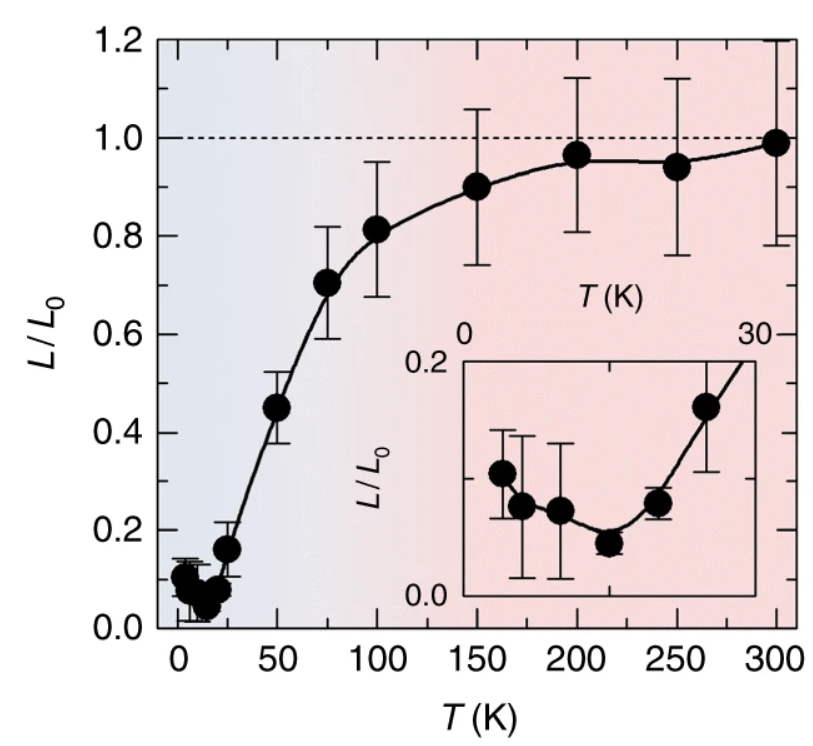}
	\end{subfigure} \hfill
	\begin{subfigure}{.32\textwidth}
		\centering
		\includegraphics[width=\linewidth]{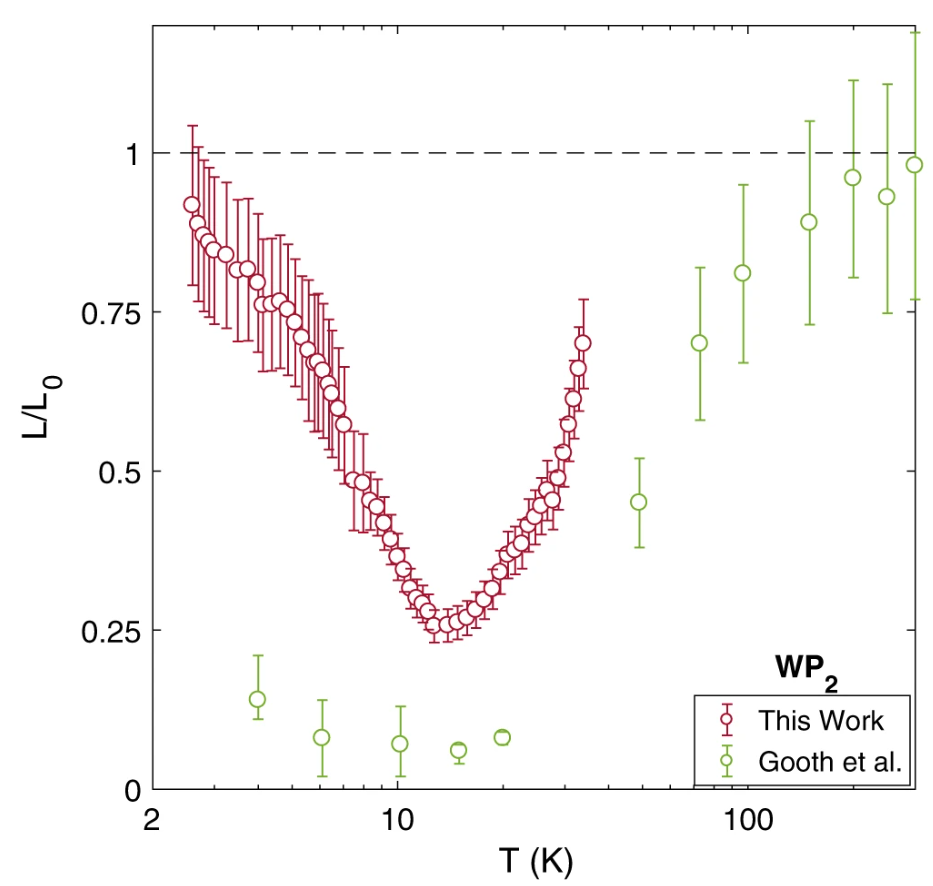}
	\end{subfigure} \hfill
	\begin{subfigure}{.32\textwidth}
		\centering
		\includegraphics[width=\linewidth]{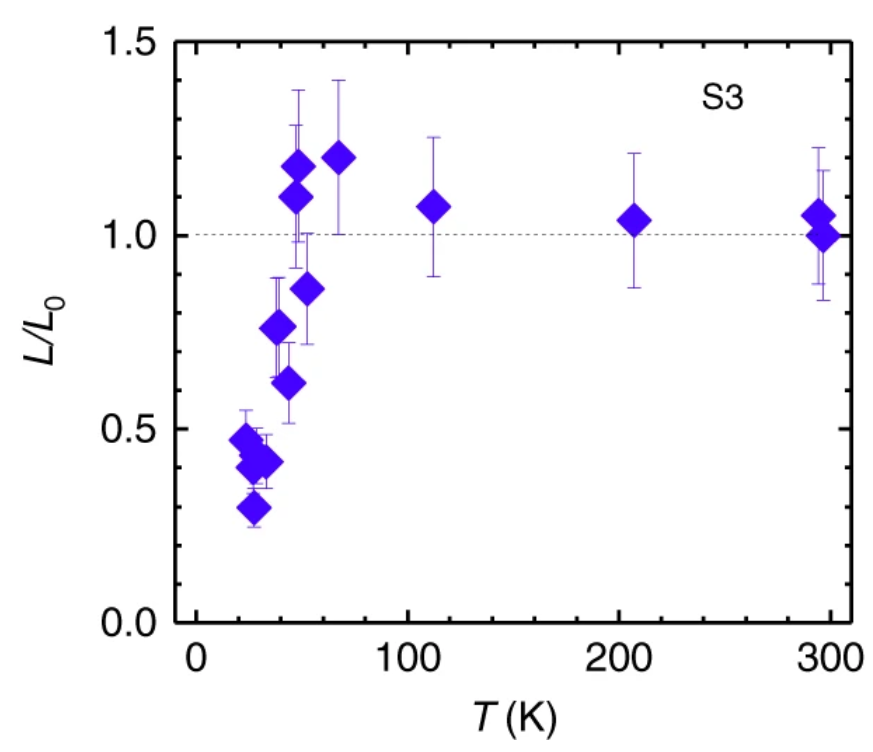}
	\end{subfigure}
	\caption{\textbf{Left:} Figure taken from \cite{Gooth:ElectricalThermalTransport}. Strong WL law violation in a micro-ribbon of $\chem{WP_2}$. \textbf{Center:} Figure taken from \cite{Jaoui:DepartureWiedemannFranzLaw}. Same violation in bulk crystals of $\chem{WP_2}$. \textbf{Right:} Figure taken from \cite{Kumar:ExtremelyHighConductivity}. WF law violation is observed also in $\chem{MoP}$.}
	\label{fig:ch1:wiedemann-franz_law}
\end{figure}

Theory predicts a very large value of the Lorenz ratio in graphene $\mathcal{L}\propto\tau_\text{ei}/\tau_\text{ee}\gg1$ in the hydrodynamic regime \cite{Lucas:HydrodynamicsElectronsGraphene}, which was indeed observed in the graphene Dirac fluid in \cite{Crossno:ObservationDiracFluid}.

Signatures of electron hydrodynamics also appear in topological materials (Dirac and Weyl semimetals) \cite{Jaoui:DepartureWiedemannFranzLaw,Gooth:ElectricalThermalTransport,Kumar:ExtremelyHighConductivity}, from the observation of strong WL law violations. In these compounds the WL law violation happens in the opposite direction compared to graphene: now the Lorenz ratio is found to be very small $\mathcal{L}\ll\mathcal{L}_0$, again attributed to the existence of hydrodynamic electronic transport, which onsets at lower temperature compared to graphene.

\section{The content of this thesis}
Hydrodynamics, as we discussed, is concerned with the dynamics of conserved charges in closed systems. Although this can be a very good approximation, real systems are rarely exactly closed and void of impurities. As a consequence, we can expect the conservation laws of hydrodynamics to be only approximate and, in general, to be broken by some effective relaxation term characterized by a timescale parameter $\Gamma\sim1/\tau_1$ which dictates the damping of fluctuations towards equilibrium. These kinds of theories are called relaxed hydrodynamics or quasihydrodynamics.

\begin{figure}
	\centering
	\includegraphics[width=\textwidth]{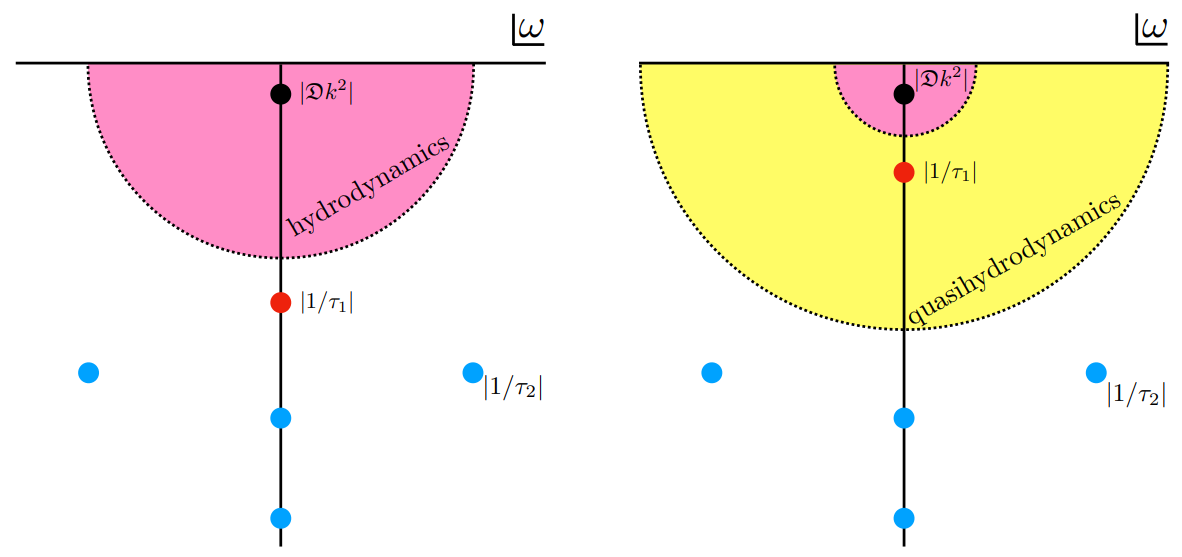}
	\caption{Figure taken from \cite{Grozdanov:HolographyHydrodynamicsWeakly}. \textbf{Left:} If the relaxation times for the non-hydrodynamic modes are all comparable, then hydrodynamics dominates in the IR and there is no quasihydrodynamics regime, since $\tau_1\sim\tau_2$. \textbf{Right:} On the other hand, if there is a mode with a much larger decay time $\tau_1\gg\tau_2$, then hydrodynamics has a reduced regime of validity, close to the origin. However, quasihydrodynamics describes the IR physics on a wider regime, by including the effect of non-hydrodynamic mode with  parametrically small decay time $\tau_1$.}
	\label{fig:ch1:quasi_hydrodynamics}
\end{figure}

Clearly, these corrections to the conservation equations must be small in amplitude, for the hydrodynamic description to still be approximately valid \cite{Grozdanov:HolographyHydrodynamicsWeakly}. Physically, this corresponds to saying that whatever is ruining the exact conservation acts on scales which are still much larger than the microscopic mean-free-path and thermalization scale (the non-hydrodynamic modes), so that a separation of scale is still possible, see Figure~\ref{fig:ch1:quasi_hydrodynamics}.

From a phenomenological perspective, we can expect the quark-gluon-plasma created in heavy-ion collisions in a vacuum to be an almost perfect fluid (\emph{the} most perfect fluid \cite{Schaefer:NearlyPerfectFluidity}), with virtually no impurities, thus quasihydrodynamics might have limited applicability in the context of nuclear physics or astrophysics\footnote{Although there are no impurities, close to thermal critical points fluctuations become parametrically large and the correlation length diverges, requiring a modification of hydrodynamics to include the presence of these slow modes \cite{Stephanov:HydroHydrodynamicsParametric}.}. On the other hand, condensed matter systems are in comparison much more complicated structure: the electron fluid always moves in the presence of a background lattice and interacts with phonons, which can degrade energy and momentum; furthermore, even clean samples have some level of impurities and Umklapp scatterings which can disrupt the electron flow and take away momentum \cite{Andreev:HydrodynamicDescriptionTransport}. The presence of boundaries and layers can also contribute to losses of would-be conserved charges, since the system is never perfectly closed. Finally, in many condensed matter systems symmetries are only approximate and accidental, thus they are not protected by any fundamental law. Consequently, the associated Noether charges are not exactly conserved, and the conservation equations are explicitly weakly broken. This is the case e.g. of the emergent Lorentz symmetry in graphene, or the chiral symmetry in Weyl semimetals, which is not exact and broken at high quasiparticle momenta, far from the Fermi surface (the band structure is not linear up to arbitrary high momenta).

In \cite{Stephanov:HydroHydrodynamicsParametric,Armas:ApproximateSymmetriesPseudoGoldstones,Delacretaz:DampingPseudoGoldstoneFields}, but also in Chapter~\ref{chapter:charge_density_waves}, the extra mode is assumed to be very light and dynamical, so that it thermalizes quickly and the regime of validity of hydrodynamics is naturally extended to include it. On the other hand, for most of this thesis, like in \cite{Grozdanov:HolographyHydrodynamicsWeakly,Baggioli:QuasihydrodynamicsSchwingerKeldyshEffective,Hartnoll:TheoryNernstEffect,Gouteraux:DrudeTransportHydrodynamic,Blake:MomentumRelaxationFluid,Lucas:MemoryMatrixTheory,Ammon:PseudospontaneousSymmetryBreaking}, we will assume that the associated microscopic non-hydrodynamical mode is integrated out and not dynamical, and its effect is that of modifying the conservation law of some quantity with a relaxation term. Weak effective breaking of symmetries, and momentum relaxations in particular, are also of great interest from a holographic perspective, see e.g. \cite{Amoretti:AnalyticDCThermoelectric,Amoretti:MagnetotransportMomentumDissipating,Davison:MomentumRelaxationHolographic,Andrade:SimpleHolographicModel} and many others.

Alternatively, another point of view is that of \cite{Grozdanov:HolographyHydrodynamicsWeakly}: hydrodynamics works on scales much larger than the mean free path $l$ and its associated microscopic time $\tau$. Normally, non-hydrodynamic operators decay quickly as $\langle O(t)O(0)\rangle\sim e^{-t/\tau_2}$, and what survives the long-wavelength/long-timescale regime is the dynamics of conserved operators. However, it can happen that there is some operator whose decay is parametrically slower $\langle P(t)P(0)\rangle\sim e^{-t/\tau_1}$ than other $\tau_2,\tau_3,\dots\ll\tau_1$. Strictly speaking, hydrodynamics applies only on scales $\tau\gtrsim\tau_1$ and breaks down when $\tau_1\partial_t\sim1$. Nonetheless, because $\tau_1$ is so much larger than the other decay rates, a separation of scales is still possible, and we can include $\langle P\rangle$ as a quasi-conserved operator, with an effective decay time $\tau_1$, which is a valid description as long as $\tau_2\partial_t\ll1$.

This is the main context of the thesis. In most of this work we will expand on these ideas, trying to learn general lessons from quasihydrodynamics that can be useful to characterize condensed matter systems in which the standard hydrodynamic charges are not exactly conserved, but instead decay to some fixed equilibrium value. The thesis is structured as follows.

In Chapter~\ref{chapter:hydrodynamics_linear_response} we review the canonical approach to hydrodynamics, we discuss how to obtain the retarded responses of fluids to external linear perturbations, and we introduce the tool of the hydrostatic generating functional.

Subsequently, in Chapter~\ref{chapter:charge_density_waves} we expand on previous works and discuss the hydrodynamic theory of electronic Charge Density Waves in terms of viscoelastic fluids which contains Goldstone bosons for broken spatial translations (spontaneous and explicit breaking), while including the effects of a strong external order-zero magnetic field and a new transport coefficient, the lattice pressure. We study the transport properties of such fluids both analytically from hydrodynamics, and numerically from holography, and find a very good match in the hydrodynamic regime between the two approaches. This work sits in the broader line of research that tries to better understand the strange-metal strongly-coupled phase of cuprates.

Chapter~\ref{chapter:electrically_driven_fluids} is devoted to constructing a theory of hydrostatic fluids in strong external electric fields for systems in which energy and momentum are not exactly conserved, but rather slowly relax. In particular, we find that momentum relaxation modifies the stationary state, hence it should be included in the hydrostatic constraints of the fluid, leading to new predictions for the thermoelectric transport.

In Chapter~\ref{chapter:onsager} we introduce the concept of generalized relaxations in linearized hydrodynamics, which are decay terms that mix the charges (e.g. a charge fluctuation which induces energy relaxation and vice versa). We obtain a set of very general constraints that these relaxations must obey, based on microscopic time-reversal invariance, positivity of entropy production and linear stability. Then, we show how to make them consistent with curved spacetime and background gauge fields by adding new source terms proportional to relaxations and imposing Onasger relations on the system. This procedure thus allows us to obtain the full spectrum of retarded correlators for the relaxed theory.

Finally, in Chapter~\ref{chapter:anomalous_hydrodynamics} first we correct previous results in the literature regarding the longitudinal magnetotransport of anomalous fluids, emphasizing the importance of working at the correct derivative order for the anomaly to appear in the conductivities. Subsequently, we use this new result to study the transport properties of Weyl semimetals in the hydrodynamic regime. We find that generalized relaxations are a necessary ingredient if we want our model to satisfy basic fundamental principles (namely Onsager relations, finite DC conductivities and electric charge conservation). However, the set of relaxations we find consistent with the constraints above do not obey positivity of entropy production, implying the system is open. Furthermore, our model predicts qualitatively different DC values for the longitudinal thermoelectric magnetoresistance compared to older works, suggesting a way to test the hydrodynamic regime in Weyl semimetals. To conclude, we also introduce a mechanism to reproduce generalized relaxations from kinetic theory, by using appropriate collision integrals with energy-dependent Relaxation Time Approximation.

\section{Notations}
We work in natural units $\hbar=c=k_B=1$. Greek indexes $\mu,\nu,\dots$ identify spacetime directions, while Latin indexes $i,j,\dots$ only the spatial ones. Latin capital indexes $I,J,\dots$, mostly in Chapter~\ref{chapter:charge_density_waves}, represent a subset of the spatial directions. The spacetime metric is taken with signature $g_{\mu\nu}=\text{diag}\left(-1,1,\dots,1\right)$. Vectors are indicated both in bold $\vect{v}$ or with the arrow $\vec{v}$. In this thesis I will use both $G^R_{ab}$ and $\langle\phi_a\phi_b\rangle$ to indicate retarded Green functions interchangeably.

\medskip
\begin{center}
	\begin{tabular}{m{7cm}|m{0.2cm}m{2cm}}
		\toprule
		Name							&	&	Acronym\\
		\midrule
		Weyl semimetal					&	&	WSM\\
		Relaxation Time Approximation	&	&	RTA\\
		Direct Current					&	&	DC\\
		Alternating Current				&	&	AC\\
		Charge Density Wave				&	&	CDW\\
		Negative MagnetoResistance		&	&	NMR\\
		Partial Differential Equation	&	&	PDE\\
		Local Thermodynamic Equilibrium	&	&	LTE\\
		Anti de Sitter					&	&	AdS\\
		Wiedemann-Franz					&	&	WF\\
		Right-Hand Side					&	&	RHS\\
		Left-Hand Side					&	&	LHS\\
		\bottomrule
	\end{tabular}
\end{center}
\chapter{Hydrodynamics and Linear Response Theory}\label{chapter:hydrodynamics_linear_response}
\epigraph{``Begin at the beginning'', the King said gravely, ``and go on till you come to the end: then stop.''}{Lewis Carroll, \emph{Alice in Wonderland}}

\section{Hydrodynamics}\label{sec:ch2:hydrodynamics}

In the previous chapter we discussed hydrodynamics in broad terms. Now we will focus on the relativistic normal fluid (without superfluid component) in $d+1$ spacetime dimensions, which will be the pedagogical example in this thesis. The approach presented here, however, applies to fluids with any symmetry content (e.g. Galilean, Carrolian, boost-agnostic, \dots). In this review we will mostly follow \cite{Kovtun:LecturesHydrodynamicFluctuations}.

\subsection{Hydrodynamic variables}
As we have already discussed, hydrodynamics is concerned with the long wavelength dynamics of conserved charges. Following Noether's theorem, we know that conservation laws are a consequence of the continuous symmetries of the fundamental theory, and this in turn implies the existence of conserved currents. In particular here we will focus on a charged relativistic fluid, which means we not only assume Poincaré symmetry (boosts, rotations and translations), but also an internal $\mathrm{U(1)}$ symmetry, such that we also have a conserved vector current (e.g. baryon number, electric charge, \dots).

Symmetry under spacetime translations gives us a conserved symmetric stress-energy tensor $T^{\mu\nu}$ that couples to the external spacetime metric $g_{\mu\nu}$ and can be defined as the variation of the action with respect to the metric \cite{Weinberg:GravitationCosmologyPrinciples,Carroll:SpacetimeGeometry}
\begin{equation}
	T^{\mu\nu}=\frac{2}{\sqrt{-g}}\frac{\delta S}{\delta g_{\mu\nu}}
\end{equation}
Although the stress-energy tensor obtained from Noether's theorem is not always symmetric, the one obtained from the above definition automatically is. Anyway, following the Belinfante's construction, it is always possible to add extra antisymmetric terms to $T^{\mu\nu}$ which are identically conserved and that lead to a fully symmetric stress-energy tensor, thus we will only consider symmetric $T^{\mu\nu}$ in this work. We remark that if spin-angular momentum is a conserved quantity relevant in hydrodynamics, one indeed needs to consider antisymmetric corrections to the stress-energy tensor \cite{Gallegos:HydrodynamicsSpinCurrents}, however spin hydrodynamics will not be covered in this thesis.

The conservation law for spacetime translations (or diffeomorphism invariance) is simply
\begin{subequations}
\begin{equation}\label{eqn:ch2:stress_energy_conservation}
	\partial_\mu T^{\mu\nu}=0
\end{equation}
We can also compute the conserved currents that correspond to the other spacetime symmetries, namely boosts and rotations, and we obtain $\mathcal{M}^{\mu\nu\alpha}=x^\mu T^{\nu\alpha}-x^\nu T^{\mu\alpha}$, which is automatically conserved thanks to \eqref{eqn:ch2:stress_energy_conservation} and the relativistic Ward Identity $T^{\mu\nu}=T^{\nu\mu}$, hence there are no other conserved quantities related to the spacetime symmetries.

We still have to discuss the internal symmetries, but for a global $\mathrm{U(1)}$ symmetry this simply implies that there exists a conserved current $J^\mu$
\begin{equation}\label{eqn:ch2:current_conservation}
	\partial_\mu J^\mu=0
\end{equation}
\end{subequations}
It is also possible to introduce in hydrodynamics higher-form symmetries, which lead to the conservation of totally antisymmetric currents with more than one index, see for example \cite{Das:HigherformSymmetriesAnomalous,Armas:ApproximateHigherformSymmetries}.

It is important to stress that \eqref{eqn:ch2:current_conservation} and \eqref{eqn:ch2:stress_energy_conservation} are to be understood as non-trivial equations of motion for the conserved charges, which are the dynamical fields in hydrodynamics. While these equations are solved identically on shell for a given microscopic action $S$, in hydrodynamics they provide the dynamical content of the theory.

It is then immediately obvious that, without further assumptions, it is not possible to solve the conservation equations. In particular, by simply counting the degrees of freedom, we see that a symmetric stress-energy tensor has $(d+1)(d+2)/2$ components and the $\mathrm{U(1)}$ current has $d+1$, however we only have $d+1$ equations from \eqref{eqn:ch2:stress_energy_conservation} and one equation from \eqref{eqn:ch2:current_conservation}, not enough to determine uniquely the evolution of the system.

The simplifying assumption of hydrodynamics is that we can express $T^{\mu\nu}$ and $J^\mu$ as functions of $d+2$ local fields, which are called \emph{hydrodynamics fields} or variables. In the context of relativistic fluid dynamics, the independent fields are usually taken to be: a local temperature $T(x)$, a local chemical potential $\mu(x)$ and a local fluid velocity, described by the four-vector $u^\mu(x)$.

To better understand this choice of variable we can consider a thermal system in the grand canonical ensemble whose equilibrium is characterized by a density operator $\hat\rho$ which is proportional to the exponential of the conserved charges \cite{Israel:ThermodynamicsRelativisticSystems}. In our present case these are described by the four-momentum $\hat P^\mu=\int\dif\Sigma_\nu \hat T^{\mu\nu}$ and particle number operator $\hat N=\int\dif\Sigma_\mu \hat J^\mu$, where $\Sigma_\mu$ is some spacelike hypersurface. Then, we can express the density operator as
\begin{equation}\label{eqn:ch2:thermal_density_operator}
	\hat\rho=\frac{1}{Z}e^{\beta_\mu \hat P^\mu+\Lambda_\beta \hat N}
\end{equation}
where $Z=\text{Tr}\ e^{\beta_\mu \hat P^\mu+\Lambda_\beta \hat N}$ is the partition function. We introduced two Lagrange multipliers: a timelike vector $\beta^\mu$, called the thermal vector, and a scalar $\Lambda_\beta$, the thermal twist \cite{Haehl:AdiabaticHydrodynamicsEightfold}. It is customary in relativistic hydrodynamics to rewrite these two parameters as $\beta^\mu=\beta u^\mu$ and $\Lambda_\beta=\beta\mu$ in terms of a four-velocity $u^\mu$ (normalized such that $u^2=-1$), a chemical potential $\mu$ and an inverse temperature $\beta=T^{-1}$.

In hydrodynamics, we move away from the uniform equilibrium state characterized by \eqref{eqn:ch2:thermal_density_operator}, and we consider temperature, chemical potential and velocity as slowly varying functions of spacetime. The idea being that, as the operators $\hat{T}^{\mu\nu}$ and $\hat J^\mu$ are always well-defined (for a given theory), their expectation values $T^{\mu\nu}=\langle\hat{T}^{\mu\nu}\rangle$ and $J^\mu=\langle\hat J^\mu\rangle$ can be expressed in terms of equilibrium quantities when these vary slowly in spacetime.

We comment here that while a specific equilibrium state breaks the boost symmetry, by picking a specific timelike vector $\beta^\mu$, there are many physically equivalent equilibrium states with different spatial velocities. It is thus only the state the breaks the symmetry, while the theory is still perfectly Lorentz covariant. Indeed, it can be shown explicitly that the Goldstone boson associated with the spontaneous breaking of the boost symmetry is not an extra dynamical field, as it must be identified with the fluid velocity \cite{Armas:CarrollianFluidsSpontaneous}.

\subsection{Constitutive relations}
Having discussed the equations of motion, i.e. the conservation laws, and the hydrodynamic variables, it is now time to turn to the constitutive relations. This means that we need to write the stress-energy tensor and the $\mathrm{U(1)}$ current as the most general expressions in terms of the hydrodynamic fields compatible with the symmetries of the theory.

In the context of relativistic hydrodynamics it is useful to decompose the stress-energy tensor and the current with respect to the timelike vector $u^\mu$, as this allows us to work with quantities that are easily recognized as Lorentz covariant objects. We work in flat Minkowski space $\eta_{\mu\nu}=\text{diag}(-1,1,\dots,1)$ and we define the explicitly symmetric projector orthogonal to the four-velocity, namely $\Delta^{\mu\nu}=\eta^{\mu\nu}+u^\mu u^\nu$.\footnote{$\Delta^{\mu\nu}$ is a projector orthogonal to $u^\mu$ in the sense that $\Delta^{\mu\nu}u_\nu=0$ and $\Delta^{\mu\lambda}\Delta_\lambda^\nu=\Delta^{\mu\nu}$.} The decomposition takes the form \cite{Eckart:ThermodynamicsIrreversibleProcesses,Kovtun:LecturesHydrodynamicFluctuations}
\begin{subequations}\label{eqn:ch2:tensor_current_standard_decomposition}
\begin{align}
	T^{\mu\nu}&=\mathcal{E}u^\mu u^\nu+\mathcal{P}\Delta^{\mu\nu}+2\mathcal{Q}^{(\mu} u^{\nu)}+\mathcal{T}^{\mu\nu}\\
	J^\mu&=\mathcal{N}u^\mu+\mathcal{J}^\mu
\end{align}
\end{subequations}
where we used the standard notation in which the brackets mean symmetrized indices. The coefficients $\mathcal{E}$, $\mathcal{P}$ and $\mathcal{N}$ are scalars, $\mathcal{J}^\mu$ and $\mathcal{Q^\mu}$ are vectors transverse with respect to $u^\mu$ (such that $\mathcal{J}^\mu u_\mu=\mathcal{Q}^\mu u_\mu=0$), while $\mathcal{T}^{\mu\nu}$ is a traceless and transverse symmetric tensor.

These quantities are explicitly defined by projecting the stress-energy tensor and current. Specifically they can be written as
\begin{subequations}\label{eqn:ch2:coefficients_decomposition}
\begin{align}
	\mathcal{E}&=u_\mu u_\nu T^{\mu\nu}		\qquad\qquad	\mathcal{P}=\frac{1}{d}\Delta_{\mu\nu}T^{\mu\nu}	\qquad	\mathcal{N}=-u_\mu J^\mu\\
	\mathcal{Q}_\mu&=-\Delta_{\mu\lambda}u_\sigma T^{\lambda\sigma}		\qquad	\mathcal{J}_\mu=\Delta_{\mu\nu}J^\nu\\
	\mathcal{T}_{\mu\nu}&=\frac{1}{2}\left(\Delta_{\mu\lambda}\Delta_{\nu\sigma}+\Delta_{\nu\lambda}\Delta_{\mu\sigma}-\frac{2}{d}\Delta_{\mu\nu}\Delta_{\lambda\sigma}\right)T^{\lambda\sigma}
\end{align}
\end{subequations}
This decomposition is an identity and holds for all symmetric tensors and vectors. The hydrodynamic assumption enters when we express the coefficients of \eqref{eqn:ch2:tensor_current_standard_decomposition} in terms of the hydrodynamic variables $\mu, T$ and $u^\mu$. In particular, we will follow the logic of a derivative expansion, common in EFT, by assuming that the hydrodynamic fields differ from equilibrium only on scales much larger than the microscopic mean free path, that plays the role of a UV cut-off. This means that first-order corrections are larger than the second-order ones, second-order terms are larger than third, and so on. Formally we can write
\begin{subequations}\label{eqn:ch2:gradient_expansion}
\begin{align}
	T^{\mu\nu}&=T^{\mu\nu}_{(0)}+T^{\mu\nu}_{(1)}+T^{\mu\nu}_{(2)}+\dots+T^{\mu\nu}_{(n)}+\mathcal{O}(\partial^{(n+1)})\\
	J^\mu&=J^\mu_{(0)}+J^\mu_{(1)}+J^\mu_{(2)}+\dots+J^\mu_{(n)}+\mathcal{O}(\partial^{(n+1)})
\end{align}
\end{subequations}
The ideal or perfect fluid contains zero derivatives, while Navier-Stokes has one-derivative corrections.

\subsection{Order zero: ideal fluid}\label{sec:ch2:ideal_fluid}
Since all the transverse vectors and transverse traceless tensor we can construct with $\mu$, $T$, $u^\mu$ and their derivatives are at least order one in gradients, we can already recognize that the ideal fluid will have $\mathcal{Q}^\mu=\mathcal{J}^\mu=\mathcal{T}^{\mu\nu}=0$. Thus, we need to write $\mathcal{E}, \mathcal{P}$ and $\mathcal{N}$ in terms of the order-zero scalars, namely our hydrodynamic variables $\mu$ and $T$. To make progress we notice that, in global thermodynamic equilibrium (in the fluid rest frame), the stress-energy tensor and the current take a universal form, namely $T^{\mu\nu}_\text{RF}=\text{diag}(\epsilon,P,\dots,P)$ and $J^\mu_\text{RF}=(n,\vect{0})$. These can be understood as the definitions of the equilibrium energy density $\epsilon$, pressure $P$ and charge density $n$. These expressions are valid in the rest frame of the fluid, thus to obtain the covariant expressions for a generic fluid moving with four-velocity $u^\mu$ we simply need to perform a Lorentz boost. The final expressions are \cite{Weinberg:GravitationCosmologyPrinciples,Landau:FluidMechanicsVolume}
\begin{subequations}\label{eqn:ch2:ideal_fluid_constitutive_relations}
\begin{align}
	T^{\mu\nu}&=\epsilon u^\mu u^\nu + P\Delta^{\mu\nu}=(\epsilon+P)u^\mu u^\nu+P\eta^{\mu\nu}\\
	J^\mu&=n u^\mu
\end{align}
\end{subequations}
By construction these equations hold in global thermodynamic equilibrium, however ideal hydrodynamics simply corresponds to assuming that the same form holds when the thermodynamic variables are promoted to slowly varying fields in local thermal equilibrium, thus we find that for an ideal fluid $\mathcal{E}(x)=\epsilon(x)$, $ \mathcal{P}(x)=P(x)$ and $\mathcal{N}(x)=n(x)$.

It is clear from this discussion that one also needs an equilibrium equation of state $P(T,\mu)$ in order to know the explicit constitutive relations in terms of the fundamental hydrodynamic fields. Thus, we assume that a sensible equation of state exists, and using the thermodynamic relations \cite{Callen:ThermodynamicsIntroductionThermostatistics,Landau:StatisticalPhysicsVolume}
\begin{subequations}\label{eqn:ch2:thermodynamic_identities}
\begin{align}
	\dif P&=\frac{\partial P}{\partial T}\biggr\rvert_\mu\dif T+\frac{\partial P}{\partial\mu}\biggr\rvert_T\dif\mu=s\dif T+n\dif\mu\\
	\epsilon&=-P+Ts+\mu n
\end{align}
\end{subequations}
we first define the entropy density $s$ and the charge density from the Gibbs–Duhem equation and subsequently the energy density via the Euler relation.

It is possible to combine the longitudinal component of \eqref{eqn:ch2:stress_energy_conservation}, namely $u_\mu \partial_\mu T^{\mu\nu}=0$, together with the equation of motion for the current \eqref{eqn:ch2:current_conservation}, expressed for the constitutive relations of an ideal fluid \eqref{eqn:ch2:ideal_fluid_constitutive_relations} to obtain
\begin{equation}
	\partial_\mu \left(s u^\mu\right)=0
\end{equation}
The term in the bracket is interpreted as the ideal fluid entropy current and this equation tells us that, for a perfect fluid, entropy is locally conserved and therefore ideal hydrodynamics is non-dissipative.

\subsection{Frame choice}\label{sec:ch2:frame_choice}
Before discussing first-order corrections to the perfect fluid, leading to the so-called relativistic Navier-Stokes equations, we must first deal with a problem that appears in all EFT constructions, related to redefinitions of the fundamental fields used to perform the gradient expansion.

Specifically, at order one in derivatives we can redefine the hydrodynamic variables in terms of order-one quantities that vanish in equilibrium, so that different definitions match in global thermodynamic equilibrium, when the gradients are zero. A specific definition of the fields $T$, $\mu$ and $u^\mu$ is usually called a \emph{frame choice}. Physically this is due to the fact that there is no microscopic definition of the hydrodynamic fields when out of equilibrium, e.g. there is no temperature operator whose expectation value would give us $T(x)$ for out-of-equilibrium states. The correct way to think about the constitutive relations is to consider $T(x)$, $\mu(x)$ and $u^\mu(x)$ as auxiliary parameters used to parametrize $T^{\mu\nu}$ and $J^\mu$, which do have a microscopic out-of-equilibrium definition. A change of frame may redefine the hydrodynamic fields, but will not change the expectation value of stress-energy tensor and the current (only how they appear in terms of the local variables) \cite{Kovtun:TemperatureRelativisticFluids,Bhattacharya:TheoryFirstOrder}.

From an EFT perspective, it is natural to have the freedom to redefine fields that differ by derivative quantities, and this is indeed the case that appears when discussing EFT approach for dissipative systems \cite{Glorioso:LecturesNonequilibriumEffective,Endlich:DissipationEffectiveField}. Since frame choice are just a redundancy of the theory, they can be interpreted as gauge-like transformation \cite{Dore:FluctuatingRelativisticDissipative}. Furthermore, exactly like gauge theories, they can be recast in a gauge invariant manner, meaning that it is possible to develop a theory of hydrodynamics without fixing a frame, only in terms of frame invariant scalars and vectors \cite{Bhattacharya:TheoryFirstOrder}.

Formally, this means that the coefficients $\mathcal{E}, \mathcal{P}$ and $\mathcal{N}$ take the generic form
\begin{subequations}
\begin{align}
	\mathcal{E}&=\epsilon(T,\mu)+f_\mathcal{E}(\partial T, \partial\mu, \partial u)\\
	\mathcal{P}&=p(T,\mu)+f_\mathcal{P}(\partial, \partial\mu, \partial u)\\
	\mathcal{N}&=n(T,\mu)+f_\mathcal{N}(\partial, \partial\mu, \partial u)
\end{align}
\end{subequations}
where the first terms are obtained by the equilibrium equation of state, as discussed in Section~\ref{sec:ch2:ideal_fluid}, while the $f$s depends on the gradients of the hydrodynamic variables and are specified by our choice of the out-of-equilibrium definitions of local temperature, chemical potential and fluid velocity.

Consider now a generic frame transformation
\begin{subequations}\label{eqn:ch2:frame_transformation}
\begin{align}
	T(x)	&\longrightarrow\quad	 T'(x)=T(x)+\delta T(x)\\
	\mu(x)	&\longrightarrow\quad 	\mu'(x)=\mu(x)+\delta\mu(x)\\
	u^\mu(x)&\longrightarrow\quad 	u'^\mu(x)=u^\mu(x)+\delta u^\mu(x)
\end{align}
\end{subequations}
where the $\delta$ represent first-order correction\footnote{Notice that $\delta u^\mu$ is a transverse vector, such that $u_\mu \delta u^\mu=0$. This is needed in order to preserve the normalization of the four-velocity at the given order in derivatives $u^2=-1+\mathcal{O}(\partial^2)$.}. One can use the new definition of $u^\mu$ to decompose the stress-energy tensor and the current as in \eqref{eqn:ch2:coefficients_decomposition}, remembering that $T^{\mu\nu}$ and $J^\mu$ are invariant under frame choice and that $\mathcal{Q}^\mu$ and $\mathcal{J}^\mu$ are themselves order one in derivatives, to determine how the coefficients of the decomposition vary under a frame transformation. The result is
\begin{subequations}\label{eqn:ch2:frame_transformation_coefficients}
\begin{align}
	\delta\mathcal{E}&=0\qquad\delta\mathcal{P}=0\qquad\delta\mathcal{N}=0\\
	\delta\mathcal{Q}^\mu&=-(\mathcal{E}+\mathcal{P})\delta u^\mu\qquad\delta\mathcal{J}^\mu=-\mathcal{N}\delta u^\mu\\
	\delta\mathcal{T}^{\mu\nu}&=0
\end{align}
\end{subequations}
From these expressions it is immediately clear that one can always pick $\delta u^\mu$ such that $\mathcal{J}^\mu=0$ or $\mathcal{Q}^\mu=0$. The first case is often called Eckart frame \cite{Eckart:ThermodynamicsIrreversibleProcesses,Weinberg:GravitationCosmologyPrinciples}, while the second choice is called Landau frame \cite{Landau:FluidMechanicsVolume}.

While frames are not physical in the hydrodynamic regime, they are simply understood as redefinitions of the fields we use to parametrize the observable quantities $T^{\mu\nu}$ and $J^\mu$, they can still have a physical interpretation. In particular the Eckart frame is defined so that there is no charge flow in the local rest frame of the fluid, while the Landau frame is such that there is no energy current in the local rest frame of the fluid. There are in principle infinite frames, however in practice only a handful of these are really used in the literature. In particular two very useful ones are the thermodynamic frame \cite{Jensen:HydrodynamicsEntropyCurrent}, which is actually a class of frame that we will discuss later on, and the recently introduced BDNK frame \cite{Kovtun:FirstorderRelativisticHydrodynamics,Bemfica:NonlinearCausalityGeneral}.

The first equation in \eqref{eqn:ch2:frame_transformation_coefficients} implies that $\epsilon(T,\mu)+f_\mathcal{E}(\partial T,\partial\mu,\partial u)=\epsilon(T',\mu')+f'_\mathcal{E}(\partial T',\partial\mu',\partial u')$ and similar expressions hold also for $\mathcal{N}$ and $\mathcal{P}$. This means that we can write
\begin{subequations}
\begin{align}
	f'_\mathcal{E}&=f_\mathcal{E}-\left(\frac{\partial\epsilon}{\partial T}\right)_\mu\delta T-\left(\frac{\partial\epsilon}{\partial\mu}\right)_T\delta\mu\\
	f'_\mathcal{P}&=f_\mathcal{P}-\left(\frac{\partial p}{\partial T}\right)_\mu\delta T-\left(\frac{\partial p}{\partial\mu}\right)_T\delta\mu\\
	f'_\mathcal{N}&=f_\mathcal{N}-\left(\frac{\partial n}{\partial T}\right)_\mu\delta T-\left(\frac{\partial n}{\partial\mu}\right)_T\delta\mu
\end{align}
\end{subequations}
to relate the primed functions in terms of the old ones plus the frame choice derivative corrections. In particular, we can always pick the out-of-equilibrium definition of $\mu$ and $T$ such that two of the three functions $f$ are zero. It is standard to set to zero $f'_\mathcal{E}$ and $f'_\mathcal{N}$, such that $\mathcal{E}=\epsilon$ and $\mathcal{N}=n$ at all orders in derivatives.

\subsection{Order one: Navier-Stokes}\label{sec:ch2:order_one}
\begin{figure}[t]
	\centering
	\includegraphics[width=0.5\textwidth]{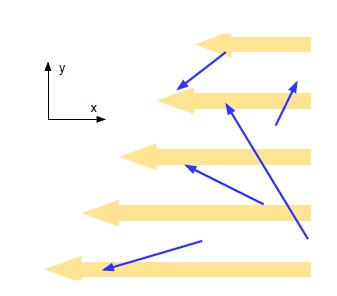}
	\caption{Figure taken from \cite{Kovtun:LecturesHydrodynamicFluctuations}.}
	\label{fig:ch2:impossible_flow}
\end{figure}

The ideal fluid we have discussed so far, however, predicts unphysical solutions: if we only consider the constitutive relations truncated at order zero in derivatives, then we find solutions to the conservation equations of hydrodynamics that describe impossible flows. For example, an equilibrated fluid flowing in the $x$ direction with a velocity gradient along $y$ is a valid solution to the ideal fluid hydrodynamic equations, but it is unphysical in actual systems, since fluid particles will transfer momentum between the fluid layers and render the flow homogeneous, see Figure~\ref{fig:ch2:impossible_flow}. Hence, it is imperative to study the first-order dissipative corrections, for which such flows are not equilibrium solutions.

We will work in the Landau frame, i.e. we will choose the out-of-equilibrium definition of $u^\mu$ to be such that $\mathcal{Q}^\mu=0$ and furthermore we also fix the definitions of $\mu$ and $T$ in such a way that $\mathcal{E}=\epsilon$ and $\mathcal{N}=n$. The Landau frame matching conditions are usually written as
\begin{equation}\label{eqn:ch2:landau_matching_conditions}
	u_\mu T^{\mu\nu}=-\epsilon u^\nu\qquad\qquad u_\mu J^\mu=-n
\end{equation}
so that the fluid velocity is an eigenvector of the stress-energy tensor with the energy as eigenvalue.

Following the hydrodynamic EFT prescription, we write the free coefficients $\mathcal{P}$, $\mathcal{J}^\mu$, $\mathcal{T}^{\mu\nu}$ (those that are not fixed by the choice of the frame) as the most general expressions in terms of the derivatives of the fundamental fields. There are only three one-derivative scalars at this order, namely the matter derivative of temperature and chemical potential $u^\alpha\partial_\alpha T$, $u^\alpha\partial_\alpha\mu$ and the expansion $\theta=\partial_\alpha u^\alpha$. There are furthermore three transverse vectors $\Delta^{\mu\nu}\partial_\nu T$, $\Delta^{\mu\nu}\partial_\nu\mu$ and $\Delta^{\mu\nu}a_\nu$ where $a_\mu=u^\alpha\partial_\alpha u_\mu$ is the acceleration. To conclude, there is only one traceless symmetric order-one tensor we can build, the shear tensor
\begin{equation}\label{eqn:ch2:shear_tensor}
	\sigma^{\mu\nu}=\Delta^{\mu\alpha}\Delta^{\nu\beta}\left(\partial_\alpha u_\beta+\partial_\beta u_\alpha-\frac{2}{d}\eta_{\alpha\beta}\partial_\lambda u^\lambda\right)
\end{equation}

Let us focus on the scalar part first. We already fixed $\mathcal{E}$ and $\mathcal{N}$ from the Landau matching conditions, so we only have one scalar left, $\mathcal{P}$, which should be written as the most generic expression in terms of the order-zero and order-one scalars listed above.
\begin{equation}
	\mathcal{P}=P+c_1u^\alpha\partial_\alpha T+c_2u^\alpha\partial_\alpha\mu+c_3\theta+\mathcal{O}(\partial^2)
\end{equation}
However, we remember that hydrodynamics should be understood as a formal series, and by using the ideal fluid constitutive relations \eqref{eqn:ch2:ideal_fluid_constitutive_relations} in the two scalars equations $\partial_\mu J^\mu=0$ and $u_\mu\partial_\nu T^{\mu\nu}=0$ we find two equations that relate the three order-one scalars. This means that, e.g. $u^\alpha\partial_\alpha T$ is proportional to $u^\alpha\partial_\alpha\mu$, up to corrections that are higher order in derivatives $\mathcal{O}(\partial^2)$, and the same holds for the expansion. Thus, at order one in derivatives there is only one independent scalar, and it is a convention to keep $\partial_\mu u^\mu$ as the independent one. We can finally write down $\mathcal{P}$ as the order-zero term plus the first-order correction
\begin{equation}
	\mathcal{P}=P-\zeta\partial_\mu u^\mu+\mathcal{O}(\partial^2)
\end{equation}
where $\zeta=\zeta(T,\mu)$, the bulk viscosity, is the first transport coefficient we encounter.

We now turn to the only transverse vector left unspecified, namely $\mathcal{J}^\mu$. As before, there are three transverse vectors at order one in derivatives, however we also have one transverse ideal-fluid equation of motion, $\Delta_{\mu\nu}\partial_\alpha T^{\alpha\nu}=0$. This relates the three order-one vectors up to higher-order corrections, which again means we only have two independent terms. Thus, we write
\begin{equation}\label{eqn:ch2:dissipative_current}
	\mathcal{J}^\mu=-\sigma T\Delta^{\mu\nu}\partial_\nu\frac{\mu}{T}+\chi_T\Delta^{\mu\nu}\partial_\nu T+\mathcal{O}(\partial^2)
\end{equation}
where we reorganized the vectors in a way that will be useful later on. For each term we introduced a new transport coefficient: the charge conductivity $\sigma$ and $\chi_T$.

To end our discussion, we observe that since the shear tensor \eqref{eqn:ch2:shear_tensor} is the only transverse traceless symmetric tensor, it must be proportional to $\mathcal{T}^{\mu\nu}$, hence we write
\begin{equation}
	\mathcal{T}^{\mu\nu}=-\eta \sigma^{\mu\nu}+\mathcal{O}(\partial^2)
\end{equation}
where $\eta$, the shear viscosity, is the last transport coefficient that appears in standard relativistic hydrodynamics.

Had we chosen Eckart frame, we would have found the same expressions for $\mathcal{T}^{\mu\nu}$ and $\mathcal{P}$, instead what we wrote for $\mathcal{J}^\mu$ would now appear for $\mathcal{Q}^\mu$. Notice that there are actually two ambiguity in hydrodynamics: one is related to the choice of frame (in this case, Eckart or Landau), while the other is related to fixing the base of independent coefficients used to write the constitutive relations. For example, we have decided not to use the acceleration in writing the constitutive relation for $\mathcal{J}^\mu$, however when working in the Eckart frame it is more common to use the acceleration instead of $\Delta^{\mu\nu}\partial_\nu\frac{\mu}{T}$. This choice might seem innocuous, however one must be careful when computing quantities in hydrodynamics to check that they are independent of these ambiguities by working at the correct derivative order \cite{Kovtun:LecturesHydrodynamicFluctuations}.

A comment on transport coefficients: these are quantities which cannot be determined from hydrodynamics alone\footnote{This is not true for certain transport coefficients related to quantum anomalies \cite{Son:HydrodynamicsTriangleAnomalies}.}, since they depend on the microscopic details of the theory, while hydrodynamics knows only about the IR macroscopic dynamics. Given a specific microscopic model one can compute these coefficients via Kubo formulae (at least in principle), however from the point of view of hydrodynamics they are just unknown functions of the scalar fields.

The Kubo formulae are independent of the choice of the frame and can be used to match the microscopic theory to the EFT of hydrodynamics. This means that the transport coefficients are frame invariant: they can appear in different places in the constitutive relations, depending on the choice of the frame, but their form in terms of the microscopic theory is unique and independent this choice.

\subsection{Entropy current}\label{sec:ch2:entropy_current}
The constitutive relations we obtained above are based only on two consistency requirements: Lorentz covariance and the formal derivative expansion. Hydrodynamics however requires more conditions, in particular when developing a theory of hydrodynamics one also imposes a local version of the second law of thermodynamics, namely we will require that the production of entropy is locally positive. Later on we will also discuss one last kind of constraint that stems from the discrete symmetries of the microscopic theory.

In equilibrium, we can write the entropy current $S^\mu=su^\mu$ for constant $u^\mu$, and we showed that the ideal fluid does not produce entropy $\partial_\mu(s u^\mu)=0$. In first-order hydrodynamics we assume that $S^\mu$ receives derivative corrections, consistent with what we would expect by covariantizing the thermodynamics. In particular, we start from the Euler relation $Ts=P+\epsilon-\mu n$, and we write it in a covariant form \cite{Eckart:ThermodynamicsIrreversibleProcesses,Israel:ThermodynamicsRelativisticSystems}
\begin{equation}
	TS^\mu=Pu^\mu-T^{\mu\nu}u_\nu-\mu J^\mu
\end{equation}
From this expression, using the standard decomposition \eqref{eqn:ch2:tensor_current_standard_decomposition}, we obtain for the entropy current
\begin{equation}\label{eqn:ch2:canonical_entropy_current}
	S^\mu=\left[s+\frac{1}{T}\left(\mathcal{E}-\epsilon\right)-\frac{\mu}{T}\left(\mathcal{N}-n\right)\right]u^\mu+\frac{1}{T}\mathcal{Q}^\mu-\frac{\mu}{T}\mathcal{J}^\mu
\end{equation}
where the terms inside the square brackets is the rest frame entropy density, while the other two terms represent the rest frame entropy flow. This is the so-called canonical entropy current, and it is frame invariant (as always, the explicit form of the current in terms of the gradients may vary depending on the frame, but the value of the current does not). There are other contributions to the entropy current, called non-canonical, which have important roles when discussing hydrostatic or non-dissipative fluids.

In the specific case of the Landau frame the entropy current simplifies to \cite{Hartnoll:TheoryNernstEffect,Kovtun:LecturesHydrodynamicFluctuations,Landau:FluidMechanicsVolume}
\begin{equation}
	S^\mu=su^\mu-\frac{\mu}{T}\mathcal{J}^\nu
\end{equation}
We must now use this expression, together with the Landau-frame constitutive relations, to impose the local form of the second law of thermodynamics
\begin{equation}
	\partial_\mu S^\mu\geq0
\end{equation}
on all possible flows that are solutions to the equations of motion \eqref{eqn:ch2:stress_energy_conservation} and \eqref{eqn:ch2:current_conservation}.

The steps are very straightforward, using the Gibbs–Duhem equation $\partial P=s\partial T+n\partial\mu$ one arrives at
\begin{align}
	\partial_\mu S^\mu&=\frac{\zeta}{T}\theta^2+\frac{\eta}{T}\sigma_{\mu\nu}\sigma^{\mu\nu}+\sigma\left(T\Delta^{\mu\nu}\partial_\nu\frac{\mu}{T}\right)^2-T\chi_T\Delta^{\mu\nu}\partial_\nu\frac{\mu}{T}\partial_\mu T\geq0
\end{align}
where we remind that $\theta$ is the expansion and $\sigma_{\mu\nu}$ the shear tensor. Because the RHS must always be non-negative, we immediately find that
\begin{equation}\label{eqn:ch2:entropy_constraints}
	\eta\geq0\qquad\zeta\geq0\qquad\sigma\geq0\qquad\chi_T=0
\end{equation}
Hence, while Lorentz covariance leaves us with four transport coefficients at first order in derivatives for relativistic hydrodynamics, positivity of entropy production reduces this number to only three.

For future reference, we write here the complete first-order constitutive relations in the Landau frame for a normal fluid that stem from the requirement of Lorentz covariance, gradient expansion and positivity of entropy production. They are
\begin{subequations}\label{eqn:ch2:landau_constitutive_relations}
\begin{align}
	T^{\mu\nu}&=\epsilon u^\mu u^\nu+P\Delta^{\mu\nu}-\eta\sigma^{\mu\nu}-\zeta\Delta^{\mu\nu}\partial_\lambda u^\lambda+\mathcal{O}(\partial^2)\\
	J^\mu&=n u^\mu-\sigma T\Delta^{\mu\nu}\partial_\nu\frac{\mu}{T}+\mathcal{O}(\partial^2)
\end{align}
\end{subequations}

There is a slightly different route to arrive at the same constitutive relations that is sometimes quicker: after fixing the frame one can skip completely the steps in Section~\ref{sec:ch2:order_one} and instead compute directly the divergence of the canonical entropy current. Then, by looking at the final expression, one can infer the form of the constitutive relations. This is the approach used e.g. in \cite{Landau:FluidMechanicsVolume}.

\subsection{Hydrostatic generating functional}\label{sec:ch2:generating_functional}
The hydrostatic generating functional \cite{Jensen:HydrodynamicsEntropyCurrent,Banerjee:ConstraintsFluidDynamics} is a powerful technique that can be used to compute the $n$-point functions (hence also the constitutive relations) from a path-integral approach in the hydrostatic regime, i.e. when everything is time independent and, in the absence of external sources, constant.

The hydrostatic flow, together with the dissipative sector described in the previous Section~\ref{sec:ch2:order_one} on Navier-Stokes fluids, are two of the eight possible class of fluid transport classified in \cite{Haehl:AdiabaticHydrodynamicsEightfold,Haehl:EightfoldWayDissipation} using the idea of adiabatic flows.

Consider a charged fluid in the presence of spacetime and gauge sources \eqref{eqn:ch2:conservation_equation_sources}. The fluid is defined to be in (time-independent) hydrostatic equilibrium if there exists a timelike Killing vector $V^\mu$ that parametrizes time-translations such that all the thermodynamic quantities and the external sources have zero Lie derivative $\mathcal{L}_V$ with respect to the Killing field $V^\mu$. Physically, this means that the sources are slowly varying and that the time derivative in the frame comoving with $V^\mu$, when $V^\mu=(1,\vect{0})$, is zero.

Because we are dealing with equilibrium configurations (in the presence of sources that are adiabatically turned on) we can formally obtain the Green functions from a generating functional by differentiating with respect to the sources $g_{\mu\nu}$ and $A_\mu$.

In the frame where $V^\mu=(1,\vect{0})$, consider the set of zero-frequency $n$-point functions truncated up to order $m$ in powers of the wavevector $\vect{k}$. We can inverse-Fourier transform to obtain a set of correlators in position space which are approximately valid on scales much larger than the microscopic correlation length of the system. By formally integrating this set of approximate Green functions we arrive to the equilibrium generating functional for the truncated correlators that is a local function of the sources
\begin{equation}
	W_m[g,A]=\int\dif^dx\ \mathcal{W}[\text{sources}(x)]
\end{equation}
where the function $\mathcal{W}$ contains terms up to order $m$ in derivatives. Because this expression is covariant, it holds also for frames which are not comoving with $V^\mu$.

In order for $W_m$ to reproduce the equations of motion \eqref{eqn:ch2:conservation_equation_sources} it must be diffeomorphism and gauge invariant, hence $\mathcal{W}$ is a function of all the possible diffeomorphism and gauge invariant scalars of the theory. These scalar observables should be local in space, but can be non-local in Euclidean time, in order to describe finite temperature systems.

Consider then the two quantities: the invariant length $L$ of the time circle in the Euclidean theory and the Polyakov loop $P_A$ of the gauge field $\mathrm{U(1)}$ around the same direction. The length $L$ can be computed simply as
\begin{equation}
	L=\int_0^\beta\dif\tau\sqrt{g_{\tau\tau}}
\end{equation}
in the frame comoving with $V^\mu$, where $\beta=T_0^{-1}$ is the period of the coordinate in the time direction. Then, covariantizing the expression and rotating back to real time, we find that $L$ and $P_A$ are written as
\begin{equation}
	L=\beta\sqrt{-V^2}\qquad\qquad \ln P_A=\beta V^\mu A_\mu
\end{equation}
We can now identify the temperature, chemical potential and fluid velocity from these invariant quantities as
\begin{subequations}\label{eqn:ch2:thermodynamic_frame_definitions}
	\begin{align}
		T&=\frac{1}{L}=\frac{T_0}{\sqrt{-V^2}}\\
		\mu&=\frac{\ln P_A}{L}=\frac{V^\mu A_\mu-\Lambda_\beta}{\sqrt{-V^2}}\\
		u^\mu&=\frac{V^\mu}{\sqrt{-V^2}}
	\end{align}
\end{subequations}
where $\Lambda_\beta$ is a gauge parameter that must be included to preserve gauge invariance. These expressions can be understood as the thermal twist and thermal vector, in the presence of sources, that appear in the equilibrium statistical mechanics perspective introduced in \eqref{eqn:ch2:thermal_density_operator}. Notice that, contrary to \eqref{eqn:ch2:thermal_density_operator}, these expressions for $T$, $\mu$ and $u^\mu$ have a dependence on spacetime coordinates inherited from the sources $g_{\mu\nu}$ and $A_\mu$.

For a given derivative order $m$ the theory has $N_m$ scalars quantities $s_{m,1},s_{m,2},\dots,s_{m,N_m}$ constructed from the invariants \eqref{eqn:ch2:thermodynamic_frame_definitions} and their derivatives. Therefore, the most general hydrostatic generating functional at order $m$ takes the form
\begin{equation}\label{eqn:ch2:generating_functional}
	W_m[g,A]=\int\dif^dx\sqrt{-g}\left[P(s_0)+\sum_{n=1}^m\sum_{i=1}^{N_n} F_{n,i}(s_0)s_{n,i}\right]
\end{equation}
In this expression: $s_0$ are the order-zero scalars and for a simple fluid they are $s_0=\{T,\mu\}$, $F_{n,i}$ are unknown functions of the order-zero scalars, while $P$ can be identified with the pressure. Indeed, at zero sources, when all the derivative corrections to $W_m$ vanish, the functional $W$ corresponds to the free energy.

Given this expression we can now compute the hydrostatic constitutive relations for the stress-energy tensor and the current, the one-point functions, by varying $W_m$ with respect to the metric and gauge field respectively
\begin{equation}
	\langle T^{\mu\nu}\rangle=\frac{2}{\sqrt{-g}}\frac{\delta W_m}{\delta g_{\mu\nu}}\quad\qquad \langle J^\mu\rangle=\frac{1}{\sqrt{-g}}\frac{\delta W_m}{\delta A_\mu}
\end{equation}
Thanks to the requirement of gauge and diffeomorphism invariance of $W_m$ the hydrostatic constitutive relations obtain from the above variations will identically satisfy the hydrodynamic equations of motion.

The constitutive relations obtained from the hydrostatic generating functional appear in a specific hydrodynamic frame called thermodynamic frame. Contrary to the Landau frame, defined by the matching conditions \eqref{eqn:ch2:landau_matching_conditions}, the thermodynamic frame physically means that the definitions of temperature, chemical potential and fluid velocity are not modified on shell with respect to their equilibrium value \eqref{eqn:ch2:thermodynamic_frame_definitions}. Notice that, while it is common to use the term thermodynamic frame, this is not a true frame, but rather a class of frame. This is because the thermodynamic frame, by definition, is a prescription for how to deal with the hydrostatic sector of the constitutive relations, but it says nothing about the dissipative parts, for which one needs further frame fixing conditions.

One might also wonder what happens to the entropy current in the hydrostatic sector. The hydrostatic solutions are, as expected, non-dissipative, hence $\partial_\mu S^\mu=0$, however in order to achieve this results the entropy current itself must be modified compared to its canonical value. The total entropy current is
\begin{equation}
	S^\mu = S^\mu_\text{c.}+S^\mu_\text{n.c.}
\end{equation}
where the first term is the canonical entropy current discussed in \eqref{eqn:ch2:canonical_entropy_current}, while the latter is the so-called non-canonical entropy current and its role is to ensure that the hydrostatic derivative corrections do not produce entropy \cite{Bhattacharyya:EntropyCurrentPartition,Bhattacharyya:EntropyCurrentEquilibrium}.

\subsection{Hydrostatic conditions}\label{sec:ch2:hydrostatic_constraints}
For the fluid to be in hydrostatic flows, its variables must obey certain conditions which in a sense define equilibrium. These constraints are obtained by requiring that the Lie derivatives of the sources and of the thermodynamic quantities \eqref{eqn:ch2:thermodynamic_frame_definitions} with respect to the Killing vector $V^\mu$ vanish. To proceed, we first decompose the $\mathrm{U(1)}$ electromagnetic field strength by projecting along the fluid velocity in ($3+1$)-dimensions
\begin{equation}\label{eqn:ch2:electric_magnetic_decomposition}
	F_{\mu\nu}=u_\mu E_\nu-u_\nu E_\mu-\varepsilon_{\mu\nu\rho\sigma}u^\rho B^\sigma
\end{equation}
where we defined the covariant electric and magnetic fields as
\begin{equation}
	E_\mu=F_{\mu\nu}u^\nu\qquad B_\mu=-\frac{1}{2}\varepsilon^{\mu\nu\alpha\beta}u_\nu F_{\alpha\beta}
\end{equation}
Notice that both $E^\mu$ and $B^\mu$ are transverse to $u^\mu$. Then, requiring $\mathcal{L}_VT=\mathcal{L}_V\mu=0$ give use the first two hydrostatic constraints  \cite{Jensen:HydrodynamicsEntropyCurrent}
\begin{subequations}\label{eqn:ch2:hydrostatic_constraints_relativistic}
	\begin{align}
		\partial_\mu T&=-Ta_\mu\\
		\partial_\mu\mu&=E_\mu-\mu a_\mu
	\end{align}
where $a^\mu=u^\lambda\nabla_\lambda u^\mu$ is the acceleration. The other constraints come from studying the Lie derivative of the fluid velocity, for which we find that an equilibrium fluid must have vanishing expansion and shear tensor
	\begin{equation}
		\theta=\nabla_\mu u^\mu=0\qquad\sigma^{\mu\nu}=0
	\end{equation}
\end{subequations}
Finally, we will require the Bianchi identity of the gauge field, $\varepsilon^{\mu\nu\alpha\beta}\nabla_\nu F_{\alpha\beta}=0$.

These are equality-type constraints that the fluid variables must obey for the fluid to be in equilibrium in the presence of external sources. In particular, any dissipative correction to the constitutive relations must be understood as fluctuating away from these constraints. For example, in equilibrium we find that the fluid must have zero expansion $\theta=0$, which agrees with the fact that the bulk viscosity is a dissipative transport coefficient and the fluid produce entropy when expansion is not zero. For the vector sector, second law of thermodynamics tells us that the two constraints \eqref{eqn:ch2:hydrostatic_constraints_relativistic} do not give rise to two independent dissipative transport coefficients, but to a single dissipative transverse vector
\begin{equation}\label{eqn:ch2:kovtun_condition}
	E^\mu-\Delta^{\mu\nu}\partial_\nu\frac{\mu}{T}=0
\end{equation}
that appears in the current constitutive relation \eqref{eqn:ch2:constitutive_relations_curved_space}.

Notice that the opposite is not true: there exist transport coefficients which are non-dissipative (they drop out of the equation when computing the positivity of entropy production), but nonetheless they vanish on hydrostatic flows. These are called non-dissipative non-hydrostatic terms and can be obtained from a generalization of the above generating functional \cite{Haehl:AdiabaticHydrodynamicsEightfold}.

\section{Linear Response Theory}\label{sec:ch2:linear_response_theory}
Hydrodynamics is a highly non-linear theory, meaning that small differences in the initial conditions can lead to drastically different evolutions of the fluid dynamics. This makes solving the full non-linear equations of hydrodynamics notoriously hard, with entire careers dedicated to numerically solving fluid flows.

As we discussed in the introduction, the problems that appear in hydrodynamics are often related to trusting the theory far beyond its regime of validity. Hydrodynamics is a low-energy EFT that works only when $\omega$ and $\vect{k}$ are small, nonetheless if one wishes to use the equations of hydrodynamics as a system of non-linear PDEs with predictive power, one needs to solve the equations even at high momenta, when we expect hydrodynamics to fail.

Things are instead better behaved if we focus on linearized hydrodynamics. This means that we trust the theory only in its regime of validity: we consider a uniform global thermodynamic equilibrium state given by a constant temperature $T=\text{const}$, chemical potential $\mu=\text{const}$ and in the rest frame $u^\mu=(1,\vect{0})$ and study the small fluctuations above this solution. Furthermore, we will only care about the small-wavevector/small-frequency expansion, which is where hydrodynamics is expected to hold.

Linearized hydrodynamics still gives plenty of information, in particular: (i) it gives access to the retarded correlators, which can then be matched against the same two-point functions obtained from other approaches (holography, kinetic theory, QFT, \dots), leading to the Kubo formulae, (ii) the same correlators give physical information about the transport properties of the system, like the thermoelectric conductivities, (iii) it allows us to find new constraints on the constitutive relations from the microscopic discrete symmetries, and (iv) it gives information about the stability of the theory to small perturbations.

For most of this section instead of working with $T$, $\mu$ and $u^\mu$ we will consider their thermodynamically conjugate charges as independent variables, namely energy, charge and momentum density. The only reason to do so is that the latter quantities have a microscopic definition in terms of operators $\hat T^{0\mu}$ and $\hat J^0$, which allows us to study how they appear in the Hamiltonian and obtain the correlators associated to these charges.

\subsection{Martin-Kadanoff method}\label{sec:ch2:martin_kadanoff}
In this section we will quickly resume the main idea behind linear response theory, that is to describe the response of a system in thermal equilibrium to small perturbations, based on the classical paper by Martin and Kadanoff \cite{Kadanoff:HydrodynamicEquationsCorrelation} from which this method takes its name. We consider a set of hydrodynamic fields $\phi_a(t,\vect{x})$ (that are microscopically well-defined) and we couple them to their (weak slowly-varying) sources $\lambda_a(t,\vect{x})$. The sources are turned on at $t=-\infty$ and are adiabatically increased; at $t=0$ they are turned off and the system is let free to evolve. In linearized hydrodynamics the classical fields $\phi_a$ are described by a set of linear equations that, in momentum space, read\footnote{Here the fields $\phi_a$ whose dynamics is described by linearized hydrodynamics are actually $\delta\phi_a$, that is the difference with respect to the equilibrium $\delta\phi_a(t,\vect{x})=\phi_a(t,\vect{x})-\phi_a$.}
\begin{equation}\label{eqn:ch2:hydrodynamic_equations_schematic}
	\partial_t\phi_a(t,\vect{k})+M_{ab}(\vect{k})\phi_b(t,\vect{k})=0
\end{equation}
where $M_{ab}$ is a matrix determined by the conservation laws of hydrodynamics and the constitutive relations. This set of equations is valid in the hydrodynamic regime (long wavelength $\vect{k}\rightarrow0$ and small frequency $\omega\rightarrow0$), and describes the free evolution of the fields at $t>0$. To study the solutions we perform a Laplace transform in time
\begin{equation}
	\phi_a(z,\vect{k})=\int_0^{\infty}\dif t\ e^{izt}\phi_a(t,\vect{k})
\end{equation}
with $z$ that has to lie on the upper half of the complex plane for the integral to converge. Transforming the set of equation we find
\begin{equation}
	(-i z\delta_{ab}+M_{ab})\phi_b(z,\vect{k})=\phi_a^0(\vect{k})
\end{equation}
where we defined $\phi^0_a(\vect{k})=\phi_a(t=0,\vect{k})$, the value of the field just when the source is turned off. The values at $t=0$ of the hydrodynamic variables are related to the initial values of the sources, that is just before we turn the sources off we have a relation that is local in space. For small fluctuations we can write
\begin{equation}\label{eqn:ch2:static_susceptibility}
	\phi_a^0(\vect{k}\rightarrow0)=\chi_{ab}\lambda_b^0(\vect{k}\rightarrow0)   \qquad\Longrightarrow\qquad \chi_{ab}=\left(\frac{\partial\phi_a}{\partial\lambda_b}\right)
\end{equation}
$\chi_{ab}$ is called static thermodynamic susceptibility, it is not a dynamical response, but rather a thermodynamic quantity that depends on the system. With this definition the hydrodynamic equations can be formally solved in terms of the initial conditions
\begin{equation}\label{eqn:ch2:phi_from_linear_equation}
	\phi_a(z,\vect{k})=(K^{-1})_{ab}\chi_{bc}\lambda_c^0(\vect{k})
\end{equation}
with $K_{ab}=-iz\delta_{ab}+M_{ab}(\vect{k})$.

We now try to relate the expectation values of the hydrodynamic fields to the correlation functions of the system. Imagine perturbing the fields using slowly-varying (with respect to both $\vect{x}$ and $t$) sources that are adiabatically turned on at $t=-\infty$ and then switched off at $t=0$, for example with $\lambda(t,\vect{x})=e^{\varepsilon t}\lambda^0(\vect{x})\theta(-t)$ ($\varepsilon>0$ is a small adiabatic parameter and $\theta(t)$ is the step function); we can then link the evolution of the system at $t>0$ to the result we found in equation \eqref{eqn:ch2:phi_from_linear_equation}. To do so we perturb the Hamiltonian with a term that couples the hydrodynamic fields to their sources
\begin{equation}
	\delta\hat H(t)=-\int \dif^d\vect{x}\ \lambda_a(t,\vect{x})\hat\phi_a(t,\vect{x})
\end{equation}
From first-order time-dependent perturbation theory in Quantum Mechanics we can find the response of an operator $\hat\phi_a(t,\vect{x})$ in the Heisenberg picture. If the system has a time-independent Hamiltonian, and we add the small perturbation above the expected value of the observable changes as
\begin{equation}
	\delta\langle\hat\phi_a(t,\vect{x})\rangle=-i\int_{-\infty}^t \dif t'\langle[\hat\phi_a(t,\vect{x}),\delta\hat H(t')]\rangle
\end{equation}
where $\langle\dots\rangle$ stands for the thermal average $\langle\hat O\rangle=\text{Tr}(\hat\rho\hat O)$ with $\hat\rho$ the density matrix (we work in the grand canonical ensemble, so the Heisenberg operators are defined with the Hamiltonian $\hat H'=\hat H-\mu \hat N$, with the chemical potential $\mu$ as the Lagrange multiplier of the charge number operator $\hat N$). Plugging together the two equations we find
\begin{equation}\label{eqn:ch2:correlation_real_space}
	\delta\langle\hat\phi_a(t,\vect{x})\rangle=-\int_{-\infty}^{\infty}\dif t'\int\dif^d x'\ G^R_{ab}(t-t',\vect{x}-\vect{x}')\lambda_b(t,\vect{x})
\end{equation}
where $G^R_{ab}$ is called retarded response function, and it is defined as
\begin{equation}
	G^R_{ab}(t-t',\vect{x}-\vect{x}')=-i\theta(t-t')\langle[\hat\phi_a(t,\vect{x}),\hat\phi_b(t',\vect{x}')]\rangle
\end{equation}
It depends only on the differences of the coordinates, due to its invariance under spacetime translations. In Fourier space the expectation value simplifies and the convolution gives
\begin{equation}
	\delta\langle\hat\phi_a(\omega,\vect{k})\rangle=-G^R_{ab}(\omega,\vect{k})\lambda_b(\omega,\vect{k})
\end{equation}
We still have to Laplace-Fourier transform equation \eqref{eqn:ch2:correlation_real_space} to link this result with the one obtained from linearized hydrodynamics.

We start by performing the Fourier transform in space; with our assumption on the form of the sources we find
\begin{equation}\label{eqn:ch2:phi_fourier}
	\langle\hat\phi_a(t,\vect{k})\rangle=-\int_{-\infty}^0\dif t' e^{\varepsilon t'}G^R_{ab}(t-t',\vect{k})\lambda_b^0(\vect{k})
\end{equation}
We now Fourier transform only the retarded function in time
\begin{equation}
	G^R(t-t',\vect{k})=\int_{-\infty}^{\infty}\frac{\dif \omega}{2\pi}G^R(\omega,\vect{k})e^{-i\omega(t-t')}
\end{equation}
Because of the step function in the definition of $G^R(t,\vect{k})$, this function is identically zero for $t<0$, this in turn means that $G^R(\omega,\vect{k})$ is an analytic function in the upper half-plane of complex $\omega$, and we are free to analytically continue $G^R(\omega,\vect{k})$ to the whole plane. After performing the $t'$ integral we are left with
\begin{equation}
	\langle\hat\phi_a(t,\vect{k})\rangle=-\lambda^0_b(\vect{k})\int\frac{\dif \omega}{2\pi}G^R_{ab}(\omega,\vect{k})\frac{e^{-i\omega t}}{i\omega+\varepsilon}
\end{equation}
where $\varepsilon$ is needed for the convergence of the integral. We can now multiply both sides by $e^{izt}$ (with the prescription that $\text{Im}(z)>0$ for the integral to convergence) and integrate over $t$ from $0$ to $\infty$ (that is, we perform the same Laplace transform we applied earlier to our linearized hydrodynamic equations)
\begin{equation}
	\langle\hat\phi_a(z,\vect{k})\rangle=-\lambda^0_b(\vect{k})\int\frac{\dif \omega}{2\pi}\frac{G^R_{ab}(\omega,\vect{k})}{(i\omega+\varepsilon)(i(\omega-z)+\varepsilon)}
\end{equation}
As we mentioned, $G^R(\omega)$ is analytic in the upper half-plane, this allows us to perform the integral by closing the contour with $\text{Im}(\omega)>0$. There are two poles inside the region, one at $\omega=i\varepsilon$ and one at $\omega=z+i\varepsilon$. The residue theorem gives us
\begin{equation}\label{eqn:ch2:martin_kadanoff_kubo_formulae}
	\langle\hat\phi_a(z,\vect{k})\rangle=-\lambda^0_b(\vect{k})\frac{G^R_{ab}(z,\vect{k})-G^R_{ab}(z=0,\vect{k})}{iz}
\end{equation}
where the argument $z=0$ is meant to be taken slightly above the real axis. We still need to find what $G^R(z=0,\vect{k})$ is, so we look at equation \eqref{eqn:ch2:phi_fourier} evaluated at $t=0$. The argument of the integral is the Laplace transform evaluated at $z=0$ (it is again meant to be slightly above the real axis, from $z=i\varepsilon$).
\begin{equation}
	\langle\hat\phi_a(t=0,\vect{k})\rangle=-\int_0^{\infty}\dif t'\ e^{-\varepsilon t'}G^R_{ab}(t',\vect{k})\lambda^0_b(\vect{k})=-G^R_{ab}(z=0,\vect{k})\lambda^0_b(\vect{k})
\end{equation}
From this equation we find that, in the long wavelength limit, $G^R(z=0,\vect{k})$ is minus the static susceptibility $\chi(\vect{k})$ defined above \eqref{eqn:ch2:static_susceptibility}, $G^R(z=0,\vect{k})=-\chi(\vect{k})$.

We can finally compare this result with our earlier expression from linearized hydrodynamics obtaining
\begin{equation}
	-\frac{1}{iz}\bigl(G^R_{ab}(z,\vect{k})+\chi_{ab}(\vect{k})\bigr)\lambda^0_b(\vect{k})=(K^{-1})_{ac}\chi_{cb}\lambda^0_b(\vect{k})
\end{equation}
and we find an expression for the retarded response function matrix
\begin{equation}\label{eqn:ch2:retarded_green_functions}
	G^R_{ab}(z,\vect{k})=-(\delta_{ac}+iz(K^{-1})_{ac})\chi_{cb}
\end{equation}

This function is always analytic in the upper half-plane of complex $z$, hence we can define $G^R(\omega,\vect{k})$ in the whole complex plane as the analytical continuation of $G^R(z,\vect{k})$ from the upper half-plane. $G^R$ has many analytic properties and can be related to other Green functions (advanced and symmetric) \cite{Kovtun:LecturesHydrodynamicFluctuations}, in particular it is possible to show that
\begin{equation}
	-\Im G^R_{aa}(\omega,\vect{k})\geq 0    \qquad \text{for}\quad\omega\geq 0
\end{equation}
This result, when applied to the hydrodynamic response functions, implies that the transport coefficients are all non-negative: $\sigma\geq0$, $\zeta\geq0$ and $\eta\geq0$, without the need to define an entropy current.

\subsection{Variational method}\label{sec:ch2:variational_method}
In the previous section we presented the classical approach to compute the retarded Green functions from hydrodynamics using linear response theory, by introducing sources for the conserved charge densities that follow from equilibrium thermodynamics. This method has the advantage of being very transparent from a physical perspective, however it has a couple of drawbacks, the most important one being that it is not possible to obtain all the correlators associated with the currents \cite{Kovtun:LecturesHydrodynamicFluctuations,Romatschke:RelativisticFluidDynamics}.

There is a second method to obtain the same quantities from a field theory perspective, which introduces sources that couple directly to the covariant form of the currents $T^{\mu\nu}$ and $J^\mu$. Given a generic theory we can always add to it external sources, in particular we will consider a curved metric $g_{\mu\nu}$ and a gauge field $A_\mu$. Then we can formally construct a generating functional $W[A,g]$ whose variations give us all the connected $n$-point functions of the theory.

By definition, varying with respect to the metric or the gauge field gives us the one-point functions associated with the stress-energy tensor and the current
\begin{equation}
	\delta W[A,g]=\int\dif x\sqrt{-g}\left(T^{\mu\nu}\delta g_{\mu\nu}+J^\mu\delta A_\mu\right)
\end{equation}
Furthermore, when the metric and gauge fields variations are associated to diffeomorphism or gauge symmetry, the invariance of $W[A,g]$ under these operations gives us the conservation equations of hydrodynamics (assuming there are no anomalies)
\begin{subequations}\label{eqn:ch2:conservation_equation_sources}
	\begin{align}
		\nabla_\mu T^{\mu\nu}&=F^{\nu\lambda}J_\lambda\\
		\nabla_\mu J^\mu&=0
	\end{align}
\end{subequations}
These are the obvious generalization of \eqref{eqn:ch2:stress_energy_conservation} and \eqref{eqn:ch2:current_conservation} in the presence of external curvature and electromagnetic fields.

Although in hydrodynamics we do not have access to a microscopic generating functional for the full dissipative theory, we can still construct the generators of the retarded Green functions in the presence of background sources as
\begin{equation}\label{eqn:ch2:green_functions_generators}
	\mathcal{T}^{\mu\nu}[A,g]=\sqrt{-g}\langle\hat T^{\mu\nu}\rangle\qquad \mathcal{J}^\mu[A,g]=\sqrt{-g}\langle\hat J^\mu\rangle
\end{equation}
where $\langle\hat T^{\mu\nu}\rangle$ and $\langle\hat J^\mu\rangle$ are the solutions to the equations of hydrodynamics in the presence of non-trivial $A_\mu$ and $g_{\mu\nu}$. Finally, we can obtain the equilibrium correlators at zero sources as
\begin{subequations}\label{eqn:ch2:green_functions_variational}
	\begin{align}
		G^R_{J^\mu J^\nu}(x)&=-\frac{\delta\mathcal{J}^\mu(x)}{\delta A_\nu(0)}\biggr\rvert_{A=h=0}	&	G^R_{T^{\mu\nu} J^\sigma}(x)&=-\frac{\delta\mathcal{T}^{\mu\nu}(x)}{\delta A_\sigma(0)}\biggr\rvert_{A=h=0}\\
		G^R_{J^\sigma T^{\mu\nu}}(x)&=-2\frac{\delta\mathcal{J}^\sigma(x)}{\delta h_{\mu\nu}(0)}\biggr\rvert_{A=h=0}	&	G^R_{T^{\mu\nu} T^{\sigma\rho}}(x)&=-2\frac{\delta\mathcal{T}^{\mu\nu}(x)}{\delta h_{\rho\sigma}(0)}\biggr\rvert_{A=h=0}
	\end{align}
\end{subequations}
where we defined the metric fluctuation as $h_{\mu\nu}=g_{\mu\nu}-\eta_{\mu\nu}$.

The way to compute these quantities is as follows: the constitutive relations in the presence of background sources in the Landau frame, at order one in derivatives, are
\begin{align}\label{eqn:ch2:constitutive_relations_curved_space}
	T^{\mu\nu}&=\epsilon u^\mu u^\nu+P\Delta^{\mu\nu}+\nonumber\\
	&\quad-\eta\Delta^{\mu\alpha}\Delta^{\nu\beta}\left(\nabla_\alpha u_\beta+\nabla_\beta u_\alpha-\frac{2}{d}g_{\alpha\beta}\nabla_\lambda u^\lambda\right)-\zeta\Delta^{\mu\nu}\nabla_\lambda u^\lambda\nonumber\\
	J^\mu&=n u^\mu+\sigma\left(E^\mu-T\Delta^{\mu\nu}\nabla_\nu\frac{\mu}{T}\right)
\end{align}
where $E_\mu=F_{\mu\nu}u^\nu$ is the covariant electric field and appears naturally from the study of the entropy current in the presence of sources. Then we can solve the equations of hydrodynamics in Fourier space by linearizing around global thermodynamic equilibrium with $T=\text{const}$, $\mu=\text{const}$ and $u^\mu=(1,\vect{0})$ in the presence of small fluctuating sources $\delta A_\mu$ and $\delta h_{\mu\nu}$. Inserting the result in the generators \eqref{eqn:ch2:green_functions_generators} and taking the functional derivatives \eqref{eqn:ch2:green_functions_variational} gives access to all the Green functions of the theory.

The Green functions obtained by this approach might differ from the ones obtain via Martin-Kadanoff by contact terms that appear because of the $\sqrt{-g}$ term in the generators \eqref{eqn:ch2:green_functions_generators}, see \cite{Herzog:LecturesHolographicSuperfluidity}.

\subsection{Discrete symmetries: Onsager relations}\label{sec:ch2:onsager_relations}
The retarded functions have to be consistent with the symmetries of the theory, in particular time-reversal turns out to impose powerful constraints on the transport coefficients. Consider a Hermitian field $\hat\phi_a(t,\vect{x})$ that transform under time reversal as $\hat\Theta\hat\phi_a(t,\vect{x})\hat\Theta^{-1}=\eta_a\hat\phi_a(-t,\vect{x})$, where $\hat\Theta$ is the anti-unitary time-reversal operator and $\eta_a=\pm 1$ is the eigenvalue of $\hat\phi_a$. If the microscopic system is time-reversal invariant, such that $[\hat H,\hat\Theta]=0$, then the retarded response functions must obey
\begin{equation}
	G^R_{ab}(t,\vect{x})=G^R_{ba}(t,-\vect{x})\eta_a\eta_b
\end{equation}

On the other hand, if time-reversal is not a symmetry of the microscopic system, for example if there is an external magnetic field $B$ that breaks the symmetry, the Hamiltonian satisfies $\hat\Theta\hat H(B)\hat\Theta^{-1}=\hat H(-B)$. Then the above equation is modified to
\begin{equation}\label{eqn:ch2:onsager_relations_general}
	G^R_{ab}(\omega,\vect{k};B)=\eta_a\eta_b G^R_{ba}(\omega,-\vect{k};-B)
\end{equation}
This equation is at the core of the so-called Onsager reciprocal relations \cite{Onsager:ReciprocalRelationsIrreversible,Onsager:ReciprocalRelationsIrreversiblea}, although older works use an approach based on thermodynamics and the relations between fluxes and forces \cite{Callen:ThermodynamicsIntroductionThermostatistics,Landau:StatisticalPhysicsVolume}.

This condition on $G^R$ imposed by time-reversal is not automatically satisfied by linearized hydrodynamics, instead it should be interpreted as a constraint on the possible terms in the constitutive equations: in the limit $\vect{k}\rightarrow0$, at $\omega=0$, this condition implies that $\chi_T=0$ in first-order hydrodynamics \eqref{eqn:ch2:dissipative_current}. This is another way to constraint the transport coefficients without resorting to the entropy current.

From \eqref{eqn:ch2:retarded_green_functions} and \eqref{eqn:ch2:onsager_relations_general} it follows that the matrix $M_{ab}$ must obey the following condition
\begin{equation}
	\chi(B)SM^T(-\vect{k};-B)=M(\vect{k};B)\chi(B)S
\end{equation}
where $S=\text{diag}\left(\eta_1,\eta_2,\dots\right)$ is the matrix of the time-reversal eigenvalues and the susceptibilities obey
\begin{equation}
	S\chi(B)S=\chi^T(-B)
\end{equation}
where $T$ stands for transpose matrix.

One could also apply the same procedure to other discrete symmetries, such as parity, however in these other cases the procedure is not as useful, since it is usually enough to keep track of the parity breaking parameters of the theory \cite{Ammon:ChiralHydrodynamicsStrong}. This is related to the fact that only time-reversal is associated with an anti-unitary operator.

\subsection{Thermoelectric transport}\label{sec:ch2:thermoelectric_transport}
In this thesis we are often interested in finding the thermoelectric transport coefficients in a linear-response-theory framework. Usually the charge and heat currents couple respectively with the electric field and a temperature gradient, but there are also thermoelectric effects where the sources create mixed responses described by the full thermoelectric matrix. In linear response we can define \cite{Callen:ThermodynamicsIntroductionThermostatistics}
\begin{equation}\label{eqn:ch2:thermoelectric_matrix}
	\begin{pmatrix}
		\delta J_i \\
		\delta Q_i
	\end{pmatrix}
	=
	\begin{pmatrix}
		\sigma_{ij} &   \alpha_{ij} \\
		T \bar{\alpha}_{ij} &   \bar{\kappa}_{ij}
	\end{pmatrix}
	\begin{pmatrix}
		\delta \mathbb{E}_j \\
		-\partial_j \delta T
	\end{pmatrix}
\end{equation}
where $\sigma$ is the electrical conductivity tensor, $\bar{\kappa}$ is the thermal conductivity tensor, $\alpha$ and $\bar{\alpha}$ are the thermoelectric tensors.

The $\sigma$ that appears in the constitutive relations is not to be confused with $\sigma_{ij}$: $\sigma$ is a transport coefficient, cannot be determined by hydrodynamics, and depends on the microscopic theory, on the other hand $\sigma_{ij}$ describes the macroscopic response of the current with respect to an external electric field and will contain a contribution due to $\sigma$.

It should be noted that $\bar{\kappa}$ is not the usual thermal conductivity measured in experiments: in the laboratory measurements of the thermal conductivity are usually performed with the boundary condition $\vect{J}=0$, while in the above definition the boundary condition is $\vect{E}=0$. The relation between $\kappa_{ij}$ and $\bar{\kappa}_{ij}$ is
\begin{equation}\label{eqn:ch2:thermal_conductivity_experiment}
	\kappa_{ij}=\bar{\kappa}_{ij}-T\bar{\alpha}_{ik}\sigma_{kl}^{-1}\alpha_{lj}
\end{equation}
Transport coefficients have to obey Onsager relations, this implies that the matrices $\alpha_{ij}$ and $\bar{\alpha}_{ij}$ are related by $\alpha_{ij}(B)=\bar{\alpha}_{ij}(B)$, with $B$ an external magnetic field that breaks time-reversal invariance \cite{Callen:ThermodynamicsIntroductionThermostatistics,Hartnoll:TheoryNernstEffect}.

In linear response theory we would like to perturb the system with a small electric field and temperature gradient, from there, find how the electric and heat currents react upon these perturbations. To do so, we have to first figure out the correct sources: for example the electric current does not couple to the electric field in the Hamiltonian, but instead it couples to the four-potential $A^{\mu}$.

A disturbance in temperature couples to the Hamiltonian, and since we are working in the grand canonical ensemble we find
\begin{equation}
	\delta H=-\int\dif^d x\left(\frac{\delta T(t,\vect{x})}{T}(\epsilon(t,\vect{x})-\mu n(t,\vect{x}))+\delta A^{\mu}(t,\vect{x})J_{\mu}(t,\vect{x})\right)
\end{equation}
We can fix the gauge by choosing the electric field to be minus the gradient of $A^0$, $\mathbb{E}_i=-\partial_iA^0$, so we can ignore the $\delta A^i J_i$ terms and focus on $\delta A^0 n$
\begin{equation}
	\delta H=-\int\dif^d x\left(\frac{\delta T(t,\vect{x})}{T}(\epsilon(t,\vect{x})-\mu n(t,\vect{x}))+\delta A^{0}(t,\vect{x})n(t,\vect{x})\right)
\end{equation}

In classic thermodynamics the heat current is defined as \cite{Callen:ThermodynamicsIntroductionThermostatistics} $\vect{J}_Q=T \vect{J}_S = \vect{J}_U - \mu \vect{J}_N$ where $\vect{J}_S$ is the entropy current while $\vect{J}_U$ and $\vect{J}_N$ are respectively the energy and charge current. In the relativistic case $\epsilon-\mu n$ is the time component of the heat current four-vector. Hence, the relativistic generalization to the heat current, in linear response theory, is simply
\begin{equation}\label{eqn_ch:2:canonical_heat_current}
	\delta Q^{i}=\delta T^{0 i}-\mu \delta J^{i}
\end{equation}

It is also possible to arrive at the same result from a variational approach perspective, by studying temperature fluctuations as changes in the Euclidean time component of the metric. This is the method followed in \cite{Hartnoll:LecturesHolographicMethods,Herzog:LecturesHolographicSuperfluidity} to arrive at the form of the canonical heat current, without resorting to thermodynamics.

The Kubo formulae for the conductivities in terms of the retarded Green functions are
\begin{subequations}\label{eqn:ch2:thermoelectric_kubo_formulae}
	\begin{align}
		\sigma_{ij}(\omega)&=-\frac{G^R_{J_iJ_j}(\omega)-G^R_{J_iJ_j}(0)}{i\omega}\\
		\alpha_{ij}(\omega)&=-\frac{G^R_{J_iQ_j}(\omega)-G^R_{J_iQ_j}(0)}{i\omega T}\\
		\bar{\alpha}_{ij}(\omega)&=-\frac{G^R_{Q_iJ_j}(\omega)-G^R_{Q_iJ_j}(0)}{i\omega T}\\
		\bar{\kappa}_{ij}(\omega)&=-\frac{G^R_{Q_iQ_j}(\omega)-G^R_{Q_iQ_j}(0)}{i\omega T}
	\end{align}
\end{subequations}
Then, from a practical perspective, if we are interested in the conductivities we can simply perturb the system with linear sources
\begin{equation}
	T\rightarrow T+\delta T-T x^i\delta\zeta_i\qquad\qquad F^{0i}\rightarrow\delta\mathbb{E}^i
\end{equation}
where $\zeta_i=\partial_iT/T$ is the heat source, and solve the equations of hydrodynamics in Laplace-Fourier space at $\vect{k}=0$ to arrive at \eqref{eqn:ch2:martin_kadanoff_kubo_formulae}, from which we can read off the conductivities \eqref{eqn:ch2:thermoelectric_kubo_formulae}.

For future reference, we report here the optical conductivities in order-one hydrodynamics without background magnetic field. These are
\begin{subequations}\label{eqn:ch2:thermoelectric_conductivities}
\begin{align}
	\sigma(\omega)&=\sigma+\frac{in^2}{\omega(\epsilon+P)}\\
	\alpha(\omega)&=-\frac{\mu}{T}\sigma+\frac{ins}{\omega(\epsilon+P)}\\
	\bar\kappa(\omega)&=\frac{\mu^2}{T}\sigma+\frac{is^2T}{\omega(\epsilon+P)}
\end{align}	
\end{subequations}
and Onsager relations constraint $\bar\alpha=\alpha$. We notice that the conductivities diverge as $\omega$ goes to zero, in particular the imaginary part has a pole, hence Kramers–Kronig relations imply that the real part has a delta function at $\omega=0$ \cite{Hartnoll:HolographicQuantumMatter}. This is because momentum is a conserved operator, which means that in the presence of a background constant electric field a charged fluid accelerates without bound. Finally, the Green functions of conserved quantities obey certain Ward identities \cite{Herzog:LecturesHolographicSuperfluidity,Kovtun:LecturesHydrodynamicFluctuations,Bailin:IntroductionGaugeField} that are reflected in relations between the conductivities. In this case it is easy to check that
\begin{subequations}\label{eqn:ch2:ward_identities_conductivities}
\begin{align}
	\alpha(\omega)&=-\frac{\mu}{T}\sigma(\omega)+\frac{in}{T\omega}\\
	\bar\kappa(\omega)&=\frac{\mu^2}{T}\sigma(\omega)+\frac{i\left(\epsilon+P-2\mu n\right)}{T\omega}
\end{align}	
\end{subequations}

\subsection{Modes and linear stability}\label{sec:ch2:modes}
The first and easiest thing one can wish to compute when he is handed a theory of hydrodynamics are the modes. These give information about how small collective fluctuations near thermal equilibrium propagate inside the fluid, and furthermore it gives important clues about the stability of the theory.

From a formal perspective, the modes $\omega(\vect{k})$ are eigenfrequencies of the linearized equations of motion. Starting from the conservation laws and the constitutive relations, we can consider fluctuations on a global thermodynamic equilibrium state of the form
\begin{equation}
	\phi_a(t,\vect{x})=e^{i\omega t-i\vect{k}\cdot\vect{x}}\phi_a
\end{equation}
The equations of hydrodynamics can be rewritten in Fourier space as
\begin{equation}
	M_{ab}(\vect{k})\phi_b=\omega\phi_a
\end{equation}
and host non-trivial solutions only if $\text{det}(M-\mathds{1}\omega)=0$, from which we obtain the modes of the system.

The requirement of causality, when applied to the analytic structure of the modes for complex frequency $\omega$, tells us that the imaginary part of the modes must be non-positive. This is related to the discussion on the analyticity of the retarded correlators $G^R$ discussed in Section~\ref{sec:ch2:martin_kadanoff}, because the poles of the Green functions are the modes themselves (this can be seen by the fact that to obtain the Green functions we need to invert the hydrodynamic matrix, which in turn leaves us with the determinant of $M-\mathds{1}\omega$ in the denominator), and the poles must lay in the lower-half of the complex $\omega$ plane.

One final comment on the terminology: by definitions the hydrodynamic modes should be such that $\omega(\vect{k}=0)=0$, however when there are impurities, magnetic fields or other slow variables that enter the hydrodynamic regime this might not be true \cite{Grozdanov:HolographyHydrodynamicsWeakly}. It is thus common to call a mode \emph{hydrodynamic} if it is the lowest lying mode (the closest one to the real axis at $\vect{k}=0$), irrespective of its actual value, and provided that the non-hydrodynamic modes are deep enough in the bottom-half of the complex plane.

\chapter[Charge Density Waves hydrodynamics]{Hydrodynamic and holographic Charge Density Waves}\label{chapter:charge_density_waves}
\epigraph{``Reality exists in the human mind, and nowhere else.''}{George Orwell, \emph{1984}}

\section{Introduction}
As we have already discussed in the introduction, High-Temperature superconductors, of which cuprates are typical examples, are believed to host strongly coupled phases of matter that do not allow a microscopic description in terms of quasiparticles, see \cite{Phillips:StrangerMetals,Greene:StrangeMetalState} for some reviews. This is usually argued based on the peculiar features of the cuprates, such as the electric resistivity that scales linear in $T$ on a very broad range of temperature (from as low as \qty{10}{\K}, up to \qty{300}{\K}), implying that the standard description in terms of phonons and Fermi-liquid cannot be accurate \cite{Amoretti:HydrodynamicalDescriptionMagnetotransport,Delacretaz:BadMetalsFluctuating,Hartnoll:TheoryNernstEffect,Hartnoll:PlanckianDissipationMetals,Lucas:ResistivityBoundHydrodynamic}. On the other hand holography, which is not based on a description in terms of quasiparticles, together with the existence of a quantum critical point, might be able to capture the Planckian transport many-body dynamics of these strongly coupled systems \cite{Balm:LinearResistivityOptical,Hartnoll:LecturesHolographicMethods,Hartnoll:StrangeMetallicHolography,Hartnoll:TheoryUniversalIncoherent}.

\begin{figure}
	\centering
	\includegraphics[width=0.6\textwidth]{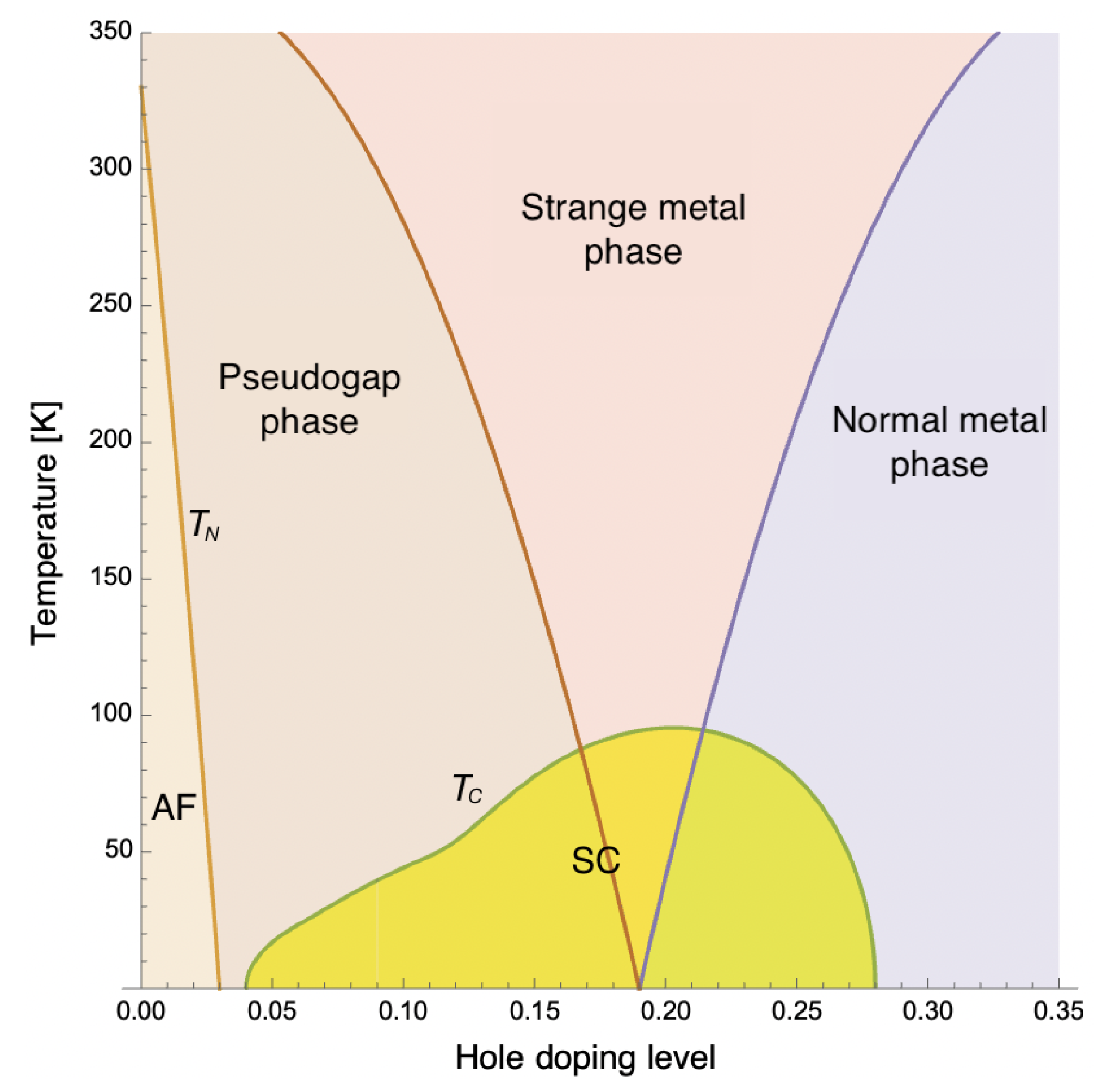}
	\caption{Typical phase diagram of cuprates, taken from \cite{Mirarchi:StrangeMetalBehaviorCuprates}. The strange metal phase above the superconducting dome is non-Fermi liquid and presents the typical linear-in-$T$ resistivity. Close to optimal doping there is often a coexistence of charge density wave and superconducting phase.}
	\label{fig:ch3:cuprate_phase_diagram}
\end{figure}

\begin{figure}
	\centering
	\includegraphics[width=0.8\textwidth]{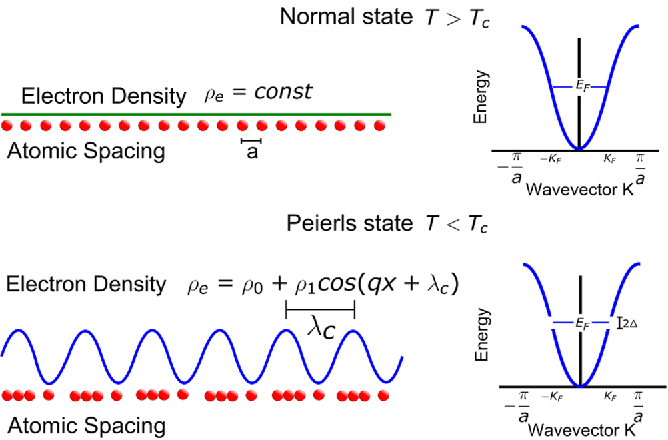}
	\caption{Figure taken from \cite{Goli:ChargeDensityWaves}. Without interactions the charge density is uniform in equilibrium. The presence of electron-phonon interactions induces a charge modulation order, while thermal fluctuations work to destroy this effect.}
	\label{fig:ch3:charge_density_waves}
\end{figure}

One common feature of cuprates is the presence of Charge Density Wave order in the phase diagram for certain values of the parameters. Charge Density Waves are the collective behaviour of charges that can sometimes arise in the presence of an ionic lattice \cite{Gruner:DynamicsChargedensityWaves}. In metals usually the charge density is uniform throughout the system, however in the presence of electron-phonon interactions the lattice is distorted to a new periodicity $\lambda_c$, and a gap opens up that leads to Charge Density Wave modulation of the electron fluid, see Figure~\ref{fig:ch3:charge_density_waves}. The long-lived CDW collective modes due to the spontaneous breaking of translations in the electron fluids, because they are (almost) gapless, can be important for the low-energy dynamics in the hydrodynamic regime, on par with other conserved quantities \cite{Chaikin:PrinciplesCondensedMatter}. Wigner crystals have a different physical origin, but nonetheless they can give rise to a low-energy dynamics which is very similar. The main difference is that Wigner crystals always break translations in all spatial directions, while CDW can break the symmetry only in certain directions, producing smectic phases. In this chapter, however, we will only focus on the case in which translations are broken in all directions.

Finally, one last ingredient we need to consider is that condensed matter experiments are usually performed in the presence of an external magnetic field. This is both to suppress superconductivity and enhance the strongly-coupled strange metal regimes, but also to study other transport coefficients such as the Hall conductivities.

For these reasons, many holographic models in which translation symmetry is broken either spontaneously or pseudo-spontaneously have been intensively studied to better understand the low energy physics of these systems and, at the same time, holography itself. This has led to many different approaches to translations breaking, from massive gravity \cite{Amoretti:AnalyticDCThermoelectric,Amoretti:MagnetotransportMomentumDissipating,Amoretti:MagnetotransportMomentumDissipating,Amoretti:ThermoelectricTransportGaugea,Baggioli:HolographicAxionModel,Baggioli:HolographicPolaronsMetalInsulator,Baggioli:MagnetophononsTypeBGoldstones,Ammon:UnifiedDescriptionTranslational,Alberte:BlackHoleElasticity,Alberte:HolographicPhonons}, to spatially modulated charge density \cite{Donos:SpatiallyModulatedInstabilities,Andrade:PinningLongitudinalPhonons,Andrade:PhaseRelaxationPattern} and Q-lattices \cite{Amoretti:HolographicPerspectivePhonons,Amoretti:EffectiveHolographicTheory,Amoretti:DCResistivityQuantum,Amoretti:UniversalRelaxationHolographic,Donos:IncoherentTransportPhases,Donos:IncoherentHydrodynamicsDensity,Amoretti:DiffusionUniversalRelaxation,Amoretti:GaplessGappedHolographic,Amoretti:HowConstructHolographic,Donos:IncoherentHydrodynamicsDensitya}.

At the same time, holography sparkled a new interest in the low energy EFT and hydrodynamic description of these phases of matter \cite{Amoretti:HydrodynamicMagnetotransportCharge,Delacretaz:BadMetalsFluctuating,Delacretaz:TheoryCollectiveMagnetophonon,Armas:ApproximateSymmetriesPseudoGoldstones,Armas:ViscoelasticHydrodynamicsHolography,Armas:HydrodynamicsChargeDensity,Armas:HydrodynamicsPlasticDeformations}. However, it is now known that in the context of broken-translation symmetry, holography and hydrodynamics do not naively match. This is because the holographic Q-lattice description is metastable, namely the vacuum does not minimize the free energy, and thus the hydrodynamic description has to be modified to include the effect of the so-called lattice pressure $P_l$ \cite{Armas:ViscoelasticHydrodynamicsHolography,Armas:HydrodynamicsChargeDensity,Ammon:LongitudinalSoundDiffusion,Ammon:HydrodynamicDescriptionHolographic}. In stable systems the lattice pressure vanishes and its thermodynamic derivatives $\partial_TP_l$, $\partial_\mu P_l$ can be absorbed by redefining certain transport coefficients.

In this chapter we will consider a model of CDW, both from a hydrodynamic and holographic perspective, in the presence of a strong magnetic field which is order zero in derivatives $B\sim\mathcal{O}(1)$. To match the two approaches, we need to also include the lattice pressure in the thermodynamic of our hydrodynamic model. We consider both the case in which translation symmetry is broken spontaneously, thus leading to a gapless Goldstone mode, and pseudo-spontaneously, i.e. in which the Goldstone acquires a parametrically-small mass. We compute the AC conductivities analytically from hydrodynamics and numerically from holography; subsequently, using a method developed in \cite{Amoretti:MagnetothermalTransportImplies} based on Ward identities, we express the AC results in terms of the DC values of the conductivities, which can be computed analytically from holography in terms of horizon data. This allows us to match the correlators obtained from the two approaches to high precision, and we observe a very good agreement between the two descriptions.

In Section~\ref{sec:ch3:symmetry_breaking} we discuss general properties of systems in which translations are broken by a scalar field. Then, in Section~\ref{sec:ch3:spontaneous_case} and~\ref{sec:ch3:pseudo_spontaneous_case}, we develop the hydrodynamic theory for the cases of spontaneous and pseudo-spontaneous breaking of translations respectively. Finally, in Section~\ref{sec:ch3:holographic_model} we briefly review our holographic model and match the conductivities obtained from numerics to the analytic hydrodynamic ones.

\section{Breaking the translation symmetry and Ward identities}\label{sec:ch3:symmetry_breaking}
The equations of motion for the hydrodynamic model are still the conservation equations for the stress-energy tensor and a $\mathrm{U(1)}$ current, however this time we must also include the presence of an extra scalar due to the symmetry breaking
\begin{subequations}\label{eqn:ch3:one_point_ward_identities}
	\begin{align}
		\partial_\mu\langle T^{\mu\nu}\rangle&=F^{\nu\lambda}\langle J_\lambda\rangle-\partial^\nu\Phi^I\langle O_I\rangle\\
		\partial_\mu\langle J^\mu\rangle&=0
	\end{align}
\end{subequations}
where $O_I$, with $I=1,\dots,d$, is a set of scalar operators that break spatial translations and can be identified with the Goldstone modes for the broken symmetry \cite{Armas:ViscoelasticHydrodynamicsHolography}, while $\Phi^I$ are their conjugate sources. The scalars $O_I$ are dynamical, which means that we need other equations of motion that will take the form of Josephson-like relations. However, because these equations are not constrained by symmetries and are obtained order by order in derivative expansion, they do not lead to Ward identities.

To achieve the spontaneous breaking of translations we consider scalars operators that take a non-zero vev proportional to the spatial coordinates in the background $\langle O_I\rangle\propto x^i\delta_{iI}$, while for the explicit and pseudo-spontaneous case instead we take the source fields to be proportional to the spatial coordinates $\Phi_I=\varphi x^i\delta_{iI}$ where $\varphi$ is some parameter, such that the spacetime derivatives of $\Phi_I$ are constants.

Tuning the value of $\varphi$ allows us to better identify two different regimes: the pseudo-spontaneous case, which is of interest for this chapter, is when $\varphi\ll\abs{\partial_i\langle O_I\rangle}$. Then the Goldstone acquires a small mass, denoted by $\omega_0^2$ and called pinning frequency, and still behaves like a pseudo-Goldstone boson, much like the pions in QCD. The second scenario is the explicit case $\varphi\geq\abs{\partial_iO_I}$ which leads to non-Goldstone dynamics \cite{Andrade:SimpleHolographicModel}.

This specific construction of translation breaking can be realized in systems that have spatial translation invariance, and in which the scalar operators have a constant internal shift symmetry $O_I\rightarrow O_I+c_I$. Then, breaking the translation symmetries leads to a diagonal subgroup of the two symmetries that remains unbroken, which in turns ensures that the equations of motion remain homogeneous. This is indeed the minimal scenario that allows for such symmetry breaking, according to the classification of phases of matter in \cite{Nicolis:ZoologyCondensedMatter}. The same setup has indeed been used for many different purposes, such as studying lattice phonons \cite{Nicolis:MutualInteractionsPhonons}, hydrodynamics \cite{Delacretaz:BadMetalsFluctuating,Delacretaz:TheoryHydrodynamicTransport} and holographic \cite{Amoretti:EffectiveHolographicTheory,Amoretti:DCResistivityQuantum,Amoretti:UniversalRelaxationHolographic,Alberte:HolographicPhonons,Alberte:BlackHoleElasticity,Baggioli:MagnetophononsTypeBGoldstones,Baggioli:HolographicAxionModel,Andrade:PinningLongitudinalPhonons,Andrade:PhaseRelaxationPattern,Ammon:HydrodynamicDescriptionHolographic} effective field theory descriptions of charge density waves states.

\begin{figure}
	\centering
	\includegraphics[width=\textwidth]{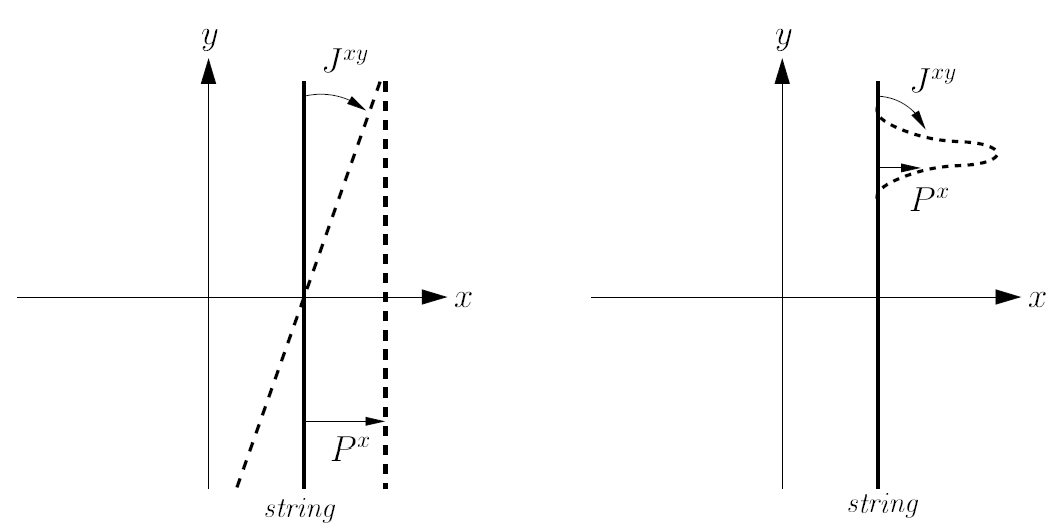}
	\caption{Figure taken from \cite{Low:SpontaneouslyBrokenSpacetime}. The presence of a string in the ground state breaks the 3-dimensional Poincaré group to the 2-dimensional one. \textbf{Left}: Clearly, global rotations and global translations are different operation that change the vacuum state. \textbf{Right:} Nonetheless, local translations can be undone with local rotations, and vice versa.}
	\label{fig:ch3:spacetime_symmetries}
\end{figure}

Following \cite{Nicolis:ZoologyCondensedMatter} we can understand this construction from an EFT point of view. Consider a microscopic theory that is symmetric under the full Poincaré group, where $P_0$, $P_i$, $\mathcal{J}_i$ and $K_i$ are respectively the generators for time and spatial translations, rotations and boosts, plus eventually other internal symmetries $\mathcal{Q}$. Now, we assume that there is a set of generators $\bar{P}_0$, $\bar{P}_i$ and $\bar{\mathcal{J}}_i$ that leave the ground state of the low-energy EFT invariant, and that are unbroken, so that the system looks homogenous. These generators are not necessarily the same of the microscopic theory, but they clearly obey the same algebra, and in general will be a combination of microscopic and internal symmetry generators. In the case of interest, for example, we can take $\bar{P}_0=P_0$, $\bar{P}_i=P_i+\mathcal{Q}_i$ and $\bar{\mathcal{J}}_i=J_i+\tilde{\mathcal{Q}}_i$ and by the requirement that $\mathcal{Q}_i$ and $\tilde{\mathcal{Q}}_i$ commute with Poincaré we find that
\begin{equation}
	\left[\tilde{\mathcal{Q}}_i,\tilde{\mathcal{Q}}_j\right]=i\varepsilon_{ijk}\tilde{\mathcal{Q}}^k\qquad\left[\mathcal{Q}_i,\mathcal{Q}_j\right]=0\qquad\left[\tilde{\mathcal{Q}}_i,\mathcal{Q}_j\right]=i\varepsilon_{ijk}\mathcal{Q}^k
\end{equation}
This is the algebra of three-dimensional Euclidean group $\mathrm{ISO(3)}$, i.e. the $\tilde{\mathcal{Q}}_i$ generate an internal $\mathrm{SO(3)}$ rotation symmetry, while the $\mathcal{Q}_i$ are the generators of internal translations.

This setup has nine broken generators (the $\mathcal{Q}_i$, $\tilde{\mathcal{Q}}_i$ and the boosts), but hosts only three Goldstone modes. This happens because in the presence of spacetime symmetries the counting of Goldstone bosons can be subtle, compared to the case with spontaneously broken internal symmetries \cite{Low:SpontaneouslyBrokenSpacetime,Watanabe:RedundanciesNambuGoldstoneBosons,Watanabe:NumberNambuGoldstoneBosons,Watanabe:EffectiveLagrangianNonrelativistic}, see Figure~\ref{fig:ch3:spacetime_symmetries} and the reviews \cite{Watanabe:CountingRulesNambuGoldstone,Beekman:IntroductionSpontaneousSymmetry}. In particular the Goldstone fields can acquire a gap \cite{Nicolis:RelativisticNonrelativisticGoldstone,Nicolis:SpontaneousSymmetryProbing,Kapustin:RemarksNonrelativisticGoldstone,Watanabe:MassiveNambuGoldstoneBosons}, so they are irrelevant for the low-energy physics, or can be removed from the spectrum using the inverse-Higgs constraint \cite{Ivanov:InverseHiggsEffect,Nicolis:MoreGappedGoldstones,Brauner:SpontaneousBreakingSpacetime}. This happens whenever the commutator between the generators of the unbroken translation of the effective theory $\bar{P}$ and the broken generators $\mathcal{Q}$ contains another multiplet of broken generators $\mathcal{Q}'$, namely
\begin{equation}
	\left[\bar{P},\mathcal{Q}\right]\supset\mathcal{Q}'
\end{equation}
When this condition is satisfied, we can then express the Goldstone modes that arise from the breaking of $\mathcal{Q}$ in terms of derivatives of those arising from the breaking of $\mathcal{Q}'$, thus reducing the total number of independent Goldstone fields. In our specific example, we have that
\begin{equation}
	\left[\bar{P}_0,K_i\right]=-i\left(\bar{P}_i-\mathcal{Q}_i\right)\qquad\text{and}\qquad\left[\bar{P}_i,\tilde{\mathcal{Q}}_j\right]=i\varepsilon_{ijk}\mathcal{Q}^k
\end{equation}
which gives six inverse-Higgs constraints and reduces the number of Goldstone bosons from nine to three. The minimal implementation of this symmetry breaking pattern describes ordinary solids, and amounts to having a triplet of scalar fields $O_I$ that transforms under the internal $\mathrm{SO(3)}$ and shifts under the internal symmetry $Q$ as $O_I\rightarrow O_I+c_I$, which can be interpreted as solid volume elements \cite{Dubovsky:NullEnergyCondition,Nicolis:RelativisticFluidsSuperfluids,Son:EffectiveLagrangianTopological}. Then, taking the expectation values $\langle O_I(x)\rangle=x_I$, realizes the desired symmetry breaking scenario and the fluctuating fields $\delta O_I=O_I-x_I$ describe the Goldstone bosons.

We now set $d=2$ and compute the Ward identities for the two-point functions of this system. We take the spatial derivatives of the scalar sources to be constants $\partial_i\Phi^I=\varphi\delta^I_i$ and expand the various quantities in Fourier modes. Then, at zero wavevector $\vect{k}=0$, we find
\begin{subequations}\label{eqn:ch3:ward_identities}
	\begin{align}
		i\omega\langle Q^iQ^j\rangle&=-\left(i\omega\mu\delta^i_k-F^i_{\ k}\right)\langle Q^kJ^j\rangle+\varphi\langle Q^iO^J\rangle\delta^j_J-i\omega\left(\chi_{\pi\pi}-\mu n\right)\delta^{ij}\\
		i\omega\langle Q^i J^j\rangle&=-\left(i\omega\mu\delta^i_k-F^i_{\ k}\right)\langle J^kJ^j\rangle+\varphi\langle J^iO^J\rangle\delta^j_J-i\omega n\delta^{ij}\\
		i\omega\langle Q^i O^J\rangle&=-\left(i\omega\mu\delta^i_k-F^i_{\ k}\right)\langle J^kO^J\rangle-\varphi\langle O^IO^J\rangle\delta^i_I+\delta^{iJ}
	\end{align}
\end{subequations}
where $Q^i=T^{i0}-\mu J^i$ is the canonical heat current and $\chi_{\pi\pi}$ is the momentum susceptibility.

There is a remarkable ladder structure in these Ward identities, in particular we can write the heat current correlators in terms of the scalars and $\mathrm{U(1)}$ current correlators. This means that our system has actually only 3 independent Green functions, which we take to be $\langle O^IO^J\rangle, \langle J^iO^J\rangle$ and $\langle J^iJ^j\rangle$, and the other ones can be obtained from these using the above Ward identities. These expressions, following \cite{Amoretti:MagnetothermalTransportImplies}, will allow us to fix the hydrodynamic transport coefficients in terms of the DC values of the conductivities, which are related to the Green functions via \eqref{eqn:ch2:thermoelectric_kubo_formulae}.

\section{Spontaneous case}\label{sec:ch3:spontaneous_case}
Here, we will focus on the simpler case of the spontaneous breaking of translation symmetry, therefore we set $\varphi=0$, and we take the fluctuations of the scalar fields $O_I$ to be the Goldstone modes associated to the breaking of spacetime translations.

\subsection{Constitutive relations}
Our construction is based on the geometric formalism developed in \cite{Armas:ViscoelasticHydrodynamicsHolography,Armas:HydrodynamicsChargeDensity}. The idea is to consider $O_I$ to be crystal fields, namely they represent coordinates of the unbroken $\mathrm{ISO(2)}$ crystal manifold\footnote{In the context of CDW and Wigner crystal, the crystal is not the standard condensed matter lattice made of positive ions, but an electronic crystal.}, such that we have both spacetime coordinates $x^\mu$ and crystal coordinates $O^I$. In this work we take $I=1,\dots,d$ with $d=2$, so that translations are broken in all directions, while systems with $I=1,\dots,m$ with $m<d$ describe smectic phases. In the context of CDW and quasicrystals, the $O_I$ field (or, better, its fluctuation), is called \emph{phason} \cite{Overhauser:ObservabilityChargeDensityWaves}.

We define the pullback from the $(2+1)$-dimensional spacetime to the 2-dimensional crystal manifold $e_\mu^I=\partial_\mu O^I$. Subsequently, we pullback the inverse spacetime metric on the manifold by defining $h^{IJ}=g^{\mu\nu}e^I_\mu e^J_\nu$, which acts as an inverse metric on the crystal manifold, allowing us to raise crystal indices. Similarly, we can also define the inverse matrix $h_{IJ}=(h^{IJ})^{-1}$ that can be used to lower the crystal indices. We take the vielbeins to be order-zero quantities in the derivative counting of our hydrodynamic formulation $e_\mu^I\sim\mathcal{O}(1)$.

We further define the non-linear strain tensor $u_{IJ}=(h_{IJ}-\mathds{h}_{IJ})/2$ that parametrizes the distortion of crystal manifold with respect to some reference rest configuration that we denote by $\mathds{h}_{IJ}$. We assume that the reference metric $\mathds{h}_{IJ}$ is fixed and time-independent, thus that we are describing elastic systems \cite{Armas:ViscoelasticHydrodynamicsHolography,Armas:HydrodynamicsChargeDensity}, while relaxing this assumption leads to the more general class of plastic materials which can be permanently deformed \cite{Armas:HydrodynamicsPlasticDeformations}. The standard choice here is to consider a flat crystal metric, hence we choose the reference configuration $\mathds{h}_{IJ}=\delta_{IJ}/\alpha^2$ which is isotropic and homogeneous. Here $\alpha$ represents the inverse crystal lattice spacing, however by redefining $O^I\rightarrow \alpha O^I$ we can always set $\alpha=1$ without loss of generality.

Until now the construction is completely arbitrary and valid for generic crystal metric. From this point, we specialize to the hydrodynamic regime, in which we require that the strain is small so that we can define an expansion in power of the strain tensor. Then the free energy $F$, that describes the equilibrium (ideal fluid) part of the system, is given by the integral of the total pressure $P=P(T,\mu,B,h_{IJ})$, which is itself the sum of the contribution of the fluid pressure $P_f$ and the crystal. Namely, we have $F=\int\dif^2\sqrt{-g}P$ where we expand the pressure up to quadratic terms in strain
\begin{equation}\label{eqn:ch3:spontaneous_free_energy}
	P=P_f-mB+P_l\left(u^I_I+u^{IJ}u_{IJ}\right)-\frac{K}{2}\left(u^I_I\right)^2-G\left(u^{IJ}u_{IJ}-\frac12\left(u^I_I\right)^2\right)+\mathcal{O}(u^3)
\end{equation}
Here $m=\partial P/\partial B$ is the magnetization density conjugate to the order-zero magnetic field $B\sim\mathcal{O}(1)$, $P_l$ is the lattice pressure, while $K$ and $G$ are respectively the bulk and shear modulus of the crystal \cite{Armas:ViscoelasticHydrodynamicsHolography}. The presence of the lattice pressure, in particular, is what makes this expression different from older results, and signals that the system is in a metastable state. Indeed, in classical elasticity theory the free energy is minimized with respect to the strain, which requires setting the linear term $P_l=0$ in equilibrium \cite{Landau:TheoryElasticityVolume}.

Given the above thermodynamic free energy, we can define the total energy, charge and entropy densities from the pressure $P$ in the usual way, according to the thermodynamic relations
\begin{equation}
	\dif P=s\dif T+n\dif\mu+m\dif B+\frac{1}{2}r_{IJ}\dif h^{IJ}\qquad\epsilon+P=sT+\mu n
\end{equation}
We also defined the elastic stress tensor $r_{IJ}$ as a thermodynamic derivative of $P$. Subsequently, we can further separate this in the contributions from the fluid, indicated by the subscript $f$, and from the lattice, identified with $l$. In particular, we assume that the fluid and lattice have a well-defined thermodynamics by themselves, and thus we can write $\dif P_f=s_f\dif T+n_f\dif\mu+m\dif B$ and equivalently $\dif P_l=s_f\dif T+n_f\dif\mu$, which also lead to the integrated forms $\epsilon_f+P_f=s_fT+n_f\mu$ and $\epsilon_l+P_l=s_lT+n_l\mu$.

With the free energy \eqref{eqn:ch3:spontaneous_free_energy} we can finally express the constitutive relations for the stress energy tensor and the $\mathrm{U(1)}$ current order-by-order in derivatives. To do this we define the projector orthogonal to the fluid velocity $\Delta^{\mu\nu}=g^{\mu\nu}+u^\mu u^\nu$, its pulled-back form $\Delta^{I\mu}=\Delta^{\mu\nu}e^I_\nu$ and, as usual, the electric field as $E_\mu=F_{\mu\nu}u^\nu$. The constitutive relations, up to order one in derivatives for an isotropic fluid in the Landau frame, take the form \cite{Armas:HydrodynamicsChargeDensity}
\begin{subequations}\label{eqn:ch3:spontaenous_constitutive_relations}
\begin{align}
	T^{\mu\nu}&=\epsilon u^\mu u^\nu+P\Delta^{\mu\nu}-r_{IJ}e^{I\mu}e^{J\nu}-\Delta^{I(\mu}\Delta^{J\nu)}\eta_{IJKL}\Delta^{K(\rho}\Delta^{L\sigma)}\nabla_\rho u_\sigma\\
	J^\mu&=nu^\mu-\Delta^{I\mu}\sigma_{IJ}\Delta^{J\nu}\left(T\partial_\nu\frac{\mu}{T}-E_\nu\right)-\Delta^{I\mu}\gamma_{IJ}u^\nu e^J_\nu
\end{align}	
\end{subequations}
The first three terms in $T^{\mu\nu}$ are the ideal fluid contribution, with an extra piece compared to the standard result due to the presence of the crystal fields $O_I$, while $\eta_{IJKL}$, $\sigma_{IJ}$ and $\gamma_{IJ}$ are transport coefficients.

As we mentioned earlier, we must also provide the equations of motion for the Goldstone fields, the Josephson-like relations. These equations are not constrained by conservation laws, hence they must be obtained order by order in derivative expansion exactly like the constitutive relations. At lowest order in the hydrodynamic expansion they constrain the scalar fields to be constant along the fluid flow, while at order one they take the form
\begin{equation}
	\sigma^\phi_{IJ}u^\mu e^I_\mu+\gamma'_{JK}\Delta^{K\mu}\left(T\partial_\mu\frac{\mu}{T}-E_\mu\right)+\nabla_\mu\left(r_{JK}e^{K\mu}\right)=K_J^\text{ext}
\end{equation}
where $\sigma^\phi_{IJ}$ and $\gamma'_{IJ}$ are two further dissipative transport coefficient matrices, while $K_J^\text{ext}$ is an external background source that couples to the scalar fields $O^I$. It can be understood as a different parametrization with respect to $\Phi_I$ used in Section~\ref{sec:ch3:symmetry_breaking}, and it enters the stress-energy tensor equations of motion as
\begin{equation}
	\nabla_\mu T^{\mu\nu}=F^{\nu\lambda}J_\lambda-K_I^\text{ext}e^{I\nu}
\end{equation}
This difference can be understood from the fact that in Section~\ref{sec:ch3:symmetry_breaking} we took $O_I$ to be the fundamental field and $\Phi_I$ its non-linear source, meaning that we used $\Phi_I$ to formally compute $n$-point functions from a generator $W=W[g,A,\Phi]$. Here, following \cite{Armas:ViscoelasticHydrodynamicsHolography}, we are considering a functional generator $W=W[g,A,O]$ which is related to the previous one via Lagrange transform, where $K^\text{ext}_I$ is the external field coupled to $O_I$. The former is better suited to compute the Ward identity, while the latter gives a more direct access to interesting thermodynamic quantities.

These constitutive relations are already the most general expressions compatible with the positivity of entropy production up to order one in derivatives, however, as we discussed in Chapter~\ref{chapter:hydrodynamics_linear_response}, we will also require time-reversal invariance of the microscopic theory. In this case the Onsager relation $\langle O^IJ^j\rangle=-\langle J^iO^J\rangle$ require us to identify $\gamma'_{IJ}=-\gamma_{IJ}$.

These expressions are valid for strain tensors $u_{IJ}$ arbitrary large, however in practice we often want to focus on the small strain fluctuations about some global thermodynamic equilibrium, thus we linearize in small amplitude $\mathcal{O}(u^2)$. In this regime, the first-order transport coefficients are independent of the strain (they can still depend on $T$ and $\mu$). However, because of the background constant magnetic field $B$, we allow for the presence of Hall transport coefficients in the constitutive relations \cite{Amoretti:MagnetothermalTransportImplies}. Then we can further decompose transport coefficients as
\begin{equation}\label{eqn:ch3:transport_coefficients_decomposition}
	\left(\gamma,\sigma,\sigma^\phi\right)_{IJ}=\left(\gamma,\sigma,\sigma^\phi\right)_{(L)}\delta_{IJ}+\left(\gamma,\sigma,\sigma^\phi\right)_{(H)}F_{IJ}
\end{equation}
where $F_{IJ}=F_{\mu\nu}e^\mu_Ie^\nu_J$ is the pullback of the electromagnetic field strength. Above, we denoted with $(L)$ the longitudinal transport coefficients, while $(H)$ stands for the Hall transverse transport coefficients, induced by the presence of the magnetic field. $\sigma_{IJ}$ is the charge conductivity, $\sigma^\phi_{IJ}$ the so-called crystal diffusivity and $\gamma_{IJ}$ is a mixed scalar-charge conductivity. Finally, $\eta_{IJKL}$ contains information about the shear and bulk viscosities, and we could in principle also decompose it in $\mathrm{SO(2)}$ terms with $\delta_{IJ}$ and $F_{IJ}$, however because we are interested in the diffusive sector, i.e. the thermoelectric transport at zero wavevector, the contribution of viscosities is not relevant.

With these definitions at hand, the constitutive relations up to order one in derivative and up to linear order in strain $u$ are
\begin{subequations}
	\begin{align}
		T^{\mu\nu}&=\left(\epsilon_f+\epsilon_l u^\lambda_{\ \lambda}\right)u^\mu u^\nu+\left(P_f+P_lu^\lambda_{\ \lambda}-mB\right)\Delta^{\mu\nu}+P_lh^{\mu\nu}-2Gu^{\mu\nu}\nonumber\\
		&\quad-\left(K-G\right)u^\lambda_{\ \lambda}h^{\mu\nu}+\mathcal{O}(u^2,\partial^2)\\
		J^\mu&=\left(n_f+n_lu^\lambda_{\ \lambda}\right)u^\mu-\sigma_{IJ}\Delta^{I\mu}\Delta^{J\nu}\left(T\partial_\nu\frac{\mu}{T}-E_\nu\right)\nonumber\\
		&\quad-\gamma_{IJ}\Delta^{I\mu}e^J_\nu u^\nu+\mathcal{O}(u^2,\partial^2)
	\end{align}
\end{subequations}
where $h_{\mu\nu}=h_{IJ}e^I_\mu e^J_\nu$ and $u_{\mu\nu}=u_{IJ}e^I_\mu e^J_\nu$, and we did not include the viscosities in the stress-energy tensor, since they are not relevant for the optical thermoelectric conductivities. The Goldstone equation in this regime is
\begin{multline}
	\sigma^{\phi}_{IJ}u^\mu e^J_\mu-h_{IJ}\nabla_\mu\left(P_le^{\mu J}-\left(K-G\right)u^\lambda_{\ \lambda}e^{\mu J}-2Gu^{\mu\nu}e_\nu^J\right)\\-\gamma_{IJ}\Delta^{J\mu}\left(T\partial_\mu\frac{\mu}{T}-E_\mu\right)=K_I^\text{ext}
\end{multline}

\subsection{AC conductivities}
With the constitutive relations and the equations of motion at hand, we can finally compute the AC conductivities analytically using the tools reviewed in Section~\ref{sec:ch2:linear_response_theory}. We consider a background with flat spacetime metric $g_{\mu\nu}=\eta_{\mu\nu}$ and constant magnetic field $F^{12}=B$, with vanishing electric field, Goldstone source $E_\mu=K^\text{ext}_J=0$ and vanishing spatial velocity $u^\mu=(1,\vect{0})$. The global thermodynamic equilibrium has constant temperature, chemical potential and uniform scalar fields $O^I=x^I$.

To proceed, we linearize the equations of motion around this background equilibrium solution by introducing fluctuations of the hydrodynamic fields
\begin{subequations}\label{eqn:ch3:fluctuations}
	\begin{align}
		T&\rightarrow T+\delta T	&	\mu&\rightarrow\mu+\delta\mu\\
		u^\mu&\rightarrow(1,v^i)	&	O^I&\rightarrow x^I-\delta O^I
	\end{align}
\end{subequations}
and of their sources $\delta\mathbb{E}^i=\delta F^{0i}$ and $\delta K^\text{ext}_I$. We can then Fourier transform all the linear fluctuations by assuming plane wave dependence $\exp(-i\omega t+i\vect{k}\cdot\vect{x})$ and solve the algebraic equations of hydrodynamics to express the hydrodynamic fields in terms of the linear sources. Plugging back in the results into the constitutive relations at $\vect{k}=0$ allows us to read off the two-point functions for the independent correlators $\langle J^iJ^j\rangle, \langle J^i O^J\rangle$ and $\langle O^IO^J\rangle$.

We can subsequently use the Ward identities in their ladder form \eqref{eqn:ch3:ward_identities} to obtain the missing correlators involving the canonical heat current. One comment here is important: in the present context the thermodynamic heat current, that couples to the source $-T\partial_i\delta\frac{1}{T}=\delta\zeta_i$ (see \eqref{eqn:ch2:thermoelectric_matrix}) is not the canonical heat current. This can be verified by also turning on a temperature gradient $T\rightarrow T+\delta T-x^i\delta\zeta_i$ and checking that the correlators obtained in this way do not obey Onsager relations or the Ward identities. Then, if one wishes to obtain the canonical heat current correlators without using the Ward identities, a metric fluctuation $\delta g_{0i}$ must be turned on. Spacetime fluctuations couple to the stress energy tensor and allows one reconstructing the heat current correlators in terms of the stress-energy tensor and current Green functions as e.g. $\langle Q^iJ^j\rangle=\langle T^{0i}J^j\rangle-\mu\langle J^iJ^j\rangle$.

In this chapter we decided to work with the canonical heat current, since it is the quantity that naturally appears in the Ward identities, and to connect with previous works \cite{Amoretti:EffectiveHolographicTheory,Amoretti:DCResistivityQuantum,Ammon:UnifiedDescriptionTranslational,Armas:ViscoelasticHydrodynamicsHolography,Armas:HydrodynamicsChargeDensity}, however it would have been possible to do the same computation using the thermodynamic heat current without major differences, see \cite{Donos:IncoherentHydrodynamicsDensity,Donos:IncoherentHydrodynamicsDensitya}.

We define the Kubo formulae for the AC conductivities in terms of the correlators \eqref{eqn:ch2:thermoelectric_kubo_formulae}
\begin{subequations}\label{eqn:ch3:spontaneous_conductivities_definitions}
	\begin{align}
		\left(\sigma^{ij},\alpha^{ij},\gamma^{iJ}\right)(\omega)&=\left(\frac{\langle J^iJ^j\rangle}{i\omega},\frac{\langle Q^iJ^j\rangle}{i\omega},\langle J^iO^J\rangle\right)\\
		\left(\kappa^{ij},X^{IJ},\theta^{iJ}\right)(\omega)&=\left(\frac{\langle Q^iQ^j\rangle}{i\omega},i\omega\langle O^IO^J\rangle,\langle Q^iO^J\rangle\right)
	\end{align}
\end{subequations}
As we will see, the DC values of the conductivities in the first line is completely fixed by symmetries, i.e. the Ward identities. The conductivities in the second line, instead, have a DC value that depends on the microscopic details of the system and is a priori unconstrained.

Using the ladder structure of the Ward identities \eqref{eqn:ch3:ward_identities}, the conductivities $\alpha(\omega)$, $\kappa(\omega)$ and $\theta(\omega)$ can always be derived from $\sigma(\omega)$ and $\gamma(\omega)$. Furthermore, in the spontaneous case $\varphi=0$, $X(\omega)$ decouples in the Ward identities and cannot be obtained from other correlators. Thus, it is enough to provide the expressions for $\sigma(\omega)$, $\gamma(\omega)$ and $X(\omega)$ to give all the information needed to describe the thermoelectric transport.

The expressions for the Green functions computed from linearized hydrodynamics are very large, even at zero wavevector, for this reason we use a matrix notation to express the results. First we define the matrices of the hydrodynamic transport coefficients and AC conductivities (always identified with an explicit $\omega$ dependence) by decomposing with respect to $\mathrm{SO(2)}$
\begin{subequations}\label{eqn:ch3:SO(2)_decomposition_spontaneous}
	\begin{align}
		(\hat\sigma,\hat\sigma_\phi,\hat\gamma)&=(\sigma,\sigma^\phi,\gamma)_{(L)}\mathds{1}_2-(\sigma,\sigma^\phi,\gamma)_{(H)}F\\
		(\hat\sigma,\hat\alpha,\hat\kappa,\hat\gamma,\hat X,\hat\theta)(\omega)&=(\sigma,\alpha,\kappa,\gamma,X,\theta)_{(L)}(\omega)\mathds{1}_2-(\sigma,\alpha,\kappa,\gamma,X,\theta)_{(H)}(\omega)F^{-1}
	\end{align}
\end{subequations}
where $F=F_{IJ}$ is the antisymmetric matrix proportional to the magnetic field. We further define the following matrix structures that appear frequently in the expressions
\begin{equation}
	\hat\sigma'=\hat\gamma^2+\hat\sigma\cdot\hat\sigma_\phi\qquad\hat\rho=2\hat\gamma+F\cdot\hat\sigma-n_f\mathds{1}_2
\end{equation}
Finally, the full analytic expressions for the three independent conductivities are
\begin{subequations}\label{eqn:ch3:conductivities_spontaneous}
	\begin{align}
		\hat\sigma(\omega)&=\hat\Lambda^{-1}\cdot\left[\omega P_l(in_f\hat\rho-\omega w_f\hat\sigma)+n_f^2\hat\sigma_\phi-(n_fF+i\omega\chi_{\pi\pi}\mathds{1}_2)\hat\sigma'\right]\\
		\hat\gamma(\omega)&=\hat\Lambda^{-1}\left(i\omega w_f\hat\gamma-n_f\hat\sigma_\phi+F\cdot\hat\sigma'\right)\\
		\hat X(\omega)&=\hat\Lambda^{-1}\cdot\left(F\cdot\hat\rho+i\omega w_f\mathds{1}_2-\hat\sigma_\phi\right)\\
		\hat\Lambda&=\omega P_l\left(iF\cdot\hat\rho-\omega w_f\mathds{1}_2\right)+\hat\sigma_\phi(Fn_f-i\omega\chi_{\pi\pi}\mathds{1}_2)-F^2\cdot\hat\sigma'
	\end{align}
\end{subequations}
with $w_f=\epsilon_f+P_f$ the enthalpy density.

The above conductivities, derived from hydrodynamics, are written in terms of the undetermined transport coefficients matrices $\hat\sigma$, $\hat\sigma_\phi$ and $\hat\gamma$. We can however use the Ward identities, by expanding at leading order in $\omega$, to express these transport coefficients in terms of the DC values of the conductivities \cite{Amoretti:MagnetothermalTransportImplies}. Assuming that they are finite as $\omega\rightarrow0$, we find
\begin{subequations}
	\begin{align}
		\sigma_{(L)}(\omega)&=-\frac{i}{B^2}\left(\mu n_f-\alpha_{(H)}(0)\right)\omega+\frac{\kappa_{(L)}(0)}{B^2}\omega^2+\mathcal{O}(\omega^3)\\
		\sigma_{(H)}(\omega)&=\sigma_{(H)}(0)+\frac{\kappa_{(H)}-\mu\left(2\chi_{\pi\pi}-\mu n_f\right)}{B^2}\omega^2+\mathcal{O}(\omega^3)\\
		\gamma_{(L)}(\omega)&=\frac{i\left(\mu+\theta_{(H)}(0)\right)}{B^2}\omega+\mathcal{O}(\omega^2)\\
		\gamma_{(H)}(\omega)&=\gamma_{(H)}(0)-i\theta_{(L)}(0)\omega+\mathcal{O}(\omega^2)
	\end{align}
\end{subequations}
where the DC values of $\sigma$, $\alpha$ and $\gamma$ are fixed by symmetries to
\begin{subequations}\label{eqn:ch3:DC_values_symmetries_spontaneous}
	\begin{align}
		\sigma_{(L)}(0)&=\alpha_{(L)}(0)=\gamma_{(L)}(0)=0	&	\sigma_{(H)}(0)&=-n_f\\
		\alpha_{(H)}(0)&=\mu n_f-\chi_{\pi\pi}	&	\gamma_{(H)}(0)&=-1
	\end{align}
\end{subequations}
However, the remaining conductivities are undetermined in DC and can be put in one-to-one correspondence with the transport coefficients \cite{Amoretti:HydrodynamicMagnetotransportCharge}. Doing so we find
\begin{subequations}\label{eqn:ch3:transport_coefficients_to_DC_spontaneous}
	\begin{align}
		\hat\sigma&=\hat\Phi^{-1}\cdot\left(\hat\kappa(0)+2\chi_{\pi\pi}\hat\theta(0)-\chi_{\pi\pi}^2\hat X(0)-\mu^2n_fF^{-1}\right)+n_fF^{-1}\\
		\hat\sigma_\phi&=\hat\Phi^{-1}\cdot\left[F\cdot(P_l^2\hat X(0)-\hat\kappa(0)-2P_l\hat\theta(0))+\mu(\mu n_f-2w_f)\mathds{1}_2\right]\cdot F\\
		\hat\gamma&=\hat\Phi^{-1}\cdot\left[F\cdot\left(P_l\chi_{\pi\pi}\hat X(0)+(w_f-2\chi_{\pi\pi})\hat\theta(0)-\hat\kappa(0)\right)+\mu(\mu n_f-w_f)\mathds{1}_2\right]\\
		\hat\Phi&=\left(\mu\mathds{1}_2-F\cdot\hat\theta(0)\right)^2+(F\cdot\hat\kappa(0)-\mu(\mu n_f-2\chi_{\pi\pi})\mathds{1}_2)\cdot F\cdot\hat X(0)
	\end{align}
\end{subequations}
This identification is particularly useful both for experiments and holography, since it allows us to fix the values of the hydrodynamic transport coefficients not via Kubo formulae or by matching the pole structure, but simply by the values of the DC conductivities. In particular in holography the DC transport can be computed analytically in terms of horizon data, therefore this approach allows us to write the analytic AC conductivities from hydrodynamics in terms of horizon quantities alone (DC conductivities and thermodynamics).

Both the AC conductivities and the transport coefficients in terms of DC data agree, at zero lattice pressure $P_l=0$, with \cite{Amoretti:HydrodynamicMagnetotransportCharge}, and at zero magnetic field $B=0$ with \cite{Armas:HydrodynamicsChargeDensity}, upon appropriate redefinitions of the transport coefficients to match the notations.

\section{Pseudo-spontaneous case}\label{sec:ch3:pseudo_spontaneous_case}
We now focus on the case of pseudo-spontaneous or explicit breaking of translation symmetry, where the difference between the two scenarios only stems from the magnitude of the source that breaks the symmetry. If $\varphi\ll\abs{\partial_i\langle O_I\rangle}$ we call it pseudo-spontaneous, otherwise if the source is large, it is an explicit breaking. Physically, pinned charge densities arise whenever there are defects or impurities in the crystal which select a preferred configuration for the ground state.

\subsection{Constitutive relations}
If the breaking of translation is small, we can take the constitutive relations to be exactly the same as in the spontaneous case \eqref{eqn:ch3:spontaenous_constitutive_relations}, however we must modify the conservation equations to also include the presence of a small parameter that explicitly breaks translations. In particular, because of the presence of a non-zero background mass term for the scalar fields, we modify the momentum conservation equation to include the effect of this coupling
\begin{equation}\label{eqn:ch3:explicit_equation_of_motion}
	\partial_t P^i+\partial_jT^{ij}=F^{i\mu}J_\mu-K^\text{ext}_Ie^{Ii}+\omega_0^2\chi_{\pi\pi}O^I\delta^i_I
\end{equation}
in which $\omega_0^2$ is called the pinning frequency. This expression comes directly from the one-point Ward identity in \eqref{eqn:ch3:one_point_ward_identities} by considering an homogeneous source $\Phi^I=\varphi x^I$ with $\varphi=\omega_0^2\chi_{\pi\pi}$.

The Josephson equations for the scalars $O_I$ also receive corrections due to the mass term. In particular, we are now allowed to add an extra term on the RHS that accounts for the possibility of the Goldstone field to relax in spacetime \cite{Delacretaz:DampingPseudoGoldstoneFields}. Thus, we modify the equation to introduce a phase relaxation term $\Omega^{IJ}O_J$, namely
\begin{equation}\label{eqn:ch3:josephson_relation_explicit}
	\sigma^\phi_{IJ}u^\mu e^I_\mu+\gamma_{JK}\Delta^{K\mu}\left(T\partial_\mu\frac{\mu}{T}-E_\mu\right)+\nabla_\mu\left(r_{JK}e^{K\mu}\right)=\Omega_{IJ}O^I+K^\text{ext}_J
\end{equation}
While the presence of pinning $\omega_0$ in \eqref{eqn:ch3:explicit_equation_of_motion} is due to defects in the crystals, phase relaxation can be understood due to the motion of topological defects, specifically dislocations, see Appendix B of \cite{Delacretaz:TheoryHydrodynamicTransport}. We can decompose the phase relaxation tensor, as we did for the other transport coefficients in \eqref{eqn:ch3:transport_coefficients_decomposition}, with respect to $\mathrm{SO(2)}$ rotation invariance and microscopic parity invariance, hence we define
\begin{equation}
	\Omega^{IJ}=\Omega_{(L)}\delta^{IJ}+\Omega_{(H)}F^{IJ}
\end{equation}

Although the phase relaxation tensor is in principle an independent transport coefficients, from the point of view of continuous symmetries, Onsager relations happen to constraint it. In particular, it can be shown that the phase relaxation is diagonal and proportional to the pinning frequency
\begin{equation}\label{eqn:ch3:phonon_relaxation}
	\Omega^{IJ}=\omega_0^2\chi_{\pi\pi}\delta^{IJ}+\dots
\end{equation}
hence $\Omega_{(H)}=0$. This result seems in contrast with previous studies that found a non-zero Hall term for the phase relaxation tensor \cite{Amoretti:HydrodynamicMagnetotransportCharge}, however the difference in the two formalisms comes from the fact that here we are normalizing the Josephson equations with an extra factor of $\sigma^\phi$. We can normalize the kinetic term in \eqref{eqn:ch3:josephson_relation_explicit} to the identity, multiplying the equation with $(\sigma^{\phi}_{IJ})^{-1}$, so that  the above Onsager constraint becomes $\Omega^{IJ}=\omega_0^2\chi_{\pi\pi}(\sigma^\phi)^{-1}$, in agreement with previous works. In the above formula the dots express the fact that the relation holds only up to the lowest order in $\omega_0^2$, and higher-order $\mathcal{O}(\omega_0^4)$ corrections in general will enter the expression \cite{Delacretaz:DampingPseudoGoldstoneFields}.

This relation between pinning frequency and the Goldstone phase relaxation was first obtained from holography in \cite{Amoretti:UniversalRelaxationHolographic} and holds generally even for other broken symmetries \cite{Grossi:TransportHydrodynamicsChiral}. Here, we obtained \eqref{eqn:ch3:phonon_relaxation} imposing microscopic time-reversal symmetry, however subsequent papers managed to find other proofs of this result based on various different approaches, such as: locality \cite{Delacretaz:DampingPseudoGoldstoneFields}, positivity of entropy production \cite{Armas:ApproximateSymmetriesPseudoGoldstones}, and EFT arguments \cite{Baggioli:EffectiveFieldTheory}, suggesting even more convincingly the universality of this relation.

Finally, we remark that if the Goldstone becomes very massive, its low-energy dynamics freezes, and it can be integrated out \cite{Amoretti:HydrodynamicMagnetotransportCharge}. In this case the Goldstone takes the equilibrium value $\langle O^I\rangle=x^I$ and from \eqref{eqn:ch3:josephson_relation_explicit} we arrive at $\Omega_{IJ}O^J\approx\sigma^\phi_{IJ} u^J$. Plugging this expression for $O^I$ into \eqref{eqn:ch3:explicit_equation_of_motion} we find
\begin{equation}\label{eqn:ch3:effective_momentum_relaxation}
	\partial_tP^i+\partial_jT^{ij}=F^{i\mu}J_\mu+\Gamma^I_{\ J}\chi_{\pi\pi}u^J\delta^i_I
\end{equation}
where $\Gamma^I_{\ J}\approx\omega_0^2\left(\Omega^{IK}\right)^{-1}\sigma_{KJ}^\phi$ acts as an effective relaxation term for the physical momentum.

\subsection{AC conductivities}
The computation of the optical conductivities goes through exactly as in the spontaneous case, without any change.

To express the results we introduce some new definitions for the conductivities, which are more appropriate for the pseudo-spontaneous case. Specifically we define
\begin{equation}
	(\varpi^{iJ},\zeta^{IJ})(\omega)=\frac{1}{i\omega}\left(\langle J^iO^J\rangle,\langle O^IO^J\rangle-\frac{\delta^{IJ}}{\varphi}\right)
\end{equation}
in place of the two spontaneous conductivities $\gamma^{iJ}(\omega)$ and $X^{IJ}(\omega)$. These conductivities have different powers of $\omega$ with respect to their spontaneous version, to reflect the fact that the two cases, spontaneous or pseudo-spontaneous, have a different low frequency expansion, as observed from the two-point functions Ward identities \eqref{eqn:ch3:ward_identities}. Furthermore, we decompose the AC conductivities with respect to $\mathrm{SO(2)}$ differently, by keeping an extra factor of $B^2$ with respect of the spontaneous case
\begin{equation}
	(\hat\sigma,\hat\alpha,\hat\kappa,\hat\varpi,\hat\zeta,\hat\theta)(\omega)=(\sigma,\alpha,\kappa,\varpi,\zeta,\theta)_{(L)}(\omega)\mathds{1}_2+(\sigma,\alpha,\kappa,\varpi,\zeta,\theta)_{(H)}F
\end{equation}
In particular, the Hall terms are now decomposed with respect to $F$ instead of $F^{-1}$ in \eqref{eqn:ch3:SO(2)_decomposition_spontaneous}, in agreement with the fact that they are smooth as $B\rightarrow0$. This is because the Goldstone mass, like the magnetic field, gaps the system. Therefore, while in the spontaneous case the $B$ to zero limit lead to divergences, in the pseudo-spontaneous case the presence of the pinning frequency prevent any blow-up as $B\rightarrow0$. Finally, we introduce one last quantity for ease of notation
\begin{equation}
	\Gamma=\omega_0^2\chi_{\pi\pi}-\omega^2P_l
\end{equation}
which should not be confused with a momentum relaxation term in \eqref{eqn:ch3:effective_momentum_relaxation}.

With this new set of definitions, and remembering the decomposition of the frequency-independent transport coefficients in \eqref{eqn:ch3:SO(2)_decomposition_spontaneous}, we can write the three independent AC conductivities. Following the same argument of the spontaneous case, we take the independent correlators to be $\sigma(\omega)$,  $\varpi(\omega)$ and $\zeta(\omega)$, while the remaining conductivities can be obtained by these from the Ward identities \eqref{eqn:ch3:ward_identities}. We find
\begin{subequations}\label{eqn:ch3:conductivities_explicit}
	\begin{align}
		\hat\sigma(\omega)&=\hat\Xi^{-1}\cdot\left[\Gamma\omega w_f\hat\sigma+\omega n_f^2\hat\sigma_\phi-i(\omega^2-\omega_0^2)\chi_{\pi\pi}\hat\sigma'-n_f(i\Gamma\hat\rho+\omega F\cdot\hat\sigma')\right]\\
		\hat\varpi(\omega)&=\hat\Xi^{-1}\cdot\left[\omega w_f\hat\gamma+i(n_f\hat\sigma_\phi-F\cdot\hat\sigma')\right]\\
		\hat\zeta(\omega)&=\frac{\hat\Xi^{-1}}{\omega_0^2\chi_{\pi\pi}}\cdot\left[\omega\chi_{\pi\pi}\hat\sigma_\phi-\omega P_l(F\cdot\hat\rho+i\omega w_f\mathds{1}_2)+iF\cdot(n_f\hat\sigma_\phi-F\cdot\hat\sigma')\right]\\
		\hat\Xi&=\Gamma(\omega w_f\mathds{1}_2-iF\cdot\hat\rho)+\omega n_fF\cdot\hat\sigma_\phi-i(\omega^2-\omega_0^2)\chi_{\pi\pi}\hat\sigma_\phi-\omega F^2-\hat\sigma'
	\end{align}
\end{subequations}

Subsequently, we can again use the Ward identities to express the undetermined hydrodynamic transport coefficients in terms of the DC values of the conductivities which are not constrained by symmetries. In the pseudo-spontaneous case, we see that the electric, thermoelectric and thermal conductivities are unconstrained in DC, thus we use $\hat\sigma(0)$, $\hat\alpha(0)$ and $\hat\kappa(0)$ to express the transport coefficients
\begin{subequations}\label{eqn:ch3:transport_coefficients_to_DC_explicit}
	\begin{align}
		\hat\sigma&=-\hat\Psi^{-1}\cdot\hat\pi(0)\\
		\hat\sigma_\phi&=\hat\Psi^{-1}\cdot\left[w_f^2\mathds{1}_2+(F\cdot\hat\pi(0)-2w_f(\hat\alpha(0)+\mu\hat\sigma(0)))\cdot F\right]+n_fF\\
		\hat\gamma&=\hat\Psi^{-1}\cdot\left[F\cdot\hat\pi(0)-w_f(\hat\alpha(0)+\mu\hat\sigma(0))\right]+n_f\mathds{1}_2\\
		\hat\Psi&=\mu^2\hat\sigma(0)+2\mu\hat\alpha(0)+\hat\kappa(0)
	\end{align}
\end{subequations}
where we have also defined a matrix $\hat\pi(0)$ that mimics the determinant of the DC thermoelectric matrix
\begin{equation}
	\hat\pi(0)=\hat\alpha^2(0)-\hat\kappa(0)\hat\sigma(0)
\end{equation}

We can again check that, as the lattice pressure goes to zero $P_l\rightarrow0$, we recover the expressions found in \cite{Amoretti:HydrodynamicMagnetotransportCharge} with perfect agreement, after an appropriate mapping of the transport coefficients.

\section{Holographic model}\label{sec:ch3:holographic_model}
In this section we describe the holographic computations of \cite{Amoretti:HydrodynamicMagnetotransportHolographic}. Because holography is not the central topic of this thesis, and because I did not personally obtain these results, I will only briefly sketch the main points, without explaining how holography works \cite{Hartnoll:HolographicQuantumMatter,Ammon:GaugeGravityDuality,Rangamani:GravityHydrodynamicsLectures} or the details of the computations, which can be found in the original paper \cite{Amoretti:HydrodynamicMagnetotransportHolographic}.

\subsection{Setup and background}
From the holographic correspondence, the large $N$, large 't~Hooft coupling $\lambda$ regime of a $d$ dimensional QFT is dual to the semiclassical limit of asymptotically-AdS Einstein gravity in $d+1$ dimensions. We consider the following Q-lattice action \cite{Donos:NovelMetalsInsulators}
\begin{equation}
	S=\int\dif^{3+1}x\sqrt{-g}\left(R-V[\phi]-\frac{1}{2}(\partial\phi)^2-\frac{Z[\phi]}{4}F^2-\frac{1}{2}Y[\phi]\sum_{i=1,2}(\partial\psi_i)^2\right)
\end{equation}
here $F=\dif A$ is the electromagnetic field strength, $\phi$ is the dilaton field and $\psi_i$ are the translation symmetry breaking scalars, dual to the $O_I$ fields in the hydrodynamic picture, which enjoy the shift symmetry $\psi_i\rightarrow\psi_i+c_i$. Following the hydrodynamic construction, we require a linear coordinate dependence on the scalars
\begin{equation}
	\psi_i=kx_i\qquad x^i=\{x,y\}
\end{equation}
which break spatial translations and shift symmetry to their diagonal $\mathrm{U(1)}$ subgroup, i.e. translations are broken homogeneously, as discussed in Section~\ref{sec:ch3:symmetry_breaking}. With this ansatz, we look for background solutions for $A_\mu$, $g_{\mu\nu}$ and $\phi$ that depend only on the radial coordinate $r$.

We consider a background with a black hole solution in the interior, which is dual to a thermal system
\begin{equation}\label{eqn:ch3:background_solution}
	\dif s^2=\frac{1}{r^2}\left(-f(r)\dif t^2+\frac{\dif r^2}{f(r)}+g(r)\dif\vect{x}^2\right)\quad A=a(r)\dif t-By\dif x\quad\phi=\phi(r)
\end{equation}
with a constant magnetic field and non-zero charge density. We want these solutions to be asymptotically AdS, this in turn constraints the form of the functions $V[\phi]$, $Z[\phi]$ and $Y[\phi]$ as $\phi\rightarrow0$. Subsequently, these asymptotics fix value of the dilaton near the boundary $r=0$
\begin{equation}
	\phi(r)=\lambda r+\phi_vr^2+\mathcal{O}(r^3)
\end{equation}
which determines if translations are broken explicitly $\lambda\neq0$, or spontaneously $\lambda=0$, in the dual theory \cite{Amoretti:EffectiveHolographicTheory,Amoretti:UniversalRelaxationHolographic,Amoretti:GaplessGappedHolographic}.

The first step now is to match the thermodynamics of the boundary QFT with the background solution \eqref{eqn:ch3:background_solution}. We need to specify the temperature, chemical potential, magnetic field and either the vev of the scalar field $\phi_v$, for the spontaneous case, or its source $\lambda$, for the explicit case. In particular, following the holographic dictionary, the boundary behaviour of $a(r)$ gives the value of the chemical potential as the leading term and of the charge density at the sub-leading order
\begin{equation}
	a(r)=\mu-n_fr+\mathcal{O}(r^3)
\end{equation}
while the near-horizon expansion gives information about the temperature and entropy density of the fluid, from the usual Gibbons–Hawking formula for black-holes horizons. Employing the equations of motion for the background one finds radially conserved quantities that can be associated with the lattice pressure $P_l$ and the magnetization density $m$. Then, matching the boundary and horizon values of these conserved quantities we find a Smarr-type relation
\begin{equation}
	\epsilon_f=2(s_fT+\mu n_f-mB-P_l+\lambda\phi_v)
\end{equation}

\subsection{Analytic DC conductivities}
We now focus on the analytic computation of the DC values of the conductivities defined in hydrodynamics, both for the spontaneous and explicit case \cite{Donos:IncoherentTransportPhases}. To do so we fluctuate about the background described above by turning on a small, constant electric field and temperature gradient in the spatial directions of the boundary QFT \cite{Donos:ThermoelectricDCConductivities}. This is implemented as
\begin{subequations}\label{eqn:ch3:holographic_fluctuations}
	\begin{align}
		\delta A_x(r)&=a_x(r)-p_x(r)t	&	\delta g_{tx}&=\frac{1}{r^2}\left(h_x(r)-\bar{p}_x(r)t\right)	&	\delta g_{rx}=\frac{1}{r}\bar{h}_x(r)\\
		\delta A_y(r)&=a_y(r)-p_y(r)t	&	\delta g_{ty}&=\frac{1}{r^2}\left(h_y(r)-\bar{p}_y(r)t\right)	&	\delta g_{ry}=\frac{1}{r}\bar{h}_y(r)
	\end{align}
\end{subequations}
In the spontaneous case we also need to turn on constant sources for the axion fields
\begin{equation}
	\delta\psi_x(r)=\frac{\chi_x(r)}{r}-k\delta V_xt\qquad\delta\psi_y(r)=\frac{\chi_y(r)}{r}-k\delta V_yt
\end{equation}
while for the explicit case these take the simpler form
\begin{equation}\label{eqn:ch3:axion_fluctuation_explicit}
	\delta\psi_x(r)=\chi_x(r)\qquad\delta\psi_y(r)=\chi_y(r)
\end{equation}
without the sliding modes $\delta V_i$. The coefficients with explicit time dependence in \eqref{eqn:ch3:holographic_fluctuations} are fixed such that the time-dependent terms drops out of the linearized equations of motion, i.e. we take
\begin{subequations}
	\begin{align}
		p_x(r)&=p_x^{(0)}+n_f\bar{E}_xa(r)	&	\bar{p}_x(r)&=-n_f\bar{E}_xf(r)\\
		p_y(r)&=p_y^{(0)}+n_f\bar{E}_ya(r)	&	\bar{p}_y(r)&=-n_f\bar{E}_yf(r)
	\end{align}
\end{subequations}
where $p_i^{(0)}$ and $\bar{E}_i$ are, for the moment, free constants. The former represent the constant electric field, the latter the temperature gradient in the boundary theory.

From the linearized equations of motions we can identify a set of four radially conserved currents, which we denote as $\delta\mathcal{J}_i(r)$ and $\delta\mathcal{Q}_i(r)$, related to charge and heat transport. In particular, we can fix the free constants $p_i^{(0)}$ and $\bar{E}_i$ such that these bulk currents correspond to the vev of the electric and canonical heat current in the boundary theory $\delta\mathcal{J}_i(0)=\lim_{r\rightarrow0}\partial_ra_i(r)=\langle J_i\rangle$.

In the spontaneous case, this requires that
\begin{subequations}
	\begin{align}
		p_x^{(0)}&=\left(s_fT-mB+P_l\right)\bar{E}_x-\frac{B}{n_f}\langle J_y\rangle+\frac{\delta s_x}{n_f}\\
		p_y^{(0)}&=\left(s_fT-mB+P_l\right)\bar{E}_y+\frac{B}{n_f}\langle J_x\rangle+\frac{\delta s_y}{n_f}\\
		\delta s_i&=k\phi^2_v\chi_i(0)
	\end{align}
\end{subequations}
where $\delta s_i$ correspond to the linear sources for the dual Goldstone fields. To compute the DC conductivities we identify the boundary electric field $E_i$ and the thermal gradients $\partial_iT/T$ from the expansion
\begin{equation}\label{eqn:ch3:electric_thermal_holographic_sources}
	E_i=-\lim_{r\rightarrow0}\partial_t\left(\delta A_i+\frac{\mu r^2}{f(r)}\delta g_{ti}\right)\qquad\frac{\partial_iT}{T}=\lim_{r\rightarrow0}\partial_t\left(\frac{r^2}{f(r)}\delta g_{ti}\right)
\end{equation}
We can rewrite the expressions for the vev of the electric and heat current at the boundary in terms of these sources while also using the properties that they are radially conserved and, from there, we can read off the DC conductivities. In the spontaneous case, for example, we find
\begin{equation}
	\sigma_{(H)}(0)=-n_f\quad\alpha_{(H)}(0)=-\left(s_fT-mB-P_l\right)\quad\gamma_{(H)}(0)=-1
\end{equation}
in agreement with the fact that these conductivities are constrained by Ward identities to have these universal values at leading order in small $\omega$ \eqref{eqn:ch3:DC_values_symmetries_spontaneous}. We are specifically interested in $\hat\kappa(0)$, $\hat\theta(0)$ and $\hat X(0)$, since these are the non-universal DC conductivities that we used to fix the hydrodynamics transport coefficients in \eqref{eqn:ch3:transport_coefficients_to_DC_spontaneous}. We can write them in terms of horizon data as
\begin{subequations}\label{eqn:ch3:holographic_DC_values_spontaneous}
	\begin{align}
		\kappa_{(L)}(0)&=\frac{Z_h(s_fT-P_l)^2}{T\left(n_f^2+B^2Z_h^2\right)}+\frac{4\pi P_l^2}{Ts_fY_h}	&	\kappa_{(H)}(0)&=-\frac{n_f(s_fT-P_l)^2}{T(n_f^2+B^2Z_h^2)}-\frac{M_Q}{n_fT}\\
		\theta_{(L)}(0)&=\frac{4\pi I_Y}{sY_h}-\frac{(s_fT-P_l)Z_h}{n_f^2+B^2Z_h^2}	&	\theta_{(H)}(0)&=\frac{n_f(s_fT-P_l)Z_h}{n_f^2+B^2Z_h^2}\\
		X_{(L)}(0)&=-\left(\frac{4\pi}{k^2s_fY_h}+\frac{Z_h}{n_f^2+B^2Z_h^2}\right)	&	X_{(H)}&=\frac{n_f}{(n_f^2+B^2Z_h^2)}
	\end{align}
\end{subequations}
where $Z_h$ and $Y_h$ are the horizon values of the functions $Z[\phi]$ and $Y[\phi]$, while $I_y$ and $M_Q$ are radial integrals computed at the horizon related to the lattice pressure and the magnetization. Further details can be found in the original paper \cite{Amoretti:HydrodynamicMagnetotransportHolographic}.

In the pseudo-spontaneous case the bulk radially-conserved electric currents are unchanged with respect to the spontaneous case, however the identification of the constants $p_i^{(0)}$ is slightly modified, to account for the different expansion of the axion fluctuations \eqref{eqn:ch3:axion_fluctuation_explicit}. This time we require
\begin{subequations}
	\begin{align}
		p_x^{(0)}&=\left(s_fT-mB+P_l\right)\bar{E}_x-\frac{B}{n_f}\langle J_y\rangle-\frac{\langle O_x\rangle}{n_f}\\
		p_y^{(0)}&=\left(s_fT-mB+P_l\right)\bar{E}_y+\frac{B}{n_f}\langle J_x\rangle-\frac{\langle O_y\rangle}{n_f}\\
		\langle O^i\rangle&=k\lambda^2\chi_i'(0)
	\end{align}
\end{subequations}
The radially conserved bulk heat currents are also slightly modified for the same reason, but we can relate boundary and horizon values for them too. Again, we express the boundary electric and heat currents in terms of the electric field and temperature gradient \eqref{eqn:ch3:electric_thermal_holographic_sources}, while also taking their values at the horizon. This allows us to obtain the zero-frequency limit of all the conductivities, e.g.
\begin{subequations}
	\begin{align}
		\sigma_{(L)}(0)&=\frac{k^2s_fY_h\left(k^2s_fY_hZ_h+4\pi(n_f^2+B^2Z_h^2)\right)}{(4\pi n_fB)^2+(k^2s_fY_h+4\pi Z_hB^2)^2}\\
		\sigma_{(H)}(0)&=-\frac{8\pi n_f\left(k^2s_fY_hZ_h+2\pi(n_f^2+B^2Z_h^2)\right)}{(4\pi n_fB)^2+(k^2s_fY_h+4\pi Z_hB^2)^2}
	\end{align}
\end{subequations}
for the electric transport, and similar results hold for the other DC conductivities too.

\subsection{AC conductivities and matching hydrodynamics}
To compute numerically the AC conductivities, both in the spontaneous and explicit case, we need to turn on linear sources for the various fields, and impose ingoing boundary conditions and regularity at the horizon. Details on the numerics and the explicit steps can be found in the original paper \cite{Amoretti:HydrodynamicMagnetotransportHolographic}.

\subsubsection{Spontaneous case}
\begin{figure}
	\centering
	\begin{subfigure}{.31\textwidth}
		\centering
		\includegraphics[width=\textwidth]{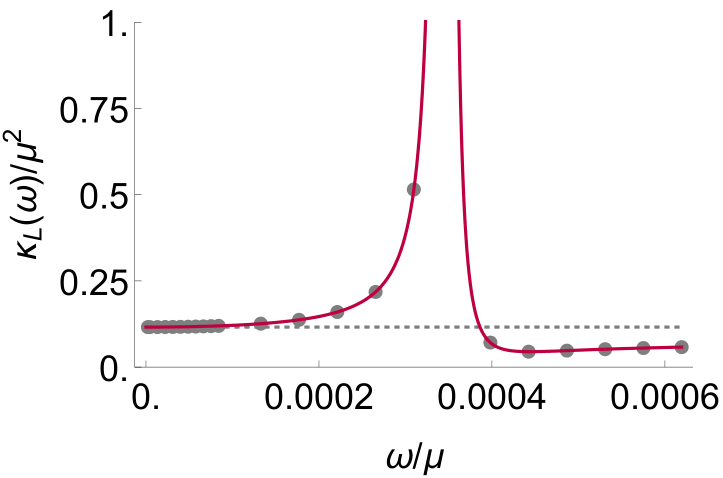}
	\end{subfigure} \hspace{.01\textwidth}
	\begin{subfigure}{.31\textwidth}
		\centering
		\includegraphics[width=\textwidth]{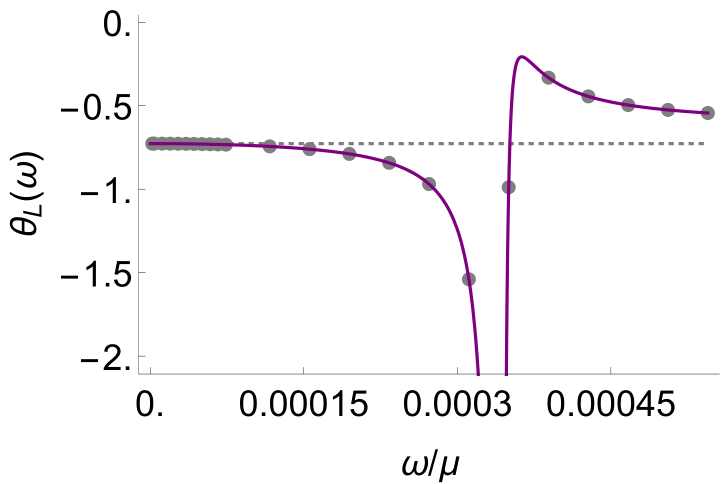}
	\end{subfigure} \hspace{.01\textwidth}
	\begin{subfigure}{.31\textwidth}
		\centering
		\includegraphics[width=\textwidth]{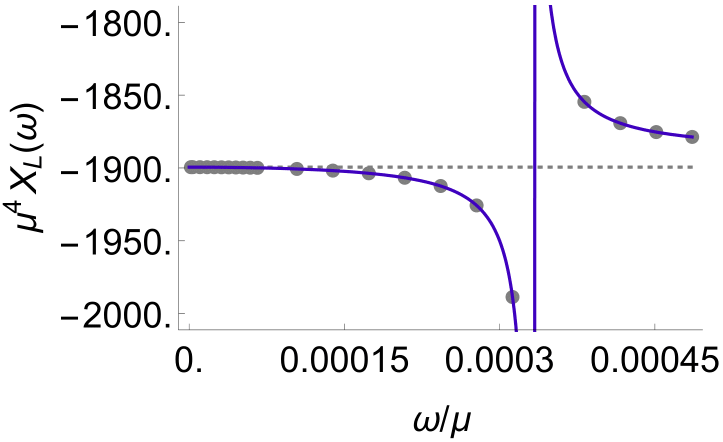}
	\end{subfigure} \hfill \\
	\begin{subfigure}{.31\textwidth}
		\centering
		\includegraphics[width=\textwidth]{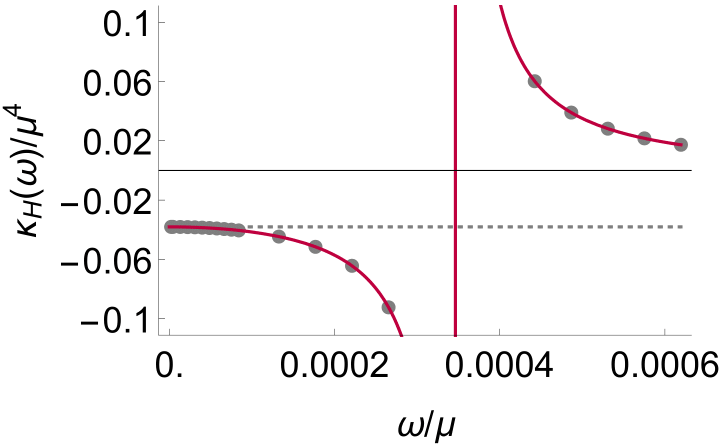}
	\end{subfigure}  \hspace{.01\textwidth}
	\begin{subfigure}{.31\textwidth}
		\centering
		\includegraphics[width=\textwidth]{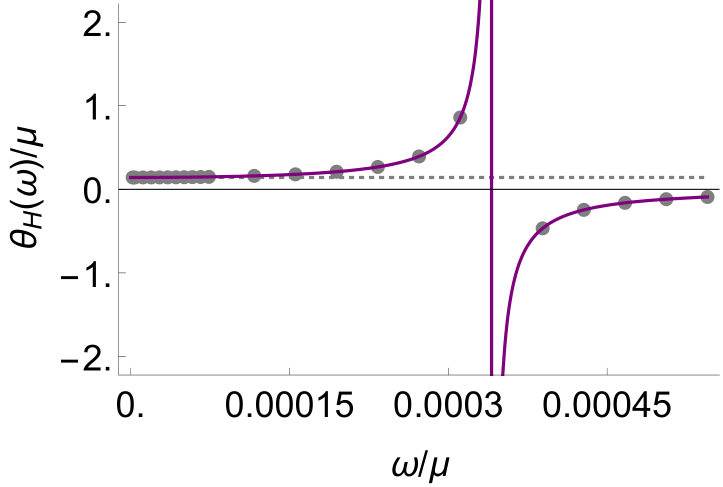}
	\end{subfigure}  \hspace{.01\textwidth}
	\begin{subfigure}{.31\textwidth}
		\centering
		\includegraphics[width=\textwidth]{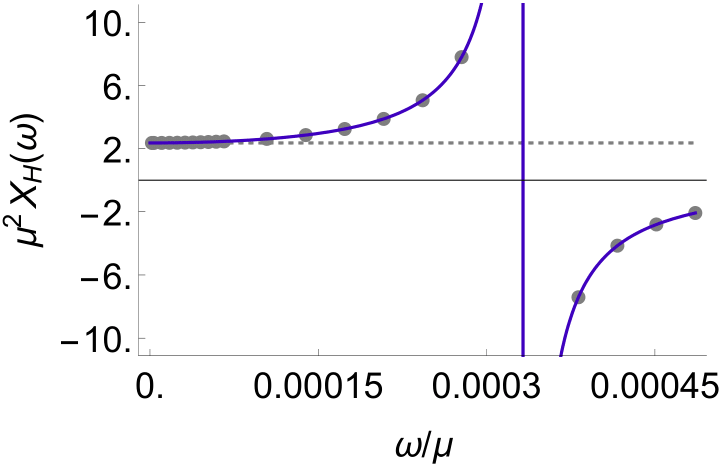} 
	\end{subfigure} \hfill
	\caption{AC correlators for the spontaneous case at $k/\mu=10^{-1}$. Grey dots are numerical data, solid lines are the analytic expressions from hydrodynamics, and the dashed grey line correspond to the DC values. \textbf{Left column:} The thermal conductivities $\hat\kappa(\omega)$ at $T/\mu = 0.06$ and $B/\mu^2 \approx 4.4 \times 10^{-4}$. \textbf{Central column:} The heat-Goldstone correlators $\hat\theta(\omega)$ at $T/\mu=0.04$ and $B/\mu^2 \approx 3.9 \times 10^{-4}$. \textbf{Right column:} The Goldstone-Goldstone correlators $\hat{X}(\omega)$ at $T/\mu=0.02$ and $B/\mu^2 \approx 3.5 \times 10^{-4}$.}
	\label{fig:ch3:spontaneous_AC_conductivities}
\end{figure}

There are in principle twelve possible conductivities \eqref{eqn:ch3:spontaneous_conductivities_definitions} we could compute for the spontaneous case, however we will focus only on the thermal $\langle Q^iQ^j\rangle$, the thermal-Goldstone $\langle Q^iO^J \rangle$ and the Goldstone-Goldstone correlators $\langle O^IO^J\rangle$. The real part of the AC hydrodynamic conductivities are computed by using the results in Section~\ref{sec:ch3:spontaneous_case} upon substituting the transport coefficients in terms of the DC values of the conductivities computed from holography \eqref{eqn:ch3:holographic_DC_values_spontaneous}. We can see the agreement in Figure~\ref{fig:ch3:spontaneous_AC_conductivities} is excellent in the shown regions.
Our hydrodynamic results seem very robust, matching with great accuracy up to very small values of the temperature and to relatively high magnetic field. The off-axis peak that appears in the thermal conductivity at $\omega>0$ is also present in the electric and thermoelectric conductivities, and corresponds to the cyclotron mode induced by the magnetic field \cite{Hartnoll:TheoryNernstEffect}.

At zero magnetic field the Goldstone-Goldstone correlator has a double pole at $\omega=0$ \cite{Amoretti:DiffusionUniversalRelaxation}, which can be observed by taking the $B$ to zero limit of our longitudinal conductivity \eqref{eqn:ch3:conductivities_spontaneous}. This means that, at zero magnetic field, the real part or the correlator has a peak of finite width in its spectrum. When the magnetic field is non-zero one of the pole becomes gapped and describes a cyclotron mode, leaving an isolated pole at $\omega=0$. The imaginary pole corresponds, via Kramers–Kronig relations, to a delta function in the real part of the conductivity, which is why the Goldstone-Goldstone conductivities $X_{(L,H)}(\omega)$ in Figure~\ref{fig:ch3:spontaneous_AC_conductivities} are smooth all the way down to $\omega=0$ without any sign of divergence.

\subsubsection{Pseudo-spontaneous case}
In the explicit and pseudo-spontaneous case the AC hydrodynamic conductivities are not completely fixed in terms of the DC values of the thermoelectric correlators. The holographic zero-frequency limits of $\hat\sigma(0)$, $\hat\alpha(0)$ and $\hat\kappa(0)$ in terms of horizon data allows us to fix the hydrodynamic transport coefficients \eqref{eqn:ch3:transport_coefficients_to_DC_explicit} (and consequently of the AC conductivities) up to the constant value of the pinning frequency $\omega_0$, which is not constrained in the hydrodynamic model and must be computed numerically.

\begin{figure}
	\centering
	\includegraphics[width=0.6\textwidth]{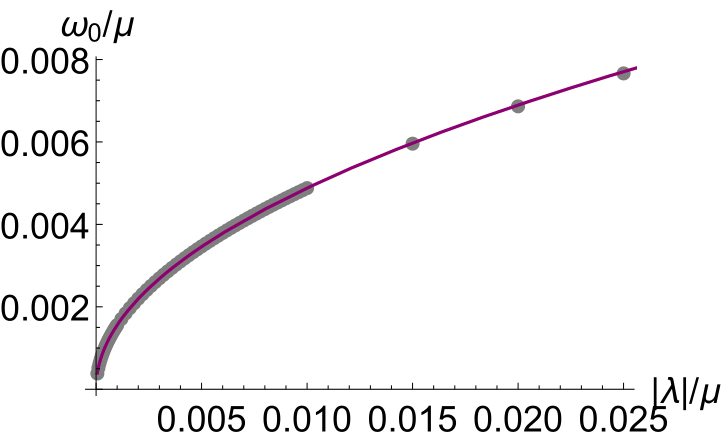}
	\caption{Pinning frequency against $\lambda/\mu$ at $k/\mu=0.1$, $B/\mu^2=10^{-2}$, and $T/\mu=5\times10^{-2}$. The solid line is a fit to the numerical points, which gives a dependence on $\sqrt{\abs{\lambda}/\mu}$.}
	\label{fig:ch3:pinning_frequency}
\end{figure}

To fix this parameter one can use two different method: either by studying the holographic spectrum of the quasinormal modes in the complex plane and solving for $\omega_0$, or by simply taking any low-frequency AC correlator and requiring that the hydrodynamic results agrees with the values computed numerically from holography. The two methods agree perfectly, and we find the dependence in Figure~\ref{fig:ch3:pinning_frequency} for the pinning frequency as a function of the holographic explicit breaking parameters $\lambda$. We found by best fit a square-root dependence of the pinning frequency on the ratio $\abs{\lambda/\mu}$, in agreement with previous computations at zero magnetic field \cite{Amoretti:UniversalRelaxationHolographic} and more general quantum field theory arguments \cite{Amoretti:HolographicPerspectivePhonons}. Furthermore, as the temperature increases, the pinning frequency becomes smaller and smaller.

\begin{figure}
	\begin{subfigure}[t]{.45\textwidth}
		\vskip 0pt
		\centering
		\includegraphics[width=\linewidth]{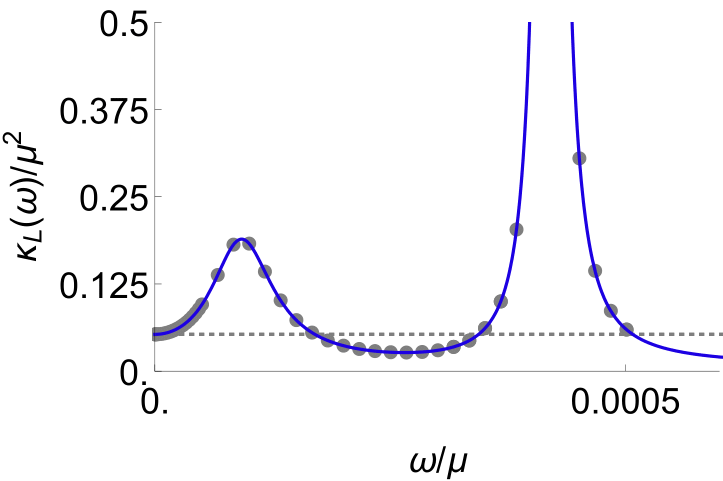}
	\end{subfigure} \hfill %
	\begin{subfigure}[t]{.45\textwidth}
		\vskip 0pt
		\centering
		\includegraphics[width=\linewidth]{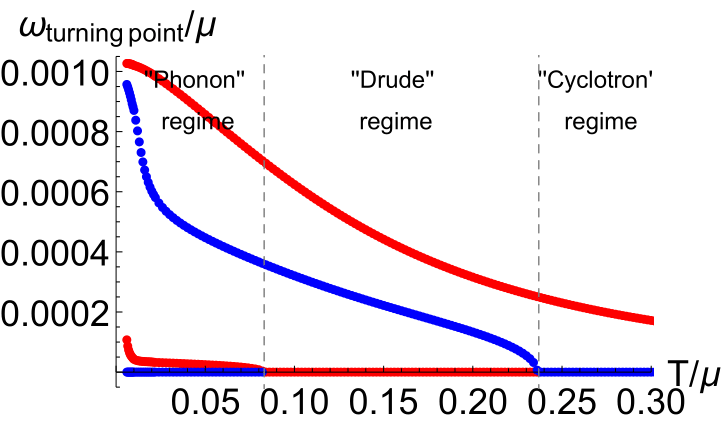}
	\end{subfigure}
	\caption{Conductivities in the (pseudo-)spontaneous regime. \textbf{Left:} AC longitudinal thermal conductivity at $\lambda/\mu=-10^{-5}$, $k/\mu=0.1$, $B/\mu^2\approx3\times10^{-4}$, and $T/\mu=10^{-2}$. Notice the two peaks at $\omega>0$. \textbf{Right:} The frequency at which the maxima (red) and minima (blue) appear in the hydrodynamic longitudinal electric charge conductivity \eqref{eqn:ch3:conductivities_explicit} as a function of $T$ at $\lambda/\mu=-10^{-5}$, $k/\mu=0.1$, and $B/\mu^2=10^{-3}$. At very low temperatures we find two peaks for $\omega>0$, and also a minimum at $\omega=0$ and a second one between the maxima. As the temperature is increased, the two pseudo-Goldstone modes merge around $T/\mu \approx 0.083$, to become a single Drude-like peak at zero frequency. This zero-frequency peak eventually drops out of the conductivity approximately at $T/\mu \approx 0.237$.}
	\label{fig:ch3:two_peaks}
\end{figure}

\begin{figure}[t]
	\begin{subfigure}{.47\textwidth}
		\centering
		\includegraphics[width=\linewidth]{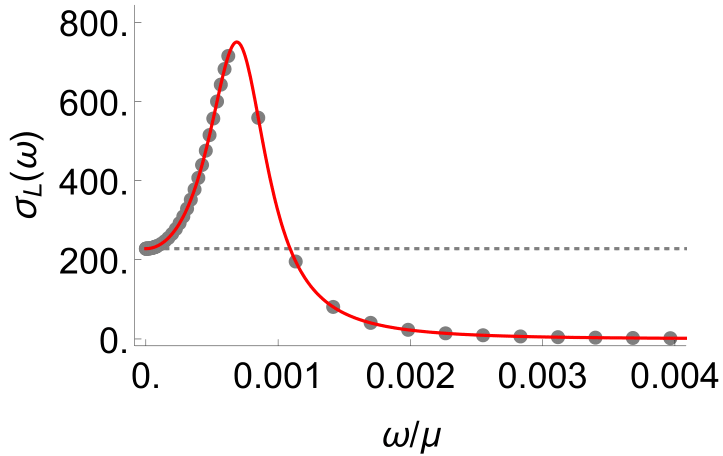}
	\end{subfigure} \hfill %
	\begin{subfigure}{.47\textwidth}
		\centering
		\includegraphics[width=\linewidth]{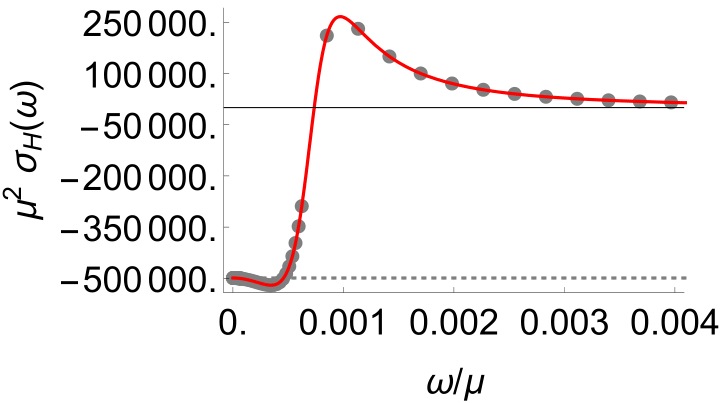}
	\end{subfigure} \\
	\begin{subfigure}{.47\textwidth}
		\centering
		\includegraphics[width=\linewidth]{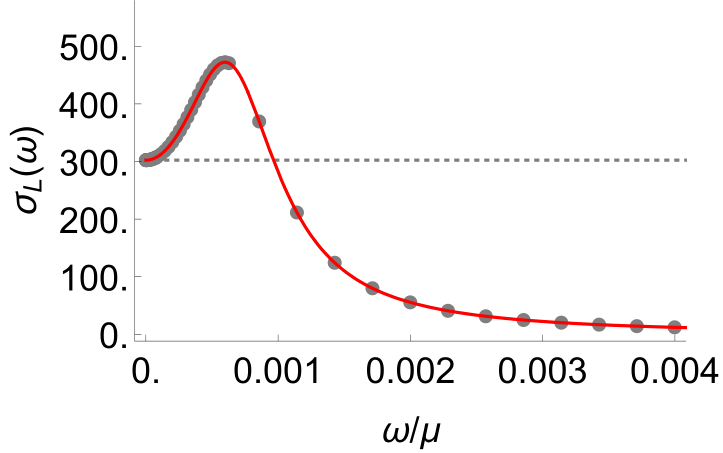}
	\end{subfigure} \hfill %
	\begin{subfigure}{.47\textwidth}
		\centering
		\includegraphics[width=\linewidth]{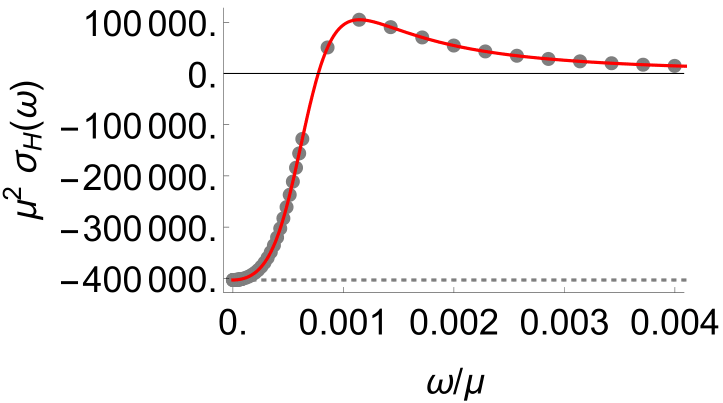}
	\end{subfigure}
	\caption{AC electric conductivities at $B/\mu^2=10^{-3}$, $T/\mu=10^{-1}$, and $k/\mu=10^{-1}$. As before, red lines are the analytic expressions obtained from hydrodynamics, while grey dots are numerical data. \textbf{Left:} The longitudinal conductivities in the pseudo-spontaneous regime (top), where $\lambda \mu/\phi_{v} \approx 0.004$, and in the strongly explicit regime (bottom), where $\lambda \mu/\phi_{v} \approx 0.95$. \textbf{Right:} The Hall conductivities in the same regimes.}
	\label{fig:ch3:explicit_conductivities}
\end{figure}

We now focus on the electric and thermal conductivities, plotted in Figure~\ref{fig:ch3:two_peaks} and \ref{fig:ch3:explicit_conductivities}, which again show a very good match between the analytic hydrodynamic computation \eqref{eqn:ch3:conductivities_explicit} and the numerical holographic results. In the pseudo-spontaneous case it is possible to find two off-axis peaks at $\omega>0$, see Figure~\ref{fig:ch3:two_peaks} for an example in the longitudinal thermal conductivity.

To better understand this point we also plotted, this time for the electric conductivity, the behaviour of the extrema (maxima in red, minima in blue) as a function of temperature. We can then loosely identify three different regimes: the low temperature regime is the phonon-dominated regime, in which we observe two off-axis peaks, specifically one is the cyclotron peak, while the low-frequency one is the contribution due to the pinned Goldstones, and also a minimum at $\omega=0$. The two peaks and their associated quasinormal modes can be understood, at least on a qualitative level, as the magnetophonon and magnetoplasmon resonance that appear in the hydrodynamic regime of a weakly-pinned Wigner crystal \cite{Delacretaz:TheoryCollectiveMagnetophonon}. As we increase the temperature we enter the Drude regime, in which the Goldstone peak moves towards $\omega=0$ to create a Drude-like peak (remember that the pinning frequency decreases with increasing $T$), while keeping the second off-axis peak qualitatively the same. At even higher temperature we are in the cyclotron regime: the second minimum moves at $\omega=0$ and washes away the Drude peak, leaving only one single off-axis maximum given by the cyclotron resonance of magnetohydrodynamics with momentum relaxation \cite{Hartnoll:TheoryNernstEffect}.

\begin{figure}
	\centering
	\includegraphics[width=0.6\linewidth]{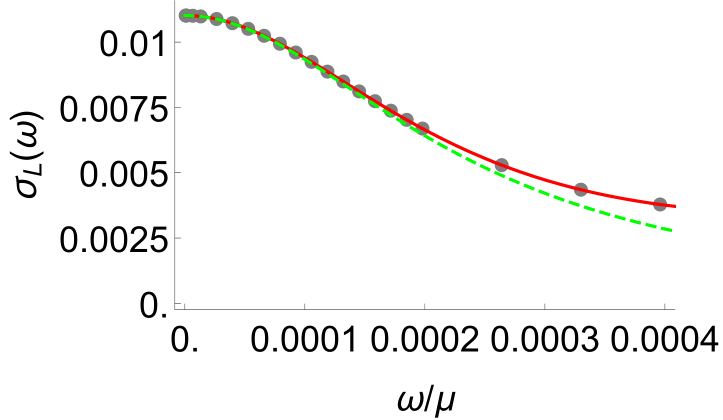}
	\caption{Longitudinal electric conductivity against frequency at small frequencies in the Drude regime for the values of the parameters $\lambda/\mu = -10^{-5}$, $k/\mu=10^{-1}$, $T/\mu = 0.3$, and $B/\mu^2 \approx 0.066$. The solid red line represents the hydrodynamic expression, the grey dots are numerical data, while the dashed Green line shows a pure Drude-like approximation $\sim 1/ (\omega - i \Gamma)$.}
	\label{fig:ch3:drude_emergence}
\end{figure}

Notice that the above definition of the Drude regime should be taken with care. In particular, it is not the standard Drude regime that appears at zero magnetic field in the presence of some effective collective momentum-relaxation rate, see Figure~\ref{fig:ch3:drude_emergence}. Indeed, while momentum decay leads to a single purely imaginary quasinormal mode, the $\omega=0$ peak in what we call Drude regime is given by the superposition of two complex quasinormal modes \cite{Amoretti:HydrodynamicMagnetotransportCharge}. In particular, if one considers hydrodynamics with momentum relaxation and a magnetic field, one does not find a Drude peak at zero frequency, which appears only when the dynamics of the Goldstone fields is relevant, but a cyclotron resonance corresponding to a pair of quasinormal modes.

\subsection{On the spurious pole}
The formalism we employed in this chapter predict an extra gapped pole \cite{Armas:ViscoelasticHydrodynamicsHolography,Armas:HydrodynamicsChargeDensity}, compared to older results \cite{Amoretti:HydrodynamicMagnetotransportCharge}. This pole is related to the fact we are working at non-zero lattice pressure and survives even at zero magnetic field, indeed looking at \eqref{eqn:ch3:conductivities_spontaneous} we see that there is an extra pole in the denominator of the spontaneous hydrodynamic conductivities, which takes the analytic form at $B=0$
\begin{equation}
	\omega=-\frac{i\left(P_f+P_l+\epsilon_f\right)\sigma_{(L)}^\phi}{P_l\left(P_f+\epsilon_f\right)}
\end{equation}
In the original papers in which this formalism was developed \cite{Armas:ViscoelasticHydrodynamicsHolography,Armas:HydrodynamicsChargeDensity} the authors considered a small-frequency expansion of the two-point functions, valid in the hydrodynamic regime at $B=0$, so that this pole effectively disappeared from the spectrum. In the present case, however, we cannot take the same low-frequency expansion, because the presence of the magnetic field gaps the system and a low-frequency expansion would wash away all the information regarding the cyclotron mode.

Instead, what we can do is to work with the weaker definition of what hydrodynamic poles are, as discussed in Section~\ref{sec:ch2:modes}. Namely, all we care about is that this pole (and eventually other non-hydrodynamic poles), appears deep in the complex frequency plane, far from the hydrodynamic cyclotron and phonon modes. Indeed, we checked numerically using holography that this pole has a large imaginary part and does not appear in the spectrum in the region predicted by hydrodynamics, furthermore other quasinormal modes become relevant before any hints of this spurious pole, hence it plays no role in characterizing the diffusive correlators in the hydrodynamic regime.

One possibility is that this spurious pole might appear in the hydrodynamic expressions as an artefact of our frame choice. Indeed, this can happen even in standard hydrodynamics, see e.g. the discussion in \cite{Kovtun:LecturesHydrodynamicFluctuations}, in which the charge-charge two-point function in the Eckart frame shows exactly this unwanted behaviour. To exclude this possibility we computed the hydrodynamic analytic correlators at zero magnetic field in two other frames, a pseudo-Eckart frame (keeping $\gamma_{(L)}$ as the only transport coefficient in the current constitutive relation) and the true Eckart frame, where $J^\mu=nu^\mu$, however even these two frames predict the same lattice-pressure-dependent spurious pole, which does not seem to be related to frame ambiguities.

\section{Summary, discussion and outlook}
In this chapter we developed a hydrodynamic theory for CDW in the presence of a strong external magnetic field $B\sim\mathcal{O}(1)$. We studied both the spontaneous case, in which translation symmetry is spontaneously broken and gives rise to a true Goldstone dynamics, and the pseudo-spontaneous/explicit case, in which spatial translations are broken explicitly by a small parameter, giving origin to pseudo-Goldstone physics.

Because we were interested in matching the hydrodynamic results against a holographic Q-lattice model, we included in our expressions a lattice pressure term, by generalizing the formalism developed in \cite{Armas:ViscoelasticHydrodynamicsHolography,Armas:HydrodynamicsChargeDensity}. Furthermore, to match more easily the numerical results from holography and the analytic hydrodynamic expressions, we followed the method of \cite{Amoretti:MagnetothermalTransportImplies}, based on Ward identities, to express the hydrodynamic transport coefficients in terms of the DC values of the conductivities. Given that the DC transport can be computed analytically from holography in terms of horizon data, we obtained a set of AC hydrodynamic conductivities which are fully determined in terms of horizon data and a single undetermined parameter (in the explicit case), the pinning frequency $\omega_0$, which must be computed numerically in our model.

Our hydrodynamic and holographic results agree excellently over a broad range of values of the parameters, up to small temperature and relatively high magnetic field. Furthermore, at zero lattice pressure and zero magnetic field, we recover all the results previously known in the literature \cite{Armas:HydrodynamicsChargeDensity,Amoretti:HydrodynamicMagnetotransportCharge}.

Since the publication of the paper this chapter is based on, many other research works have been published on the topic \cite{Andrade:ThermoelectricTransportProperties,Armas:ApproximateSymmetriesPseudoGoldstones,Armas:HydrodynamicsPlasticDeformations,Delacretaz:DampingPseudoGoldstoneFields,Ammon:PseudospontaneousSymmetryBreaking}, see in particular the reviews \cite{Baggioli:ColloquiumHydrodynamicsHolography,Amoretti:HydrodynamicsDimensionalStrongly}. These works generalized our results regarding the relation between phonon relaxation and pinning frequency \eqref{eqn:ch3:phonon_relaxation} using locality and the second law of thermodynamics. Furthermore, \cite{Armas:ApproximateSymmetriesPseudoGoldstones,Armas:HydrodynamicsPlasticDeformations} found new transport coefficients that are missing from our pseudo-spontaneous description using an approach based on spurion fields. Nonetheless, the accuracy of our hydrodynamic description in matching the Q-lattice holographic model we considered suggests that these missing transport coefficients are very small in our system.
\chapter{Electrically-driven fluids}\label{chapter:electrically_driven_fluids}
\epigraph{``Nature doesn't care what we call it, she just keeps on doing it whatever ways she wants''}{Feynman, \emph{Feynman Lectures on Physics}}

\section{Introduction}
In this chapter we focus on a very simple physical system comprised of a charged fluid that moves under an externally applied electric field; this is the typical case discussed in the Drude model for a gas of weakly-coupled electrons to describe the conductivity of metals \cite{Ashcroft:SolidStatePhysics,Chaikin:PrinciplesCondensedMatter}. Here we revisit a similar construction, but from the point of view of hydrodynamics: we consider the transport of a charged fluid across a piece of wire with an external linear electric field $\vec{\mathbb{E}}$, we wait a long time until the system reaches a steady states, and we study the stationary regime.

When momentum is an exactly conserved quantity, hydrodynamics predicts that the only stationary solutions in the presence of an external electric field, i.e. the hydrostatic time-independent flows we discussed in Section~\ref{sec:ch2:generating_functional}, are such that spatial gradient of the chemical potential cancel the external field \cite{Kovtun:ThermodynamicsPolarizedRelativistic,Hernandez:RelativisticMagnetohydrodynamics}, see \eqref{eqn:ch2:kovtun_condition}. In this case the electric field effectively disappears from the dynamics, cancelled by the derivative of the chemical potential, while on the other hand the velocity of the fluid is totally unconstrained in magnitude and can take on arbitrary values independent on $\vec{\mathbb{E}}$.

This is very different to what we would expect to find in real devices, which are open systems and, when an electric field is applied externally so that a current starts to flow, the system exchanges heat and momentum with the environment to reach a steady state. In this paper we try to describe this real-world situation, which standard hydrodynamics fails to capture, by focussing on fluids without boost symmetry \cite{deBoer:PerfectFluids,deBoer:NonBoostInvariantFluid,Novak:HydrodynamicsBoosts,Armas:NewtonCartanSubmanifoldsFluid,Armas:EffectiveFieldTheory,deBoer:HydrodynamicModesHomogeneous} and considering stationary time-independent states in which an arbitrary external electric field and the fluid polarization co-exist.

We find a new class of hydrostatic theories that incorporate the presence of an order zero external electric field, without having it constrained to balance the chemical potential gradient, while also including the effect of relaxations in the fluid theory which pin the fluid velocity. These stationary states present a non-zero ground-state DC conductivity and heat transport, suggesting that they are a more accurate representation of what happens in standard DC measurements.

We have already claimed that the only stationary state in standard hydrodynamics, in the presence of a background electric field, can be achieved when the electric field cancels the chemical potential gradient. This is because if this condition fails, then the electric field keeps adding momentum and energy to the system without limit, until the sample melts. To look for more generic solutions we must relax some of the assumptions of hydrodynamics, in particular we will assume that the system can exchange heat and momentum with external sinks, so that some steady state can be reached.

We are agnostic on the possible microscopic origin of the sinks to keep  our hydrodynamic description as general as possible (could be some UV degrees of freedom which, upon integrating out, lead to momentum decay, as was shown in the last chapter in \eqref{eqn:ch3:effective_momentum_relaxation}, or simply the system being open). We take the sinks to be order zero in derivatives and weak, in the sense that they do not affect the thermodynamics of the fluid and are to be understood as small corrections to the equations of hydrodynamics. Furthermore, we will assume that the corresponding susceptibilities, which should be understood as varying with respect to the hidden degrees of freedom, are small and can be ignored, so that the corresponding degrees of freedom are frozen, and their dynamics is not relevant in the hydrodynamic regime. We will however assume that, in general, the relaxation rates might depend on the other thermodynamic quantities, it is only their dependence on the UV degrees of freedom that is hidden (see \cite{Landry:DynamicalChemistryNonequilibrium} for a similar setup related to charge relaxations in the presence of chemical reactions).

Among other things, one of the most important effect of the relaxations/sinks, is that their presence break the Lorentz symmetry, in particular they break boost invariance. To explain this better, consider the case of standard hydrodynamics without momentum relaxation: in this case stationary states (that correspond to global equilibrium, in the absence of external sources) are obtained when the thermodynamic quantities are constant in spacetime $\mu=\text{const}$, $T=\text{const}$ and the velocity too is constant and arbitrary, namely all solutions with different velocities are equally good equilibria, related by boosts (and eventually rotations). However, in the absence of external sources, momentum relaxation pins the equilibrium velocity of the system to zero, which is then the only solution and cannot be related to other solutions via boosts. The same argument holds even with external sources: momentum relaxation and the source still constraint the velocity to take a specific stationary value, breaking the boost symmetry.

We are thus forced to use the boost-agnostic formalism, which means that the spatial velocity is now part of the equilibrium thermodynamics of the system. This should be contrasted with the more standard Galilean or Lorentz symmetry, in which the velocity can be set to zero via boosts, so that is not part of the thermodynamics. In this context, following Drude intuition, we expect to find a stationary state in which the fluid velocity is (implicitly) constrained by the other parameters of the system
\begin{equation}\label{eqn:ch4:velocity_electric_field_implicit}
	v^i=\Omega_\mathbb{E}\left(T,\mu,\vec{v}^2,\vec{\mathbb{E}}^2,\vec{v}\cdot\vec{\mathbb{E}}\right)\mathbb{E}^i-\Omega_\mu\left(T,\mu,\vec{v}^2,\vec{\mathbb{E}}^2,\vec{v}\cdot\vec{\mathbb{E}}\right)\partial^i\mu
\end{equation}
where $\Omega_{\mathbb{E},\mu}$ are, for the moment, undetermined functions. The chemical potential term appears because its gradient acts as an effective electric field for the system that can also drive the fluid.

Although the relaxation rates are taken to be order zero in derivatives, they are still relatively small in amplitudes. This is necessary not to completely destroy the hydrodynamic picture \cite{Amoretti:HydrodynamicsDimensionalStrongly}. This is similar to what happens with order-zero magnetic fields, which must not be so large to confine the dynamics only to the lowest Landau levels, otherwise the hydrodynamic description is not applicable. We can quantify this assumption by looking at the decay of the modes, a feature of many hydrodynamic systems \cite{Grozdanov:HolographyHydrodynamicsWeakly}. In the present context it is due to the effective relaxations, but mode decay also appears in the presence of external magnetic fields (cyclotron modes) \cite{Hartnoll:TheoryNernstEffect}, charge density waves (as discussed in the last Chapter~\ref{chapter:charge_density_waves}) or Wigner crystals \cite{Delacretaz:TheoryCollectiveMagnetophonon,Armas:HydrodynamicsPlasticDeformations}. We remind here the same argument discussed in Section~\ref{sec:ch2:modes}: what is important for the applicability of hydrodynamics is not that the modes are weakly decaying, but rather that the non-hydrodynamic modes are far enough in the complex frequency plane so that they can be ignored for the low-energy dynamics. Because in practice it is not easy to know the position of the first non-hydrodynamic mode, we can set for a more pragmatic approach and simply check whether our hydrodynamic models match with experiments.

We decided to focus on the hydrostatic regime for three reasons: the first one is practical, because it is easier, since we can use simple tools such as the hydrostatic generating functional \cite{Jensen:HydrodynamicsEntropyCurrent}, secondly because of the strong interest in describing the physical situation of a system that reaches a steady state upon application of a constant external electric field, and lastly because the hydrostatic constraints can be regarded as the building blocks to then construct the full hydrodynamic theory. Indeed, dissipative corrections in hydrodynamics can always be understood as fluctuations away from the hydrostatic regime, see the discussion in Section~\ref{sec:ch2:hydrostatic_constraints}.

In this chapter we will follow the standard approach for non-conservative systems: first in Section~\ref{sec:ch4:boost_agnostic_fluids} we will compute the constitutive relations for the ideal boost-agnostic fluid in the presence of a strong electric field in the absence of relaxation rates. To achieve this we can use the hydrostatic generating functional, suitably generalized for the current symmetries. Then, under the assumptions that the only effect of the relaxations is to modify the equations of motion, in Section~\ref{sec:ch4:adding_relaxations} we will add sinks to the system, and we will check how the theory must be modified to account for their presence. Finally, in Section~\ref{sec:ch4:dissipative_corrections} we will sketch the effect of dissipative corrections on our formalism and discuss how the thermoelectric transport of the fluid is modified.

\section{Boost-agnostic fluids}\label{sec:ch4:boost_agnostic_fluids}
We argued for the need of boost-agnostic hydrodynamics to properly describe fluids with momentum relaxation. In this section we explain the formalism and its covariant description in terms of geometric structures. These quantities can then be used to construct the hydrostatic generating functional presented in Section~\ref{sec:ch2:generating_functional} for fluids without any boost symmetry, which allows us to obtain the hydrostatic constitutive relations at order zero and one in derivative expansion.

We will consider the simple case of a fluid with (almost-)conserved stress-energy tensor and $\mathrm{U(1)}$ current, in the presence of a strong background electric field $\mathbb{E}\sim\mathcal{O}(1)$ which polarizes the fluid.

\subsection{Aristotelian geometry}
Aristotelian geometry was first introduced in \cite{Penrose:StructureSpacetime} to describe gravity without boost symmetry. This geometry naturally couples to boost-agnostic fluids and leads, via variational principle, to a generating functional for such hydrodynamic theories. This should be compared to the pseudo-Riemann description that leads to Lorentz-symmetric fluids.

In this geometry, spacetime is a manifold on which two different geometric quantities live: a clock 1-form $\tau_\mu$ and a symmetric covariant tensor $h_{\mu\nu}$ with signature $(0,1,1,\dots,1)$, which acts as a spatial metric. This geometry naturally incorporates space and time on different ground, thus allowing the description of continuous systems which do not have boost symmetry. The physical meaning of $\tau_\mu$ is that of defining a clock in the laboratory frame, which allows us to separate the spatial part of the fluid velocity, as we will see.

This geometry contains Carrollian, Galilean or Lorentzian geometries as suitable limits. In particular, the geometric structures defined above do not have tangent space transformations rule. In the Carrollian case \cite{Hartong:GaugingCarrollAlgebra} it is $\tau_\mu$ which transforms under local boosts, and $h_{\mu\nu}$ does not, while in Newton-Cartan it is the opposite \cite{Jensen:AspectsHotGalilean,Jensen:CouplingGalileaninvariantField}. Riemann geometry instead appears if we require that both $\tau_\mu$ and $h_{\mu\nu}$ transform, but in such a way that the object $\gamma_{\mu\nu}=-\tau_\mu\tau_nu+h_{\mu\nu}$ is invariant under Lorentz boosts.

We can decompose the spatial metric into vielbeins
\begin{equation}
	h_{\mu\nu}=\delta_{ab}e^a_\mu e^b_\nu\qquad a,b=1,\dots,d
\end{equation}
where $d$ is the number of spatial dimensions. This means that, contrary to Riemann geometry, the vielbeins are not square matrices and cannot be inverted directly to reconstruct the inverse metric. This is obvious by the fact that $h_{\mu\nu}$ has a zero eigenvalue and is not a real metric. We then construct the square matrix $(\tau_\mu, e^a_\mu)$, which is instead invertible, and we denote its inverse as $(-\nu^\mu,e^\mu_a)$. The vielbeins, vectors and covectors obey the following relations
\begin{equation}
	\nu^\mu\tau_\mu=-1\qquad\nu^\mu e^a_\mu=0\qquad e^\mu_a\tau_\mu=0\qquad e^\mu_a e^b_\mu=\delta^b_a
\end{equation}
together with the completeness relation, i.e. the requirement that $(-\nu^\mu,e^\mu_a)$ is indeed the inverse of $(\tau_\mu, e^a_\mu)$
\begin{equation}\label{eqn:ch4:completeness_relation}
	-\nu^\mu\tau_\nu+e^\mu_a e^a_\nu = \delta^\mu_\nu
\end{equation}
From the inverse vielbeins we can define a $(2,0)$ tensor
\begin{equation}
	h^{\mu\nu}=\delta^{ab}e^\mu_a e^\nu_b
\end{equation}
Contrary to our Lorentzian intuition, this is not the inverse of $h_{\mu\nu}$, but it is a related object that satisfies
\begin{equation}
	h_{\mu\rho}h^{\rho\nu}=\delta^\nu_\mu+\nu^\nu\tau_\mu
\end{equation}
This last equation shows that $h_{\mu\rho}h^{\rho\nu}$ acts as a projector orthogonal to the clock form $\tau_\mu$. For this reason we can still think of $h^{\mu\nu}$ as the inverse metric of $h_{\mu\nu}$, but only on the spatial hypersurfaces normal to the clock form $\tau_\mu$.

To describe changes in this geometry, we need to provide a notation of covariant derivative. Following our intuition based on Riemannian geometry, we introduce a covariant derivative which is metric compatible with respect to certain spacetime quantities, namely we constrain it to obeys
\begin{equation}
	\nabla_\mu\tau_\nu=0\qquad\nabla_\mu h^{\nu\rho}=0
\end{equation}
This condition fixes the connection to take the form
\begin{equation}
	\Gamma^\lambda_{\mu\nu}=-\nu^\lambda\partial_\mu\tau_\nu+\frac{1}{2}h^{\lambda\kappa}\left(\partial_\mu h_{\nu\kappa}+\partial_\nu h_{\nu\kappa}-\partial_\kappa h_{\mu\nu}\right)+\frac{1}{2}h^{\lambda\sigma}Y_{\sigma\mu\nu}
\end{equation}
where the tensor $Y_{\sigma\mu\nu}$ is arbitrary, but it must satisfy the condition
\begin{equation}
	\left(h^{\rho\sigma}h^{\lambda\nu}-h^{\lambda\sigma}h^{\rho\nu}\right)Y_{\sigma\mu\nu}=0
\end{equation}
This should be compared to pseudo-Riemannian geometry, in which metric compatibility $\nabla_\mu g_{\alpha\beta}=0$ and torsion freedom $\Gamma^\lambda_{[\mu\nu]}=0$ uniquely fix the connection to be the Levi-Civita one.

To proceed and fix the connection uniquely, we impose metric compatibility also on the dual objects $\nu^\mu$ and $h_{\mu\nu}$
\begin{equation}
	\nabla_\mu\nu^\nu=0\qquad\nabla_\mu h_{\nu\rho}=0
\end{equation}
For this particular choice, the affine connection can be written as
\begin{equation}
	\Gamma^\lambda_{\mu\nu}=-\nu^\lambda\partial_\mu\tau_\nu+\frac{1}{2}h^{\lambda\kappa}\left(\partial_\mu h_{\nu\kappa}+\partial_\nu h_{\nu\kappa}-\partial_\kappa h_{\mu\nu}\right)-h^{\lambda\kappa}\tau_\nu K_{\mu\kappa}+C^\lambda_{\mu\nu}
\end{equation}
where $K_{\mu\nu}$ is the extrinsic curvature, defined as
\begin{equation}
	K_{\mu\nu}=-\frac{1}{2}\mathcal{L}_\nu h_{\mu\nu}
\end{equation}
in terms of the Lie derivative along the direction of $\nu^\mu$. The other undefined object is $C^\lambda_{\mu\nu}$, which obeys
\begin{equation}
	C^\lambda_{\mu\nu}\tau_\lambda=0\qquad C^\lambda_{\mu\nu}h_{\lambda\rho}+C^\lambda_{\mu\rho}h_{\nu\lambda}=0
\end{equation}
In particular, without loss of generality, we can now pick $C^\lambda_{\mu\nu}=0$, which is indeed a solution of the above constraints. Our final choice of the connection is thus
\begin{equation}
	\Gamma^\lambda_{\mu\nu}=-\nu^\lambda\partial_\mu\tau_\nu+\frac{1}{2}h^{\lambda\kappa}\left(\partial_\mu h_{\nu\kappa}+\partial_\nu h_{\nu\kappa}-\partial_\kappa h_{\mu\nu}\right)-h^{\lambda\kappa}\tau_\nu K_{\mu\kappa}
\end{equation}
which is not torsion free $\Gamma^\lambda_{[\mu\nu]}\neq0$, hence torsion can appear as a tensor structure to construct hydrostatic scalars.

We will use this curved-space Aristotelian geometry to obtain the hydrostatic constitutive relations, however because we are interested in flat spacetime, at the end of the computations we take the tensor structures to take the following flat-space values
\begin{equation}
	\tau_\mu=\delta^0_\mu\qquad h_{\mu\nu}=\delta^i_\mu\delta^j_\nu\delta_{ij}\qquad\nu^\mu=-\delta^\mu_0\qquad h^{\mu\nu}=\delta^\mu_i\delta^\nu_j\delta^{ij}
\end{equation}
This means that the laboratory is stationary with time form $(1,0,\dots,0)$ and that $h^{\mu\nu}$ is understood as an Euclidean spatial metric. In Cartesian coordinates furthermore, $\nabla_\mu=\partial_\mu$ as the covariant derivative reduces to the partial one.

\subsection{Geometry, thermodynamics and hydrostatic flows}
To construct the hydrostatic generating functional, we first need to express the thermodynamic fields in terms of the basic geometric objects in the theory. In our case we consider the Aristotelian geometry of the previous section, together with a $\mathrm{U(1)}$ gauge field $A_\mu$, hence the full set of geometric quantities which we can use is $(\tau_\mu,h_{\mu\nu},A_\mu)$.

To define the notion of stationarity we need to introduce a timelike Killing vector $\beta^\mu$ that parametrizes time translations, which are an isometry of spacetime. Timelike and future-directed, in this context, means that $\tau_\mu\beta^\mu>0$. This means we must impose that the Lie derivative along the thermal Killing vector must vanish when acting on the geometric quantities of the theory, which will later be interpreted as the sources for the hydrodynamical fields. Namely, we require
\begin{subequations}\label{eqn:ch4:killing_condition_geometry}
	\begin{align}
		\mathcal{L}_\beta\tau_\mu&=0\\
		\mathcal{L}_\beta h_{\mu\nu}&=0\\
		\mathcal{L}_\beta A_\mu&=0
	\end{align}
\end{subequations}
which means that the sources are time-independent and stationary, with respect to $\beta^\mu$. Furthermore, we require that the field $F=\dif A$ obeys the Bianchi identity
\begin{equation}
	\partial_{[\mu}F_{\nu\rho]}=0
\end{equation}

Having identified the geometric objects as sources of the theory, we can now define the thermodynamic variables in terms of the sources and of the Killing vector. Following the usual prescription we define the temperature $T$ and chemical potential $\mu$ as
\begin{subequations}\label{eqn:ch4:geometric_thermodynamics}
	\begin{align}
		T&=\frac{1}{\tau_\mu\beta^\mu}\\
		\mu&=T\left(A_\mu\beta^\mu+\Lambda_\beta\right)
	\end{align}
\end{subequations}
where, exactly as in Section~\ref{sec:ch2:generating_functional}, $\Lambda_\beta$ is a gauge parameter that preserves $\mathrm{U(1)}$ gauge invariance. Finally, we also define the fluid velocity to be proportional to the thermal vector, and we normalize the fluid velocity with the clock form such that
\begin{equation}\label{eqn:ch4:velocity_normalization}
	u^\mu\tau_\mu=1
\end{equation}
which uniquely fixes the proportionality coefficient between the fluid four-velocity and the Killing vector as
\begin{equation}
	u^\mu=T\beta^\mu
\end{equation}
We can also use the completeness relation \eqref{eqn:ch4:completeness_relation}, together with the normalization condition \eqref{eqn:ch4:velocity_normalization}, to separate the spatial part of the fluid velocity
\begin{equation}
	u^\mu=-\nu^\mu+v^ae^\mu_a\qquad v^\mu=u^\mu e_\mu^a
\end{equation}
where we defined the spatial velocity as $\vec{v}=v^ae^\mu_a\partial_\mu$. In the flat spacetime limit the fluid four-velocity simply reduces to $u^\mu=(1,v^i)$. We emphasize that the velocity $v^a$ is in general non-zero due to the presence of the fixed background clock-form $\tau_\mu$, which forbid changes of frame that set the spatial velocity to zero, contrary to what happens in the relativistic or Galilean case.

We also need to properly define the electric field in terms of the field strength $F_{\mu\nu}$. We take the relations between the two to be
\begin{equation}\label{eqn:ch4:electric_field_decomposition}
	\mathbb{E}_\mu=-F_{\mu\nu}\nu^\nu\qquad F_{\mu\nu}=2\partial_{[\mu}A_{\nu]}=\mathbb{E}_\mu\tau_\nu-\mathbb{E}_\nu\tau_\mu\qquad\mathbb{E}_a=e^\mu_a\mathbb{E}_\mu
\end{equation}
which in flat spacetime reduces to $\mathbb{E}_\mu=(0,\mathbb{E}_i)$, as it should. In this definition we decomposed the field strength with respect to the laboratory velocity $\nu^\mu$, therefore the electric field $\mathbb{E}_\mu$ is measured in the laboratory frame. This is different from the usual decomposition in terms of the covariant $E_\mu$ and $B_\mu$ \eqref{eqn:ch2:electric_magnetic_decomposition} used for relativistic systems: usually $E_\mu=F_{\mu\nu}u^\nu$ and $B_\mu$ are the electromagnetic fields in the comoving frame of the fluid, and indeed for non-zero fluid velocity $E_\mu$ can be expressed in terms of both the electric and magnetic field in the laboratory, due to Lorentz symmetry. In particular, in the relativistic decomposition, $E_\mu$ is always normal to the fluid four-velocity $E_\mu u^\mu=0$, which is not true in our laboratory-frame decomposition.

We now have all the tools to recast the Killing conditions \eqref{eqn:ch4:killing_condition_geometry} as constraints that the thermodynamic fields must obey in order for the fluid to be in hydrostatic equilibrium. In particular, from the first equation $\mathcal{L}_\beta \tau_\mu=0$ we obtain
\begin{equation}\label{eqn:ch4:temperature_Killing_condition}
	\mathcal{L}_\beta \tau_\mu=0\qquad\Longrightarrow\qquad\frac{\partial_\mu T}{T}-u^\nu\left(\partial_\nu\tau_\mu-\partial_\mu\tau_\nu\right)=0
\end{equation}
We can also rewrite this expression in an explicitly covariant form as
\begin{equation}
	2u^\nu\Gamma^\rho_{[\nu\mu]}\tau_\rho=\frac{\nabla_\mu T}{T}
\end{equation}
in terms of covariant derivative and torsion\footnote{Indeed, torsion is related to the translation algebra as $[\nabla_\mu,\nabla_\nu]=-T^\rho_{\mu\nu}\nabla_\rho$, hence the result above makes intuitive sense.}. This relation means that certain components of the torsion tensor can be written, in the hydrostatic limit, as temperature derivatives. In the flat spacetime limit, in Cartesian coordinates, this simply reduces to
\begin{equation}
	\partial_\mu T=0
\end{equation}
which means that temperature must be constant in spacetime for the fluid to be stationary.

We can now repeat the same computation for the chemical potential as defined in \eqref{eqn:ch4:geometric_thermodynamics}. Using the Killing equations we find the hydrostatic constraint
\begin{equation}\label{eqn:ch4:curved_space_chemical_constraint}
	\mathcal{L}_\beta A_\mu=0\qquad\Longrightarrow\qquad\mathbb{E}_\mu-T\partial_\mu\frac{\mu}{T}=\mathbb{E}_\nu u^\nu\tau_\mu
\end{equation}
This expression is reminiscent of the corresponding hydrostatic constraint in the Lorentzian case \eqref{eqn:ch2:kovtun_condition}, which we rewrite here for comparison purpose
\begin{equation}
	E_\mu-T\partial_\mu\frac{\mu}{T}=0
\end{equation}
We see that in both cases the external electric field must be compensated by the derivatives of temperature and chemical potential in order for the fluid to be stationary. This makes intuitively sense, since otherwise an imbalanced electric field would drive the fluid out of equilibrium by continuously adding energy and momentum to it. In the flat spacetime limit the above Killing condition simply reduces to
\begin{subequations}
	\begin{align}
		\partial_t\mu+v^i\partial_i\mu&=0\\
		\mathbb{E}_iv^i+\partial_t\mu&=0\\
		\mathbb{E}_i-\partial_i\mu&=0
	\end{align}
\end{subequations}
in which we used that $\partial_\mu T=0$. The first constrain can be obtained from the other two, hence they are not all independent. The first one is simply the usual conservation of scalar quantities along the fluid flow, while the third one expresses the fact that the external electric field must be balanced against the gradient of the chemical potential.

Expression \eqref{eqn:ch4:curved_space_chemical_constraint} and its flat spacetime limit inherit an unexpected feature from its Lorentz counterpart. Specifically, one of the implication of the above formula is that the electric field and the chemical potential gradient must always have the same derivative counting, otherwise they could not cancel on hydrostatic flows \cite{Kovtun:ThermodynamicsPolarizedRelativistic}. This condition is straightforward when the electric field is assumed order one in derivative, because then the chemical potential gradient keeps its naive scaling and must also be order one in derivatives. However, in the present case, as in \cite{Kovtun:ThermodynamicsPolarizedRelativistic}, the electric field is taken to be large and order zero in derivatives. This means that $\partial\mu\sim\mathcal{O}(1)$ must also be order zero in derivative counting, contrary to the naive expectation. We remark that although the two quantities are order zero in derivatives, when the fluid is not in a stationary state the difference is always at least order one in derivatives
\begin{equation}\label{eqn:ch4:derivative_counting_chemical_potential}
	\mathbb{E}_i-\partial_i\mu\sim0+\mathcal{O}(\partial)
\end{equation}
Which means that any dissipative correction, hence any fluctuation away from the hydrostatic regime, must be at least order one in derivatives, following the common expectation.

We can now consider the implication of the Killing conditions on the electric field, and we find
\begin{equation}
	\mathcal{L}_\beta\mathbb{E}_\mu=0\qquad\Longrightarrow\qquad0=u^\nu\partial_\nu\mathbb{E}_\mu+\mathbb{E}_\nu\partial_\mu u^\nu-\frac{\mathbb{E}_\nu u^\nu}{T}\partial_\mu T
\end{equation}
In the flat spacetime limit, in Cartesian coordinates, this reads as
\begin{equation}
	0=\partial_t\mathbb{E}_i+v^j\partial_j\mathbb{E}_i+\mathbb{E}_j\partial_iv^j\qquad\mathbb{E}_i\partial_tv^i=0
\end{equation}
As we mentioned earlier, the electric field is also always constrained by the Bianchi identity $\dif F=0$, which in the decomposition \eqref{eqn:ch4:electric_field_decomposition} reads
\begin{equation}
	0=\partial_{[\mu}\mathbb{E}_{\nu]}\tau_\rho+\mathbb{E}_\nu\partial_{[\mu}\tau_{\rho]}+\partial_{[\nu}\mathbb{E}_{\rho]}\tau_\mu+\mathbb{E}_\mu\partial_{[\rho}\tau_{\nu]}+\partial_{[\rho}\mathbb{E}_{\mu]}\tau_\nu+\mathbb{E}_\rho\partial_{[\nu}\tau_{\mu]}
\end{equation}
and in flat spacetime reduces to the simple requirement
\begin{equation}
	\partial_j\mathbb{E}_i-\partial_i\mathbb{E}_j=0
\end{equation}
namely, the electric field must be irrotational.

Finally, the last quantity we have to worry about is the velocity field. Constraints on $u^\mu$ can be more easily obtained from imposing the Killing condition on the $\nu^\mu$ and $h_{\mu\nu}$. The first one gives
\begin{equation}
	\mathcal{L}_\beta\nu^\mu=0\qquad\Longrightarrow\qquad0=u^\nu\partial_\nu\nu^\mu-\nu^\nu\partial_\nu u^\mu+u^\mu\nu^\nu\frac{\partial_\nu T}{T}
\end{equation}
which, in flat space, becomes
\begin{equation}
	\partial_tv^i=0
\end{equation}
The last constraint comes from imposing the Killing condition on $h_{\mu\nu}$ and reads
\begin{equation}
	\mathcal{L}_\beta h_{\mu\nu}=0\qquad\Longrightarrow\qquad0=\mathcal{L}_uh_{\mu\nu}-\frac{u^\rho}{T}h_{\rho\nu}\partial_\mu T-\frac{u^\rho}{T}h_{\rho\mu}\partial_\nu T=0
\end{equation}
In flat space it reduces to
\begin{equation}
	\partial_iv_j+\partial_jv_i=0
\end{equation}
which corresponds to the vanishing of the shear tensor.

All the conditions we found above are obvious generalizations of the constraints obtained in \ref{sec:ch2:hydrostatic_constraints} for relativistic fluids. Exactly like in the Lorentz case, we can readily see (at least intuitively), that any fluctuation away from these hydrostatic constraints will give rise to dissipative corrections.

\paragraph{Summary of the hydrostatic conditions in flat spacetime.} For ease of reference, we report here all the hydrostatic conditions that we found, specifically in the flat spacetime, Cartesian-coordinates limit.

Any scalar quantity must not change along the fluid flow, hence all scalars are such that $\mathcal{L}_u(\dots)=\left(\partial_t+v^i\partial_i\right)(\dots)=0$. Furthermore, we showed that
\begin{subequations}\label{eqn:ch4:hydrostatic_constraints_flat_space}
	\begin{align}
		\partial_\mu T&=0	&	\partial_i\mathbb{E}_j-\partial_j\mathbb{E}_i&=0	&	\partial_iv_j+\partial_jv_i&=0\\
		\partial_tv^i&=0	&	\partial_t\mathbb{E}_i+v^i\partial_j\mathbb{E}_i+\mathbb{E}_j\partial_iv^j&=0	&	\mathbb{E}_i-\partial_i\mu&=0
	\end{align}
\end{subequations}

\subsection{Constitutive relations and thermodynamics}
We can finally turn to the computation of the constitutive relations for our fluid using the hydrostatic generating functional. To proceed, following the general prescription, we write the most generic functional that depends on the sources $W=W[\tau_\mu,h_{\mu\nu},A_\mu]$ in terms of diffeomorphism and gauge invariant scalars quantities of the theory, constructed from the diffeomorphism and gauge invariant hydrodynamic fields defined in \eqref{eqn:ch4:geometric_thermodynamics}. At order zero in derivatives the generating functional $W_{(0)}$ can be written as
\begin{align}\label{eqn:ch4:order_zero_generating_functional}
	W_0[\tau,h,A,F]&=\int\dif^{d+1}x\ eP(T,\mu,\mathbb{E}^2,u^2,\mathbb{E}\cdot u)\nonumber\\
	&=\int\dif^{d+1}x\ P(T,\mu,\vec{\mathbb{E}}^2,\vec{v}^2,\vec{v}\cdot\vec{E})
\end{align}
where the second line is the flat spacetime limit and $e=\text{det}(\tau_\mu,e_\mu^a)$. Notice that we also included $F=\dif A$ as a source: although obviously it is not independent of $A_\mu$, $F$ (which is order zero in derivatives with our counting) can be understood as the source term for the polarization tensor. The polarization can also be obtained from varying with respect to $A_\mu$, but the computation becomes much more involved.

In the above expression we defined $\mathbb{E}^2=\mathbb{E}_\mu\mathbb{E}_\nu h^{\mu\nu}$ and $u^2=u^\mu u^\nu h_{\mu\nu}$. In general, we will take $h_{\mu\nu}$ and $h^{\mu\nu}$ to raise and lower indices of $\mathbb{E}_\mu$ and $u^\mu$, therefore $u_\nu = u^\mu h_{\mu\nu}$ and $\mathbb{E}^\nu=h^{\mu\nu}\mathbb{E}_\mu$\footnote{We remind that $h^{\mu\nu}u_\nu=h^{\mu\nu}h_{\nu\alpha}u^\alpha\neq u^\mu$, because $h^{\mu\nu}$ is not the inverse matrix of $h_{\mu\nu}$, and similarly for $\mathbb{E}^\mu$.}.

By varying the above generating functional with respect to the sources we find the one-point functions
\begin{equation}\label{eqn:ch4:one_point_functions}
	\delta W_0[\tau,h,A,F]=\int\dif^{d+1}xe\left(-T^\mu\delta\tau_\mu+\frac{1}{2}T^{\mu\nu}\delta h_{\mu\nu}+J^\mu\delta A_\mu+\frac{1}{2}M^{\mu\nu}\delta F_{\mu\nu}\right)
\end{equation}
where we identify $T^\mu$ with the energy current, $T^{\mu\nu}$ with the symmetric stress-momentum tensor, $J^\mu$ is the usual $\mathrm{U(1)}$ charge current and $M^{\mu\nu}$ the antisymmetric polarization-density tensor. As mentioned below \eqref{eqn:ch4:order_zero_generating_functional}, integrating by part the last term gives a contribution to the current proportional to the derivative of the polarization tensor, however keeping them separate in this form is more convenient.

Following \cite{deBoer:NonBoostInvariantFluid} we can construct the standard stress-energy-momentum tensor from a combination of the energy current $T^\mu$ and the stress-momentum tensor $T^{\mu\nu}$
\begin{equation}\label{eqn:ch4:stress_energy_momentum_tensor}
	T^\mu_{\ \nu}=-T^\mu\tau_\nu+T^{\mu\rho}h_{\rho\nu}
\end{equation}
This stress-energy tensor is symmetric only in its spatial indices, because we are assuming isotropy and rotational invariance of the microscopic theory, however it is not symmetric in the time indices, because Lorentz boosts are not a symmetry and the associated Ward identity $T^{\mu0}=T^{0\mu}$ fails.

We can also decompose the polarization tensor following the electromagnetic field strength decomposition \eqref{eqn:ch4:electric_field_decomposition} as \cite{Kovtun:ThermodynamicsPolarizedRelativistic}
\begin{equation}
	M^{\mu\nu}=\nu^\mu\mathbb{P}^\nu-\nu^\mu\mathbb{P}^\mu
\end{equation}
where $\mathbb{P}^\mu$ is the polarization vector, and we are assuming that there are no contributions due to the magnetization, since we are working at zero external magnetic field.

One final comment on the generating functional is in order: in the theory there are no parity-breaking parameters, hence all the terms (and the generating functional itself) are $\mathcal{P}$-even. However, the same is not true for the time-reversal symmetry. Namely, the velocity $v^i$ which is now part of the thermodynamics is $\mathcal{T}$-odd, and consequently the scalar $\vec{\mathbb{E}}\cdot\vec{v}$ is also $\mathcal{T}$-odd, thus the pressure and all other derived thermodynamic quantities do not have a definite sign under time-reversal symmetry. This is similar to the case discussed in \cite{Kovtun:ThermodynamicsPolarizedRelativistic}, in which the scalar $B\cdot E$ breaks time-reversal (and parity). Usually imposing $\mathcal{T}$-reversal symmetry constraints the number of possible terms in the constitutive relations, however in the present case this does not happen and requiring $\mathcal{T}$-symmetry only makes the expressions longer, without any real gain, therefore we will not take the square of the $\mathcal{T}$-odd scalar.

The equations of motion on generic background are obtained by simply imposing that the generating functional $W_{(0)}$ should be diffeomorphism and gauge invariant. We arrive at
\begin{subequations}
	\begin{align}
		e^{-1}\partial_\mu\left(e T^\mu_{\ \rho}\right)+T^\mu\partial_\rho\tau_\mu-\frac{1}{2}T^{\mu\nu}\partial_\rho h_{\mu\nu}&=F_{\rho\mu}J^\mu\\
		e^{-1}\partial_\mu\left(eJ^\mu\right)&=0
	\end{align}
\end{subequations}
which, in the flat spacetime limit, reduce to the standard hydrodynamic conservation equations without relaxations
\begin{equation}
	\partial_\mu T^\mu_{\ \nu}=F_{\nu\lambda}J^\lambda\qquad\qquad\partial_\mu J^\mu=0
\end{equation}

To compute the ideal-fluid constitutive relations, we need to vary the generating functional \eqref{eqn:ch4:order_zero_generating_functional} with respect to the sources as in \eqref{eqn:ch4:one_point_functions}. We employ the following variations for the hydrodynamic and geometric fields in terms of the sources
\begin{subequations}
	\begin{align}
		\delta T&=-Tu^\mu\delta\tau_\mu\\
		\delta u^\mu&=-u^\mu u^\nu\delta\tau_\nu\\
		\delta\mathbb{E}_\mu&=\mathbb{E}_\mu\nu^\rho\delta\tau_\rho-\tau_\mu\mathbb{E}_\nu h^{\nu(\rho}\nu^{\sigma)}\delta h_{\rho\sigma}-\left(\partial_\mu\delta A_\nu-\partial_\nu\delta A_\mu\right)\nu^\nu\\
		\delta h^{\mu\nu}&=-h^{\mu\rho}\delta h_{\rho\sigma}h^{\sigma\nu}+2\nu^{(\mu}h^{\nu)\rho}\delta\tau_\rho\\
		\delta\nu^\mu&=\nu^\mu\nu^\nu\delta\tau_\nu-h^{\mu(\nu}\nu^{\rho)}\delta_{\nu\rho}\\
		\delta\mu&=-\mu u^\mu\delta\tau_\mu+u^\mu\delta A_\mu\\
		\delta e&=e\left(-\nu^\mu\delta\tau_\mu+\frac{1}{2}h^{\mu\nu}\delta h_{\mu\nu}\right)
	\end{align}
\end{subequations}
Doing so, we obtain the following expressions for the constitutive relations of an ideal fluid on a generic curved spacetime
\begin{subequations}
	\begin{align}
		T^\mu&=\epsilon u^\mu+\left(P-\mathbb{P}\cdot\mathbb{E}\right)h^{\mu\rho}u_\rho\\
		T^{\mu\nu}&=Ph^{\mu\nu}+\rho_m u^\mu u^\nu-\kappa_\mathbb{E}\mathbb{E}^\mu \mathbb{E}^\nu-2\beta_\mathbb{P}\mathbb{E}^{(\mu)}\nu^{\nu)}
	\end{align}
\end{subequations}
which, upon using \eqref{eqn:ch4:stress_energy_momentum_tensor}, give the final result for the stress-energy tensor and the current
\begin{subequations}\label{eqn:ch4:ideal_fluid_constitutive_relations}
	\begin{align}
		T^\mu_{\ \nu}&=-\epsilon u^\mu\tau_\nu-\left(P-\mathbb{P}\cdot\mathbb{E}\right)h^{\mu\rho}u_\rho\tau_\nu+P h^{\mu\rho}h_{\rho\nu}+\rho_m u^\mu u_\nu \nonumber\\
		&\quad-\kappa_\mathbb{E}\mathbb{E}^\mu\mathbb{E}^\rho h_{\rho\nu}-\beta_\mathbb{P}\mathbb{E}^\rho\nu^\mu h_{\rho\nu}\\
		J^\mu&=n u^\mu+\frac{1}{e}\partial_\nu\left(2e\nu^{[\mu}\mathbb{P}^{\nu]}\right)
	\end{align}
\end{subequations}
Observe that, because the electric field is order zero in derivatives, the current is separated in two pieces: the first term proportional to the velocity is the standard free current, related to the transport of free charges; the second term is the bound current and, because it is antisymmetric in its indices, it identically solves the conservation equation for the current. This means that free and bound charges are separately conserved, and is because of the hidden winding symmetry that implies the polarization current is conserved identically off shell, in the QFT context. The polarization current is peculiar, because it is of derivative order, even if we are considering only ideal fluids. This feature appears every time we have an order zero electric or magnetic field, see e.g. \cite{Kovtun:ThermodynamicsPolarizedRelativistic,Hernandez:RelativisticMagnetohydrodynamics} for an example in relativistic hydrodynamics.

In equation \eqref{eqn:ch4:ideal_fluid_constitutive_relations}, we define the fluid pressure $P$, the energy density $\epsilon$, the momentum susceptibility (or kinetic mass density in \cite{deBoer:PerfectFluids}) $\rho_m$, the density of free charges $n$, and $s$ the entropy density. In equilibrium these are defined in terms of the derivatives of the grand-canonical potential $P$ as
\begin{subequations}\label{eqn:ch4:thermodynamics_derivatives}
	\begin{align}
		n&=\frac{\partial P}{\partial\mu}\biggr|	&	\rho_m&=2\frac{\partial P}{\partial\vec{v}^2}\biggr|	&	s&=\frac{\partial P}{\partial T}\biggr|\\
		\beta_\mathbb{P}&=\frac{\partial P}{\partial(\vec{\mathbb{E}}\cdot\vec{v})}\biggr|	&	\kappa_\mathbb{E}&=2\frac{\partial P}{\partial\vec{\mathbb{E}}^2}\biggr|
	\end{align}
\end{subequations}
where the vertical bar means the derivatives are computed keeping all the other thermodynamic variables fixed. The free energy is defined so that the following thermodynamic equation is obeyed
\begin{equation}
	\epsilon+P=sT+\mu n+\rho_m\vec{v}^2+\kappa_\mathbb{E}\vec{\mathbb{E}}^2+2\beta_\mathbb{P}\vec{\mathbb{E}}\cdot\vec{v}
\end{equation}

The two thermodynamic quantities $\kappa_\mathbb{E}$ and $\beta_\mathbb{P}$ were discussed for the first time in \cite{Amoretti:NondissipativeElectricallyDriven}, and we can understand their physical meaning from \eqref{eqn:ch4:thermodynamics_derivatives}. The function $\kappa_\mathbb{E}$ is interpreted as the non-linear electric susceptibility, and is different from the standard thermodynamic susceptibilities, which are second derivatives of the grand canonical potential $\chi=\frac{\partial^2P}{\partial\lambda^2}$, see \eqref{eqn:ch2:static_susceptibility}. On the other hand, the role of $\beta_\mathbb{P}$ can be understood by remembering that $\vec{J}\cdot\vec{\mathbb{E}}\sim \vec{v}\cdot\vec{\mathbb{E}}$ is the work done on the system by the electric field. Then, the inverse derivative $\frac{\partial(\vec{v}\cdot\vec{\mathbb{E}})}{\partial P}$ is a quantifier of how the work done on the system per unit charge changes as we change the pressure. Thus, we can interpret $\beta_\mathbb{P}$ as compressibility caused by polarization.

We can express, in thermodynamic equilibrium, the momentum density $\vec{P}$ and polarization $\vec{\mathbb{P}}$ as
\begin{subequations}
	\begin{align}
		\vec{\mathbb{P}}&=\frac{\partial P}{\partial\vec{\mathbb{E}}}\biggr|=\kappa_\mathbb{E}\vec{\mathbb{E}}+\beta_\mathbb{P}\vec{v}\\
		\vec{P}&=\frac{\partial P}{\partial\vec{v}}\biggr|=\rho_m\vec{v}+\beta_\mathbb{P}\vec{\mathbb{E}}
	\end{align}
\end{subequations}
from which we see that the system can have a non-zero momentum at zero velocity, and similarly a non-zero polarization at vanishing external electric field. This happens because our system is not time-reversal invariant: had we kept $(\vec{\mathbb{E}}\cdot\vec{v})^2$ in the pressure, thus making it $\mathcal{T}$-even, we would have found that the polarization always vanishes at zero electric field. This is again similar to what happens in \cite{Kovtun:ThermodynamicsPolarizedRelativistic}, in which the scalar $E\cdot B$ breaks the time-reversal symmetry and for this reason the system can have a non-zero polarization at zero electric field, because the magnetic field induces a polarization too.

We can rewrite the Euler relation as
\begin{equation}\label{eqn:ch4:euler_relation}
	\epsilon+P=sT+\vec{\mathbb{E}}\cdot\vec{\mathbb{P}}+\vec{v}\cdot\vec{P}+n\mu
\end{equation}
And from here, obtain the Gibbs-Duhem equation in the form
\begin{equation}\label{eqn:ch4:gibbs_duhem_relation}
	\dif P=s\dif T+n\dif\mu+P_i\dif v^i+\mathbb{P}^i\dif\mathbb{E}_i
\end{equation}

We remark that the constitutive relations \eqref{eqn:ch4:ideal_fluid_constitutive_relations} are peculiar, in the sense that usually the velocity is an eigenvector of the stress-energy tensor for ideal fluids, and the total energy density its eigenvalue. Indeed, one of the matching conditions of the Landau frame \eqref{eqn:ch2:landau_matching_conditions} is to extend this ideal fluid relation to all orders in derivative expansion. In the present case however one finds
\begin{equation}
	T^\mu_{\ \nu}u^\nu=-\left(\epsilon-\rho_m u^2-\mathbb{E}\cdot\mathbb{P}\right)u^\mu+\left(\mathbb{P}\cdot\mathbb{E}-\beta_\mathbb{P}\mathbb{E}\cdot u\right)\nu^\mu-\kappa_\mathbb{E}(\mathbb{E}\cdot u)\mathbb{E}^\mu
\end{equation}
in which the velocity is an eigenvector with the energy density as eigenvalue only at zero electric field. Thus, for an order-zero in derivatives electric field, the Landau frame does not exist, since it is violated already at the level of the ideal fluid.

In the flat spacetime, Cartesian-coordinates limit the constitutive relations \eqref{eqn:ch4:ideal_fluid_constitutive_relations} reduce to
\begin{subequations}\label{eqn:ch4:ideal_fluid_flat_spacetime_constitutive_relation}
\begin{align}
	T^0_{\ 0}&=-\epsilon\\
	T^0_{\ i}&=\rho_mv_i+\beta_\mathbb{P}\mathbb{E}_i=P_i\\
	T^i_{\ 0}&=-\left(\epsilon+P-\vec{\mathbb{P}}\cdot\vec{\mathbb{E}}\right)v^i\\
	T^i_{\ j}&=P\delta^i_j+\rho_m v^iv_j-\kappa_\mathbb{E}\mathbb{E}^i\mathbb{E}_j\\
	J^0&=n-\partial_j\mathbb{P}^j\\
	J^i&=nv^i+\partial_t\mathbb{P}^i
\end{align}	
\end{subequations}
in which we see explicitly that the energy flux $T^i_{\ 0}$ is different from the momentum density $T^0_{\ i}$.

These constitutive relations, and their respective curved spacetime form \eqref{eqn:ch4:ideal_fluid_constitutive_relations}, identically solve the equations of hydrodynamics after imposing the hydrostatic constraints \eqref{eqn:ch4:hydrostatic_constraints_flat_space}. This means that any fluid flow that is stationary will be a solution to the equations of hydrodynamic without further constraints.

\section{Adding relaxations}\label{sec:ch4:adding_relaxations}
As we showed explicitly in \eqref{eqn:ch4:curved_space_chemical_constraint}, for a fluid to be stationary in the presence of a background external electric field, the chemical potential must be arranged in such a way that its gradient counter-balances the electric field. This, however, means that the velocity, provided that it obeys the hydrostatic constraints \eqref{eqn:ch4:hydrostatic_constraints_flat_space}, is completely free and unrelated to the applied electric field, hence we do not find any stationary state driven by the electric field. To reach such steady states we need to include relaxations in the equations of hydrodynamics, which we will do in this Section, and modify the hydrostatic constraints accordingly.

\subsection{Order zero}
Following the EFT prescription, we add the most general relaxation terms consistent with the ideal fluid data, such as energy and charge density, momentum, etc\dots\ At order zero in derivatives all the terms we can construct do not produce entropy, since they do not vanish on hydrostatic configurations (it is only terms which do vanish on the hydrostatic constraints that are associated with entropy production). It is only at order one in derivatives that we can consider dissipative corrections, which could then also appear in the relaxation terms, see the recent work \cite{Gouteraux:DrudeTransportHydrodynamic}. This means that at order zero in derivatives we can have relaxation terms which do not vanish in the hydrostatic regime, while at order one we can have relaxations proportional to the hydrostatic constraints, e.g. $\partial_\mu T$, and that are zero when the fluid is in a stationary state. The former kind can, in principle, lead to a modification of the hydrostatic constraints, as we will see.

We now focus on the constitutive relations \eqref{eqn:ch4:ideal_fluid_flat_spacetime_constitutive_relation} and (non-)conservation equations in the flat spacetime limit and modify them accordingly to include the effect of the sinks
\begin{subequations}\label{eqn:ch4:equations_of_motion_relaxations}
	\begin{align}
		0&=\partial_t\epsilon+\partial_iJ^i_\epsilon-\mathbb{E}_iJ^i+\hat\Gamma_\epsilon\\
		0&=\partial_tP_i+\partial_jT^j_{\ i}-nJ_i+\hat\Gamma_{P,i}\\
		0&=\partial_tJ^0+\partial_iJ^i
	\end{align}
\end{subequations}
Here we defined the energy current $J^i_\epsilon=T^i_{\ 0}$. Notice that we are not considering charge relaxation effects.

As a comment, we remind that our fluid is boost-agnostic only because of the presence of relaxation rates, as discussed in the introduction of this chapter. Therefore, the system could initially be Galilean or Lorentzian, but stop fulfilling the relevant Ward identities after relaxation rates are introduced. Because of this, it is reasonable to expect that the thermodynamic parameters specific to boost-agnostic fluid should be small, of the scale of the relaxation rates $\Gamma$.

The most general choice compatible with the EFT prescription is to expand the momentum relaxation $\hat\Gamma_P^i$ in terms of the order zero vector of the theory, while $\hat\Gamma_\epsilon$ can be a generic function of the scalars
\begin{equation}
	\hat\Gamma^i_P=\Gamma_\mathbb{P}\mathbb{P}^i+\Gamma_PP^i+\mathcal{O}(\partial)\qquad\hat\Gamma_\epsilon=\Gamma_\epsilon+\mathcal{O}(\partial)
\end{equation}
In principle, we also have another vector at order zero, namely $\partial^i\mu$, but as we discussed earlier it can always be exchanged for the electric field on stationary states, and this fact survives even in the presence of relaxations. We included decay terms up to order zero in derivatives, since we are considering ideal fluids. This means that the relaxation terms appear at order zero in the equations of motion, while the other terms, coming from the gradients of the fluid four-currents, can be order zero or one in derivatives.

If we naively apply the constitutive relations to our new modified equations of motion \eqref{eqn:ch4:equations_of_motion_relaxations}, while imposing the hydrostatic constraints \eqref{eqn:ch4:hydrostatic_constraints_flat_space}, we clearly find that the relaxation terms must vanish for the equations of motion to be satisfied identically, as expected\footnote{This is because the equations of motions for ideal fluids can be written as linear combinations of Lie derivatives \cite{deBoer:NonBoostInvariantFluid}, so  that they are identically satisfied when we impose the Killing conditions \eqref{eqn:ch4:killing_condition_geometry}.}. However, we can consider a second possibility, by looking at the equations of motion order by order in derivatives: at order one there are no relaxation rates in the equations of motion (remember, the sinks are order zero terms), hence the terms coming from the currents must cancel with themselves in the hydrostatic limit. This is indeed what happens, provided that we use all the hydrostatic constraints in \eqref{eqn:ch4:hydrostatic_constraints_flat_space} except for the last one, which is not needed.

Next, we can look at the equations of motion at order zero in derivatives, the order at which our relaxation rates appear. We use all the hydrostatic constraints already employed to solve the order-one equations identically, in particular the vanishing of the Lie derivatives of scalar quantities along the fluid flow. Thus, at the end of the computation what remains is
\begin{subequations}\label{eqn:ch4:order_zero_equations_motion_relaxations}
	\begin{align}
		nv^i\left(\mathbb{E}_i-\partial_i\mu\right)&=\Gamma_\epsilon+\mathcal{O}(\partial)\\
		n\left(\mathbb{E}_i-\partial_i\mu\right)&=\Gamma_\mathbb{P}\mathbb{P}_i+\Gamma_PP_i+\mathcal{O}(\partial)
	\end{align}
\end{subequations}
respectively for the energy and momentum equations of motion. As we remarked above, it is immediately clear that if we also impose the last hydrostatic requirement $\mathbb{E}_i-\partial_i\mu=0$, then the relaxation rates must all cancel with each other. However, this is an unphysical solution: we would find that
\begin{subequations}
	\begin{align}
		\Gamma_\mathbb{P}\mathbb{P}_i+\Gamma_PP_i&=0\\
		\Gamma_\epsilon&=0
	\end{align}
\end{subequations}
indeed leads to a relation that implicitly links the fluid velocity to the external electric field
\begin{equation}
	v^i=-\left(\frac{\kappa_\mathbb{E}\Gamma_\mathbb{P}+\beta_\mathbb{P}\Gamma_P}{\beta_\mathbb{P}\Gamma_\mathbb{P}+\rho_m\Gamma_P}\right)\mathbb{E}^i
\end{equation}
So the fluid would be dragged along by the electric field, but would somehow produce no heat in the process, as the energy sink is set to zero $\Gamma_\epsilon=0$.

For this reason we focus on a second possibility, specifically that the last hydrostatic constraint $\mathbb{E}_i-\partial_i\mu=0$ must be modified to accommodate the presence of relaxation rates. First, we observe that the energy and momentum relaxation rates are not independent in \eqref{eqn:ch4:order_zero_equations_motion_relaxations}: by projecting the momentum equation along the velocity we find that the LHS is the same for the two equations and so
\begin{align}\label{eqn:ch4:energy_momentum_relaxations_relation}
	\Gamma_\epsilon&=v^i\left(\Gamma_PP_i+\Gamma_\mathbb{P}\mathbb{P}_i\right)\nonumber\\
	&=\left(\rho_m\Gamma_P+\beta_\mathbb{P}\Gamma_\mathbb{P}\right)\vec{v}^2+\left(\beta_\mathbb{P}\Gamma_P+\kappa_\mathbb{E}\Gamma_\mathbb{P}\right)\vec{v}\cdot\vec{\mathbb{E}}
\end{align}
This relation between energy and momentum relaxation is a kinematic constraint that must be obeyed for the system to reach hydrostatic equilibrium, and can be interpreted as a requirement of stationarity for the sinks. We will see later that this condition can also be obtained by the requirement that the ideal fluid does not generate entropy, and thus holds more generally than just hydrostatic flows. Notice that a similar relation holds for the simple Drude model, if we require the energy of the system not to blow up.

Assuming the validity of the above condition, we modify the hydrostatic constraints involving the electric field to also include the momentum relaxation
\begin{equation}\label{eqn:ch4:hydrostatic_condition_relaxations}
	\mathbb{E}_i-\partial_i\mu=0\qquad\Longrightarrow\qquad n\left(\mathbb{E}_i-\partial_i\mu\right)=\Gamma_\mathbb{P}\mathbb{P}_i+\Gamma_PP_i
\end{equation}
We can arrange \eqref{eqn:ch4:hydrostatic_condition_relaxations} in the same form of \eqref{eqn:ch4:velocity_electric_field_implicit}, explicitly
\begin{equation}
	\vec{v}=\left(\frac{n-\kappa_\mathbb{E}\Gamma_\mathbb{P}-\beta_\mathbb{P}\Gamma_P}{\beta_\mathbb{P}\Gamma_\mathbb{P}+\rho_m\Gamma_P}\right)\vec{\mathbb{E}}-\frac{n}{\beta_\mathbb{P}\Gamma_\mathbb{P}+\rho_m\Gamma_P}\vec{\partial}\mu+\mathcal{O}(\partial)
\end{equation}
leading to the identifications
\begin{subequations}
	\begin{align}
		\Omega_\mathbb{E}&=\left(\frac{n-\kappa_\mathbb{E}\Gamma_\mathbb{P}-\beta_\mathbb{P}\Gamma_P}{\beta_\mathbb{P}\Gamma_\mathbb{P}+\rho_m\Gamma_P}\right)\\
		\Omega_\mu&=\frac{n}{\beta_\mathbb{P}\Gamma_\mathbb{P}+\rho_m\Gamma_P}
	\end{align}
\end{subequations}
Notice in particular that $\Omega_\mu\neq0$ only if $n\neq0$.

In the presence of external sources the system never reaches global thermodynamic equilibrium, but rather relax to hydrostatic solutions. Remember that, in the context of General Relativity \cite{Carroll:SpacetimeGeometry}, static means that the system is time-independent and symmetric under time reversal. On the other hand, stationary refers only to the first property, and does not require the system to look the same under $\mathcal{T}$-symmetry. From this perspective our solution is perfectly stationary (there exists a timelike Killing vector field), although not static.

The condition \eqref{eqn:ch4:energy_momentum_relaxations_relation} can be understood in the same way to what happens to any external source in hydrostatic equilibrium: the sources cannot vary arbitrarily, but are constrained by the Killing condition, if a stationary state is to be reached at all. Once the external sources are time independent, meaning $\mathcal{L}_\beta(\dots)=0$, then the fluid can finally relax to a hydrostatic flow and the fluid variables too will slowly start obeying the hydrostatic constraints.

We can also understand better \eqref{eqn:ch4:hydrostatic_condition_relaxations} by following the hydrodynamic approach of Chapter~\ref{chapter:hydrodynamics_linear_response}. Without any external source, the hydrodynamic EFT prescription leads us to the constitutive relations \eqref{eqn:ch2:landau_constitutive_relations} in the Landau frame. There, the chemical potential sits alone in the current constitutive relations and stationary states are reached when $\partial_\nu\mu=0$. We can now ask what happens when we also include an electric field in the theory, and the answer are the constitutive relations in \eqref{eqn:ch2:constitutive_relations_curved_space}: the external source field $E_\mu$ now sits together with the gradient of the chemical potential in the current one-point function, thus a steady state is reached only if the dissipative term vanishes $E_\mu-T\partial_\mu\frac{\mu}{T}=0$, leading us to the hydrostatic constraint. It should not come as a surprise then that if we add other sources (or sinks in this case) that couple to the fluid degrees of freedom the hydrostatic conditions should be modified.

There are however a couple of differences with respect to adding a $\mathrm{U(1)}$ gauge field: first, the external gauge field does not depend on the thermodynamics variables, while the relaxation rates $\Gamma$ might do. Secondly, the $\mathrm{U(1)}$ gauge field is a conservative field, which can be included in the theory via exact arguments based on an action principle, while the relaxation rates are non-conservative effects that break diffeomorphism invariance. This means that to include them in the theory we must either add them by hand and then study them via heuristic means, like in this chapter, or consider a specific realization and subsequently freeze the dynamic of the field that break diffeomorphism to obtain an effective relaxation, as was done in the last chapter around \eqref{eqn:ch3:effective_momentum_relaxation}.

\subsection{Order one}
In standard hydrodynamics, as in Chapter~\ref{chapter:hydrodynamics_linear_response}, if the constitutive relations are at order $\mathcal{O}(\partial^n)$ in derivative expansion, then the equations of motion are up to order $\mathcal{O}(\partial^{n+1})$ in derivatives, and the formal derivative series of hydrodynamic means that each order should be solved independently. This is what happens for example when applying the hydrostatic conditions to stationary flows, in which each order in derivative expansion identically obeys the conservation equations.

However, in the presence of order zero external fields this is not true, and we have mixing of derivatives. Constitutive relations at order $\mathcal{O}(\partial^n)$ contains terms up to order $\mathcal{O}(\partial^{n+1})$ in derivatives, due to polarization effects, while the equations of motion contain terms from $\mathcal{O}(1)$ to $\mathcal{O}(\partial^{n+1})$.

When we consider first-order corrections to the ideal fluid, we must also take into account first-order terms into the relaxation rates, in particular both $\mathbb{P}_i$ and $P_i$ receive derivative contributions at higher orders. These corrections to the relaxation terms will appear at order one in the equations of motion; hence, provided they obey certain integrability constraints, they can be in principle cancelled by appropriate redefinitions of the order-zero stress-energy tensor. For this reason, (some of) the order-one relaxations are unphysical, since they can be redefined away. If they cannot be reabsorbed in the constitutive relations, then they might lead to further modifications of the hydrostatic conditions \eqref{eqn:ch4:hydrostatic_condition_relaxations} or the other constraints \eqref{eqn:ch4:hydrostatic_constraints_flat_space}.

We can start the analysis following the same steps used to study the ideal fluid: we first focus on the constitutive relations in the absence of relaxation rates at order one in derivatives, then we add the relaxation rate, and check for consistency. At order one in derivatives there are 14 scalars that we can use to construct the hydrostatic generating functional in terms of the geometric sources $(\tau_\mu,h_{\mu\nu},A_\mu)$. One possible choice for these scalars is given by
\begin{align}
	s_{(1)}&=\left\{\nu^\mu\partial_\mu\left(T,u^2,\mathbb{E}^2,u\cdot\mathbb{E}\right),\mathbb{E}_\nu h^{\nu\mu}\partial_\mu\left(T,u^2,\mathbb{E}^2,u\cdot\mathbb{E}\right),\right.\nonumber\\
	&\qquad u^\mu u^\nu\nabla_\mu\mathbb{E}_\nu,\nu^\mu u^\nu\nabla_\mu\mathbb{E}_\nu,\mathbb{E}^\mu u^\nu\nabla_\mu\mathbb{E}_\nu,\nabla_\mu\mathbb{E}^\mu,\nonumber\\
	&\qquad\left.\partial_\mu\tau_\nu\nu^{[\mu}h^{\nu]\sigma}\mathbb{E}_\sigma,\partial_\mu\left(h_{\nu\sigma}u^\sigma\right)\nu^{[\mu}h^{\nu]\rho}\mathbb{E}_\rho\right\}
\end{align}
which lead to the following form for the first order corrections to the generating functional \eqref{eqn:ch2:generating_functional}
\begin{equation}\label{eqn:ch4:order_one_generating_functional}
	W_1[\tau,h,A]=\int\dif^{d+1}x\ e\sum_{i=1}^{14}F_i\left(T,\mu,\mathbb{E}^2,u^2,\mathbb{E}\cdot u\right)s_{(1)}^i
\end{equation}
that should be added to $W_{(0)}$, used to compute the ideal fluid constitutive relations \eqref{eqn:ch4:order_zero_generating_functional}. Notice that we did not include $\nu^\mu\partial_\mu\mu$ and $\mathbb{E}^\nu\partial_\nu\mu$ in the list of scalars, since these can always be traded for terms related to the electric field \cite{Kovtun:ThermodynamicsPolarizedRelativistic}.

We vary this functional with respect to the sources according to \eqref{eqn:ch4:one_point_functions} to obtain the first order corrections to the stress-energy tensor and the current. At the end we take the flat spacetime limit, which is the regime we are interested in when we add the relaxations. In flat spacetime the 14 scalars reduce significantly
\begin{subequations}
	\begin{align}
		s_1&=s_2=s_5=s_{13}=s_{14}=0\\
		s_3&=2s_8=2s_{11}=v^i\partial_i\vec{\mathbb{E}}^2\equiv \bar{s}_1\\
		s_4&=s_{10}=v^i\partial_i\left(\vec{\mathbb{E}}\cdot\vec{v}\right)\equiv\bar{s}_2\\
		s_6&=2s_9=\mathbb{E}^i\partial_i\vec{v}^2\equiv\bar{s}_3\\
		s_7&=\mathbb{E}^i\partial_i\vec{\mathbb{E}}^2\equiv\bar{s}_4\\
		s_{12}&=\partial_i\mathbb{E}^i\equiv\bar{s}_5
	\end{align}
\end{subequations}
with only 5 independent order-one scalars left. For the non-composite vectors at order one in derivatives we take the following basis
\begin{subequations}
	\begin{align}
		\bar{v}_1&=\partial_i\vec{v}^2\\
		\bar{v}_2&=\partial_i\vec{\mathbb{E}}^2\\
		\bar{v}_3&=\partial_i\left(\vec{v}\cdot\vec{\mathbb{E}}\right)\\
		\bar{v}_4&=v^j\partial_i\mathbb{E}_j
	\end{align}
\end{subequations}
The final result for the order-one constitutive relations are long and not very illuminating, so we refer to the original paper for further details \cite{Amoretti:NondissipativeElectricallyDriven}.

We comment here on a problem related to derivatives of the chemical potential. Given the order-one generating functional \eqref{eqn:ch4:order_one_generating_functional} we expect to find in the constitutive relations terms which are order zero in derivatives, e.g. $\frac{\partial F_i}{\partial\mu}\partial_j\mu$, since the derivative of the chemical potential is order zero $\partial\mu\sim\mathcal{O}(1)$, as already discussed near \eqref{eqn:ch4:derivative_counting_chemical_potential}. This means that the constitutive relations at order zero are influenced by the generating functional at order one, and more generally the constitutive relations at order $n$ will contain terms coming from the generating functional at order $n+1$, rendering the formal gradient expansion of hydrodynamics problematic.

This issue already arises in the relativistic case \cite{Kovtun:ThermodynamicsPolarizedRelativistic}, where we remind that the gradient of $\mu$ must be order zero to balance the order-zero electric field. To solve this problem, in \cite{Kovtun:ThermodynamicsPolarizedRelativistic} it was assumed that $\frac{\partial F}{\partial\mu}\sim\mathcal{O}(\partial)$, so that these terms do not affect the ideal fluid constitutive relations. However, in the presence of relaxations there is another possibility we can explore, namely we can consider the more natural approach to take $\partial\mu\sim\mathcal{O}(\partial)$. This can be done thanks to our modified hydrostatic constrain
\begin{equation}
	n\left(\mathbb{E}_i-\partial_i\mu\right)-\Gamma_\mathbb{P}\mathbb{P}_i-\Gamma_PP_i=0
\end{equation}
Without relaxations the gradient of the chemical potential must be of the same derivative order of the electric field, but with relaxations we can take $\partial\mu$ order one in derivatives and balance the electric field against the sinks.

In \cite{Amoretti:NondissipativeElectricallyDriven} we considered the most general expansion for the order-one momentum relaxation as
\begin{equation}
	\hat\Gamma^i_P=\left(\Gamma_P+\sum_{j=1}^5\Gamma_{P,j}\bar{s}_j\right)P^i+\left(\Gamma_\mathbb{P}+\sum_{j=1}^5\Gamma_{\mathbb{P},j}\bar{s}_j\right)\mathbb{P}^i+\sum_{j=1}^4\Gamma_j(\bar{v}_j)^i+\mathcal{O}(\partial^2)
\end{equation}
where we remind that both $\mathbb{P}^i$ and $P^i$ receives derivative corrections. We then assumed that even at order one, all the hydrostatic conditions were preserved. Thus, we still need to modify only the constraint \eqref{eqn:ch4:hydrostatic_condition_relaxations}, and took the above expression to appear on the RHS of the hydrostatic constraint \eqref{eqn:ch4:hydrostatic_condition_relaxations}. Had we chosen $v^i$ and $\mathbb{E}^i$ instead of $P^i$ and $\mathbb{P}^i$ as our basis vectors to write the relaxations, we would not have found order-one corrections to the relaxations. This might be related to frame transformations, however it is not clear at the present stage.

Finally, we remark that there are other possibilities for how to deal with the order-one corrections to the relaxations rates: a full study of the first-order corrections is still in progress, and we refer to the original paper for some further preliminary results and discussions \cite{Amoretti:NondissipativeElectricallyDriven}.

\subsection{Entropy current}
Because our constitutive relations and equations of motion do not simply follow from the hydrostatic generating functional, since we added the relaxations by hand, positivity of entropy production of the ideal fluid is not automatically true and must be checked. We proceed in the standard way to compute the canonical entropy current from the hydrodynamic equations: we contract the momentum conservation equation with $v^i$ and subtract it to the energy conservation equation
\begin{multline}
	\left(\partial_t+v^i\partial_i\right)\left(\epsilon-\vec{\mathbb{E}}\cdot\vec{\mathbb{P}}\right)+\left(\epsilon+P-\vec{\mathbb{P}}\cdot\vec{\mathbb{E}}-\vec{v}\cdot\vec{P}\right)\partial_iv^i+\nonumber\\
	-v^i\left(\partial_t+v^j\partial_j\right)P_i+v^i\mathbb{P}^j\partial_j\mathbb{E}_i+\mathbb{P}^j\partial_t\mathbb{E}_j=\hat\Gamma^\epsilon-v_i\hat\Gamma^i_P+\mathcal{O}(\partial^2)
\end{multline}
From the thermodynamic relation \eqref{eqn:ch4:gibbs_duhem_relation}, it follows that
\begin{multline}
	\left(\partial_t+v^i\partial_i\right)\left(\epsilon-\vec{\mathbb{P}}\cdot\vec{\mathbb{E}}\right)=T\left(\partial_t+v^i\partial_i\right)s+\\
	+\mu\left(\partial_t+v^i\partial_i\right)n-\mathbb{P}^j\left(\partial_t+v^i\partial_i\right)\mathbb{E}_j+v_j\left(\partial_t+v^i\partial_i\right)P^j
\end{multline}
and together with the charge conservation equation, we arrive at the following expression for the divergence of the entropy current
\begin{equation}
	\left(\partial_t+v^i\partial_i\right)s+s\partial_iv^i=\hat\Gamma^\epsilon-v_i\hat\Gamma^i_P+\mathcal{O}(\partial^2)
\end{equation}
Without relaxations the RHS is identically zero, as discussed in Chapter~\ref{chapter:hydrodynamics_linear_response} for relativistic ideal fluids. Because the relaxations we are considering are hydrostatic, i.e. they can be non-zero on stationary states, we require that the ideal fluid does not produce entropy even when decay rates are present. This leads to the constraint \eqref{eqn:ch4:energy_momentum_relaxations_relation} between energy and momentum sinks that we found requiring the conservation equations to be compatible with relaxations on stationary flows.

It is possible to have ideal fluids that produce entropy in the presence of relaxations, as in \cite{Landry:DynamicalChemistryNonequilibrium} or \cite{Gouteraux:DrudeTransportHydrodynamic}, however the relaxation considered there are of a different kind, namely they must always disappear in hydrostatic solutions. Notice, however, that for an open system entropy must not always grow: if the sinks parametrize the effect of some weak coupling with a bath it might be possible to give up on the positivity of entropy production.

Finally, if we insist that the ideal fluid must not produce entropy, we can give an intuitive remark on the relation we found between the sinks \eqref{eqn:ch4:energy_momentum_relaxations_relation}. Consider the Euler equation in the form
\begin{equation}
	s=\frac{1}{T}\left(\epsilon-n\mu-\vec{v}\cdot\vec{P}+P-\vec{\mathbb{E}}\cdot\vec{\mathbb{P}}\right)
\end{equation}
then, if the momentum $\vec{P}$ decreases due to the relaxation term, we can keep the entropy constant by also reducing the energy via $\hat\Gamma_\epsilon$.

\subsection{Linearised stability and DC conductivities}
To properly analyse the most important aspects of our theory we want to make use of the modified form of the hydrostatic constraint \eqref{eqn:ch4:hydrostatic_condition_relaxations}. For this reason, we consider a (2+1)-dimensional fluid in a background with constant temperature, chemical potential\footnote{This is possible because with our prescription \eqref{eqn:ch4:hydrostatic_condition_relaxations} we can balance the electric field with the $\Gamma v_x$ while still keeping $\partial_x\mu=0$.} and electric field in the $x$-direction $\vec{\mathbb{E}}=(\mathbb{E}_x,0)$. Because we want to study the modes around a stationary state, we employ our hydrostatic condition and find that, for this background, the velocity is
\begin{equation}
	v_x=\left(\frac{n-\kappa_\mathbb{E}\Gamma_\mathbb{P}-\beta_\mathbb{P}\Gamma_P}{\beta_\mathbb{P}\Gamma_\mathbb{P}+\rho_m\Gamma_P}\right)\mathbb{E}_x
\end{equation}

Subsequently, we linearize around this background, assuming for simplicity that the sinks are independent on the thermodynamic variables, and we compute the modes at zero wavevector. We find four modes: two of them are gapless and vanish at zero wavevector $\omega_{1,2}=0$. A third mode instead is gapped and decaying as
\begin{equation}
	\omega_3=-i\Gamma_\text{eff}\qquad\text{with}\qquad\Gamma_\text{eff}=\frac{1}{\rho_m}\left(\beta_\mathbb{P}\Gamma_\mathbb{P}+\rho_m\Gamma_P\right)
\end{equation}
The last mode, even at zero wavevector, is very complicated and depends on many thermodynamic susceptibilities, hence a complete study of its stability depends also on the specific details of the equations of state. However, for small electric field we can expand the mode finding
\begin{equation}
	\omega=-i\Gamma_\text{eff}+\mathcal{O}(\mathbb{E}_x^2)
\end{equation}

Hence, the system is stable with respect to the background we considered if $\Gamma_\text{eff}\geq0$. In particular, this means that the single relaxation rates can be negative, provided that $\Gamma_\text{eff}$ is positive.

Finally, observe that the system we have described has a finite DC electric and thermoelectric conductivity in the hydrostatic limit. Specifically we can evaluate the electric and heat currents for the ideal fluid as
\begin{subequations}
	\begin{align}
		\vec{J}&=\frac{n^2\tau}{\rho_m}\vec{\mathbb{E}}+\mathcal{O}(\partial)\\
		\vec{Q}&=\frac{sTn\tau}{\rho_m}\vec{\mathbb{E}}+\mathcal{O}(\partial)
	\end{align}
\end{subequations}
where we defined an effective relaxation time $\tau$
\begin{equation}
	\tau=\Gamma_p^{-1}\left(\frac{1-\frac{\kappa_\mathbb{E}\Gamma_\mathbb{P}}{n}-\frac{\beta_\mathbb{P}\Gamma_P}{n}}{1+\frac{\beta_\mathbb{P}\Gamma_\mathbb{P}}{\rho_m\Gamma_P}}\right)
\end{equation}
to make contact with the standard Drude formula. Then, remembering that the thermodynamic parameters can all depend on the external electric field, we can define the non-linear DC conductivities as
\begin{equation}\label{eqn:ch4:non_linear_conductivities}
	\sigma_\text{DC}=\frac{\vec{J}}{\vec{\mathbb{E}}}=\frac{n^2\tau}{\rho_m}\qquad\qquad\alpha_\text{DC}=\frac{\vec{Q}}{\vec{\mathbb{E}}}=\frac{sTn\tau}{\rho_m}
\end{equation}
which correspond to linear response conductivities for small electric fields.

To conclude, we can thoroughly consider the analogy between our approach and the Drude model. Consider a system composed of charged, weakly-interacting particles, like electrons in ordinary metals. If we ignore interactions between the carriers and only focus on scatterings with impurities, Newton's second law reads
\begin{equation}\label{eqn:ch4:drude_model}
	m\frac{\dif}{\dif t}\langle \vec{v}\rangle=q\mathbb{\vec{E}}-\frac{m}{\tau}\langle\vec{v}\rangle
\end{equation}
where $\langle\vec{v}\rangle$ is the average velocity, while $m$ and $q$ are respectively the mass and charge of the particle. The last term represents the interactions with impurities as an effective drag force in terms of a single phenomenological parameter $\tau$. The system is in a steady state only when $\tau^{-1}\neq0$, then requiring that the velocity is not time-dependent we obtain $\langle\vec{v}\rangle=q\tau/m\mathbb{\vec{E}}$, which is the drift velocity of the electron gas and is a fundamental property of the system: it does not depend on initial conditions or the geometry of the sample, exactly like in our hydrodynamic formalism. Furthermore, we can write the electric current as $\vec{J}=nq\langle\vec{v}\rangle$ for some uniform charge density $n$, which gives us the DC conductivity
\begin{equation}
	\vec{J}=\frac{nq^2\tau}{m}\vec{\mathbb{E}}\qquad\Rightarrow\qquad\sigma_\text{DC}=\frac{nq^2\tau}{m}
\end{equation}

An applied external electric field generally heats up the system, as it happens also in hydrodynamics. This means that if the system reaches a steady state the heat produced must be dissipated away, otherwise the system melts. We can see this effect also in the Drude model: from \eqref{eqn:ch4:drude_model} we arrive to an equation for the kinetic energy of the gas particles
\begin{equation}
	\frac{m}{2}\frac{\dif}{\dif t}\langle\vec{v}\rangle^2=q\langle\vec{v}\rangle\vec{\mathbb{E}}-\frac{m}{\tau}\langle\vec{v}\rangle^2
\end{equation}
which implies that if the velocity is time-independent (steady state), then the energy provided by the electric field is lost to impurities with a rate proportional to $\langle\vec{v}\rangle^2$. The last term is effectively an energy relaxation rate $\Gamma_\epsilon$ like the one we introduced in our hydrodynamic formalism.

\section{Dissipative corrections and optical conductivity}\label{sec:ch4:dissipative_corrections}
This final section is based on work still in progress, and is not published research yet.

The obvious next step forward is to start discussing dissipative corrections. However, this becomes quickly very complicated with the setup discussed in this chapter, because having two different vectors at order zero in derivatives, $v^i$ and $\mathbb{E}^i$, means that we can construct many tensor structures $\sim\mathcal{O}(10^2)$ that can appear in the constitutive relations at order one, each with its own associated transport coefficient\footnote{Even without order-zero electric fields, boost-agnostic hydrodynamics has 30 transport coefficients at order one in derivatives \cite{Armas:EffectiveFieldTheory}.}.

To make our life simpler, we can then demote the electric field to order one in derivative counting, hence dropping all the terms that depend on the electric field in the thermodynamics \eqref{eqn:ch4:thermodynamics_derivatives} and ideal fluid \eqref{eqn:ch4:ideal_fluid_flat_spacetime_constitutive_relation}. Furthermore, we also demote to order one the relaxation rate $\Gamma_P$ and $\Gamma_\epsilon$, while $\Gamma_\mathbb{P}$ disappears, since we do not have an order-zero polarization any more. We remark that this does not modify the result of \eqref{eqn:ch4:hydrostatic_condition_relaxations}, since hydrostatic constraints are exact results which do not depend on the choice of derivative counting for the external fields.

With these simplifications, the most interesting term we may wish to study are the dissipative corrections to the $\mathrm{U(1)}$ current. This is because usually it receives derivative corrections in the form $J_i\supset\sigma\left(\mathbb{E}_i-\partial_i\mu\right)$, but now that we modified the hydrostatic constraint \eqref{eqn:ch4:hydrostatic_condition_relaxations} we expect to find
\begin{equation}\label{eqn:ch4:dissipative_current}
	J_i\supset \sigma\left(\mathbb{E}_i-\partial_i\mu-\Gamma\frac{P_i}{n}\right)
\end{equation}
where we dropped the subindex $P$ from the relaxation rate, since there is now only one decay rate. In particular, we are interested in studying how this correction to the constitutive relations affects the usual linear-response charge transport.

To understand the physics, consider a relativistic constitutive relation expanded at small velocity. If we add dissipative correction to the current $J^\mu=J^\mu_{(0)}+J^\mu_{(1)}$ we find that second law of thermodynamics becomes
\begin{equation}
	\partial_ts+\partial_i\left(sv^i-\frac{\mu}{T}J^i_{(1)}\right)+J^i_{(1)}\partial_i\frac{\mu}{T}=-\frac{\Gamma_\epsilon\epsilon}{T}+\frac{\Gamma(P\cdot v)}{T}+\frac{E\cdot J_{(1)}}{T}\geq0
\end{equation}
The first two terms in the LHS are the divergence of the entropy current, thus we rewrite it as
\begin{equation}
	\partial_\mu S^\mu=-\frac{\Gamma_\epsilon}{T}+\Gamma\frac{P\cdot v}{T}+\frac{J^i_{(1)}}{T}\left(E_i-T\partial_i\frac{\mu}{T}\right)\geq0
\end{equation}
If we now impose the relations we found for ideal fluids \eqref{eqn:ch4:energy_momentum_relaxations_relation} to cancel the two relaxation terms, we are left with a quantity on the RHS that does not vanish in equilibrium if we use our modified condition \eqref{eqn:ch4:hydrostatic_condition_relaxations}.

To address the problem, the relation between energy and momentum relaxation must receive dissipative derivative corrections. Namely, we identify
\begin{equation}
	\Gamma_\epsilon=\Gamma P\cdot v+\Gamma\frac{P\cdot J_{(1)}}{n}=\Gamma\frac{P_i\left(nv^i+J^i_{(1)}\right)}{n}=\Gamma\frac{P\cdot J}{n}
\end{equation}
where in the last step we introduced the total free current, meaning that if polarization was present, it would not appear there. Obviously, on hydrostatic solutions the new term proportional to $J_{(1)}$ vanishes and we find again \eqref{eqn:ch4:energy_momentum_relaxations_relation}.

Indeed, with this identification, we see that the divergence of the entropy current takes the desired form
\begin{equation}
	\partial_\mu S^\mu=\frac{J^i_{(1)}}{T}\left(E_i-T\partial_i\frac{\mu}{T}-\Gamma\frac{P_i}{n}\right)\geq0
\end{equation}
and the RHS is positive definite only if the dissipative part of the current is of the form \eqref{eqn:ch4:dissipative_current}
\begin{equation}\label{eqn:ch4:dissipatve_current_relativistic}
	J^i_{(1)}=\sigma\left(E_i-T\partial_i\frac{\mu}{T}-\Gamma\frac{P_i}{n}\right)
\end{equation}

From this result, we can now compute the optical conductivities on a background with zero velocity and zero electric field using the standard Martin-Kadanoff method \ref{sec:ch2:martin_kadanoff}. We find that
\begin{equation}
	\sigma(\omega)=\frac{n^2-i\chi_{\pi\pi}\sigma\omega}{\Gamma\chi_{\pi\pi}-i\chi_{\pi\pi}\omega}
\end{equation}
with $\chi_{\pi\pi}$ the momentum-momentum susceptibility. Notice, in particular, that in DC at $\omega=0$ the incoherent conductivity $\sigma$ drops out and we find
\begin{equation}
	\sigma_\text{DC}=\frac{n^2}{\Gamma\chi_{\pi\pi}}
\end{equation}
in agreement with \eqref{eqn:ch4:non_linear_conductivities}.

There are ways to reintroduce the transport coefficient $\sigma$ in the DC conductivity, by considering non-hydrostatic momentum relaxations $\Gamma=\Gamma_{(0)}+\Gamma_{(1)}$ and modifying the relation between $\Gamma_\epsilon$ and $\Gamma$ appropriately. This gives rise to interesting results that renormalize the incoherent conductivity $\sigma$ similarly to what is found in \cite{Gouteraux:DrudeTransportHydrodynamic}, however we will not discuss them further.

\section{Summary, discussion and outlook}
In this chapter we characterized the stationary flows of a charged fluid immersed in an electric field, which is also subject to the effect of relaxations that remove energy and momentum from the system, thus behaving as sinks connected with external baths.

Since momentum relaxation breaks boost, we employed a boost-agnostic formalism and computed the constitutive relations up to order one in derivatives using the hydrostatic generating functional. Subsequently, we added by hand the relaxation terms and studied the minimal corrections to the hydrostatic conditions \eqref{eqn:ch4:hydrostatic_constraints_flat_space} consistent with the equations of motion. We found that only one single order-zero constraint had to be modified, taking the new form \eqref{eqn:ch4:hydrostatic_condition_relaxations}, and we also obtained a kinematic relation between energy and momentum relaxation \eqref{eqn:ch4:energy_momentum_relaxations_relation}.

As for future perspectives, it would be fascinating to see if it is possible to implement relaxation rates as in our setup from a more formal and geometrical approach, so to include the effect of relaxations in a (suitably generalized) generating functional. This could be done using the modified diffeomorphisms of \cite{Abbasi:MagnetotransportAnomalousFluid}, following the approach of \cite{Gouteraux:DrudeTransportHydrodynamic}, from the Schwinger-Keldysh effective actions as in \cite{Baggioli:QuasihydrodynamicsSchwingerKeldyshEffective}, or, finally, following the thermodynamic approach of \cite{Jain:SchwingerKeldyshEffectiveField} to include a fixed vector in the theory.

In the line of Section~\ref{sec:ch4:dissipative_corrections}, it would also be interesting to systematically study dissipative corrections to discuss how the modified hydrostatic relation \eqref{eqn:ch4:hydrostatic_condition_relaxations} changes the AC thermoelectric transport.

Finally, we could discuss a similar setup in holographic models \cite{Fu:ChargeTransportProperties}, following e.g. the result of \cite{Blake:MomentumRelaxationFluid}, which finds a contribution due to the velocity in the spatial current as in \eqref{eqn:ch4:dissipative_current}.
\chapter{Onsager relations in relaxed hydrodynamics}\label{chapter:onsager}
\epigraph{``Nature does nothing uselessly.''}{Aristotle, \emph{Politics}}

\section{Introduction}
A common feature of all hydrodynamic theories is that a charged fluid in an external electric field gives rise to optical conductivities which are divergent in the DC limit $\omega\rightarrow0$ \cite{Hartnoll:TheoryNernstEffect}. The physical reason for this divergence is very simple: in hydrodynamics momentum is a conserved quantity, therefore an external electric field accelerates the fluid increasing momentum without bound, since there are no mechanism to dissipate it. On a more technical level, this can be traced back to the exact translation symmetry of the theory, or more directly to the existence of a conserved momentum operator that couples to the charged degrees of freedom.

Exactly like in Chapter~\ref{chapter:electrically_driven_fluids}, the standard approach to avoid this issue is to include a small effective phenomenological relaxation parameter $\tau_m^{-1}$ which breaks translation invariance and relaxes the total momentum of the fluid, so that a steady state with finite DC conductivities can be reached. One assumption behind this approach is that, although momentum is no longer conserved, it can still be a relevant hydrodynamic charge if the parameter $\tau_m$ is large enough \cite{Amoretti:HydrodynamicsDimensionalStrongly,Amoretti:MagnetotransportMomentumDissipating,Amoretti:ThermoelectricTransportGauge,Amoretti:ThermoelectricTransportGaugea,Grozdanov:HolographyHydrodynamicsWeakly,Gouteraux:DrudeTransportHydrodynamic,Blake:MomentumRelaxationFluid,Andrade:SimpleHolographicModel}, i.e. if the associated gapped mode is closer to the real line in frequency space compared to the other non-hydrodynamic modes (see the discussion about the hydrodynamic modes in Section~\ref{sec:ch2:modes} and in the Introduction~\ref{chapter:introduction}).

As we will see better in the next chapter, there are circumstances in which momentum relaxation is not enough to obtain finite DC conductivities. This is particularly relevant for the case of anomalous hydrodynamics, i.e. chiral fluids in the presence of a $\mathrm{U(1)}$ axial anomaly. In this case, momentum relaxation $\tau_m$ is not enough to obtain finite DC conductivities, and one also needs to relax energy and axial charge \cite{Landsteiner:NegativeMagnetoresistivityChiral,Abbasi:MagnetotransportAnomalousFluid,Lucas:HydrodynamicTheoryThermoelectric,Jimenez-Alba:AnomalousMagnetoconductivityRelaxation,Rogatko:MagnetotransportWeylSemimetals}. Furthermore, from a physical perspective, even without chiral anomalies, energy and charge relaxations can be important ingredients of the low energy dynamics of condensed matter systems, as argued in \cite{Pongsangangan:HydrodynamicsChargedTwodimensional,Narozhny:HydrodynamicApproachElectronic,Fritz:HydrodynamicElectronicTransport,Gall:ElectronicViscosityEnergy,Narozhny:HydrodynamicApproachTwodimensional} from a kinetic theory perspective.

On the other hand, from a computational perspective there are two main approaches to obtain the retarded response functions (thus the conductivities) from hydrodynamics, as reviewed in Section~\ref{sec:ch2:linear_response_theory}: the Martin-Kadanoff \cite{Kadanoff:HydrodynamicEquationsCorrelation}, see Section~\ref{sec:ch2:martin_kadanoff} and the variational method \cite{Kovtun:LecturesHydrodynamicFluctuations,Romatschke:RelativisticFluidDynamics}, Section~\ref{sec:ch2:variational_method}. In particular, the former is based on thermodynamics and linear response theory, and works by perturbing the fluid with thermodynamic sources, while the latter is rooted in field theory and uses metric and gauge field fluctuations as sources.

However, because relaxation rates break the Lorentz and $U(1)$ symmetry, the variational approach is usually difficult to implement in quasihydrodynamics, since there is no unique proper way to covariantize the decay terms. On the other hand, the canonical approach is more robust against modifications of hydrodynamics, and is applicable also to quasihydrodynamic models, however it is lacking under other aspects. Specifically it cannot be used to compute all the Green functions in the theory \cite{Kovtun:LecturesHydrodynamicFluctuations,Romatschke:FluidDynamicsFar} and, if the thermodynamics becomes more complicated (e.g., in the presence of order zero electromagnetic fields), it might be hard to identify the susceptibilities and the correct conjugate sources.

In this chapter we are going to explore models of linearized quasihydrodynamics to compute Green functions using both the Martin-Kadanoff and the variational method. In particular, we find that it is possible to use the variational method in the presence of relaxations too. We start by including extra source terms in the linearised equations of motion, and what we find is that the coefficients of these extra terms are uniquely fixed by the requirement of time-reversal covariance of the response functions alone, i.e. Onsager reciprocal relations.

\section{Martin-Kadanoff for quasihydrodynamics}\label{sec:ch5:martin_kadanoff}
We now focus specifically on a simple charged relativistic fluid with conserved stress-energy tensor and a $\mathrm{U(1)}$ current. We linearize the theory and add the most general relaxations terms to the equations of motion; subsequently we impose Onsager relations on the Green functions, positivity of entropy production, and linear stability from the study of the modes. What we find is a set of constraints and relations that must be obeyed very generally by all fluids.

As we discussed in Section~\ref{sec:ch2:linear_response_theory}, we can write the equations of linearized hydrodynamics in the form
\begin{equation}
	\partial_t\phi_a(t,\vect{k})+M_{ab}(\vect{k})\phi_b(t,\vect{k})=0
\end{equation}
where we Fourier transformed in space, while keeping the explicit time dependence. Here, $\phi_a$ are the fluctuations of the (almost-)conserved charges, while $M_{ab}$ is the hydrodynamic matrix that depends on the constitutive relations and equations of motion. We also rewrite here the definition of the susceptibility matrix as
\begin{equation}
	\chi_{ab}=\frac{\partial\phi_a}{\partial\lambda_b}
\end{equation}
in which $\lambda_a$ are the sources canonically conjugate to the charges.

Following the arguments of Section~\ref{sec:ch2:martin_kadanoff} we can use these expressions to compute the Green functions using the Martin-Kadanoff method \ref{eqn:ch2:retarded_green_functions}. Furthermore, we have also discussed how time-reversal covariance of the microscopic theory, namely the requirement that
\begin{equation}
	G_{ab}^R(\omega,\vect{k})=\eta_a\eta_bG^R_{ab}(\omega,-\vect{k})
\end{equation}
imposes a constraint on the hydrodynamic matrix
\begin{equation}\label{eqn:ch5:onsager_constraint_general}
	\chi SM^T(-\vect{k})=M(\vect{k})\chi S
\end{equation}
in which we set the magnetic field to zero and remind here that $S=\text{diag}(\eta_1,\eta_2,\dots)$ is the matrix of the time-reversal eigenvalues $\eta_a$.

Consider now a relativistic fluid in the Landau frame in $(3+1)$-dimensional flat spacetime and without external electromagnetic fields, so that the order one constitutive relations are given in \eqref{eqn:ch2:landau_constitutive_relations}, and the non-linear conservation equations are simply $\partial_\mu T^{\mu\nu}=\partial_\mu J^\mu=0$. We now linearize these equations around a global equilibrium state with constant energy and charge density, and zero spatial velocity
\begin{equation}
	u^\mu=(1,\delta v^i)\qquad \epsilon=\epsilon_0+\delta\epsilon\qquad n=n_0+\delta n
\end{equation}
The linearised equations of motion take the following form
\begin{subequations}\label{eqn:ch5:equations_of_motion_flat_space}
	\begin{align}
		\partial_t\delta\epsilon+\left(\epsilon_0+P_0\right)\partial_i\delta v^i&=-\left(\frac{\delta\epsilon}{\tau_{\epsilon\epsilon}}+\frac{\delta n}{\tau_{\epsilon n}}\right)\\
		\partial_t\delta n+n_0\partial_i\delta v^i-\sigma T_0\partial_i^2\delta\frac{\mu}{T}&=-\left(\frac{\delta\epsilon}{\tau_{n\epsilon}}+\frac{\delta n}{\tau_{nn}}\right)\\
		\partial_t\delta\pi^i+\partial^i\delta P-\eta\left(\partial_j^2\delta v^i+\frac{1}{3}\partial^i\partial_j\delta v^j\right)-\zeta\partial_i\partial_j\delta v^j&=-\frac{1}{\tau_m}\delta\pi^i
	\end{align}
\end{subequations}
where the linear momentum is $\delta\pi^i=(\epsilon_0+P_0)\delta v^i$ for a relativistic fluid. These correspond to the (non-)conservation of energy, charge and momentum respectively. In the above equations we also included on the RHS the most general relaxation terms at order one in fluctuations and order zero in derivatives\footnote{The reason for this is that we take the relaxation rates $\tau$s to be large, namely $\tau^{-1}\sim\mathcal{O}(\partial)$ in the linearized regime.} compatible with isotropy and rotational invariance (hence only a single $\tau_m$ appears, instead of a momentum-relaxation matrix). Specifically, $\tau_m$ is the standard momentum relaxation, while $\tau_{\epsilon\epsilon}$ and $\tau_{nn}$ are commonly known as energy and charge relaxation respectively. The other two off-diagonal relaxations $\tau_{\epsilon n}$ and $\tau_{n\epsilon}$, however, are mostly new (see \cite{Lucas:HydrodynamicTheoryThermoelectric}) and represent mixed relaxations which do not have units of time: $\tau_{\epsilon n}$ parametrizes the energy loss due to charge fluctuations and vice versa for $\tau_{n\epsilon}$. Notice that in principle, because of the relaxations, the constitutive relations might have more and different transport coefficients \cite{Baggioli:QuasihydrodynamicsSchwingerKeldyshEffective}, however because we are assuming the $\tau$s are very large these corrections are taken to be very small.

In the context of hydrodynamics, \emph{relaxation} usually means that the fluctuation of some conserved charge decays exponentially to equilibrium, e.g. consider the simple charge conservation equation on a uniform background
\begin{equation}
	\partial_t\delta n=-\frac{1}{\tau_{nn}}\delta n
\end{equation}
Real time solutions to this equation take the form
\begin{equation}
	\delta n(t)=\delta n(0)e^{-t/\tau_{nn}}
\end{equation}
However, even a simple example like this one, cannot be properly covariantized, for example writing $u_\mu J^\mu/\tau$ would not give the proper relaxation term at non-zero velocity. Indeed, $J^0$ represent the charge density in the laboratory frame, while $n=-u_\mu J^\mu$ is the charge density in the fluid rest frame, thus $\partial_t J^0=\frac{1}{\tau}u_\mu J^\mu$ would not relax the right charge. Furthermore, if we try to covariantize in this way, we will never be able to write a momentum relaxation term, since the comoving momentum is zero by definition.

We then realize that to write any relaxation in a covariant non-linear form we need some extra information beyond the fluid data alone. This can be done e.g. in the boost-agnostic formalism by using the laboratory-frame clock form $\tau_\mu=(1,\vect{0})$, in the relativistic case following \cite{Stephanov:NodragFrameAnomalous} by considering a second four-velocity vector $U^\mu$ which pins the fluid velocity, or introducing extra fields as in \cite{Jain:SchwingerKeldyshEffectiveField,Gouteraux:DrudeTransportHydrodynamic,Delacretaz:DampingPseudoGoldstoneFields}.

The second important thing to observe is that relaxations act as constraints on the system, by selecting a specific background. This is seen very clearly for the momentum relaxation, in which $1/\tau_m\delta \pi^i$ naturally selects the equilibrium solution with zero velocity as the only equilibrium solution of the system. This should be contrasted with the case without momentum relaxation, in which any equilibrium with constant non-zero velocity is a valid equilibrium, thanks to the boost symmetry.

However, even without relaxations, although the theory has Lorentz symmetry, the equilibrium solution does not \cite{Kovtun:LecturesHydrodynamicFluctuations,Armas:CarrollianFluidsSpontaneous}, and because we are interested in computing two-point functions about some global thermodynamic equilibrium any choice of the background will break the symmetry. For these reasons we are agnostic on possible non-linear completions of the theory, thus we focus only on the linearized version of hydrodynamics, which is enough for the purpose of computing the Green functions. Then, we simply write the relaxation terms as the most general expressions, following the EFT prescription, of all the scalars and vectors in the linearized theory. At the end, we expect that any covariant version of the non-linear relaxations should lead to the same linearized expressions we work with, as we will see in the next chapter with specific examples.

Having clarified the reasoning behind our choice of relaxations, we proceed by imposing Onsager relations in \eqref{eqn:ch5:onsager_constraint_general} on our linearized hydrodynamic theory \eqref{eqn:ch5:equations_of_motion_flat_space}. We take $\phi(t,\vect{k})=\left(\delta\epsilon,\delta n,\delta\pi^i\right)$ and, setting $\vect{k}=(k_x,0,0)$ without loss of generality, the matrix $M_{ab}$ becomes
\begin{multline}
	M=\\
	\begin{pmatrix}
		\frac{1}{\tau_{\epsilon\epsilon}}   &   \frac{1}{\tau_{\epsilon n}} &   ik_x    &   0   &   0\\
		k_{x}^2\sigma\beta_\epsilon+\frac{1}{\tau_{n\epsilon}} &  k_{x}^2\sigma\beta_n+\frac{1}{\tau_{nn}}    &   \frac{ik_x n_0}{P_0+\epsilon_0}   &   0   &   0\\
		ik_x\frac{\partial P}{\partial\epsilon} &   ik_x\frac{\partial P}{\partial n}   &   \frac{k_{x}^2(3\zeta+4\eta)}{3(P_0+\epsilon_0)}+\frac{1}{\tau_p} &   0   &   0   \\
		0   &   0   &   0   &   \frac{k_{x}^2\eta}{P_0+\epsilon_0}+\frac{1}{\tau_m}  &   0   \\
		0   &   0   &   0   &   0   &   \frac{k_{x}^2\eta}{P_0+\epsilon_0}+\frac{1}{\tau_m}
	\end{pmatrix}
\end{multline}
where $\beta_\epsilon=\frac{\partial\mu}{\partial\epsilon}-\frac{\mu_0}{T_0}\frac{\partial T}{\partial\epsilon}$ and $\beta_n=\frac{\partial\mu}{\partial n}-\frac{\mu_0}{T_0}\frac{\partial T}{\partial n}$. Then we find that Onsager relations are obeyed iff the relaxation rates satisfy the condition
\begin{equation}\label{eqn:ch5:onsager_condition}
	\frac{\chi_{\epsilon\epsilon}}{\tau_{n\epsilon}}-\frac{\chi_{\epsilon n}}{\tau_{\epsilon\epsilon}}+\frac{\chi_{n\epsilon}}{\tau_{nn}}-\frac{\chi_{nn}}{\tau_{\epsilon n}}=0
\end{equation}
where the susceptibilities are
\begin{equation}
	\chi_{n\epsilon} =\chi_{\epsilon n}=\frac{\partial\epsilon}{\partial\mu}= T_0 \frac{\partial n}{\partial T} + \mu_0 \frac{\partial n}{\partial \mu}\qquad
	\chi_{nn} =\frac{\partial n}{\partial\mu}\qquad
	\chi_{\epsilon\epsilon} =T_0\frac{\partial\epsilon}{\partial T}+\mu_0\frac{\partial\epsilon}{\partial\mu}
\end{equation}
The above thermodynamic derivatives are in the grand canonical ensemble, and should be understood at fixed $\mu$ or $T$ respectively.

Notice that this result is not in contrast with the previous chapter, in particular the $\tau_{\epsilon\epsilon}$ that appears here is not related to the $\Gamma_\epsilon$ introduced in \eqref{eqn:ch4:equations_of_motion_relaxations}. This is because, due to the relation \eqref{eqn:ch4:energy_momentum_relaxations_relation}
\begin{equation}\label{eqn:ch5:energy_momentum_relaxation_relation}
	\Gamma_\epsilon=\Gamma_PP^iv_i
\end{equation}
between energy and momentum relaxations on hydrostatic solutions, $\Gamma_\epsilon$ vanishes when linearizing around a zero-velocity background.

The constraint \eqref{eqn:ch5:onsager_condition} is very general and must be obeyed by any fluid which preserves microscopic time-reversal invariance. In particular, we can see that if the mixed relaxations are zero $\tau_{n\epsilon}^{-1}=\tau_{\epsilon n}^{-1}=0$, then it implies $\tau_{\epsilon\epsilon}=\tau_{nn}$, i.e. energy and charge must relax at the same rate as was already observed in \cite{Abbasi:MagnetotransportAnomalousFluid}.

We can also study the effect of generalized relaxations on the second law of thermodynamics. The entropy current, at the order we are interested in, is just given by the canonical entropy current
\begin{equation}
	S^\mu=\frac{1}{T}\left(Pu^\mu+T^{\mu\nu}u_\nu-\mu J^\mu\right)
\end{equation}
We can linearize the entropy current and compute the divergence. Using the linearized equations of motion \eqref{eqn:ch5:equations_of_motion_flat_space} and following the usual steps, we arrive at the result
\begin{equation}\label{eqn:ch5:entropy_production_order_one}
	T_0\partial_\mu \delta S^\mu=\delta\epsilon\left(\frac{\mu_0}{\tau_{n\epsilon}}-\frac{1}{\tau_{\epsilon\epsilon}}\right)+\delta n\left(\frac{\mu_0}{\tau_{nn}}-\frac{1}{\tau_{\epsilon n}}\right)+\mathcal{O}(\partial^2,\delta^2)
\end{equation}
The RHS should be positive definite on any solution of the hydrodynamic equations of motion. In this case, because the fluctuations of $\delta\epsilon$ and $\delta n$ are not positive definite, we must impose that the coefficients multiplying the fluctuations vanish, namely
\begin{equation}\label{eqn:ch5:entropy_constraint}
	\frac{1}{\tau_{\epsilon\epsilon}}=\frac{\mu_0}{\tau_{n\epsilon}}\qquad\qquad\frac{1}{\tau_{\epsilon n}}=\frac{\mu_0}{\tau_{nn}}
\end{equation}
These are two further constraints on the relaxation rates that stem from the local form of the second law of thermodynamics, together with the constraint \eqref{eqn:ch5:onsager_condition} required by Onsager relations. For a system that is constrained by both second law and Onsager relations there are then three constraints and four total relaxation rates, which means that the whole relaxation dynamics is actually characterized by a single parameter (in addition to momentum decay $\tau_m$). Furthermore, because of the structure of these constraints, not all relaxation parameters must be positive.

There is an intuitive thermodynamic argument that we can give, following \cite{Landsteiner:NegativeMagnetoresistivityChiral}, to explain why the mixed relaxations are related by the positivity of entropy production in \eqref{eqn:ch5:entropy_constraint}. Consider a fluid element which undergoes a microscopic scattering process that relaxes the energy. Then, because of the scattering, the fluid will lose some energy $\delta\epsilon$ in a time $\tau_{\epsilon\epsilon}$, on average. However, if it loses energy, this implies that it is also losing charge by $\delta n=\frac{\partial n}{\partial\epsilon}\delta\epsilon=\delta\epsilon/\mu_0$ in the same interval. The argument in reverse, for a scattering that relaxes the charge, tells us that the system must also lose energy as $\delta\epsilon=\mu_0\delta n$. We are thus let to the expressions \eqref{eqn:ch5:entropy_constraint} that come from the second law.

Indeed, both the above argument based on thermodynamics and the second law hold only for closed systems. However, if the relaxation rates parametrize the fact that our system is open and loses energy and charge to the environment, then the second law might fail if applied to the fluid alone, as we will see in the next chapter. The total entropy should still increase, but we have to account for the entropy of the fluid plus of its environment for this to happen.

In \eqref{eqn:ch5:entropy_production_order_one} we computed the entropy production at first order in fluctuations, the order in which the relaxation rates first appear. However, linearized hydrodynamics is usually discussed at order two in fluctuations, from the point of view of the entropy current, since only at order $\mathcal{O}(\delta^2)$ we can bound the dissipative transport coefficients. Therefore, we check the second law at order two in fluctuations, and we find
\begin{align}
	T_0\partial_\mu\delta s^\mu&=\frac{\delta\epsilon\delta T}{T_0\tau_{\epsilon\epsilon}}+\frac{\delta n\delta T}{T_0\tau_{\epsilon n}}+\frac{\delta\mu\delta n}{\tau_{nn}}+\frac{\delta\mu\delta\epsilon}{\tau_{n\epsilon}}-\mu_0\frac{\delta T\delta n}{T_0\tau_{nn}}-\mu_0\frac{\delta T\delta\epsilon}{T_0\tau_{n\epsilon}}\nonumber\\
	&\quad+\frac{1}{\tau_m}(\epsilon_0+P_0)\delta v^2+\sigma\left(\frac{\mu_0}{T_0}\partial\delta T-\partial\delta\mu\right)^2\nonumber\\
	&\quad+\eta\left(\delta\sigma^{ij}\right)^2+\zeta\left(\partial_i\delta v^i\right)^2+\mathcal{O}(\delta^3,\partial^3)
\end{align}
where $\delta\sigma^{ij}$ are the fluctuations of the shear tensor. Thus, we find that positivity of entropy production at order two in fluctuations gives the usual constraints on the transport coefficients $\sigma\geq0$, $\eta\geq0$ and $\zeta\geq0$, furthermore, as expected, we find $\tau_m\geq0$ for the momentum relaxation rate.

Then, we can focus only on the first line in the above formula, the one depending on the energy and charge relaxations. We rewrite it in terms of fluctuations of temperature and chemical potential alone, by expanding the energy and charge fluctuations as
\begin{equation}
	\delta\epsilon=\frac{\partial\epsilon}{\partial T}\delta T+\frac{\partial\epsilon}{\partial\mu}\delta\mu\qquad\delta n=\frac{\partial n}{\partial T}\delta T+\frac{\partial n}{\partial\mu}\delta\mu
\end{equation}
and arrive at the following expression
\begin{align}
	T_0\partial_\mu \delta s^\mu
	&=\delta\mu\delta T\left(\frac{\partial\epsilon}{\partial\mu}\frac{1}{T_0\tau_{\epsilon\epsilon}}+\frac{\partial n}{\partial\mu}\frac{1}{T_0\tau_{\epsilon n}}+\frac{\partial n}{\partial T}\frac{1}{\tau_{nn}}+\frac{\partial\epsilon}{\partial T}\frac{1}{\tau_{n\epsilon}} -\frac{\mu_0}{T_0}\frac{\partial n}{\partial\mu}\frac{1}{\tau_{nn}} \right. \nonumber \\
	&\left.\quad-\frac{\mu_0}{T_0}\frac{\partial\epsilon}{\partial\mu}\frac{1}{\tau_{n\epsilon}}\right)+\frac{\delta T^2}{T_0}\left(\frac{\partial\epsilon}{\partial T}\frac{1}{\tau_{\epsilon\epsilon}}+\frac{\partial n}{\partial T}\frac{1}{\tau_{\epsilon n}}-\frac{\partial n}{\partial T}\frac{\mu_0}{\tau_{nn}}-\frac{\partial\epsilon}{\partial T}\frac{\mu_0}{\tau_{n\epsilon}}\right) \nonumber\\
	&\quad+\delta\mu^2\left(\frac{\partial n}{\partial\mu}\frac{1}{\tau_{nn}}+\frac{\partial\epsilon}{\partial\mu}\frac{1}{\tau_{n\epsilon}}\right) + \ldots
\end{align}
where the dots represent the positive definite terms we already discussed. Employing both Onsager relations \eqref{eqn:ch5:onsager_condition} and the constraints \eqref{eqn:ch5:entropy_constraint} we can simplify the above expression to
\begin{equation}
	T_0\partial_\mu\delta S^\mu=\delta\mu^2\left(\frac{\partial n}{\partial\mu}\frac{1}{\tau_{nn}}+\frac{\delta\epsilon}{\delta\mu}\frac{1}{\tau_{n\epsilon}}\right)+\dots
\end{equation}
which can finally be rearranged in terms of susceptibilities as
\begin{equation}\label{eqn:ch5:thermodynamic_constraint}
	\frac{\delta\mu^2}{\tau_{nn}}\left(\chi_{\epsilon\epsilon}\chi_{nn}-\chi_{\epsilon n}^2\right)\lessgtr 0
\end{equation}
The sign of the inequality depends on the sign of $\partial\epsilon/\partial T$. The susceptibility matrix is positive semi-definite, which means that its determinant is also non-negative and so is the quantity in the bracket above. Thus, we see that the sign of $\tau_{nn}$ is constrained by the sign of $\partial\epsilon/\partial T$: when $\partial\epsilon/\partial T\geq0$ then $\tau_{nn}\geq0$ and vice versa. It is important to observe that this condition is not of equality-type, like \eqref{eqn:ch5:entropy_constraint} and \eqref{eqn:ch5:onsager_condition}, therefore, even including terms which are order two in fluctuations in the second law, we still have a one-parameter family of relaxations\footnote{Notice that we could also include relaxation terms which are order two in fluctuations directly in the equations of motion, to cancel $\mathcal{O}(\delta^2)$ terms that appear in the positivity of entropy production.}.

Finally, we can also check if linear stability, namely the requirement that hydrodynamic modes must decay and not grow with time, imposes further constraints on the relaxation rates. In $d+1$ spacetime dimensions we find $d+2$ modes, one for each equation of motion, which at $\vect{k}=0$ are given by
\begin{equation}\label{eqn:ch5:modes}
	\omega_1=-\frac{i}{\tau_m}\qquad\omega_2=-\frac{i}{2}\left(\frac{1}{\tau_{\epsilon\epsilon}}+\frac{1}{\tau_{nn}}\right)\pm\frac{i}{2}\sqrt{\left(\frac{1}{\tau_{\epsilon\epsilon}}-\frac{1}{\tau_{nn}}\right)^2+\frac{4}{\tau_{n\epsilon}\tau_{\epsilon n}}}
\end{equation}
where $\omega_1$ has multiplicity $d$. Stability requires that the imaginary part of the modes should be negative, thus the first mode simply tells us that $\tau_m\geq0$, exactly the same constraint that we find from the second law of thermodynamics. The two other $\omega_2$ modes instead can be propagating or not, and their stability depends on the thermodynamics and the relations between the $\tau$s. However, if we also impose all the other constraints \eqref{eqn:ch5:entropy_constraint} and \eqref{eqn:ch5:onsager_condition}, then we can rewrite $\omega_2$ as a null mode and a decaying mode
\begin{equation}
	\omega_2=0\qquad\omega_2=-i\left(\frac{1}{\tau_{nn}}+\frac{1}{\tau_{\epsilon\epsilon}}\right)=-\frac{i}{\tau_{nn}}\left(\frac{\frac{\partial\epsilon}{\partial T}-\mu_0\frac{\partial n}{\partial T}}{\frac{\partial\epsilon}{\partial T}}\right)
\end{equation}
In particular, the stability of the second mode depends again on the relative sign of thermodynamic derivatives and $\tau_{nn}$, as in \eqref{eqn:ch5:thermodynamic_constraint}.

Albeit it is not the main goal of this chapter, to conclude the section we can ask about possible non-linear completion of our generalized relaxations (we will see more in the next chapter). In particular, we can focus on relaxations which obey both the second law of thermodynamics and Onsager relations, and see how these constraints restrict the possible non-linear expressions.

To proceed, we change the basis of fluctuations from the charges $\delta n$ and $\delta\epsilon$ to the conjugate sources $\delta T$ and $\delta\mu$, specifically
\begin{equation}
	\frac{\delta\epsilon}{\tau_{\epsilon\epsilon}}+\frac{\delta n}{\tau_{\epsilon n}}=\frac{\delta T}{\tau_{\epsilon T}}+\frac{\delta\mu}{\tau_{\epsilon\mu}}\qquad\qquad\frac{\delta\epsilon}{\tau_{n\epsilon}}+\frac{\delta n}{\tau_{nn}}=\frac{\delta T}{\tau_{nT}}+\frac{\delta\mu}{\tau_{n\mu}}
\end{equation}
We can now impose Onsager relations and positivity of entropy production on this new basis, obtaining the constraints
\begin{equation}
	\frac{1}{\tau_{\epsilon T}}=0\qquad\frac{1}{\tau_{nT}}=0\qquad\tau_{\epsilon\mu}=\frac{\tau_{nn}}{\mu_0}\frac{\frac{\partial\epsilon}{\partial T}}{\frac{\partial n}{\partial\mu}\frac{\partial\epsilon}{\partial T}-\frac{\partial n}{\partial T}\frac{\partial\epsilon}{\partial\mu}}\qquad\tau_{n\mu}=\frac{\tau_{nn}\frac{\partial\epsilon}{\partial T}}{\frac{\partial n}{\partial\mu}\frac{\partial\epsilon}{\partial T}-\frac{\partial n}{\partial T}\frac{\partial\epsilon}{\partial\mu}}
\end{equation}
Consider now a pair of differentiable functions\footnote{This $\Gamma_\epsilon$ must not be confused with the one of the previous chapter \eqref{eqn:ch5:energy_momentum_relaxation_relation}.} $\Gamma_\epsilon$ and $\Gamma_n$ which, upon linearization, give rise to generalized relaxations
\begin{subequations}
	\begin{align}
		\delta\Gamma_\epsilon&=\frac{\partial\Gamma_\epsilon}{\partial\mu}\delta\mu+\frac{\partial\Gamma_\epsilon}{\partial T}\delta T=\frac{\delta\mu}{\tau_{\epsilon\mu}}+\frac{\delta T}{\tau_{\epsilon T}}\\
		\delta\Gamma_n&=\frac{\partial\Gamma_n}{\partial\mu}\delta\mu+\frac{\partial\Gamma_n}{\partial T}\delta T=\frac{\delta\mu}{\tau_{n\mu}}+\frac{\delta T}{\tau_{nT}}
	\end{align}
\end{subequations}
If we want this to be consistent, then the linearized relaxations must satisfy integrability conditions
\begin{equation}
	\frac{\partial}{\partial T}\left(\frac{1}{\tau_{\epsilon\mu}}\right)=\frac{\partial}{\partial\mu}\left(\frac{1}{\tau_{\epsilon T}}\right)\qquad\qquad\frac{\partial}{\partial T}\left(\frac{1}{\tau_{n\mu}}\right)=\frac{\partial}{\partial\mu}\left(\frac{1}{\tau_{nT}}\right)
\end{equation}
because of the commutativity of the second derivatives of $\Gamma_\epsilon$ and $\Gamma_n$. This leads to
\begin{equation}
	\frac{\partial}{\partial\mu}\left(\frac{1}{\tau_{\epsilon T}}\right)=0\qquad\qquad\frac{\partial}{\partial\mu}\left(\frac{1}{\tau_{nT}}\right)=0
\end{equation}
thus, the most general $\tau_{nn}$ compatible with Onsager relations, second law of thermodynamics and non-linear completion takes the form
\begin{equation}
	\frac{1}{\tau_{nn}(T,\mu)}=\frac{f(\mu)\frac{\partial\epsilon}{\partial T}}{\frac{\partial n}{\partial\mu}\frac{\partial\epsilon}{\partial T}-\frac{\partial n}{\partial T}\frac{\partial\epsilon}{\partial\mu}}
\end{equation}
where $f$ is an arbitrary function of $\mu$. The remaining relaxations are uniquely fixed in terms of $\tau_{nn}$ from \eqref{eqn:ch5:entropy_constraint} and \eqref{eqn:ch5:onsager_condition}.

\section{Variational approach and Onsager relations}\label{sec:ch5:variational_approach}
In the previous section we considered a charged relativistic fluid in the presence of the most general linear relaxations, and found a set of very generic constraints that the relaxation parameters must obey to satisfy positivity of entropy production, Onsager relations and linear stability. Now we investigate the same problem, but from the perspective of the variational method, and give a prescription for how to obtain Onsager reciprocal Green functions in flat spacetime using this approach.

To proceed we must find the current and stress-energy tensor which solve the linearized equations of motion on a curved background $g_{\mu\nu}=\eta_{\mu\nu}+\delta h_{\mu\nu}$ and with a gauge field perturbation $\delta A_\mu$. The linearized equations of motion for an order one fluid in the Landau frame then take the form \eqref{eqn:ch2:constitutive_relations_curved_space}
\begin{subequations}\label{eqn:ch5:equations_of_motion_curved_space}
	\begin{align}
		&\tau^\mu\partial_\mu\delta\epsilon + \left( \epsilon_0 + P_0 \right) \nabla^{(0)}_{\mu} \delta v^{\mu} = - \left( \frac{1}{\tau_{\epsilon \epsilon}} \delta \epsilon +  \frac{1}{\tau_{\epsilon n}} \delta n \right) \\
		&\tau^\mu\partial_\mu\delta n + n_0 \nabla^{(0)}_{\mu} \delta v^{\mu} - \sigma \left( P^{\mu \nu} \nabla^{(0)}_{\mu} \partial_{\nu} \delta \mu - \frac{\mu_0}{T_0} P^{\mu \nu} \nabla^{(0)}_{\mu} \partial_{\nu} \delta T -\nabla^{(0)}_{\mu} \delta E^{\mu}\right) \nonumber \\
		&= - \left( \frac{1}{\tau_{n \epsilon}} \delta \epsilon +  \frac{1}{\tau_{nn}} \delta n \right)\\
		&P^{\mu \nu} \partial_{\nu} \delta P
		-  \zeta P^{\mu \nu} \nabla_{\nu}^{(0)} \nabla_{\rho}^{(0)} \delta v^{\rho} + \left( \epsilon_0 + P_0 \right) P^\mu_{\ \rho} \left( \tau^{\nu} \nabla_{\nu}^{(0)} \delta v^{\rho}  \right) - 2 \eta P^{\mu \rho} \nabla_{\sigma}^{(0)} \delta \sigma^\sigma_{\ \rho} \nonumber \\
		&= - \frac{1}{\tau_{m}} \left( \epsilon_0 + P_0 \right) \delta v^{\mu} + n_0 \delta E^{\mu} - P_0 P^\mu_{\ \nu} \nabla_{\rho}^{(0)} \delta h^{\rho \nu} - \left( \epsilon_0 + P_0 \right) P^\mu_{\ \rho}  \tau^{\nu} \delta \Gamma_{\nu \sigma}^{\rho} \tau^{\sigma}
	\end{align}
\end{subequations}
where $P^{\mu\nu}=\eta^{\mu\nu}+\tau^\mu\tau^\nu$ is the projector orthogonal to the background velocity $\tau_\mu=(-1,\vect{0})$ and $\nabla^{(0)}_\mu$ is the covariant derivative computed with respect to the background metric (if, like in the present case, we expand around Minkowski metric, then $\nabla^{(0)}_\mu=\partial_\mu$).

As we did in the previous section, we also included decay terms on the RHS. We remark that because the relaxation rates break Lorentz symmetry they cannot be uniquely written in a covariant form. This means that, depending on the arbitrary choice of non-linear covariantizations, it is possible to find different source terms associated to the relaxations, i.e. terms in which a source fluctuation, $\delta h_{\mu\nu}$ or $\delta A_\mu$, multiplies a relaxation parameter. In order to avoid making this choice, in \eqref{eqn:ch5:equations_of_motion_curved_space} we have decided to keep the relaxation terms as they appear in flat spacetime, without any possible source contribution.

One can then solve the above linear equations to express the fluctuations of the hydrodynamic fields in terms of the sources, plug the results into the definitions of the generators $\mathcal{J}^\mu$ and $\mathcal{T^{\mu\nu}}$ in \eqref{eqn:ch2:green_functions_generators} in the presence of (linear) background source fields, and compute the complete set of Green functions using \eqref{eqn:ch2:green_functions_variational}, reducing to flat spacetime, zero gauge field at the end. Proceeding in this way, we find that not all correlators are time-reversal covariant, and some of them violate Onsager relations by terms explicitly dependent on the relaxation rates, e.g.
	\begin{multline}
	\left. \langle T^{tt} T^{xx} \rangle - \langle T^{xx} T^{tt} \rangle \right|_{\vect{k}=0} = \\
	=-\frac{\left(\left(\epsilon_0+P_0\right)\left(\tau_{n\epsilon}\tau_{\epsilon n}-\tau_{nn}\tau_{\epsilon\epsilon}\right)\right)+i\tau_{nn}\tau_{n\epsilon}\left((\epsilon_0+P_0)\tau_{\epsilon n}+n_0\tau_{\epsilon\epsilon}\right)\omega}{\tau_{nn}\tau_{\epsilon\epsilon}+\tau_{n\epsilon}\tau_{\epsilon n}\left(i+\tau_{nn}\omega\right)\left(i+\tau_{\epsilon\epsilon}\omega\right)}
\end{multline}
The above expression is non-zero even after we impose \eqref{eqn:ch5:entropy_constraint} and \eqref{eqn:ch5:onsager_condition} on the relaxation parameters, furthermore some of the two-point functions are different when computed from the variational and Martin-Kadanoff approach.

All these issues are related to the presence of relaxations in our quasihydrodynamic model. Therefore, we modify the linearized theory \eqref{eqn:ch5:equations_of_motion_curved_space} by including the most general set of extra source terms. Because we want to make contact with the flat spacetime, zero gauge field limit of the previous section, we will take these extra source terms to be proportional to the fluctuations $\delta h_{\mu\nu}$ and $\delta A_\mu$, so that they vanish in the appropriate limit. Furthermore, because the relaxation terms contain only ideal fluid data (the $\tau$ parameters are never multiplied by derivatives of the hydrodynamic fields), we restrict ourselves to extra source terms without any derivatives. Schematically, we write
\begin{equation}\label{eqn:ch5:extra_sources}
	\text{sources of \eqref{eqn:ch5:equations_of_motion_curved_space}}\quad\rightarrow\quad\text{sources of \eqref{eqn:ch5:equations_of_motion_curved_space}}+c_a^{\mu\nu}\delta h_{\mu\nu}+r_a^\mu\delta A_\mu
\end{equation}
and, with this new set of equations of motion, we compute again all the correlators in the theory using the variational approach, which will now depend on these extra coefficients $c_a^{\mu\nu}$ and $r_a^\mu$. In the expression above we take $a=(\epsilon,n,x,y,z)$ to identify the corresponding equation of motion (energy, charge or momentum along $x$, $y$ or $z$).

From the above prescription we add in total 70 source terms, 14 for each equation, however we can use symmetry arguments to reduce the number of independent extra source terms. Specifically, our theory does not have any scalar parameter that breaks parity, hence all the $c_a^{\mu\nu}$ and $r_a^\mu$ must be parity even. Furthermore, the background flat spacetime theory has rotational invariance, which, together with $\mathcal{P}$-symmetry, allows us to impose parity with respect to a single axis $\mathcal{P}_i:i\rightarrow-i$, with $i=x,y,z$ and reduce the number of non-zero coefficients to only 13. Finally, because we take $\tau_m$ to be isotropic, we also assume isotropy of the sources too, which reduces the total number of independent non-vanishing coefficients to 9.

We can now take the full set of Green functions, that depends on these extra parameters $c_a^{\mu\nu}$ and $r_a^\mu$, and impose Onsager reciprocal relations \eqref{eqn:ch5:onsager_constraint_general}, trying to fix the free coefficients in such a way that the correlators preserve time-reversal covariance of the microscopic theory. The expressions are very large, for this reason we first impose Onsager relations at $\vect{k}=0$ and then at $\omega=k_y=k_z=0$. We found that this is enough to fix all nine parameters in terms of the relaxations $\tau$s, and subsequently we checked that these expressions lead to Green functions which obey Onsager relations at arbitrary $\omega$ and $\vect{k}$.

Of the nine extra source terms, we found that five of them are zero, while the non-vanishing ones are
\begin{subequations}\label{eqn:ch5:modified_equations_of_motion}
	\begin{align}
		\text{energy:}&	&	&-\left(\frac{\delta\epsilon}{\tau_{\epsilon\epsilon}}+\frac{\delta n}{\tau_{\epsilon n}}\right)-c_\epsilon^{tt}\delta h_{tt}-r^t_\epsilon \delta A_t\\
		\text{charge:}&	&	&-\left(\frac{\delta\epsilon}{\tau_{n\epsilon}}+\frac{\delta n}{\tau_{nn}}\right)-c_n^{tt}\delta h_{tt}-r^t_n\delta A_t
	\end{align}
\end{subequations}
The value of the non-vanishing extra coefficients is given in terms of the susceptibilities and relaxation rates
\begin{subequations}\label{eqn:ch5:extra_sources_final}
	\begin{align}
		c_\epsilon^{tt}&=\frac{1}{2}\left(\frac{\chi_{\epsilon n}}{\tau_{\epsilon n}}+\frac{\chi_{\epsilon\epsilon}}{\tau_{\epsilon\epsilon}}\right)\\
		r^t_\epsilon&=\frac{\chi_{\epsilon\epsilon}}{\tau_{n\epsilon}}+\frac{\chi_{\epsilon n}}{\tau_{nn}}=2c_n^{tt}\\
		r_n^t&=\frac{\chi_{\epsilon n}}{\tau_{n\epsilon}}+\frac{\chi_{nn}}{\tau_{nn}}
	\end{align}
\end{subequations}
With these corrections the correlators computed using the variational approach agree with the ones obtained via Martin-Kadanoff, up to the usual contact terms. Furthermore, we have not found other constraints on the relaxation parameters other than the one found already in the Martin-Kadanoff framework \eqref{eqn:ch5:onsager_condition}.

It is worth pointing out that the value of the coefficients is not important, what matters is that the final equations of motion \eqref{eqn:ch5:modified_equations_of_motion} have the right form after we substitute in the explicit values of $r_a^\mu$ and $c_a^{\mu\nu}$. In particular, we explained above that we decided to write the relaxation terms as they appear in flat space, agnostic on possible covariant completions. However, starting from a specific covariant expression for the relaxation terms, we could find relaxations-dependent source terms already in the undeformed equations of motion \eqref{eqn:ch5:equations_of_motion_curved_space}. Then, adding all possible relaxations as in \eqref{eqn:ch5:modified_equations_of_motion}, we expect different values for the coefficients $r_a$ and $c_a$, but such that they lead to the same final result.

Finally, we can also consider positivity of entropy production for the linearized hydrodynamic theory with background source, particularly taking into account the new metric and gauge field fluctuation terms discovered. In this case, the divergence of the entropy current gives
\begin{subequations}
	\begin{align}
		T_0\nabla_\mu^{(0)}\delta S^\mu&=\delta\epsilon\left(\frac{\mu_0}{\tau_{n\epsilon}}-\frac{1}{\tau_{\epsilon\epsilon}}\right)+\delta n\left(\frac{\mu_0}{\tau_{nn}}-\frac{1}{\tau_{\epsilon n}}\right)\nonumber\\
		&\quad-\left(r_\epsilon^t-\mu_0r_n^t\right)\delta A_t-\left(c_\epsilon^{tt}-\mu_0c_n^{tt}\right)\delta h_{tt}+\mathcal{O}(\partial^2,\delta^2)
	\end{align}
\end{subequations}
However, it is easy to check that with the expressions in \eqref{eqn:ch5:extra_sources_final}, positivity of entropy production is satisfied if the relaxation rates obey the Martin-Kadanoff constraints \eqref{eqn:ch5:entropy_constraint}, specifically the new terms proportional to $\delta A_t$ and $\delta h_{tt}$ in the equation above identically vanish and do not lead to further constraints.

\section{Summary, discussion and outlook}
In this chapter we studied relativistic hydrodynamics in the presence of relaxation terms in the linearized regime \eqref{eqn:ch5:equations_of_motion_flat_space}. We added all the possible relaxations up to order zero in derivatives and order one in fluctuations, and computed all the constraints the relaxation rates must satisfy that arise from microscopic time-reversal symmetry \eqref{eqn:ch5:onsager_condition}, positivity of entropy production \eqref{eqn:ch5:entropy_constraint} and linear stability \eqref{eqn:ch5:modes}.

Subsequently, we included the same set of relaxations in a theory placed on a weakly curved spacetime $\delta h_{\mu\nu}$ and in the presence of a linear gauge field $\delta A_\mu$. The response functions computed from the variational approach in general are not Onsager reciprocal and differ from the one obtained from the Martin-Kadanoff procedure. To amend the problem, we added all the possible terms linear in the external sources to the equations of motion \eqref{eqn:ch5:extra_sources}, each multiplied with a different unknown coefficient. What we found is that the simple requirement of time-reversal invariance of the theory allows us to uniquely fix all the newly added transport coefficients in terms of relaxation rates and susceptibilities \eqref{eqn:ch5:extra_sources_final}.

Although we tested this approach only for a relativistic charged fluid on a flat spacetime without background magnetic fields, we expect the procedure outlined in Section~\ref{sec:ch5:variational_approach} to hold very generally, for fluids with different symmetry content (boost-agnostic, Carrollian, Galilean, \dots), with different degrees of freedom (such as extra scalar fields), and for different backgrounds (curved spacetime or constant magnetic fields).

The outlook for this paper is closely connected to the previous Chapter~\ref{chapter:electrically_driven_fluids}: it would be interesting to study the generalized relaxations we introduced in a broader context, particularly looking for realizations beyond the linearized regime. Then, we could try to understand if the relaxation can enter the constitutive relations by modifying certain terms as in \cite{Amoretti:NondissipativeElectricallyDriven}, inducing new transport coefficients \cite{Baggioli:QuasihydrodynamicsSchwingerKeldyshEffective} or renormalizing existing ones \cite{Gouteraux:DrudeTransportHydrodynamic}.

Finally, in the next chapter we are going to use certain results obtained here to study the anomalous DC transport in Weyl semimetals. We will see how generalized relaxations can be obtained from a kinetic theory perspective, however it remains an open question how to arrive at the same result from holography or Schwinger-Keldysh formalism.
\chapter{Anomalous hydrodynamics and Weyl semimetals}\label{chapter:anomalous_hydrodynamics}
\epigraph{``I usually solve problems by letting them devour me.''}{Franz Kafka, \emph{Letter to Max Brod}}

\section{Introduction}
One of the most peculiar features of quantum field theories is the presence of anomalies: the breakdown of some classical symmetry due to quantum fluctuations. One of the most well-known examples is the so-called chiral/axial anomaly (or ABJ anomaly, after Adler, Bell and Jackiw \cite{Adler:AxialVectorVertexSpinor,Bell:PCACPuzzleP0}), namely the classical system enjoys a chiral symmetry, so that left- and right-handed fermions are separately conserved, but the quantum system is anomalous and there is mixing between left- and right-handed particles in the presence of external electromagnetic fields \cite{Bertlmann:AnomaliesQuantumField,Arouca:QuantumFieldTheory,Landsteiner:NotesAnomalyInduced}.

More than a decade ago, following hints from holography \cite{Erdmenger:FluidDynamicsRcharged,Banerjee:HydrodynamicsChargedBlack}, it was discovered that quantum anomalies could leave an imprint on the macroscopic dynamics of fluids, not only by modifying the equations of motion for the anomalous current, but also by allowing for extra transport coefficients in the constitutive relations \cite{Son:HydrodynamicsTriangleAnomalies,Sadofyev:ChiralMagneticEffect,Neiman:RelativisticHydrodynamicsGeneral}. The anomalous terms are non-dissipative and parity-odd, since they are proportional to the external magnetic fields and vorticity. See some of the reviews \cite{Landsteiner:NotesAnomalyInduced,Chernodub:ThermalTransportGeometry,Arouca:QuantumFieldTheory} and original works \cite{Son:HydrodynamicsTriangleAnomalies,Jensen:AnomalyInflowThermal,Jensen:TriangleAnomaliesThermodynamics,Jensen:ThermodynamicsGravitationalAnomalies,Landsteiner:GravitationalAnomalyTransport} for more details, also about the role of the mixed-gravitational anomaly and the generating functional.

What was quickly realized is that anomalous hydrodynamics could play an important role, not only for the description of heavy-ion collisions, but also to understand the transport properties of Weyl semimetals (WSMs) \cite{Lucas:HydrodynamicTheoryThermoelectric}. WSMs are a novel class of $3d$ topological quantum materials that host gapless chiral excitations. In these compounds the valence and conductance bands have linear crossing points (the Weyl nodes) close to the Fermi surface, and the system enjoys an accidental chiral symmetry, such that the Weyl nodes must always come in pairs of opposite chirality \cite{Nielsen:AbsenceNeutrinosLattice,Nielsen:NogoTheoremRegularizing}. If the Weyl cones lie on top of each other, then the material is a Dirac semimetal, but in the presence of parity or time-reversal symmetry breaking the cones will be separated in momentum space, in which case we speak of a WSM. Because of their chiral symmetry, many of their properties are strongly influenced by the presence of the axial anomaly, thus allowing us to study QFT anomalies in a controlled tabletop setting, contrary to what happens with heavy-ion-collisions experiments. For some reviews on WSMs, see e.g., \cite{Armitage:WeylDiracSemimetals,Burkov:WeylMetals,Hosur:RecentDevelopmentsTransport}.

\begin{figure}
	\centering
	\includegraphics[width=0.6\textwidth]{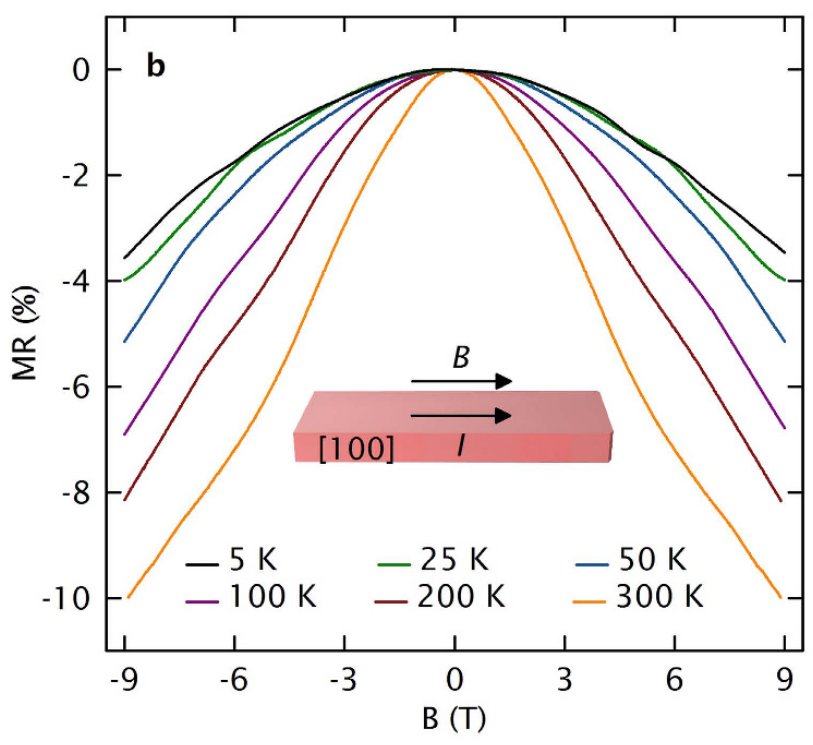}
	\caption{Figure taken from \cite{Niemann:ChiralMagnetoresistanceWeyl}. Observation of longitudinal NMR in a sample of doped $\chem{NbP}$ Weyl semimetal.}
	\label{fig:ch6:magneto_resistance}
\end{figure}

In particular, one of the most prominent features expected in WSMs due to the anomaly is a large longitudinal negative magnetoresistance (NMR), first argued in \cite{Nielsen:AdlerBellJackiwAnomalyWeyl}. This means that, contrary to ordinary metals in which the conductivity decreases with the applied magnetic field \cite{Pippard:MagnetoresistanceMetals}, in WSMs we expect the longitudinal electrical conductivity (measured parallel to the applied external magnetic field) to be strongly enhanced by the presence of a background electric field. Indeed, longitudinal NMR has been observed both for the electric and thermal transport in different compounds \cite{Niemann:ChiralMagnetoresistanceWeyl,Vu:ThermalChiralAnomaly,Xiong:EvidenceChiralAnomaly,Huang:ObservationChiralAnomaly,Gooth:ExperimentalSignaturesMixed,Jia:ThermoelectricSignatureChiral}. In practice, real materials can have multiple source of longitudinal NMR due to other effects, and distinguishing the various contributions from experiments is generally difficult.

The NMR can be predicted from many different approaches, based on semiclassical models (such as diffusion or chiral kinetic theory \cite{Stephanov:ChiralKineticTheory}) or on microscopic quantum models \cite{Abbasi:MagnetoTransportChiralFluid,Gorbar:NonlocalTransportWeyl,Son:ChiralAnomalyClassical,Sukhachov:AnomalousGurzhiEffect,Gorbar:AnomalousThermoelectricPhenomena,Gorbar:AnomalousTransportProperties,Kim:AnomalousTransportPhenomena,Das:NonlinearMagnetoconductivityWeyl,Mandal:NonlinearMagnetotransportWeyl}, but also from hydrodynamic arguments \cite{Lucas:HydrodynamicTheoryThermoelectric,Landsteiner:NegativeMagnetoresistivityChiral,Abbasi:MagnetotransportAnomalousFluid}. Furthermore, like other Dirac materials, WSMs could enjoy an extended hydrodynamic regime, as observed in \cite{Gooth:ElectricalThermalTransport,Kumar:ExtremelyHighConductivity,Jaoui:DepartureWiedemannFranzLaw}. Consequently, anomalous hydrodynamics might apply reasonably well to study the macroscopic transport dynamics of these systems.

In this chapter, we are going to discuss some questions related to anomalous fluids and possible applications to models of WSMs. First, following \cite{Amoretti:LeadingOrderMagnetic}, we review the usual computations of the anomalous optical thermoelectric conductivities and show that the NMR obtained in previous works is not physical, since that different hydrodynamic frames lead to different results. Specifically, the NMR depends quadratically on the applied magnetic field, which means that order-one hydrodynamics with an order-one in derivatives magnetic field is not appropriate to study the magnetic-field dependence of the conductivities. We discuss this issue and suggest a way to properly compute the NMR in Section~\ref{sec:ch6:negative_magneto_resistance}.

Next, in Section~\ref{sec:ch6:dc_conductivities} we discuss some problems related to the DC values of the thermoelectric conductivities. We already discussed in the previous chapter that hydrodynamics leads to finite DC conductivities only if there is some effective mechanism to relax momentum \cite{Hartnoll:TheoryNernstEffect}, but in the presence of an axial anomaly energy and axial charge should relax too for the conductivities to be finite in the $\omega\rightarrow0$ limit \cite{Landsteiner:NegativeMagnetoresistivityChiral,Abbasi:MagnetotransportAnomalousFluid}. However, we show that previous models of anomalous relaxed hydrodynamics necessarily predict that electric charge, energy and momentum should all relax at the same rate of the axial charge, which we deem unphysical. Therefore, to build a better model, we employ the generalized relaxations introduced in Chapter~\ref{chapter:onsager} and obtain a theory of relaxed hydrodynamics which conserves the electric charge of the system.

Finally, in Section~\ref{sec:ch6:kinetic_theory}, we show how these relaxations can be obtained from kinetic theory, using an appropriate Relaxation-Time-Approximation (RTA) ansatz for the scattering of electrons with impurities and phonons in which the microscopic relaxation rate depends on the quasiparticle energy.

\section{Anomalous hydrodynamics and transport}\label{sec:ch6:negative_magneto_resistance}
Remember that hydrodynamics, as an effective field theory, presents an ambiguity in the choice of the hydrodynamic variables at derivative order, see Section~\ref{sec:ch2:frame_choice}. By changing the definitions of temperature $T$, fluid velocity $u^\mu$ and chemical potential $\mu$, we can change how the constitutive relations look. This choice however is not physical, but a simple re-parametrization: all frames, in the hydrodynamic regime, should give the same values for any observable, therefore the conductivities, i.e. the retarded Green functions, should take the same form in all hydrodynamic frames. Nevertheless, the choice of frame can have huge impact on certain aspects of the theory (such as stability and causality) at high-frequency/high-momenta, see the discussion in the Introduction~\ref{chapter:introduction}.

In this section we consider order-one hydrodynamics with a $\mathrm{U(1)}$ anomalous chiral current, and we show that the optical thermoelectric conductivities computed in the presence of a background order-one magnetic field $B\sim\mathcal{O}(\partial)$ are explicitly frame-dependent, which means that the contribution of the anomaly to the conductivities is not a physical effect and should be discarded.

To resolve this problem, one could work with hydrodynamics at order two in derivatives, however this would just shift the issue to the $\mathcal{O}(B^3)$ subleading corrections to the conductivities. Thus, to avoid the issue altogether, we suggest working with an external magnetic field which is order zero in derivatives $B\sim\mathcal{O}(1)$, so that the expressions for the conductivities can be trusted at any value of $B$ without resorting to order-two hydrodynamics.

\subsection{Standard approach to anomalous hydrodynamics}
We consider now a hydrodynamic theory with a conserved stress-energy tensor and a single $\mathrm{U(1)}$ anomalous chiral current in $(3+1)$-dimensional spacetime. For the moment, we will not consider the effects of the gravitational anomaly or other setups, however in later sections we will instead consider a theory with symmetry $\mathrm{U(1)}_V\times\mathrm{U(1)}_A$, which is the relevant symmetry content to describe WSMs.

The equations of hydrodynamics for a covariant anomalous current\footnote{For a short discussion on covariant and consistent anomalies we refer to Appendix~\ref{appendix:currents}.} and stress-energy tensor in an external electromagnetic field are
\begin{subequations}\label{eqn:ch6:anomalous_equations_of_motion_one_current}
	\begin{align}
		\partial_\mu T^{\mu\nu}&=F^{\nu\lambda}J_\lambda\\
		\partial_\mu J^\mu&=cE^\mu B_\mu
	\end{align}
\end{subequations}
where $c$ is the anomaly coefficient and its explicit value depends on the microscopic theory. For the moment, following the standard approach \cite{Son:HydrodynamicsTriangleAnomalies,Neiman:RelativisticHydrodynamicsGeneral,Jensen:AnomalyInflowThermal}, we are considering both $E_\mu$ and $B_\mu$ to be order one in derivatives $\mathcal{O}(\partial)$.

To write the constitutive relations for the anomalous fluid we will follow the original work \cite{Son:HydrodynamicsTriangleAnomalies}. While there are now other methods, based on the equilibrium generating functional \cite{Jensen:TriangleAnomaliesThermodynamics,Jensen:AnomalyInflowThermal,Banerjee:ConstraintsAnomalousFluid}, which are easier to generalize to arbitrary dimensions or different anomalies (while also providing insights on the appearance of the gravitational anomaly at order one in derivatives \cite{Jensen:ThermodynamicsGravitationalAnomalies}), the original approach gives a clearer picture on the issue of hydrodynamic frames.

To begin, consider the constitutive relations in the Landau frame, and add the most generic parity-odd terms which are order one in derivatives and that depend on the external magnetic field $B^\mu$ and vorticity $\Omega^\mu=\varepsilon^{\mu\nu\rho\sigma}u_\nu\partial_\rho u_\sigma$. Thus, we arrive at
\begin{subequations}\label{eqn:ch6:constitutive_relations_generic_frame}
	\begin{align}
		T^{\mu\nu}&=\epsilon u^\mu u^\nu+P\Delta^{\mu\nu}+\xi^\epsilon_B\left(u^\mu B^\nu+u^\nu B^\mu\right)+\xi^\epsilon_\Omega\left(u^\mu\Omega^\nu+u^\nu\Omega^\mu\right)\nonumber\\
		&\quad-\eta\Delta^{\mu\alpha}\Delta^{\nu\beta}\sigma_{\alpha\beta}-\zeta\Delta^{\mu\nu}\partial_\alpha u^\alpha +\mathcal{O}(\partial^2)\\
		J^\mu&=nu^\mu+\sigma\Delta^{\mu\nu}\left(E_\nu-T\partial_\nu\frac{\mu}{T}\right)+\xi_\Omega \Omega^\mu+\xi_B B^\mu+\mathcal{O}(\partial^2)
	\end{align}
\end{subequations}
where $\sigma_{\mu\nu}$ is the shear tensor. The quantities $\xi$ are the anomalous parity-odd dissipationless transport coefficients. Like $\sigma$, $\zeta$ and $\eta$ their values depend on the order-zero thermodynamic variables $\xi(\mu,T)$, however, unlike the standard dissipative transport coefficients, this dependence is almost entirely fixed by the anomaly and the choice of hydrodynamic frame.

From a physical perspective, the $\xi_\Omega$ and $\xi_B$ terms represent the well-known Chiral Vortical Effect, Chiral Magnetic Effect (or Chiral Separation Effect, if multiple currents are present) discussed in \cite{Fukushima:ChiralMagneticEffect,Landsteiner:AnomalousTransportKubo,Landsteiner:GravitationalAnomalyTransport}.

\subsection{Positivity of entropy production}
The constitutive relations must be constrained by the second law of thermodynamics. To begin, we write the total entropy current $S^\mu$ as the sum of the canonical entropy current (that is related to the standard non-anomalous fluid) and a second contribution $S^\mu_\text{n.c.}$ that is related to the anomaly
\begin{equation}
	S^\mu=\frac{1}{T}\left(Pu^\mu-T^{\mu\nu}u_\nu-\mu J^\mu\right)+S^\mu_\text{n.c.}
\end{equation}
This second term is needed to ensure positivity of entropy production holds, and we can express it in a generic form as
\begin{equation}
	S^\mu_\text{eq}=\xi^s_BB^\mu+\xi^s_\Omega\Omega^\mu
\end{equation}
Subsequently, we can compute the divergence of the entropy current, using the equations of motion and the constitutive relations in the usual way, to arrive at the following expression
\begin{align}
	\partial_{\mu} S^{\mu}
	&=	\frac{\zeta}{T} \theta^2 + \frac{\eta}{T} \sigma_{\mu \nu} \sigma^{\mu \nu} + \sigma \left( E_{\mu} - T \partial_{\mu}^{\perp} \frac{\mu}{T} \right)^2\nonumber \\
	&\quad+ \left(  \frac{\partial \xi^{s}_{B}}{\partial T} + \frac{\mu}{T} \frac{\partial \xi^{s}_{B}}{\partial \mu} - \frac{\xi^{\epsilon}_{B}}{T^2} \right) B^{\mu} \left( \partial_{\mu}^{\perp} T + T a_{\mu} \right) \nonumber \\
	&\quad +  \left(  \frac{\partial \xi^{s}_{\Omega}}{\partial T} + \frac{\mu}{T} \frac{\partial \xi^{s}_{\Omega}}{\partial \mu} - \frac{\xi^{\epsilon}_{\Omega}}{T^2} \right) \Omega^{\mu} \left( \partial_{\mu}^{\perp} T + T a_{\mu} \right)  \nonumber \\
	&\quad - \left( \frac{\partial \xi^{s}_{B}}{\partial \mu} - \frac{\xi_{B}}{T} \right) B^{\mu} \left(  E_{\mu}  - T \partial_{\mu}^{\perp} \left( \frac{\mu}{T} \right) \right) \nonumber \\
	&\quad - \left( \frac{\partial \xi^{s}_{\Omega}}{\partial \mu} - \frac{\xi_{\Omega}}{T} \right) \Omega^{\mu} \left(  E_{\mu} - T \partial_{\mu}^{\perp} \left( \frac{\mu}{T} \right) \right) \nonumber \\
	&\quad + \left( \xi^{s}_{B} - T \frac{\partial \xi^{s}_{B}}{\partial T} - \mu \frac{\partial \xi^{s}_{B}}{\partial \mu} \right) B\cdot a+ \left( 2 \xi^{s}_{\Omega} - T \frac{\partial \xi^{s}_{\Omega}}{\partial T} - \mu \frac{\partial \xi^{s}_{\Omega}}{\partial \mu} \right) \Omega\cdot a \nonumber \\
	&\quad + \left( \frac{\partial \xi^{s}_{B}}{\partial \mu} - \frac{c \mu}{T} \right) E\cdot B + \left( \frac{\partial \xi^{s}_{\Omega}}{\partial \mu}  - \xi^{s}_{B} \right) E\cdot\Omega\geq0
\end{align}
where we defined $\partial_\mu^\perp=\Delta_{\mu\nu}\partial^\nu$.

When studying the positivity of entropy production, we can encounter two different kinds of conditions: inequality type constraints, which are related to the terms with positive definite signature in the first line, thus requiring $\sigma,\zeta,\eta\geq0$, and equality type constraints, which originate from terms without definite sign and therefore their coefficients must vanish. This is for example what we found in Section~\ref{sec:ch2:hydrodynamics} when discussing $\chi_T$, see \eqref{eqn:ch2:entropy_constraints}.

We can employ the equations of motion of the ideal fluid to relate the divergences of the vorticity and the magnetic field to other quantities as
\begin{subequations}
	\begin{align}
		\partial_\mu\Omega^\mu&=-\frac{2}{\epsilon+P}\Omega^\mu\left(\partial_\mu P-nE_\mu\right)\\
		\partial_\mu B^\mu&=-2\Omega^\mu E_\mu+\frac{1}{\epsilon+P}\left(nE_\mu B^\mu-B^\mu\partial_\mu P\right)
	\end{align}
\end{subequations}
which we can use to simplify the constraints that stem from the second law of thermodynamics. The requirement that $\partial_\mu S^\mu\geq0$ means that we should set to zero all the anomaly-related terms, which corresponds to
\begin{subequations}
	\begin{align}
		0&=\partial_\mu\xi^s_\Omega-\frac{2\partial_\mu P}{\epsilon+P}\xi^s_\Omega-\xi_\Omega\partial_\mu\frac{\mu}{T}+\left(\frac{2\partial_\mu P}{\epsilon+P}-a_\mu-\partial_\mu\right)\xi^\epsilon_\Omega\\
		0&=\partial_\mu\xi^s_B-\frac{\partial_\mu P}{\epsilon+P}\xi^s_B-\xi_B\partial_\mu\frac{\mu}{T}+\left(\frac{\partial_\mu P}{\epsilon+P}-a_\mu-\partial_\mu\right)\xi^\epsilon_B\\
		0&=\frac{2n\xi^s_\Omega}{\epsilon+P}-2\xi^s_B+\frac{\xi_\Omega}{T}+2\xi^\epsilon_B-2\xi^\epsilon_\Omega \frac{n}{\epsilon+P}\\
		0&=\frac{n\xi^s_B}{\epsilon+P}+\frac{\xi_B}{T}-c\frac{\mu}{T}-\xi^\epsilon_B\frac{n}{\epsilon+P}
	\end{align}
\end{subequations}

The above set of equations is not closed: there are six different $\xi$s, but only four equations. This is because we kept $\xi^\epsilon_B$ and $\xi^\epsilon_\Omega$ non-zero in the above expressions, contrary to what is done in \cite{Son:HydrodynamicsTriangleAnomalies}. It is then clear that we need some other constraints to uniquely determine the $\xi$s, which come from the choice of frame. In particular, the Landau frame matching conditions require us to set $\xi^\epsilon_B=\xi^\epsilon_\Omega=0$, then we can integrate the above differential equations to obtain, up to integration constants related to the gravitational anomaly and CPT-violating terms \cite{Neiman:RelativisticHydrodynamicsGeneral}, the following expressions
\begin{subequations}\label{eqn:ch6:landau_frame_anomalous_coefficients}
	\begin{align}
		\xi^\epsilon_B&=0	&	\xi_B&=c\left(\mu-\frac{1}{2}\frac{n\mu^2}{\epsilon+p}\right)\\
		\xi^\epsilon_\Omega&=0		&	\xi_\Omega&=c\left(\mu^2-\frac{2}{3}\frac{n\mu^3}{\epsilon+p}\right)
	\end{align}
\end{subequations}
in agreement with \cite{Son:HydrodynamicsTriangleAnomalies}.

Another possibility is to look for a frame in which the entropy current receives no corrections \cite{Loganayagam:AnomalyInducedTransport}, or one might prefer to work in the thermodynamic frame, that keeps the definition of temperature, fluid velocity and chemical potential the canonical ones \eqref{eqn:ch2:thermodynamic_frame_definitions}.

To move from one frame to another, as we discussed in Section~\ref{sec:ch2:frame_choice}, we have to change the definitions of the fluid fields by derivative corrections which vanish in equilibrium. For the purpose of this discussion, i.e. frame transformations that shift around the anomaly-related terms, the only frame transformations we are interested in are of the form
\begin{equation}\label{eqn:ch6:frame_transformation}
	u^\mu\rightarrow u^\mu+f_B(\mu,T)B^\mu+f_\Omega(\mu,T)\Omega^\mu
\end{equation}
which act on the constitutive relations by changing the values of the anomalous coefficients as
\begin{subequations}
	\begin{align}
		\xi^{\epsilon}_{B,\Omega} &\rightarrow \xi^{\epsilon}_{B,\Omega} + (\epsilon + P) f_{B,\Omega}(\mu,T)\\
		\xi_{B,\Omega} &\rightarrow \xi_{B,\Omega} + n f_{B,\Omega}(\mu,T)
	\end{align}
\end{subequations}
Notice that the above frame transformations are allowed because both $B^\mu$ and $\Omega^\mu$ are $\mathcal{O}(\partial)$ and orthogonal to $u^\mu$. In particular, we can use
\begin{equation}
	f_B=\frac{c\mu^2}{2\left(\epsilon+P\right)}\qquad\qquad f_\Omega=\frac{c\mu^3}{3\left(\epsilon+P\right)}
\end{equation}
to move between the constitutive relations in the Landau frame \eqref{eqn:ch6:landau_frame_anomalous_coefficients} and the thermodynamic frame, in which the anomalous transport coefficients take the simplest form
\begin{subequations}\label{eqn:ch6:anomalous_transport_coefficients_thermodynamic_frame}
	\begin{align}
		\xi^\epsilon_B&=\frac{1}{2}c\mu^2&	\xi_B&=c\mu\\
		\xi^\epsilon_\Omega&=\frac{1}{3}c\mu^3&	\xi_\Omega&=\frac{1}{2}c\mu^2
	\end{align}
\end{subequations}

\subsection{Frame-dependent conductivities}
Having discussed the constitutive relations, we now study how the choice of frame affects the value of conductivities, at order one in derivatives. To proceed, we can analyse more carefully the general argument from Section~\ref{sec:ch2:linear_response_theory} that is used to solve for the retarded Green functions in hydrodynamics. Consider the linearized hydrodynamic equations for a fluid order one in derivatives, schematically
\begin{equation}
	D_{ab}\phi_b=\lambda_b+\mathcal{O}(\partial^3)
\end{equation}
where $D_{ab}$ is an operator which is at least order one in derivatives, $\phi_a$ is a vector of the fluctuating hydrodynamic fields $\phi_a=\left(\delta T,\delta\mu,\delta v^i\right)$, and $\lambda_a$ are the external sources, like gauge fields, metric fluctuations, but also their derivatives. If we consider order-one fluids, then both $\lambda_a$ and $D_{ab}$ contain, at most, terms with two derivatives.

In Fourier space we can introduce a formal parameter $\varepsilon$ to count the derivative order of the various terms. With the setup we are using we have $\omega,\vect{k},B\sim\varepsilon$. Then, a generic Green function is obtained by inverting the operator $D$ and extracting the corresponding entry on the RHS \eqref{eqn:ch2:retarded_green_functions}, so that schematically we arrive at
\begin{equation}
	G^R\sim\frac{a(\omega,\vect{k})+\mathcal{O}(\varepsilon^2)}{b(\omega,\vect{k})+\mathcal{O}(\varepsilon^3)}
\end{equation}
where $a$ and $b$ are functions that depend on the correlator of interest. The numerator is thus determined only up to order one in the counting parameter $\varepsilon$, as a consequence of the fact that the constitutive relations for the currents are known only up to order one in derivatives.

We can study the consequences of this argument on the anomalous transport: consider the constitutive relations for the chiral fluid in a generic frame \eqref{eqn:ch6:constitutive_relations_generic_frame}. To compute the longitudinal magnetoresistance we need to select a background with non-zero magnetic field $B$, which we take along $\hat{z}$, $F^{12}=B$, moreover we will have $T$ and $\mu$ constant and zero spatial velocity. With this choice of background the equilibrium configuration for the fluid is given by
\begin{subequations}\label{eqn:ch6:equilibrium_order_one}
	\begin{align}
		T^{\mu\nu}&=\begin{pmatrix}
			\epsilon    &   0   &   0   &   \xi_B^\epsilon B\\
			0   &   P   &   0   &   0\\
			0   &   0   &   P   &   0\\
			\xi_B^\epsilon B  &  0   &   0   &   P
		\end{pmatrix}\\
		J^\mu&=\left(n,0,0,\xi_B B\right)
	\end{align}
\end{subequations}
We can now add linearized fluctuations to the hydrodynamic fields and sources $F^{0z}=\delta\mathbb{E}^z$, and solve the equations of motion \eqref{eqn:ch6:anomalous_equations_of_motion_one_current} in Fourier space to obtain the optical conductivities using the Martin-Kadanoff method as discussed in Section~\ref{sec:ch2:martin_kadanoff}.

Generally, we find that the fluctuations of the hydrodynamic fields depend on the sources as, e.g.
\begin{equation}
	\delta u_{z} \bigr|_{\vect{k}=\vect{0}} = \left[ \frac{- i n + \varepsilon^2 \alpha B^2 + \mathcal{O}(\varepsilon^3)}{(\epsilon + p) \omega \varepsilon + \mathcal{O}(\varepsilon^3)} \right] \delta \mathbb{E}_{z}
\end{equation}
with $\alpha$ some thermodynamic function. From the argument above, we know that the $\mathcal{O}(\varepsilon^2)$ dependence on the numerator should be discarded: it is not physical and is an artefact of our equations and choice of the background.

Indeed, we can compute the leading-order dependence on the magnetic field of the longitudinal electric conductivity (similar results hold for the thermal and thermoelectric ones too) and check directly that this expression depends explicitly on the hydrodynamic frame, proving that it is not physical
\begin{align}\label{eqn:ch6:conductivity_generic_frame}
	\sigma(\omega)&=\sigma+\frac{i n^2}{\omega w}+\frac{i B^2}{\omega w^2(\frac{\partial\epsilon}{\partial T} \frac{\partial n}{\partial\mu}-\frac{\partial\epsilon}{\partial\mu}\frac{\partial n}{\partial T})}\left[\left(w\frac{\partial\xi_B}{\partial\mu}-n\frac{\partial\xi_B^\epsilon}{\partial\mu}\right)\right.\nonumber\\
	&\quad\left(w\left(c\frac{\partial\epsilon}{\partial T}-\frac{\partial n}{\partial T}\xi_B\right)-\frac{\partial\epsilon}{\partial T}n\xi_B+2\frac{\partial n}{\partial T}n\xi_B^\epsilon\right)-\left(w\frac{\partial\xi_B}{\partial T}-n\frac{\partial\xi_B^\epsilon}{\partial T}\right)\nonumber\\
	&\left.\quad\left(w\left(c\frac{\partial\epsilon}{\partial\mu}-\frac{\partial n}{\partial\mu}\xi_B\right)-\frac{\partial\epsilon}{\partial\mu}n\xi_B+2\frac{\partial n}{\partial\mu}n\xi_B^\epsilon\right)\right]
\end{align}
where $w=\epsilon+P$ is the enthalpy density. The $\mathcal{O}(B^2)$ term is explicitly frame dependent, as it depends on the values of the anomalous transport coefficients $\xi$s and their derivatives.

We can compute the above formula in the Landau frame at $\xi^\epsilon_B=0$ to recover the results of \cite{Landsteiner:NegativeMagnetoresistivityChiral}, or we can plug in the thermodynamic-frame transport coefficients \eqref{eqn:ch6:anomalous_transport_coefficients_thermodynamic_frame} to get a different result \cite{Abbasi:MagnetotransportAnomalousFluid}.

Again, we emphasize that this observed frame dependence, and thus the $B^2$ correction in the conductivity, are the signature that we are incorrectly keeping terms which are higher order in derivatives than our constitutive relations allow us to. The $B^2$ term should be discarded, it is not physical, which means that the conductivities do not depend on the anomaly at all, at this order in derivatives.

There are other issues with the thermoelectric conductivities computed in different frames at order one in derivatives, which hint to the fact that the magnetic-field dependent terms should be ignored. For example, in general the correlators are not Onsager reciprocal $\alpha\neq\bar{\alpha}$, moreover computing the conductivities using the gradient of the chemical potential as a source, instead of $\delta\mathbb{E}_z$, leads to different results.

Notice that in anomalous hydrodynamics the equilibrium around which we fluctuate also depend on the frame \eqref{eqn:ch6:equilibrium_order_one}, while usually the equilibrium configuration is frame independent. Although even in standard hydrodynamics response functions computed in different frames can lead to apparently non-equivalent expressions, they can always be related to one another by mapping the transport coefficients. In the anomalous case \eqref{eqn:ch6:conductivity_generic_frame} such map does not exist, because the parity-odd anomalous transport coefficients take a specific value for each given frame, see \eqref{eqn:ch6:landau_frame_anomalous_coefficients} and \eqref{eqn:ch6:anomalous_transport_coefficients_thermodynamic_frame}.

\subsection{Killing the ambiguity}
It should be clear by now that the longitudinal conductivities in order-one anomalous hydrodynamics do not have any dependence on the magnetic field. To restore the NMR, i.e. to observe the effect of the anomaly on transport, one way is to work at order two in derivatives \cite{Abbasi:MagnetotransportAnomalousFluid,Megias:AnomalousTransportSecond,Kharzeev:AnomaliesTimeReversal,Bhattacharyya:SecondOrderTransport}. Then the $\mathcal{O}(B^2)$ term in the conductivity is an honest physical effect and can be trusted, however the same ambiguity would now arise at $\mathcal{O}(B^3)$.

To avoid these problems altogether we suggest to simply consider the magnetic field to be order zero in derivatives $B\sim\mathcal{O}(1)$, so that it is an integral part of the thermodynamics. With this choice, we can study the anomalous transport without resorting to order-two hydrodynamics, and furthermore we are allowed to trust our results at large values of $B$, since we do not have to truncate the results at leading order in $B^2$. Morally, we can think of this procedure as a partial re-summation at all orders of the hydrodynamic theory with $B\sim\mathcal{O}(\partial)$ (partial in the sense that we are not including derivatives of $B$ in the re-summation).

When $B\sim\mathcal{O}(1)$ it cannot be used in frame transformations \eqref{eqn:ch6:frame_transformation}, then the equilibrium configuration on a background with a fixed magnetic field will be unique for all frames and the conductivities will be well-defined.

A theory with order-zero magnetic field has been studied in \cite{Ammon:ChiralHydrodynamicsStrong} for the case of a single $\mathrm{U(1)}$ axial current. We can then start from their constitutive relations to compute the thermoelectric transport of an anomalous fluid. We will focus only on the ideal fluid: on one hand, because the expressions are simpler (and even the conductivities of an ideal fluid can quickly become complicated with relaxation terms, as we will see), on the other hand, because this is already enough to observe anomalous transport phenomena.

We decompose the stress-energy tensor and the current with respect to $u^\mu$ in the usual way \eqref{eqn:ch2:tensor_current_standard_decomposition}
\begin{subequations}
	\begin{align}
		T^{\mu\nu}&=\mathcal{E}u^\mu u^\nu+\mathcal{P}\Delta^{\mu\nu}+\mathcal{Q}^\mu u^\nu+\mathcal{Q}^\nu u^\mu+\mathcal{T}^{\mu\nu}\\
		J^\mu&=\mathcal{N}u^\mu+\mathcal{J}^\mu
	\end{align}
\end{subequations}
and, because we are interested in the ideal fluid, we can use the hydrostatic generating functional. We define the hydrodynamic fields in terms of the metric, gauge field and Killing vector $V^\mu$ in agreement with \eqref{eqn:ch2:thermodynamic_frame_definitions}, which at order one and higher defines the thermodynamic frame.

The non-anomalous part of the constitutive relations can be obtained from the standard generating functional
\begin{equation}
	W[g,A]=\int\dif^{3+1}x\ \sqrt{-g}P(T,\mu,B^2)
\end{equation}
where the equilibrium pressure now depends also on $B^2$. Varying with respect to metric and gauge fluctuations as explained in Section~\ref{sec:ch2:generating_functional}, we arrive at the constitutive relations
\begin{subequations}
	\begin{align}
		\mathcal{E}&=-P+sT+\mu n \\
		\mathcal{P}&=P-\frac{2}{3}\chi_BB^2 \\
		\mathcal{Q}^\mu&=-\chi_B\varepsilon^{\mu\nu\rho\sigma}u_\nu E_\rho B_\sigma\\
		\mathcal{T}^{\mu\nu}&=\chi_B\left(B^\mu B^\nu-\frac{1}{3}\Delta^{\mu\nu}B^2\right) \\
		\mathcal{N}&=n \\
		\mathcal{J}^\mu&=\varepsilon^{\mu\nu\rho\sigma}u_\nu\nabla_\rho\mathfrak{m}_\sigma+\varepsilon^{\mu\nu\rho\sigma}u_\nu a_\rho\mathfrak{m}_\sigma
	\end{align}
\end{subequations}
where we introduced the magnetization and magnetic susceptibility as
\begin{equation}
	\mathfrak{m}^\mu=\chi_BB^\mu\qquad\qquad\chi_B=2\frac{\partial P}{\partial B^2}
\end{equation}
This completely characterizes the ideal fluid constitutive relations without anomaly with an order-zero magnetic field.

Finally, we need to introduce the parity-odd contributions due to the anomaly. The details can be found in \cite{Ammon:ChiralHydrodynamicsStrong,Jensen:AnomalyInflowThermal}. In $3+1$-dimensions, the anomalous generating functional for the covariant current takes the form
\begin{multline}\label{eqn:ch6:generating_functional_anoamly}
	W_\text{anom}=\int\mathrm{d}^{3+1}x\ \sqrt{-g}\ \frac{c}{3}\mu\ B^\mu A_\mu	-\frac{c}{24}\int\mathrm{d}^{4+1}x\ \sqrt{-G}\varepsilon^{mnopq}A_mF_{no}F_{pq}\ .
\end{multline}
The first term is responsible for generating the consistent-anomaly constitutive relations \cite{Bardeen:ConsistentCovariantAnomalies,Wess:ConsequencesAnomalousWard}, the latter term instead is the anomaly-inflow Chern-Simons functional \cite{Callan:AnomaliesFermionZero}, whose boundary contributions are the Bardeen-Zumino currents that make the total current covariant \cite{Landsteiner:NotesAnomalyInduced}, see Appendix~\ref{appendix:currents}. By varying this functional with respect to metric and gauge fluctuations we find the two anomalous transport coefficients for the covariant current as
\begin{equation}
	\xi^\epsilon_B=\frac{1}{2}c\mu^2\qquad\qquad\xi_B=c\mu
\end{equation}
These coefficients agree with the thermodynamic frame, obviously: they are obtained from the same generating functional \eqref{eqn:ch6:generating_functional_anoamly} in which the magnetic field has been promoted to be $B\sim\mathcal{O}(1)$. Nevertheless, even if the contribution to the constitutive relations due to the anomaly is the same as in the thermodynamic frame, the longitudinal optical conductivities computed on a background at constant $B$ are different
\begin{subequations}\label{eqn:ch6:anomalous_conductivities_one_current}
	\begin{align}
		\sigma(\omega)&= \frac{i}{\omega} \left[ \frac{n^2}{(p+\epsilon)} + \Xi B^2 \right] + \mathcal{O}(\omega^{0})\\
		\alpha(\omega) &= \frac{i}{\omega} \left[ \frac{ns}{(p+\epsilon)} - \mu \Xi B^2 \right] + \mathcal{O}(\omega^{0})\\
		\bar\kappa(\omega) &=  \frac{i}{\omega} \left[ \frac{s^2T}{(p+\epsilon)} + \frac{\mu^2 \Xi B^2}{T} \right] + \mathcal{O}(\omega^{0})
	\end{align}
where
\begin{equation}
	\Xi = \frac{c^2s^2 T^2( \frac{\partial n}{\partial T} \mu - \frac{\partial \epsilon}{\partial T})}{(p+\epsilon)(\frac{\partial\epsilon}{\partial\mu} \frac{\partial n}{\partial T} -  \frac{\partial \epsilon}{\partial T} \frac{\partial n}{\partial\mu}) +B^2c^2\mu^2( \frac{\partial \epsilon}{\partial T} - \frac{\partial n}{\partial T} \mu)}
\end{equation}
\end{subequations}
This is because now the expressions must not be truncated at $\mathcal{O}(B^0)$, but can be taken to be valid also for large values of $B$. In the above expressions, all thermodynamic quantities are now functions of $T$, $\mu$ and $B^2$.

The conductivities \eqref{eqn:ch6:anomalous_conductivities_one_current} are finally well-defined: the anomalous magnetoresistance contribution is a physical effect, they do not change upon frame transformations, and are Onsager reciprocal. Furthermore, they obey the standard Ward identities that hold also for non-anomalous hydrodynamics \eqref{eqn:ch2:ward_identities_conductivities}.

With this simple computation we correct a derivative-counting mistake that was unnoticed and easy to miss in the literature. In particular, we characterize the full magnetic-field dependence of the longitudinal magnetoresistance for an anomalous fluid in arbitrary background magnetic field. It is possible, by following \cite{Ammon:ChiralHydrodynamicsStrong}, to also include order-one corrections to such anomalous fluid. What we expect to find is that, contrary to what is usually believed, other transport coefficients should enter the conductivities together with $\sigma$\footnote{We still expect $\sigma$ to be the only relevant transport coefficient when expanding at small magnetic field.}.

\section{The problem with DC conductivities}\label{sec:ch6:dc_conductivities}
\subsection{Weyl semimetals, hydrodynamics and transport}
In this section we consider a hydrodynamic model for WSMs. We focus on the simplest case of a WSM with two Weyl nodes of opposite chirality separated in momentum space by a vector $b^\mu$, thus the low-energy excitations are massless Weyl fermions.

Classically, we would expect two separate conserved currents $\mathrm{U(1)}_L\times\mathrm{U(1)}_R$, for the left- and right-handed particles, however due to quantum fluctuations these currents become anomalous. It is then convenient to describe the system in terms of vector and axial currents
\begin{equation}\label{eqn:ch6:chiral_to_axial_currents}
	J^\mu=J^\mu_++J^\mu_-\qquad\qquad J^\mu_5=J^\mu_+-J^\mu_-
\end{equation}
so that the vector electric current, protected by the gauge symmetry, is conserved, while the axial one is anomalous. The hydrodynamic equations of motion for the covariant currents are simply
\begin{subequations}
	\begin{align}
		\partial_\mu T^{\mu\nu}&=F^{\nu\lambda}J_\lambda\\
		\partial_\mu J^\mu&=0\\
		\partial_\mu J^\mu_5&=cE\cdot B
	\end{align}
\end{subequations}
where $c$ is again the axial anomaly coefficient.

In our setup we are considering a single fluid with two different species of particles, which is the simplest and more realistic setup for interacting Weyl nodes. However, if the two Weyl nodes are very far separated in momentum space $b^\mu\gg T$, one could also consider the Weyl nodes to give rise to two separate, weakly-interacting fluids at different temperatures and chemical potentials, such that each fluid conserves energy and momentum \cite{Lucas:HydrodynamicTheoryThermoelectric}.

Using the insight of the previous section, we study the WSM in the hydrodynamic regime using ideal fluid dynamics with an order zero magnetic field $B\sim\mathcal{O}(1)$. Then the constitutive relations (neglecting magnetization terms which are irrelevant for the longitudinal transport) read
\begin{subequations}
	\begin{align}
		T^{\mu\nu}&=\epsilon u^\mu u^\nu + P\Delta^{\mu\nu}+\xi_\epsilon \left(B^\mu u^\nu+B^\nu u^\mu\right)\\
		J^\mu&=n u^\mu+\xi B^\mu\\
		J^\mu_5&=n_5 u^\mu+\xi_5 B^\mu
	\end{align}
\end{subequations}
where $n_5=n_5(T,\mu,\mu_5,B^2)$ is the axial charge density and, like every other thermodynamic quantity, in general depends on the temperature, the two chemical potentials, and the external magnetic field. The $\xi$s are the anomalous transport coefficients and take the value \cite{Landsteiner:NotesAnomalyInduced}
\begin{equation}\label{eqn:ch6:anomalous_coefficients_two_currents}
	\xi=c\mu_5\qquad\xi_5=c\mu\qquad\xi_\epsilon=c\mu\mu_5
\end{equation}
Interestingly, there are no contributions due to the mixed axial-gravitational anomaly in the anomalous transport coefficients multiplying the magnetic field, which would appear as a term proportional to the temperature. This fact, as we will see, has the important consequence that the DC thermal and thermoelectric conductivities are not anomalous.

From this setup we can now simply compute the full longitudinal thermoelectric matrix \eqref{eqn:ch2:thermoelectric_matrix} in linear response theory. We consider a background with zero spatial velocity, constant $T$, $\mu$ and $\mu_5$, and an external magnetic field along the $\hat z$ axis. Around this background we add fluctuations and linear sources parallel to the magnetic field
\begin{align}
	T&\rightarrow T+\delta T-Tz\delta\zeta_z	&   \mu&\rightarrow\mu+\delta\mu    &   \mu_5&\rightarrow\mu_5+\delta\mu_5\nonumber\\
	u^\mu&\rightarrow(1,\delta\vect{v})  &   F^{0z}&\rightarrow\delta \mathbb{E}^z
\end{align}
where $\delta\zeta^i$ and $\delta\mathbb{E}^i$ are, as usually, the thermal and electric source respectively, see Section~\ref{sec:ch2:thermoelectric_transport}. Finally, the linearized canonical heat current is defined as $\delta Q^i=\delta T^{0i}-\mu\delta J^i-\mu_5\delta J^i_5$.

\subsection{DC limits and relaxation terms}
Unexpectedly, when we compute the full thermoelectric matrix at non-zero frequency using the prescription above, we find that the DC limits of the conductivities diverge.

We have already discussed in Chapter~\ref{chapter:electrically_driven_fluids} and \ref{chapter:onsager} the importance of momentum relaxation to obtain finite DC conductivities: the electric field keeps adding momentum to the fluid without bound, so a mechanism to balance this effect and dissipate momentum on a timescale $\tau_m$ is needed. In the case of anomalous fluids, however, we also need energy and axial charge relaxation. The former comes from the fact that the covariant current $\vect{J}\propto\vect{B}$ is not zero in the background with a constant magnetic field, due to the anomalous current, subsequently the joule heating term $\vect{E}\cdot\vect{J}\sim\vect{E}\cdot\vect{B}$ is not zero either and the energy of the fluid keeps increasing without bound even at zero velocity, unless energy relaxation is present. Finally, axial charge decay is also needed, to balance the effect of the anomaly itself $\vect{E}\cdot\vect{B}$, which otherwise would lead to an infinite axial charge. These terms act as soft cut-off for the associated conserved charge, rendering the conductivities finite.

From a physical perspective, the presence of these relaxations in WSMs can also be argued from a microscopic point of view. Momentum relaxation is a feature of all metals, due to the presence of impurities and phonons that can take away momentum \cite{Ashcroft:SolidStatePhysics,Lifshitz:PhysicalKineticsVolume,Andreev:HydrodynamicDescriptionTransport}. A finite axial charge relaxation time $\tau_5$ is also expected for Weyl semimetals, because the chiral symmetry is not exact, but only approximate. In particular, the electronic bands are not linear at arbitrary-high momenta, and instead they flatten, leading to an effective mass for the chiral particles which destroy the anomaly. Furthermore, arguments from topology requires WSMs to have multiple Weyl cones in each Brillouin zone \cite{Nielsen:NogoTheoremRegularizing}, such that the total Berry curvature is zero, but this allows particles of opposite chirality to interact with each other via impurities (the so-called inter-valley scattering) and this leads to a finite axial charge depletion rate $\tau_5$. Lastly, energy relaxation $\tau_\epsilon$ can happen in the presence of phonons, that can take away energy from the electron fluid \cite{Lifshitz:PhysicalKineticsVolume,Andreev:HydrodynamicDescriptionTransport}.

Although finite momentum, energy and axial charge relaxations can be present in Weyl semimetals, they are caused by different microscopic phenomena, as we have just discussed, hence we would anticipate the various relaxation rates to have different values in real materials. Specifically, we expect electric charge to be exactly conserved (it is protected by the gauge symmetry), while energy relaxation should be a subleading effect compared to momentum and axial charge relaxation, which can also be rather strong effects $\tau_\epsilon\gg\tau_5,\tau_m$.

Keeping in mind this discussion, we now study some possible approaches to relaxations in anomalous fluids. Specifically, we constrain our model from three fundamental and phenomenological assumptions: the fluid must have finite DC conductivities in real samples, electric charge is conserved (or, at least, the decay rate $\tau^{-1}$ can be tuned to be arbitrary small), and finally the Green functions must obey Onsager relations.

\subsubsection{Option 1: Canonical charge relaxation}
The first and simplest case is the one used in \cite{Landsteiner:NegativeMagnetoresistivityChiral,Abbasi:MagnetotransportAnomalousFluid,Rogatko:MagnetotransportWeylSemimetals}, and is the most straightforward generalization of the momentum relaxation term in \cite{Hartnoll:TheoryNernstEffect}. Namely, this corresponds to relaxing the total charge densities in the frame of the laboratory. In this approach we modify the linearized equations of motion to take the following form
\begin{subequations}\label{eqn:ch6:landsteiner_eom_total_charge}
	\begin{align}
		\partial_\mu\delta T^{\mu0}&=\delta(F^{0\lambda}J_\lambda)-\frac{1}{\tau_\epsilon}\delta T^{00}~,\\
		\partial_\mu\delta T^{\mu i}&=\delta(F^{i\lambda}J_\lambda)-\frac{1}{\tau_m}\delta T^{0i}~,\\
		\partial_\mu\delta J^\mu&=-\frac{1}{\tau_n}\delta J^0~,\\
		\partial_\mu\delta J^\mu_5&=c\delta E\cdot B-\frac{1}{\tau_5}\delta J^0_5~.
	\end{align}
\end{subequations}
Notice that we also included a possible electric-charge relaxation term, not present in the original works, and which is not necessary to obtain finite DC results.

First, we notice that all the longitudinal DC conductivities computed from this model are anomalous, meaning that they all depend on $c^2B^2$ in a non-trivial way. Specifically the scaling is not simply quadratic in the magnetic field, since it also appears in the denominator (the linear $B^2$ dependence is recovered only in the limit of small magnetic field). Although the expressions are too large to report here, we can write the results at zero axial chemical potential $\mu_5=n_5=\frac{\partial n_5}{\partial\mu}=\frac{\partial n_5}{\partial T}=0$ obtaining
\begin{subequations}\label{eqn:ch6:conductivities_zero_mu5_total_relaxation}
	\begin{align}
		\sigma&=\frac{n^2\frac{\partial n_5}{\partial\mu_5}\tau_m+B^2c^2\left(sT\tau_5-n\mu\tau_m\right)}{\frac{\partial n_5}{\partial\mu_5}(P+\epsilon)-B^2c^2\mu^2}~,\\
		\alpha&=\frac{s^2\frac{\partial n_5}{\partial\mu_5}T\tau_m}{\frac{\partial n_5}{\partial\mu_5}(P+\epsilon)-B^2c^2\mu^2}~,\\
		\bar\kappa&=\frac{s\left(n\frac{\partial n_5}{\partial\mu_5}-B^2c^2\mu\right)\tau_m}{\frac{\partial n_5}{\partial\mu_5}(P+\epsilon)-B^2c^2\mu^2}~.
	\end{align}
\end{subequations}
Notice that in the non-chiral regime, only axial charge and momentum relaxation are needed to obtain finite DC conductivities. This is expected, since at $\mu_5=0$ there is no background current $\vect{J}=0$, see \eqref{eqn:ch6:anomalous_coefficients_two_currents}, and thus no Joule heating happens.

There is however one problem with the above approach: in general, on a background with $\mu_5\neq0$, $\alpha$ and $\bar\kappa$ depend only on $\tau_m$, while $\sigma$ and $\bar\alpha$ also on $\tau_5$ and $\tau_\epsilon$. Thus, when we impose Onsager relations $\alpha=\bar\alpha$, we find that time-reversal invariance implies that all relaxation rates must be equal $\tau_\epsilon=\tau_5=\tau_n=\tau_m$. This result is against our working assumptions (electric charge is not conserved) and contradicts phenomenological expectations, for which we expect the relaxation rates to be different, since they should come from various microscopic processes.

Hence, we are forced to look for other approaches to obtain finite conductivities, while also preserving charge conservation.

\subsubsection{Option 2: Normal charge relaxation}\label{sec:ch6:normal_charge_relaxation}
Looking at the relaxation terms of the previous paragraph, we see that the linearized charge densities observed in the laboratory are always composed of two terms: one is the normal charge density, respectively $\delta\epsilon$, $\delta n$, $\delta n_5$, $(P+\epsilon)\delta v^i$, while the second one is the anomalous contribution that comes from the effect of Lorentz boost on the anomaly-related terms $\xi B^\mu$.

However, we know that the anomalous part of the fluid does not produce entropy, create drag or heat \cite{Stephanov:NodragFrameAnomalous,Sadofyev:DragSuppressionAnomalous,Copetti:AnomalousTransportHolographic} and has superfluid-like behaviour\footnote{Of course, it is not a real superfluid, since it carries entropy, but it shares some phenomenological similarities, such as a persistent dissipationless current.}. Particularly, a chiral fluid in a constant magnetic field in the presence of impurities will relax the momentum until its equilibrium velocity is zero, but the anomalous momentum will keep flowing past the impurities. Furthermore, even arguments from kinetic theory suggest that only the normal part of the fluid should relax \cite{Gorbar:ConsistentHydrodynamicTheory,Dantas:MagnetotransportMultiWeylSemimetals}.

Then, following these lines of reasoning, we propose that the relaxation rates of the previous section should be modified in order to relax only the normal fluid components
\begin{subequations}\label{eqn:ch6:normal_charge_relaxation}
	\begin{align}
		\frac{\delta T^{00}}{\tau_\epsilon}&=\frac{1}{\tau_\epsilon}\left(\delta\epsilon+2c\mu\mu_5\vect{B}\cdot\delta \vect{v}\right)&\longrightarrow& & &\frac{\delta\epsilon}{\tau_\epsilon}\\
		\frac{\delta T^{0i}}{\tau_m}&=\frac{1}{\tau_m}\left[(\epsilon+P)\delta v^i+B^ic(\mu_5\delta\mu+\mu\delta\mu_5)\right]&\longrightarrow& & &\frac{\epsilon+P}{\tau_m}\delta v^i\\
		\frac{\delta J^0}{\tau_n}&=\frac{1}{\tau_n}\left(\delta n+c\mu_5\vect{B}\cdot\delta\vect{v}\right)&\longrightarrow& & &\frac{\delta n}{\tau_n}\\
		\frac{\delta J^0_5}{\tau_5}&=\frac{1}{\tau_5}\left(\delta n_5+c\mu\vect{B}\cdot\delta\vect{v}\right)&\longrightarrow& & &\frac{\delta n_5}{\tau_5}
	\end{align}
\end{subequations}
This partially solves the problems of the previous approach, namely with this choice momentum relaxation ($\tau_m\geq0$ for stability) now decouples from the other relaxations and is arbitrary, however Onsager relations \eqref{eqn:ch2:onsager_relations_general} still require that all other relaxation rates must take the same value $\tau_n=\tau_5=\tau_\epsilon$, which is again in strong disagreement with phenomenological and theoretical arguments.

The conductivities computed from this approach are qualitatively and quantitatively different from the previous case: all conductivities now depend on energy, momentum and axial charge relaxation, however in DC only the electric conductivity remains anomalous, and with a simple $B^2$ dependence, while the other ones take the standard hydrodynamic form \eqref{eqn:ch2:thermoelectric_conductivities}. This, as we have already argued above, is due to the fact that the linearized canonical heat current does not have any anomaly-related term, and that the anomalous transport coefficients $\xi$s do not contain $T$-dependent contributions induced by the mixed-gravitational anomaly\footnote{We remind that the thermal conductivity measured in experiments $\kappa$ is related to the one obtained from linear response $\bar\kappa$ via $\kappa=\bar\kappa-T\alpha^2/\sigma$, see the discussion around \eqref{eqn:ch2:thermal_conductivity_experiment}. Thus, even if $\bar\kappa$ is not anomalous in DC, the same is not true for $\bar\kappa$, which inherits a magnetoresistance from $\sigma$.}.

Finally, at zero chemical potential  $\mu_5=n_5=\frac{\partial n_5}{\partial\mu}=\frac{\partial n_5}{\partial T}=0$ we find that the DC electric conductivity is
\begin{equation}\label{eqn:conductivity_zero_axial}
	\sigma=\frac{n^2\tau_m}{P+\epsilon}+\frac{B^2c^2\tau_5}{\frac{\partial n_5}{\partial\mu_5}}~.
\end{equation}
while the other conductivities take on their standard hydrodynamic form that depends only on $\tau_m$.

This approach already leads to an interesting result, specifically a different prediction for the thermoelectric transport compared to other studies. In older works \cite{Chernodub:ThermalTransportGeometry,Lucas:HydrodynamicTheoryThermoelectric,Abbasi:MagnetotransportAnomalousFluid,Das:NonlinearMagnetoconductivityWeyl,Spivak:MagnetotransportPhenomenaRelated,Lundgren:ThermoelectricPropertiesWeyl,Sharma:NernstMagnetothermalConductivity} the anomaly seems to appear also in the thermoelectric conductivity from hydrodynamics, lattice simulations and kinetic theory approaches. This difference can be understood because the authors either relax the total charge, use different assumptions on the model (e.g. having two separate fluids for each Weyl cone) or simply are not in the hydrodynamic regime. This prediction is then open to experimental verification \cite{Vu:ThermalChiralAnomaly,Gooth:ExperimentalSignaturesMixed,Jia:ThermoelectricSignatureChiral} and could be used as a probe of single-fluid hydrodynamic anomalous transport, although practically it can be difficult to disentangle the various contribution to the magnetoresistance and to isolate the effect of the anomaly.

\subsubsection{Option 3: Generalised relaxations}\label{sec:ch6:generalised_relaxations}
In the previous section we managed to decouple momentum relaxation from the other decay rates, however electric charge must still relax for the system to obey Onsager relations, and moreover the relaxations must all be equal. To improve the model, we propose to use the generalized relaxation of Chapter~\ref{chapter:onsager}.

Following the previous chapter, we then consider the most general set of relaxations
\begin{equation}\label{eqn:ch6:generic_relaxations}
	\begin{rcases*}
		\text{energy:}\hspace{1.3cm} \frac{1}{\tau_{\epsilon\epsilon}}\delta\epsilon+\frac{1}{\tau_{\epsilon n}}\delta n+\frac{1}{\tau_{\epsilon n_5}}\delta n_5\\
		\text{charge:}\hspace{1.3cm}\frac{1}{\tau_{n\epsilon}}\delta\epsilon+\frac{1}{\tau_{nn}}\delta n+\frac{1}{\tau_{nn_5}}\delta n_5\\
		\text{axial charge:}\quad\frac{1}{\tau_{n_5\epsilon}}\delta\epsilon+\frac{1}{\tau_{n_5n}}\delta n+\frac{1}{\tau_{n_5n_5}}\delta n_5
	\end{rcases*} =\hat{\tau}\cdot\varphi~,
\end{equation}
where $\varphi=\left(\delta\epsilon,\delta n, \delta n_5\right)$ and $\hat\tau$ is the $3\times3$ matrix of inverse relaxation times. We can again employ Onsager relations \eqref{eqn:ch2:onsager_relations_general} on these relaxed equations of motion to find the constraints that the relaxation times must obey for the system to preserve time-reversal invariance \eqref{eqn:ch5:onsager_constraint_general}. Thus, we find
\begin{subequations}\label{eqn:ch6:onsager_constraints}
	\begin{align}
		0&=\frac{\chi_{nn_5}}{\tau_{\epsilon n_5}}+\frac{\chi_{nn}}{\tau_{\epsilon n}}-\frac{\chi_{\epsilon n_5}}{\tau_{nn_5}}+\frac{\chi_{\epsilon n}}{\tau_{\epsilon\epsilon}}-\frac{\chi_{\epsilon n}}{\tau_{nn}}-\frac{\chi_{\epsilon\epsilon}}{\tau_{n\epsilon}}~,\\
		0&=\frac{\chi_{n_5n_5}}{\tau_{\epsilon n_5}}+\frac{\chi_{nn_5}}{\tau_{\epsilon n}}-\frac{\chi_{\epsilon n_5}}{\tau_{n_5n_5}}+\frac{\chi_{\epsilon n_5}}{\tau_{\epsilon\epsilon}}-\frac{\chi_{\epsilon n}}{\tau_{n_5n}}-\frac{\chi_{\epsilon\epsilon}}{\tau_{n_5\epsilon}}~,\\
		0&=\frac{\chi_{n_5n_5}}{\tau_{nn_5}}-\frac{\chi_{nn_5}}{\tau_{n_5 n_5}}+\frac{\chi_{n n_5}}{\tau_{nn}}-\frac{\chi_{nn}}{\tau_{n_5n}}+\frac{\chi_{\epsilon n_5}}{\tau_{n\epsilon}}-\frac{\chi_{\epsilon n}}{\tau_{n_5\epsilon}}~,
	\end{align}    
\end{subequations}
or in matrix formulation
\begin{equation}
	\hat\tau\cdot\hat\chi-\hat\chi\cdot\hat\tau^T=0~,
\end{equation}
where $\hat\chi$ is the $3\times3$ susceptibility matrix for the energy, electric and axial charge, while $T$ denotes the transpose matrix. In agreement with the previous case, these constraints tell us that if we set the off-diagonal relaxations to zero, then Onsager relations imply $\tau_{nn}=\tau_{n_5n_5}=\tau_{\epsilon\epsilon}$. 

The claim is that we now have enough parameters to satisfy all our requirements, namely: finite DC conductivities, charge conservation and Onsager relations (microscopic time-reversal invariance). To begin the analysis of the parameter space, first we set to zero the charge relaxations $\tau_{nn}^{-1}=\tau_{nn_5}^{-1}=\tau_{n\epsilon}^{-1}=0$ so that the electric charge is exactly conserved. Thus, starting from 9 relaxation (plus momentum $\tau_m$, which we decoupled in the previous section) we are left with only 6 non-zero relaxations. There are 3 Onsager conditions, which reduce the number of independent parameters to only 3, which we assume to be $\tau_{\epsilon\epsilon}, \tau_{n_5n_5}$ and $\tau_{n_5\epsilon}$, and express the other relaxations rates as functions of them.

This setup is guaranteed to lead to conductivities which are Onsager reciprocal, however the only way to check if they are also finite in DC is to perform the computation explicitly, and we find that they are indeed finite. Notice that this is not trivial: from Onsager relations alone one might be tempted to set to zero also $\tau_{\epsilon n}^{-1}=\tau_{n_5n}^{-1}=0$, leaving $\tau_{n_5n_5}$ as the only free parameter\footnote{From 9 initial relaxations, 5 are set to zero $\tau_{nn}^{-1}=\tau_{nn_5}^{-1}=\tau_{n\epsilon}^{-1}=\tau_{\epsilon n}^{-1}=\tau_{n_5n}^{-1}=0$, thus only 4 are non-vanishing, but with 3 Onsager conditions we are left with a 1-parameter family of solutions.}, however the conductivities computed on this setup are not DC finite.

From a qualitative perspective the conductivities computed with generalized relaxations are similar to the ones analysed in the previous section: all conductivities depend on all relaxation rates, however in DC only the electric conductivity is anomalous, while the other conductivities are not. Furthermore, at zero axial chemical potential we have the freedom to also conserve energy, namely $\tau_{\epsilon n}^{-1}=\tau_{\epsilon\epsilon}^{-1}=\tau_{\epsilon n_5}^{-1}=0$. Following the argument below \eqref{eqn:ch6:conductivities_zero_mu5_total_relaxation}, this is expected: energy relaxation is needed to balance the Joule heating term $\vect{E}\cdot\vect{J}$ with $\vect{J}\propto\vect{B}$, however at $\mu_5=0$ the electric current is zero and not anomalous, thus we expect no Joule heating.

Remarkably, with the above setup we cannot also impose positivity of entropy production. Following the approach of Chapter~\ref{chapter:onsager} on the linearized second law of thermodynamics, we find the following set of constraints
\begin{subequations}\label{eqn:ch6:entropy_production}
	\begin{align}
		0&=\frac{1}{\tau_{\epsilon\epsilon}}-\frac{\mu}{\tau_{n\epsilon}}-\frac{\mu_5}{\tau_{n_5\epsilon}}~,\\
		0&=\frac{1}{\tau_{\epsilon n_5}}-\frac{\mu}{\tau_{nn_5}}-\frac{\mu_5}{\tau_{n_5n_5}}~,\\
		0&=\frac{1}{\tau_{\epsilon n}}-\frac{\mu}{\tau_{nn}}-\frac{\mu_5}{\tau_{n_5n}}~.
	\end{align}
\end{subequations}
However, it is not possible to have electric charge conservation $\tau_{nn}^{-1}=\tau_{nn_5}^{-1}=\tau_{n\epsilon}^{-1}=0$, positivity of entropy production \eqref{eqn:ch6:entropy_production} and time-reversal covariance \eqref{eqn:ch6:onsager_constraints} with generalized relaxations. Indeed, charge conservation reduces the number of relaxation parameters from 9 to 6, Onsager relations and positivity of entropy production each correspond to three equations, but the only solution to the whole set of equations is the trivial one that sets all the relaxations to zero. We remark that this conclusion, however, is not too problematic, since generalized relaxations can be understood in terms of the fluid interacting with the environment and thus describing an open system (it can lose energy, charge and also entropy).

Finally, we can briefly comment that in the presence of charged impurities, or multiple gates, one might have an arbitrary small effective electric charge relaxation. However, this can be easily incorporated using generalized relaxations as an extra parameter that can be tuned at will.

To conclude, generalized relaxations allow us to obtain finite DC conductivities, which obey Onsager relations and that conserve the electric charge. Furthermore, the six non-zero relaxations are constrained by only three equations, meaning that the parameter space can accommodate for different phenomenological timescales for which e.g., energy relaxation can be parametrically smaller than axial charge decay rate.

\section{Kinetic theory}\label{sec:ch6:kinetic_theory}
In Chapter~\ref{chapter:onsager} we introduced the generalized relaxations with arguments uniquely based on the EFT prescription, while above we argued how they are necessary to properly describe the DC transport in models of Weyl semimetals. However, we still have to show that generalized relaxation can actually happen in condensed matter systems. This is exactly what we are going to do in this section: we will show that if charge relaxations are present at all in the system, then we should always expect generalized relaxations that take exactly the form described in Chapter~\ref{chapter:onsager}, and obey Onsager relations \eqref{eqn:ch6:onsager_constraints}. We will do this by using suitable energy-dependent modifications of the Relaxation Time Approximation (RTA for short), which are supported by microscopic collision integrals with impurities and phonons.

Notice that Onsager constraints \eqref{eqn:ch6:onsager_constraints} and the structure of generalized relaxations \eqref{eqn:ch6:generic_relaxations} do not depend on the anomaly when we relax only the non-anomalous charges, thus they hold true for all fluids (even if their relevance becomes more apparent in anomalous hydrodynamics). This means that we can use non-chiral kinetic theory to justify their presence. Although not necessary, we will consider a linear dispersion relation such as in Weyl semimetals, so that we can perform certain computations analytically. First, we will focus on a single-current ideal fluid, and we will explain how to obtain generalized relaxations from modifications of the RTA in this simpler case, matching the expressions of the previous Chapter~\ref{chapter:onsager}. After that we will move to the case relevant to Weyl semimetals, namely that of two chiral currents as in Section~\ref{sec:ch6:dc_conductivities}, and we will discuss possible scattering mechanisms there.

\subsection{A primer on kinetic theory and hydrodynamics}\label{sec:ch6:hydrodynamics_kinetic_theory}
Consider the relativistic Boltzmann equation for the one-particle distribution function $f(\vect{x},\vect{p},t)=f_{\vect{p}}$ in the absence of external forces\footnote{External forces correspond, in the hydrodynamic picture, to the Maxwell term in the stress-energy tensor conservation equation.}
\begin{equation}
	\partial_tf_{\vect{p}}+\vect{p}\cdot\vect{\nabla}f_{\vect{p}}=I_\text{coll}[f_{\vect{p}}]
\end{equation}
with $\vect{p}$ the quasiparticle momentum and $I_\text{coll}$ the collision integral, which is a functional of $f_{\vect{p}}$.

In the simplest case, which is often used to construct hydrodynamics from kinetic theory, we consider only a single species of particles (in our case, dressed electrons) for which 2-to-2 particles scatterings dominate. Then $I_\text{coll}=I_\text{ee}$ and corresponds to the usual textbook electron-electron collision integral \cite{Lifshitz:PhysicalKineticsVolume,Huang:StatisticalMechanics,Cercignani:MathematicalMethodsKinetic}. Under standard assumptions about the microscopic scattering, such as unitarity, conservation of energy and momentum, time-reversal and parity symmetry, and the hypothesis of molecular chaos, the electron-electron collision integral is written as
\begin{multline}\label{eqn:ch6:electron_collision_integral}
	I_\text{ee}=\int\dif^3\vect{p}_2\dif^3\vect{p}_1'\dif^3\vect{p}_2'W_{\vect{p}_1'\vect{p}_2'\rightarrow\vect{p}\vect{p}_2}\left(f_{\vect{p}_1'}f_{\vect{p}_2'}[1+af_{\vect{p}}][1+af_{\vect{p}_2}]\right.\\
	\left.-f_{\vect{p}}f_{\vect{p}_2}[1+af_{\vect{p}_1'}][1+af_{\vect{p}_2'}]\right)
\end{multline}
where $a=\pm1$ for bosons/fermions and $a=0$ for classical particles, while $W_{\vect{p}_1'\vect{p}_2'\rightarrow\vect{p}\vect{p}_2}$ is the scattering rate.

From here, it can be shown that the collision integral vanishes when $f_{\vect{p}}$ satisfies the detailed balance equation
\begin{equation}\label{eqn:ch6:detailed_balance}
	\log\left(\frac{f_{\vect{p}_1'}}{1+af_{\vect{p}_1'}}\right)+\log\left(\frac{f_{\vect{p}_2'}}{1+af_{\vect{p}_2'}}\right)=\log\left(\frac{f_{\vect{p}}}{1+af_{\vect{p}}}\right)+\log\left(\frac{f_{\vect{p}_2}}{1+af_{\vect{p}_2}}\right)
\end{equation}
which is understood as a conservation equation ($\vect{p}_1'$ and $\vect{p}_2'$ are after the collision, $\vect{p}$ and $\vect{p}_2$ before). Hence, the logarithm solves the detailed balance when it is proportional to the conserved quantities in the system: energy, momentum and particle number. The solution is then given by the local thermodynamic equilibrium (LTE) form of $f_{\vect{p}}$ as a Fermi-Dirac (for $a=-1$) distribution
\begin{equation}\label{eqn:ch6:local_thermodynamic_equilibrium}
	f_{\vect{p}}^\text{eq}=\frac{1}{1+e^{(\epsilon_p-\vect{u}\cdot\vect{p}-\mu)/T}}
\end{equation}
where the Lagrange multipliers $T$, $\mu$ and $\vect{u}$ are functions of space. The LTE distribution function solves the collision integral, however it does not vanish on the LHS of the Boltzmann equation (only global thermodynamic equilibrium solves the full equation), nonetheless in the hydrodynamic regime the non-vanishing of the LHS corresponds to small gradient corrections. For this reason, the ideal fluid corresponds in the kinetic-theory picture to LTE \eqref{eqn:ch6:local_thermodynamic_equilibrium}, while higher derivative corrections in hydrodynamics take us away from LTE.

Given the LTE form $f_{\vect{p}}^\text{eq}$ and the dispersion relation $\epsilon(p)=\epsilon_p$ we can compute thermodynamic quantities in local equilibrium, by integrating over momentum space and summing over particles and holes contributions
\begin{subequations}\label{eqn:energy_charge_equilibrium}
	\begin{align}
		\epsilon^\text{eq}(T,\mu)&=\sum_{p,h}\int\frac{\dif^3\vect{p}}{(2\pi)^3}\epsilon_pf_{\vect{p}}^\text{eq}\\
		n^\text{eq}(T,\mu)&=\sum_{p,h}\int\frac{\dif^3\vect{p}}{(2\pi)^3}f_{\vect{p}}^\text{eq}
	\end{align}    
\end{subequations}
Finally, to obtain the hydrodynamic equations of motions we simply multiply the Boltzmann equation by $\epsilon_p$, $\vect{p}$ or 1 and then integrate over the quasiparticle momentum. In particular, given the electron-electron collision integral \eqref{eqn:ch6:electron_collision_integral}, it can be shown that it vanishes when integrated over conserved quantities, independently on the form of the distribution function $f_{\vect{p}}$. Specifically
\begin{equation}\label{eqn:ch6:electron_collision_integral_integration}
	\int\frac{\dif^3\vect{p}}{(2\pi)^3} A(\vect{x},\vect{p})I_\text{ee}[f_{\vect{p}}]=0\qquad\qquad\text{for}\quad A=\{\epsilon_p,\vect{p},1\}
\end{equation}
Thus, multiplying with $\epsilon_p$, $\vect{p}$, or 1 and integrating the Boltzmann equation over momentum space, we recover the equations of ideal hydrodynamics.

If one wishes to go beyond order zero in the hydrodynamic derivative expansion the distribution function receives out-of-equilibrium corrections, which can be computed perturbatively (e.g. in Knudsen number) \cite{Denicol:MicroscopicFoundationsRelativistic,Huang:StatisticalMechanics}
\begin{equation}
	f_{\vect{p}}=f_{\vect{p}}^\text{eq}+\delta f_{\vect{p}}
\end{equation}
However, the non-equilibrium distribution function $f_{\vect{p}}$ does not solve the electron-electron collision integral, hence to make progress we linearize it and employ the standard RTA \cite{Lifshitz:PhysicalKineticsVolume}\footnote{We are not using the relativistic form of the RTA \cite{Rocha:NovelRelaxationTime,Anderson:RelativisticRelaxationtimeModel} since this form is enough for the present discussion.}
\begin{equation}\label{eqn:ch6:electron_RTA}
	I_\text{ee}\approx-\frac{f_{\vect{p}}-f_{\vect{p}}^\text{eq}}{\tau}=-\frac{\delta f_{\vect{p}}}{\tau}
\end{equation}
with $\tau$ some constant that represents the timescale it takes for the system to relax to equilibrium.

Importantly, the RTA is only a crude model for the real linearized collision integral, and does not inherit all of its properties \cite{Cercignani:MathematicalMethodsKinetic}: the only property that survives this approximation is the vanishing of $I_\text{coll}$ when $f_{\vect{p}}$ takes the Fermi-Dirac form \eqref{eqn:ch6:local_thermodynamic_equilibrium}, $I_\text{coll}[f^\text{eq}_{\vect{p}}]=0$. Indeed, the RTA model does not vanish when integrated in momentum space against conserved quantities \eqref{eqn:ch6:electron_collision_integral_integration}, this means that the equations of hydrodynamics receive corrections as decay terms at order one in derivatives
\begin{subequations}\label{eqn:ch6:electron_collision_integral_hydrodynamics}
	\begin{align}
		\partial_t\epsilon+\dots&=-\frac{\epsilon-\epsilon^\text{eq}}{\tau}\\
		\partial_t n+\dots&=-\frac{n-n^\text{eq}}{\tau}
	\end{align}
\end{subequations}
Then, enforcing the equations of hydrodynamics to hold true, we are led to the requirements $\epsilon=\epsilon^\text{eq}$ and $n=n^\text{eq}$, which are part of the Landau matching conditions used to fix the hydrodynamic frame\footnote{Notice that, using the method developed in \cite{Rocha:NovelRelaxationTime}, it is possible to obtain hydrodynamics from kinetic theory in any frame.}.

Because in what follows we will only be interested in perfect fluids to match the discussion of the previous section, $f_{\vect{p}}$ will always be in LTE. Thus, we drop the \emph{eq} superscript from now on and simply call $f_{\vect{p}}$ the LTE form of the Fermi-Dirac distribution \eqref{eqn:ch6:local_thermodynamic_equilibrium}. Similarly, we will avoid using \emph{eq} to define the equilibrium energy and charge, since they are always assumed to be in LTE.

\subsection{Adding momentum relaxation}
We will now focus on ideal fluid dynamics and, following \cite{Gorbar:ConsistentHydrodynamicTheory,Narozhny:HydrodynamicApproachTwodimensional}, we will see how to modify kinetic theory to incorporate a momentum relaxation effect in the equations of hydrodynamics.

Given the Fermi-Dirac distribution \eqref{eqn:ch6:local_thermodynamic_equilibrium} we can expand it at small velocity
\begin{equation}\label{eqn:ch6:small_velocity_expansion}
	f_{\vect{p}}\approx f^{(0)}-(\vect{p}\cdot\vect{u})\frac{\partial f^{(0)}}{\partial\epsilon_p}\qquad\text{with}\qquad f^{(0)}=\frac{1}{1+e^{(\epsilon_p-\mu)/T}}
\end{equation}
where $f^{(0)}$ indicates the zero velocity distribution function. We can compute the energy and charge density for massless fermions by employing a linear dispersion relation $\epsilon_p=p$ (we are setting the Fermi velocity to unity). Because of the linearization at small velocity, the second term linear in $\vect{u}$ does not contribute to any thermodynamic quantity, since it vanishes when integrated in momentum space, see the integrals in the Appendix~\ref{appendix:useful_integrals}. Then we find
\begin{subequations}\label{eqn:ch6:energy_charge_one_current}
	\begin{align}
		\epsilon&=\frac{15\mu^4+30\pi^2T^2\mu^2+7\pi^4T^4}{120\pi^2}\\
		n&=\mu\left(\frac{\mu^2}{6\pi^2}+\frac{T^2}{6}\right)
	\end{align}
\end{subequations}
while the pressure is $P=\epsilon/3$.

To include momentum relaxation effects in the hydrodynamic picture, we now assume the RTA ansatz given by
\begin{equation}\label{eqn:ch6:momentum_relaxation_RTA}
	I_\text{imp}\approx-\frac{f_{\vect{p}}-f^{(0)}}{\tau_m}
\end{equation}
The form of the RTA is such that it vanishes when integrated against the quasiparticle energy and the identity, while it gives the momentum relaxation term we seek when integrated against the quasi-momentum $\vect{p}$
\begin{equation}
	\partial_t(\epsilon+P)v^i+\dots=-\frac{(\epsilon+P)v^i}{\tau_m}
\end{equation}
This RTA is physically very different from the standard one \eqref{eqn:ch6:electron_RTA}: first, here both $f_{\vect{p}}$ and $f^{(0)}$ are LTE solutions, hence they both make the electron-electron collision integral \eqref{eqn:ch6:electron_collision_integral} vanish, while this is not true in \eqref{eqn:ch6:electron_RTA}. Furthermore, this RTA has physical effects and leads to momentum relaxation, while the electron-electron RTA model was only used to impose Landau matching conditions. We remark that momentum relaxation behaves as an extra constraint on the theory, and acts in such a way that of all the possible equilibrium solutions with different constant velocities (which are related by boosts), only the zero-velocity equilibrium is actually picked by the system.

The expression in \eqref{eqn:ch6:momentum_relaxation_RTA} can be justified from electron-impurity scatterings. Consider a collision integral that takes the following form
\begin{equation}
	I_\text{coll}=I_\text{ee}+I_\text{imp}
\end{equation}
The first term is the electron-electron scattering of \eqref{eqn:ch6:electron_collision_integral}: in the hydrodynamic regime it dominates (its associated timescale is very short) and its role is to force the distribution function to be Fermi-Dirac. The second term instead describes interaction of electrons with impurities and is assumed to be a weak correction (so that a quasihydrodynamic description still applies, and momentum is still a relevant degree of freedom). The electron-impurity collision integral takes the simple form \cite{Lifshitz:PhysicalKineticsVolume}
\begin{equation}\label{eqn:ch6:impurities_collision_integral}
	I_\text{imp}=\int\dif^3\vect{p}'W_{\vect{p}\rightarrow\vect{p}'}[f_{\vect{p}}-f_{\vect{p}'}]\delta(\epsilon_p-\epsilon_{p'})
\end{equation}
where again $W_{\vect{p}\rightarrow\vect{p}'}$ is the scattering rate. Following the usual approach of the detailed balance \eqref{eqn:ch6:detailed_balance}, we should look for solutions to this collision integral (among the spectrum of all the LTE solutions, required by the vanishing of $I_\text{ee}$). However, these are exactly the zero-velocity distribution functions $f^{(0)}$ we introduced in \eqref{eqn:ch6:small_velocity_expansion}, therefore a good ansatz for an RTA model of this collision integral should vanish when $f_{\vect{p}}=f^{(0)}$, and we are led to \eqref{eqn:ch6:momentum_relaxation_RTA}.

\subsection{Adding charge relaxations}
Having understood how to include momentum relaxation in our hydrodynamic model from kinetic theory, we can now follow the same logic to include energy and charge relaxation in the system.

In particular, the momentum-relaxing RTA suggests we can add charge relaxation by including an additional collision integral into $I_\text{coll}$ that depends on $\bar{f}^{(0)}$, so that the system relaxes to a reference LTE solution with fixed energy $\bar\epsilon$ and charge $\bar n$ respectively. With this new quantity at hand, the most generic RTA is
\begin{equation}\label{eqn:ch6:charge_relaxation_RTA}
	I_\text{coll}-I_\text{ee}\approx-\frac{f_{\vect{p}}-f^{(0)}}{\tau'_m}-\frac{f_{\vect{p}}-\Bar{f}^{(0)}}{\tau_n}=-\frac{f_{\vect{p}}-f^{(0)}}{\tau_m}-\frac{f^{(0)}-\bar{f}^{(0)}}{\tau_n}
\end{equation}
where $\bar{f}^{(0)}$ has the same form of $f^{(0)}$
\begin{equation}
	\bar{f}^{(0)}=\frac{1}{1+e^{(\epsilon_p-\bar\mu)/\bar T}}
\end{equation}
but now computed at fixed reference values of temperature $\bar T$ and chemical potential $\bar\mu$. From now on, we will assume that all barred thermodynamic quantities are computed with respect to $\bar{f}^{(0)}$ and are thus functions of $\bar T$ and $\bar\mu$. In the second equality we redefined $\tau_m$ to separate the RTA in terms which induce different relaxations: the former is zero on the charge and energy equations, and only gives momentum relaxation, while the latter relaxes energy and charge at the same rate $\tau_n$ and vanishes in the momentum equation, see again the integrals in the Appendix~\ref{appendix:useful_integrals}. With the collision integral \eqref{eqn:ch6:charge_relaxation_RTA} the equations of hydrodynamics become
\begin{subequations}
	\begin{align}
		\partial_t\epsilon+\dots&=-\frac{\epsilon-\Bar{\epsilon}}{\tau_n}\\
		\partial_t(P+\epsilon)v^i+\dots&=-\frac{(P+\epsilon)v^i}{\tau_m}\\
		\partial_tn+\dots&=-\frac{n-\bar{n}}{\tau_n}
	\end{align}
\end{subequations}
When linearizing around $\bar\epsilon$ and $\bar n$ (which are the reference equilibrium values), we immediately recover the one-current analogue of equations \eqref{eqn:ch6:normal_charge_relaxation}, while also obtaining the Onsager constraint $\tau_\epsilon=\tau_n$ for free, directly from kinetic theory.

We can again justify the RTA in \eqref{eqn:ch6:charge_relaxation_RTA} from a microscopic model. The total collision integral now takes the form
\begin{equation}
	I_\text{coll}=I_\text{ee}+I_\text{imp}+I_\text{ph}
\end{equation}
We have already discussed the role of $I_\text{ee}$, that dominates and forces the system to be in LTE, and of $I_\text{imp}$, which relaxes momentum via scattering with impurities. Provided that our solution is in LTE, these first two collision integrals vanish in the energy and charge equation. The third term instead is the collision integral that describes scattering between electrons and phonons. It is given by \cite{Lifshitz:PhysicalKineticsVolume}
\begin{align}\label{eqn:ch6:phonon_collision_integral}
	I_\text{ph}&=\int\dif^3\vect{q}W_{\vect{p}'\vect{q}\rightarrow\vect{p}}\left[f_{\vect{p}'}(1-f_{\vect{p}})n_{\vect{q}}-f_{\vect{p}}(1-f_{\vect{p}'})(1+n_{\vect{q}})\right]\delta\left(\epsilon_p-\epsilon_{p'}-\omega_q\right)\nonumber\\
	&+\int\dif^3\vect{q}W_{\vect{p}'\rightarrow\vect{p}\vect{q}}\left[f_{\vect{p}'}(1-f_{\vect{p}})(1+n_{\vect{q}})-f_{\vect{p}}(1-f_{\vect{p}'})n_{\vect{q}}\right]\delta\left(\epsilon_p+\omega_k-\epsilon_{p'}\right)
\end{align}
where $\omega_q$ and $n_{\vect{q}}$ are respectively the energy and one-particle distribution functions for the phonons, while $W_{\vect{p}'\vect{q}\rightarrow\vect{p}}$ is, as usual, the scattering rate. The first term represents the case in which a quasi-electron of momentum $\vect{p}$ emits a phonon with momentum $\vect{q}$ and is scattered to a state with momenta $\vect{p}'$, and its inverse process, in which an electron of momentum $\vect{p}'$ absorbs the phonon. In this first case, the momenta obey the conservation equation $\vect{p}=\vect{p}'+\vect{k}+\vect{b}$, where $\vect{b}$ accounts for Umklapp processes, and the electron $\epsilon_p$ is scattered to lower-energy states. On the other hand, the second term represents the case in which an electron of momentum $\vect{p}$ absorbs a phonon of momentum $\vect{q}$, and its inverse process of emission, so that momentum conservation reads $\vect{p}+\vect{k}=\vect{p}'+\vect{b}$ and the electron $\epsilon_p$ is of lower energy. Notice that the above collision integral identically vanishes when integrated in momentum space against the identity, thanks to the fact that electron-phonon scatterings conserves the electric charge.

Furthermore, the above electron-phonon collision integral also vanishes when both electrons and phonons are in global thermodynamic equilibrium \cite{Lifshitz:PhysicalKineticsVolume}. Following \cite{Narozhny:HydrodynamicApproachElectronic}, we assume that the temperature is high enough so that the phonons are in a global thermodynamic equilibrium state defined by $\bar T$ and $\bar\mu$
\begin{equation}
	n_{\vect{q}}=\bar{n}^{(0)}=\frac{1}{e^{\left(\omega_q-\bar\mu\right)/\bar T}-1}
\end{equation}
and behave as a thermal bath for the electrons. Having fixed the phonon distribution, then $I_\text{ph}$ vanishes only when $f_{\vect{p}}$ is also in global thermodynamic equilibrium with the phonons, i.e. only when $f_{\vect{p}}=\bar{f}^{(0)}$.

If we now wish to write the electron-phonon collision integral in a RTA form, this should vanish when $f_{\vect{p}}$ is equal to the global thermodynamic equilibrium induced by the phonon bath $\bar{f}^{(0)}$, and we are thus led to the second term in \eqref{eqn:ch6:charge_relaxation_RTA}.

\subsection{Generalized relaxations from RTA}
We have shown how to introduce energy and charge relaxation in the system, without mixed relaxations, hence it is time to discuss how to include generalized relaxations from a kinetic theory perspective. In the RTA \eqref{eqn:ch6:charge_relaxation_RTA}, the first term is related to momentum relaxation and will not be affected by this discussion, however we will modify the second term in such a way to obtain generalized relaxations.

To begin, we simply need to assume that the scattering rate $\tau$ is a function of the momentum or, to be more specific, of the quasiparticle energy $\tau=\tau(\epsilon_p)$, assuming isotropy. From a microscopic perspective, this is believed to be universally true for most scattering mechanism \cite{Lifshitz:PhysicalKineticsVolume,Cercignani:MathematicalMethodsKinetic}: the RTA is a very simple approximation of the true collision integral that tries to capture the complicated non-linear scattering dynamics in a single constant parameter $\tau$, however in realistic cases the effective scattering rate does depend on the energy of the quasiparticles \cite{Pal:ResistivityNonGalileaninvariantFermi,Ochi:ElectronholeDichotomyThermoelectric} and, as we will see, this induces generalized relaxations on the equations of hydrodynamics.

Consider the second term in \eqref{eqn:ch6:charge_relaxation_RTA} due to electron-phonon scattering, and assume that $\tau_n(\epsilon_p)$ can be expanded in the generic power-series form
\begin{align}\label{eqn:ch6:generalised_relaxations_RTA}
	I_\text{ph}&\approx\sum_{N\geq-2}\epsilon_p^N\frac{f^{(0)}-\Bar{f}^{(0)}}{\tau_{N+2}}=\nonumber\\
	&=\frac{1}{\epsilon_p^2}\frac{f^{(0)}-\Bar{f}^{(0)}}{\tau_0}+\frac{1}{\epsilon_p}\frac{f^{(0)}-\Bar{f}^{(0)}}{\tau_1}+\frac{f^{(0)}-\bar{f}^{(0)}}{\tau_2}+\epsilon_p\frac{f^{(0)}-\Bar{f}^{(0)}}{\tau_3}+\dots
\end{align}
The first two terms, with negative powers of $\epsilon_p$, are the only ones with $N<0$ that lead to finite quantities when integrated in momentum space for the linear dispersion relation we are considering $\epsilon_p=p$. We define with $M_{N+2}$ the $N$-th energy-moment of $f^{(0)}$
\begin{equation}
	M_{N+2}=\sum_{p,h}\int\frac{\dif^3\vect{p}}{(2\pi)^3}\epsilon_p^N f^{(0)}
\end{equation}
such that, in particular, $M_2=n$ and $M_3=\epsilon$. Then, inserting \eqref{eqn:ch6:generalised_relaxations_RTA} in the Boltzmann equation, multiplying with the identity and the quasiparticle energy, and subsequently integrating over momentum space, we obtain the equations of relaxed hydrodynamic. The expressions for energy and charge are
\begin{subequations}\label{eqn:ch6:generalised_relaxations_RTA_hydrodynamic_equations}
	\begin{align}
		\partial_t\epsilon+\dots&=-\frac{M_1-\bar{M}_1}{\tau_0}-\frac{n-\bar n}{\tau_1}-\frac{\epsilon-\bar\epsilon}{\tau_2}-\frac{M_4-\bar{M}_4}{\tau_3}+\dots\\
		\partial_t n+\dots&=-\frac{M_0-\bar{M}_0}{\tau_0}-\frac{M_1-\bar{M}_1}{\tau_1}-\frac{n-\bar n}{\tau_2}-\frac{\epsilon-\bar\epsilon}{\tau_3}+\dots
	\end{align}
\end{subequations}
where the dots on the RHS represent higher moments. The $M_N$ are thermodynamic functions of $\mu$ and $T$ that can be analytically computed, e.g. the first few terms for massless relativistic fermions are
\begin{subequations}
	\begin{align}
		M_0&=\frac{\mu}{2\pi^2}\\
		M_1&=\frac{T^2}{4\pi^2}\left(\frac{\mu^2}{T^2}+\frac{\pi^2}{3}\right)\\
		M_4&=\frac{T^4\mu}{30\pi^2}\left(\pi^2+\frac{\mu^2}{T^2}\right)\left(7\pi^2+3\frac{\mu^2}{T^2}\right)
	\end{align}
\end{subequations}
while $M_2$ and $M_3$ are given in \eqref{eqn:ch6:energy_charge_one_current}. In \eqref{eqn:ch6:generalised_relaxations_RTA_hydrodynamic_equations}, the non-bar quantities are dynamical functions of $\mu$ and $T$ that obey the equations of hydrodynamics, while the bar quantities are understood as constraints without dynamics.

As they appear now, they do not resemble the generalized relaxations of Chapter~\ref{chapter:onsager}. This is because, as we have already discussed (see Chapter~\ref{chapter:onsager}), the generalized relaxations added to the equations of linearized hydrodynamics are agnostic on the full non-linear form of the relaxations. For this reason, if we want to match the expressions in \eqref{eqn:ch6:generalised_relaxations_RTA_hydrodynamic_equations} with the relaxations in \eqref{eqn:ch5:equations_of_motion_flat_space}, we must first linearize the equations. Clearly, the equilibrium solution is chosen so that all the collision integrals, and thus the relaxations terms, vanish. Consequently, the equilibrium will be given by the zero-velocity state with $\epsilon=\bar\epsilon$ and $n=\bar n$, to which we add linear fluctuations as $\epsilon=\bar{\epsilon}+\delta\epsilon$ and $n=\bar{n}+\delta n$ to obtain the linearized equations of hydrodynamics
\begin{subequations}
	\begin{align}
		\partial_t\delta\epsilon+\dots&=-\frac{\delta\epsilon}{\tau_{\epsilon\epsilon}}-\frac{\delta n}{\tau_{\epsilon n}}\\
		\partial_t\delta n+\dots &=-\frac{\delta\epsilon}{\tau_{n\epsilon}}-\frac{\delta n}{\tau_{nn}}
	\end{align}
\end{subequations}
where in writing the above non-conservation equations we defined the usual macroscopic relaxation times in terms of the microscopic $\tau_N$ of \eqref{eqn:ch6:generalised_relaxations_RTA}. For example, considering the first few terms shown in \eqref{eqn:ch6:generalised_relaxations_RTA_hydrodynamic_equations}, they are
\begin{subequations}\label{eqn:ch6:generalised_relaxations_microscopic}
	\begin{align}
		\frac{1}{\tau_{nn}}&=\frac{\partial M_0}{\partial n}\frac{1}{\tau_0}+\frac{\partial M_1}{\partial n}\frac{1}{\tau_1}+\frac{1}{\tau_2}+\dots\\
		\frac{1}{\tau_{n\epsilon}}&=\frac{\partial M_0}{\partial\epsilon}\frac{1}{\tau_0}+\frac{\partial M_1}{\partial\epsilon}\frac{1}{\tau_1}+\frac{1}{\tau_3}+\dots\\
		\frac{1}{\tau_{\epsilon n}}&=\frac{\partial M_1}{\partial n}\frac{1}{\tau_0}+\frac{1}{\tau_1}+\frac{\partial M_4}{\partial n}\frac{1}{\tau_3}+\dots\\
		\frac{1}{\tau_{\epsilon\epsilon}}&=\frac{\partial M_1}{\partial\epsilon}\frac{1}{\tau_0}+\frac{1}{\tau_2}+\frac{\partial M_4}{\partial\epsilon}\frac{1}{\tau_3}+\dots
	\end{align}
\end{subequations}
where again the dots represent terms associated with larger powers of $\epsilon_p$ in \eqref{eqn:ch6:generalised_relaxations_RTA_hydrodynamic_equations}. These final expressions are reminiscent of Matthiessen rule, with the difference that the microscopic times $\tau_N$ can be understood as different energy scaling of the same scattering process and are weighted by thermodynamic functions.

We can now check that Onsager relations, i.e. the constraint \eqref{eqn:ch5:onsager_condition}
\begin{equation}
	\frac{\chi_{\epsilon\epsilon}}{\tau_{n\epsilon}}-\frac{\chi_{\epsilon n}}{\tau_{\epsilon\epsilon}}+\frac{\chi_{n\epsilon}}{\tau_{nn}}-\frac{\chi_{nn}}{\tau_{\epsilon n}}=0
\end{equation}
is solved identically by the expressions \eqref{eqn:ch6:generalised_relaxations_microscopic}, upon using thermodynamic identities. Although time-reversal covariance is a property that emerges from our kinetic theory approach, we find the same is not true for the second law of thermodynamics \eqref{eqn:ch5:entropy_constraint}, which fails in general, as we argued in Section~\ref{sec:ch6:generalised_relaxations} for the case with two currents. The reason is clear from our kinetic theory model: we introduced generalized relaxations as interactions with a bath (in our case, of phonons), thus we are describing a weakly open system that must not obey positivity of entropy production.

The above computation shows that Onsager-reciprocal generalized relaxations can be obtained from a microscopic model, starting from a simple energy-dependent RTA ansatz which can be understood as interaction with a bath in equilibrium. Therefore, generalized relaxations are actually a rather universal feature, and we can expect them to appear in many different cases, since they only require the scattering rate to depend on the quasiparticle energy.

Some comments we think are necessary. First, although the results above are obtained for a linear dispersion relation and for an analytic function $\tau(\epsilon_p)$ only, we can expect the same conclusions to hold even for more exotic models, at the cost of giving up analytic control over some of the computations.

Furthermore, it is clear from the expressions in \eqref{eqn:ch6:generalised_relaxations_RTA_hydrodynamic_equations} and \eqref{eqn:ch6:generalised_relaxations_microscopic} that each separate microscopic scattering time $\tau_N$ leads to generalized relaxations which satisfy Onsager relations. What this means is that if a specific microscopic model suggests that only one or two $\tau_N$ in the expansion \eqref{eqn:ch6:generalised_relaxations_RTA} are relevant, the a-priori unrelated four macroscopic generalized relaxations will depend only on these relevant microscopic times, reducing the number of independent parameters. These relations cannot be detected by means of hydrodynamics and require a fully microscopic approach to derive them.

We can also consider the effect of energy dependence on the momentum RTA term in \eqref{eqn:ch6:charge_relaxation_RTA}. What happens in this case is that we find new thermodynamics functions that multiply the velocity, however when linearizing around $\vect{v}=0$ we can always write the momentum relaxation term as $(P+\epsilon)/\tau_m$, by appropriately redefining $\tau_m$ to absorb thermodynamic factors. Thus, an energy-dependent relaxation time in the first term of \eqref{eqn:ch6:charge_relaxation_RTA} does not alter the momentum (non-)conservation equation, it simply changes the effective value of $\tau_m$.

Finally, we comment on the issue of charge conservation. From a microscopic point of view the electron-phonon collision integral \eqref{eqn:ch6:phonon_collision_integral} conserves the total electric charge, even if it relaxes energy and momentum. This property, however, is not shared by the RTA we wrote in \eqref{eqn:ch6:charge_relaxation_RTA} and its subsequent generalization in \eqref{eqn:ch6:generalised_relaxations_RTA}, that do not conserve charge. Collision integrals are complicated functionals, therefore solving exactly the Boltzmann equation is very hard, hence to make progress we often exchange the true collision integral for a simpler model. Any such approximation will generally lose certain properties of the scattering process, while trying to preserve other ones, in the hope that the approximation is good enough to describe the dynamics one is interested in \cite{Cercignani:MathematicalMethodsKinetic}. From this point of view, it is clear that the RTA is the simplest approximation, in which all the complicated scattering process is captured by a single function $\tau(\epsilon_p)$: the only property it preserves is that the collision integral vanishes when the distribution function reaches equilibrium $f_{\vect{p}}=\bar{f}^{(0)}$, however it gives up on other important properties, such as charge conservation. Nonetheless, we can impose that electric charge is conserved using generalized relaxations and their dependence on a set of common microscopic times $\tau_N$, namely we require that
\begin{equation}
	\partial_tn+\dots=\int\frac{\dif^3p}{(2\pi)^3}\frac{f^{(0)}-\bar{f}^{(0)}}{\tau(\epsilon_p)}=\sum_{N\geq-2}\int\frac{\dif^3p}{(2\pi)^3}\epsilon_p^N\frac{f^{(0)}-\bar{f}^{(0)}}{\tau_N}\stackrel{?}{=}0
\end{equation}
This might not always be possible, e.g. if the only relevant terms in the expansion of $\tau(\epsilon_p)$ in \eqref{eqn:ch6:generalised_relaxations_RTA} are $\tau_2$ and $\tau_3$, then we find that $\tau_{nn}=\tau_2$ and $\tau_{n\epsilon}=\tau_3$, thus enforcing charge conservation amounts to requiring $\tau_2^{-1}=\tau_3^{-1}=0$, washing away any energy relaxation too. However, if more or different $\tau_N$ are relevant, we can in general expect to be able to solve $\tau_{nn}^{-1}=\tau_{n\epsilon}^{-1}=0$ in terms of the microscopic times $\tau_N$, thus obtaining charge conservation as desired. It is not clear at the moment whether these conditions would necessarily arise from a microscopic calculation, or should be regarded as a computational trick to fine-tune charge conservation.

\subsection{Two-currents model}
We can generalize the results of the previous sections to the case of interest for Weyl semimetals, namely a fluid with two species of particles that thermalize together. To proceed, we consider two distinct distribution functions $f_{\vect{p},\lambda}$ with $\lambda=\pm$ denoting the chirality of each species of particles, located around its corresponding Weyl cone \cite{Dantas:MagnetotransportMultiWeylSemimetals,Gorbar:ConsistentChiralKinetic}. Then, in the absence of impurities and phonons, the two coupled Boltzmann equations take the standard form \cite{Lifshitz:PhysicalKineticsVolume,Pongsangangan:HydrodynamicsChargedTwodimensional,Fotakis:MulticomponentRelativisticDissipative,Bianca:BoltzmannGasMixture}
\begin{equation}\label{eqn:ch6:boltzmann_equation_two_currents}
	\partial_t \vec{f}_{\vect{p}}+\vect{p}\cdot\vect{\nabla}\vec{f}_{\vect{p}}=I_{ee}[\vec{f}_{\vect{p}}]=\begin{pmatrix}
		I_{++}[f_{\vect{p},+}]  +   I_{+-}[f_{\vect{p},\lambda}]\\
		I_{-+}[f_{\vect{p},\lambda}]  +   I_{--}[f_{\vect{p},-}]
	\end{pmatrix}
\end{equation}
where $\vec{f}_{\vect{p}}=\left(f_{\vect{p},+},f_{\vect{p},-}\right)^T$. Exactly like in the one-current case, it can be shown that the electron-electron collision integral vanishes when the distribution functions are in LTE. In particular, because of the off-diagonal interaction terms between particles of different chirality $I_{\pm\mp}\neq0$, the two species can exchange energy and momentum, thus only the total energy and momentum are conserved, while the chiral charges are separately conserved (not considering the anomaly) \cite{Bianca:BoltzmannGasMixture,Fotakis:MulticomponentRelativisticDissipative}. Consequently, they thermalize to LTE distribution functions defined at same temperature $T$ and velocity $\vect{u}$, but at different chemical potentials $\mu_\lambda=\mu+\lambda\mu_5$
\begin{equation}\label{eqn:ch6:local_thermodynamic_equilibrium_two_currentsz}
	f_{\vect{p},\lambda}=\frac{1}{1+e^{(\epsilon_p-\vect{u}\cdot\vect{p}-\mu_\lambda)/T}}~.
\end{equation}
From here, we again linearize at small velocity
\begin{equation}
	f_{\vect{p},\lambda}\approx f^{(0)}_\lambda-(\vect{p}\cdot\vect{u})\frac{\partial f^{(0)}_\lambda}{\partial\epsilon_p}\qquad\text{with}\qquad f^{(0)}_\lambda=\frac{1}{1+e^{(\epsilon_p-\mu_\lambda)/T}}
\end{equation}
which allows us to decouple charge and momentum relaxations in a block diagonal form. Finally, we can compute the thermodynamic quantities (energy, electric and axial charge) in LTE for a linear dispersion relation $\epsilon_p=p$ that describes Type I Weyl semimetals \cite{Gorbar:ConsistentHydrodynamicTheory}
\begin{subequations}\label{eqn:ch6:two_currents_thermodynamics}
	\begin{align}
		\epsilon&=\sum_\lambda\sum_{p,h}\int\frac{\dif^3\vect{p}}{(2\pi)^3}\epsilon_p f_{\vect{p},\lambda}=\frac{7\pi^2T^4}{60}+\frac{T^2\left(\mu^2+\mu_5^2\right)}{2}+\frac{\mu^4+6\mu^2\mu_5^2+\mu_5^4}{4\pi^2}~,\\
		n&=\sum_\lambda\sum_{p,h}\int\frac{\dif^3\vect{p}}{(2\pi)^3}f_{\vect{p},\lambda}=\frac{\mu\left(\pi^2T^2+\mu^2+3\mu_5^2\right)}{3\pi^2}~,\\
		n_5&=\sum_\lambda\lambda\sum_{p,h}\int\frac{\dif^3\vect{p}}{(2\pi)^3}f_{\vect{p},\lambda}=\frac{\mu_5\left(\pi^2T^2+3\mu^2+\mu_5^2\right)}{3\pi^2}~.
	\end{align}    
\end{subequations}
where the first sum is over the two chiralities (notice that $n_5$ is computed weighting with $\lambda$), and the second one on particles and holes contributions. These thermodynamic quantities obey the Euler relation
\begin{equation}
	P+\epsilon=sT+n\mu+n_5\mu_5
\end{equation}
where $P=\epsilon/3$ and $s=\partial P/\partial T$.

To include momentum relaxation we can proceed exactly as before: we add an elastic electron-impurity collision integral that preserves chirality, which takes the usual form \eqref{eqn:ch6:impurities_collision_integral} for each species $\lambda$. Then, we arrive at the same RTA as in \eqref{eqn:ch6:momentum_relaxation_RTA}, one for each chirality, and computing the momentum equation we obtain the momentum-relaxation term as desired.

Regarding the relaxations of the other charges, however, there are different choices we can make to write the RTA, as we discuss now.

\subsubsection{Intra-valley scattering} Consider the simplest RTA that relaxes the charges, while not mixing the chirality, namely
\begin{equation}\label{eqn:ch6:intra_valley_RTA}
	I_\text{ph}\approx-\frac{f_\lambda^{(0)}-\bar f_\lambda^{(0)}}{\tau}~,
\end{equation}
where, as before, the bar stands for the global thermodynamic equilibrium distribution function. This collision integral relaxes all the charges at the same rate $\tau_\epsilon=\tau_n=\tau_{n_5}$, which is indeed the correct solution to the Onsager constraints \eqref{eqn:ch6:onsager_constraints} without off-diagonal relaxations, as discussed in Section~\ref{sec:ch6:normal_charge_relaxation}.

Following the one-current case, we consider the effect of an energy-dependent relaxation rate $\tau=\tau(\epsilon_p)$: expanding as in \eqref{eqn:ch6:generalised_relaxations_RTA}, computing the equations of hydrodynamics, and linearizing around an equilibrium with $\epsilon=\bar\epsilon$, $\ n=\bar n$ and $n_5=\bar{n}_5$, we arrive at the generalized relaxations of \eqref{eqn:ch6:generic_relaxations}. The resulting macroscopic decay rates, written in terms of the microscopic times $\tau_N$ as in \eqref{eqn:ch6:generalised_relaxations_microscopic}, identically satisfy Onsager relations \eqref{eqn:ch6:onsager_constraints}.

From a physical perspective, this collision integral mimics intra-valley scatterings, i.e. quasi-electrons that interact only within each Weyl node and do not change chirality. The microscopic justification of this RTA comes again from electron-phonon collisions, in which we assume that phonons in equilibrium near each cone do not carry chirality and the electrons are scattered back in their node of origin.

\subsubsection{Inter-valley scattering} We can also consider more generic collision integrals in the presence of multiple species, in particular we can analyse the effect of inter-valley scatterings mediated by impurities.

Inter-valley scatterings are usually implemented  as out of equilibrium processes (not in the ideal-fluid hydrodynamic regime we are in) by subtracting to $f_{\vect{p}}$ its angular average \cite{Dantas:MagnetotransportMultiWeylSemimetals,Zyuzin:MagnetotransportWeylSemimetals}. This cannot be done in our present case, since the distribution functions $f_\lambda^{(0)}$ are isotropic, nonetheless we can write an RTA form that represents inter-valley scatterings by including distribution functions with different chirality. Namely, we can write
\begin{equation}\label{eqn:ch6:inter_valley_RTA}
	I_{\text{imp},\lambda}\approx-\frac{f^{(0)}_\lambda-f^{(0)}_{-\lambda}}{2\tau}
\end{equation}
Notice that we are subtracting to $f_\lambda^{(0)}$ the dynamical distribution function $f_{-\lambda}^{(0)}$ associated to the other chirality, instead of the fixed equilibrium value $\bar{f}^{(0)}_{-\lambda}$.

The above collision integral, when $\tau$ is constant, vanishes in the energy, momentum and electric charge equation, however it appears in the axial one and its form is such that it relaxes the axial charge to zero. More specifically, it takes the following integrated form in the axial charge equation
\begin{equation}
	\partial_t n_5+\dots=-\frac{n_5}{\tau}
\end{equation}
so that in the decay term only $n_5$ appears, instead of the difference $n_5-\bar{n}_5$. Thus, as already apparent from \eqref{eqn:ch6:inter_valley_RTA}, we see that this kind of collision integral tries to completely destroy any imbalance between the left- and right-handed distribution functions, driving the axial charge density to zero.

Naively, this collision integral seems to not satisfy Onsager relations: indeed, Onsager constraints \eqref{eqn:ch6:onsager_constraints} indicate that if $\tau_{n_5n_5}$ is non-zero, then we must also relax energy and charge at the same rate (we are not considering off-diagonal relaxations yet), see Section~\ref{sec:ch6:normal_charge_relaxation}. However, this does not happen from \eqref{eqn:ch6:inter_valley_RTA}, since it only enters the axial charge equation.

The solution to this problem comes from the observation that we should always linearize on the equilibrium solution that sets to zero all the collision integrals. It is then clear that RTA above \eqref{eqn:ch6:inter_valley_RTA} forces the chemical potentials of the two chiralities to be equal, thus the equilibrium solution is given by $\mu_5=n_5=0$, instead of $n_5=\bar{n}_5$. This means we should check Onsager relations on this equilibrium at zero axial chemical potential too. Doing so we find that Onsager relations are indeed obeyed and time-reversal invariance is identically preserved even in this case.

As usual, we can consider a generic energy dependence of the relaxation time to induce generalized relaxations. With $\tau=\tau(\epsilon_p)$ the inter-valley collision integral \eqref{eqn:ch6:inter_valley_RTA} still vanishes in the energy, charge and momentum equation, however now it induces non-zero generalized relaxations in the axial charge equation $\tau_{n_5\epsilon}\neq0\neq\tau_{n_5n}$. Again, when computed in equilibrium with $\mu_5=n_5=0$, these relaxations identically satisfy the constraints coming from Onsager relations \eqref{eqn:ch6:onsager_constraints}.

To justify the collision integral \eqref{eqn:ch6:inter_valley_RTA} from a microscopic perspective we can follow \cite{Lucas:HydrodynamicTheoryThermoelectric} and consider an impurity-mediated inter-valley scattering event of the form
\begin{equation}
	I_{\text{imp},\lambda}=\int\dif^3\vect{p}'W_{\vect{p}\rightarrow\vect{p}'}[f_{\vect{p},\lambda}-f_{\vect{p}',-\lambda}]\delta(\epsilon_p-\epsilon_{p'})~.
\end{equation}
that mixes the two chiralities. Then, the RTA model of this collision integral not only will set the velocity of the fluid to zero (contributing to momentum relaxation), but will also drive the chiral distribution functions to be equal, by depleting $\mu_5$ to zero, leading us to \eqref{eqn:ch6:inter_valley_RTA} as desired.

\bigskip
Finally, we see that, exactly like in the single-current case, the intra-valley RTA model \eqref{eqn:ch6:intra_valley_RTA} does not conserve charge in general. The same arguments we used before are valid now too: this is not a property of the full collision integral (we expect electron-phonon scattering to be unitary), but rather a spurious effect specific to the RTA model. Nonetheless, we can enforce charge conservation on the system by requiring that the effective charge-relaxation rates, written in terms of the microscopic times $\tau_N$, vanish $\tau_{n\epsilon}^{-1}=\tau_{nn_5}^{-1}=\tau_{nn}^{-1}=0$.

\subsection{Modified RTA: charge conservation}\label{sec:ch6:BGK_model}
We have already discussed how the standard RTA and its energy-dependent generalization \eqref{eqn:ch6:generalised_relaxations_RTA} violate charge conservation at the linear order in fluctuations \eqref{eqn:ch6:generalised_relaxations_RTA_hydrodynamic_equations}. We also argued that it is possible to solve this problem by enforcing charge conservation at the level of the microscopic $\tau_N$, which however requires some fine-tuning between the parameters. To conclude this chapter, we want to address whether it is possible to find suitable modifications of the RTA model that preserve charge conservation identically, while not spoiling Onsager relations or fine-tuning the relaxation rates. We find that in general this is possible, by employing a generalization of the so-called BGK model \cite{Bhatnagar:ModelCollisionProcesses,Cercignani:MathematicalMethodsKinetic,Rocha:NovelRelaxationTime}. As we discussed in Section \ref{sec:ch6:hydrodynamics_kinetic_theory}, the standard RTA \eqref{eqn:ch6:electron_RTA} induces the Landau frame on the hydrodynamic constitutive relations by ruining the conservation of the charges, for this reason in \cite{Rocha:NovelRelaxationTime} the BGK model (which conserves the charges identically) was used to obtain the hydrodynamic constitutive relations in a generic frame. To be more concrete, we will consider a model with a single conserved current, which can be straightforwardly generalized to models with multiple conserved currents.

First, we review the standard BGK model, before generalizing it for our needs. Consider the electron-electron collision integral in \eqref{eqn:ch6:electron_collision_integral}: it vanishes identically in LTE $f_{\vect{p}}=f_{\vect{p}}^\text{eq}$, thus to compute the hydrodynamic equations we can linearize around the LTE solution as
\begin{equation}
	f_{\vect{p}}=f_{\vect{p}}^\text{eq}+\delta f_{\vect{p}}=f_{\vect{p}}^\text{eq}(1+h_{\vect{p}})
\end{equation}
where we introduced the auxiliary function $h_{\vect{p}}$. With respect to $h_{\vect{p}}$, the linearized collision integral $I_\text{ee}$ takes the form \cite{Cercignani:MathematicalMethodsKinetic}
\begin{multline}
	I_\text{ee}[f_{\vect{p}}]\simeq L_\text{ee}h_{\vect{p}}=\\
	\int\dif^3\vect{p}_2\dif^3\vect{p}'_1\dif^3\vect{p}'_2W_{\vect{p}\vect{p}'\rightarrow\vect{p}'_1\vect{p}'_2}f^\text{eq}_{\vect{p}}f^\text{eq}_{\vect{p}'_1}\left(h_{\vect{p}}+h_{\vect{p}_2}-h_{\vect{p}'_1}-h_{\vect{p}'_2}\right)
\end{multline}
The linearized collision integral $L_\text{ee}$ inherits many properties from the full non-linear integral. In particular, because $I_\text{ee}$ conserve energy, charge and momentum, it vanishes when integrated in momentum space over conserved charges \eqref{eqn:ch6:electron_collision_integral_integration}. Thus, the linearized operator has three eigenfunctions with vanishing eigenvalue (zero modes)
\begin{subequations}\label{eqn:ch6:linearised_collision_integral_properties_electron}
\begin{equation}
	L_\text{ee}1=0\qquad L_\text{ee}\vect{p}=0\qquad L_\text{ee}\epsilon_p=0
\end{equation}
which correspond to the conservation of charge, momentum and energy respectively. Furthermore, like in \eqref{eqn:ch6:electron_collision_integral_integration}, it vanishes when integrated in momentum against conserved quantities, irrespective of the distribution function $h_{\vect{p}}$
\begin{equation}
	\int\dif^3\vect{p}\ L_\text{ee}h_{\vect{p}}=0\qquad\int\dif^3\vect{p}\ \vect{p}L_\text{ee}h_{\vect{p}}=0\qquad\int\dif^3\vect{p}\ \epsilon_pL_\text{ee}h_{\vect{p}}=0
\end{equation}

From a more formal perspective, $L_\text{ee}$ acts as a linear operator on the Hilbert space spanned by the real functions $h_{\vect{p}}$. The inner product is defined as
\begin{equation}
	(h,g)=\langle hg\rangle=\int\dif^3\vect{p}\ f^\text{eq}_{\vect{p}}h_{\vect{p}}g_{\vect{p}}
\end{equation}
such that $L_\text{ee}$ is self-adjoint and negative semi-definite
\begin{equation}
	(g,L_\text{ee}h)=(L_\text{ee}g,h)\qquad\text{and}\qquad(h,L_\text{ee}h)\leq0
\end{equation}
\end{subequations}
In particular, the last inequality is saturated only when $h$ is a collision invariant, i.e. $L_\text{ee}h=0$ for conserved quantities.

From this point of view, it is clear that the RTA approximation \eqref{eqn:ch6:electron_RTA}, described by the operator $L_\text{RTA}$
\begin{equation}
	L_\text{ee}h_{\vect{p}}\approx L_\text{RTA}h_{\vect{p}}=-f^\text{eq}_{\vect{p}}\frac{h_{\vect{p}}}{\tau}
\end{equation}
is proportional to the identity operator in the Hilbert space $L_\text{RTA}\sim-\mathds{1}$. Notice that it does not have all the properties of the original linearized operator $L_\text{ee}$, specifically it does not vanish when integrated in momentum space against conserved quantities, which led to the Landau matching conditions in \eqref{eqn:ch6:electron_collision_integral_hydrodynamics}.

We can approximate $L_\text{ee}$ with an improved RTA model, by projecting away the states related to the conserved charges. We write
\begin{equation}\label{eqn:ch6:BGK_model_electron}
	L_\text{ee}h_{\vect{p}}\approx L_\text{BGK}h_{\vect{p}}=\frac{f^\text{eq}_{\vect{p}}}{\tau}\left(-h_{\vect{p}}+\sum_{n=1}^5\psi^a_{\vect{p}}(\psi^a,h)\right)
\end{equation}
where the $\psi^a_{\vect{p}}$ are the orthonormal zero modes of $L_\text{ee}$
\begin{equation}
	\left(\psi^a,\psi^b\right)=\delta^{ab}
\end{equation}
and they are in one-to-one correspondence with collision invariants. It is then possible to check that the BGK model satisfies all the properties listed above for the complete linear collision integral \eqref{eqn:ch6:linearised_collision_integral_properties_electron}, while also rendering the Boltzmann equation more manageable \cite{Cercignani:MathematicalMethodsKinetic}.

We would now like to apply the same construction to the electron-phonon collision integral in ideal hydrodynamics. We expect that the linearized scattering should conserve charge, while relaxing energy and momentum. In this case, the function $h$ is defined by linearizing around the global equilibrium state
\begin{equation}
	f^{(0)}=\bar{f}^{(0)}(1+h)
\end{equation}
where we have $f^{(0)}$ instead of $f_{\vect{p}}$ on the LHS, because we already know that this allows us to decouple the RTA into separate relaxations for the momentum and the charges \eqref{eqn:ch6:charge_relaxation_RTA}. Furthermore, in the present context, the angled brackets refer to the integral in the Hilbert space against $\bar{f}^{(0)}$
\begin{equation}
	\langle g\rangle_0=\int\dif^3\vect{p}\ \bar{f}^{(0)}g_{\vect{p}}
\end{equation}
Because we are interested in charge conservation, we should naively project away only the normalized state associated with charge density, i.e. $\psi^1=1/\sqrt{\langle1\rangle_0}=1/\sqrt{\bar{n}}$ where $\bar n$ is the global thermodynamic equilibrium charge density. From \eqref{eqn:ch6:BGK_model_electron} we then find that
\begin{subequations}\label{eqn:ch6:BGK_model_phonons_naive}
	\begin{align}
		L_\text{ph}h\approx L_\text{BGK}h&=\frac{\bar{f}^{(0)}}{\tau}\left(-h+\psi\langle\psi h\rangle_{0}\right)\nonumber\\
		&=\frac{\bar{f}^{(0)}}{\tau}\left(\frac{n}{\bar{n}}-1\right)-\frac{f^{(0)}-\bar{f}^{(0)}}{\tau}
	\end{align}	
\end{subequations}
To obtain the equations of hydrodynamics we proceed in the usual way, by multiplying with the identity and the quasiparticle energy and integrating over momentum the Boltzmann equation. We find that charge is indeed conserved, as desired, while the energy relaxes as
\begin{equation}\label{eqn:ch6:BGK_model_phonons_naive_hydrodynamics}
	\partial_t\epsilon+\dots=\int\dif^3\vect{p}\ \epsilon_{\vect{p}}L_\text{BKG}h=\frac{\bar\epsilon}{\bar n}\frac{n-\bar n}{\tau}-\frac{\epsilon-\bar\epsilon}{\tau}
\end{equation}
We see that the above decay term vanishes in equilibrium defined by $n=\bar n$ and $\epsilon=\bar\epsilon$. Then, fluctuating around this background, we can check Onsager relations \eqref{eqn:ch6:onsager_constraints} following the steps of the previous sections. Interestingly, however, this procedure now fails: the generalized relaxations obtained from the above energy equation do not obey Onsager relations.

To understand this problem, remember that we should always fluctuate around a background that sets to zero all the collision integral. Although the solution $n=\bar n$ and $\epsilon=\bar\epsilon$ sets to zero the energy relaxation term, there is also a second spurious solution given by $n$ arbitrary and $\epsilon=n\bar\epsilon/\bar n$. Indeed, if we compute the susceptibilities on this equilibrium and subsequently check Onsager relations for the relaxation found in \eqref{eqn:ch6:BGK_model_phonons_naive_hydrodynamics}, we find that time-reversal holds identically, as expected.

Even if the above approach works, indeed charge is conserved and Onsager relations are obeyed, the result is unsatisfactory. We would like to obtain the same results, but for an equilibrium background in which the energy is not constrained to be proportional to the charge. To do so we must further modify the BGK model, as we will now see. Consider a modified BGK operator of the form
\begin{equation}
	L_* = -\frac{\bar{f}^{(0)}}{\tau}\sum_{i,j}a_{i,j}\psi^i\tilde{\psi}^j =  -\frac{\bar{f}^{(0)}}{\tau}\left[1+ \sum_{i,j}(a_{i,j}-\delta_{ij})\psi^i\tilde{\psi}^j\right]
\end{equation}
where $\tilde{\psi}^a$ is a linear functional of the $\psi$ (a bra in Dirac notation, dual to $\psi^a$) defined by $\left(\psi^a,\psi^b\right)=\tilde{\psi}^a(\psi^b)$, and $a_{i,j}$ is a matrix of momentum-independent coefficients. The above $L_*$ means that we are assuming that the full linear collision integral $L_\text{ph}$ and its RTA form cannot be made diagonal on the same basis, thus we had to introduce extra coefficients $a_{i,j}$ which tell us how the basis vectors transform.

If we want $L_*$ to preserve charge identically, exactly like $L_\text{ph}$ does, then we must enforce that $\left(\psi^1,L_*h\right)=0$ for any function $h$, where $\psi^1$ is the charge zero mode of $L_\text{ph}$. To do so, we require $a_{1,i}=0$ on the $a$-matrix of coefficients. Notice, however,that we might have $a_{i,1}\neq0$ in general, while also having the other entries non-zero. This implies that $\psi^1$ is not a zero mode of $L_*$, hence the phase-space charge density is not conserved. We can understand it as an indication that a system described by $L_*$ is weakly open, and this is further supported by the fact that $L_*$ may not have any other property of the standard linearized collision integral, such as Hermiticity or negative semi-definiteness. This perspective suggests that it might violate the second law of thermodynamics and not conserve any other charge, except for the electric one.

In the hydrodynamic regime we are only interested in the first moments of the conserved quantities, thus we restrict ourselves to $a_{i,j>5}=a_{i>5,j}=0$, and denote with $i,j=1,2,3,4,5$ the modes related to charge, energy and momentum conservation. Subsequently, we can decouple momentum relaxation from charge and energy, as we did in \eqref{eqn:ch6:charge_relaxation_RTA}, by assuming that $a_{i,j}$ takes a block-diagonal form, so that we can ignore momentum relaxation from now on and focus on charge and energy relaxations. With this in mind, $a_{i,j}$ is now a $2\times2$ matrix
\begin{equation}
	a_{i,j} = \begin{pmatrix}
		0 & 0
		\\
		\alpha_2 & a_1
	\end{pmatrix}
	\Rightarrow \alpha_{i,j} \equiv a_{i,j}-\delta_{ij} = 
	\begin{pmatrix}
		-1 & 0
		\\
		\alpha_2 & \alpha_1
	\end{pmatrix}
\end{equation}
that, when expanded, leads to the following linearized collision integral
\begin{align}\label{eqn:ch6:modified_BGK_operator}
	L_* &= -\frac{\bar{f}^{(0)}}{\tau}\sum_{i,j}\left(\alpha_2 \psi^2\tilde{\psi}^1 + a_1 \psi^2\tilde{\psi^2}\right)=\nonumber \\
	&=  -\frac{\bar{f}^{(0)}}{\tau}\left[1-\psi^1\tilde{\psi}^1 + \alpha_2 \psi^2\tilde{\psi}^1 + \alpha_1 \psi^2\tilde{\psi^2}\right]
\end{align}
In particular, for $\alpha_2=0$ and $\alpha_1=-1$ we recover the standard diagonal expression used for the BGK model of electron-electron scatterings (modulo the momentum terms) \cite{Cercignani:MathematicalMethodsKinetic,Rocha:NovelRelaxationTime}. The new coefficients $\alpha_1,\alpha_2$ are phenomenological parameters used to identically enforce charge conservation and Onsager relations on the background we desire, by tuning their values appropriately. We can show how this works explicitly: for our purpose, we only need the modes $\psi^1$ and $\psi^2$, associated with charge and energy respectively, which take the form
\begin{equation}\label{eqn:ch6:zero_modes}
	\psi^{1} = \frac{1}{\sqrt{\bar{n}}}\quad\qquad\psi^2 = \frac{\bar{\epsilon}-\bar{n}\epsilon_p}{\sqrt{\bar{n}^2\bar{\epsilon^2}-\bar{\epsilon}^2\bar{n}}}  
\end{equation}
Subsequently, using the expression for $L_*$ \eqref{eqn:ch6:modified_BGK_operator} in the Boltzmann equation, together with the above values for the modes $\psi^i$ \eqref{eqn:ch6:zero_modes}, we arrive at
\begin{subequations}\label{eqn:ch6:BGK_modified}
\begin{equation}
	\partial_t f^{(0)} + \dots = -\frac{1}{\tau}\left[f^{(0)} - \frac{n}{\bar{n}}\bar{f}^{(0)} + \tilde{\alpha}_2 \bar{f}^{(0)} (n-\bar{n}) + \tilde{\alpha}_1 \bar{f}^{(0)}(\bar{\epsilon}n-\bar{n}\epsilon)\right]
\end{equation}
where
\begin{equation}
	\tilde{\alpha}_1 = \alpha_1 \frac{\bar{\epsilon}-\bar{n}\epsilon_p}{\bar{n}^2\bar{\epsilon^2} - \bar{\epsilon}^2\bar{n}}\qquad
	\tilde{\alpha}_2 = \alpha_2 \frac{\bar{\epsilon}-\bar{n}\epsilon_p}{\sqrt{\bar{n}^3\bar{\epsilon^2} - \bar{\epsilon}^2\bar{n}^2}}
\end{equation}
\end{subequations}
Notice that, contrary to $\alpha_1$ and $\alpha_2$, we defined $\tilde{\alpha}_1$ and $\tilde{\alpha}_2$ to be functions of momenta, so we will need to integrate them to compute the equations of hydrodynamics. From here, we can already see that the RHS vanishes identically when $f^{(0)}=\bar{f}^{(0)}$, i.e. $n=\bar n$ and $\epsilon=\bar\epsilon$, as desired.

We can compute the hydrodynamic (non-)conservation equations from the Boltzmann equation above. As expected, integrating over momentum we see that electric charge is identically conserved (thanks to the fact that $\langle\tilde\alpha_1\rangle_0=\langle\tilde\alpha_2\rangle_0=0$), while the energy equation becomes
\begin{equation}
	\partial_t\epsilon+\dots=-\frac{\epsilon-\bar\epsilon}{\tau}+\frac{\bar\epsilon}{\bar n}\frac{n-\bar n}{\tau}-\frac{\alpha_2(\bar\epsilon^2-\bar n\bar{\epsilon^2})}{\sqrt{\bar{n}^3\bar{\epsilon^2} - \bar{\epsilon}^2\bar{n}^2}}\frac{n-\bar n}{\tau}+\\
	-\frac{\alpha_1(\bar\epsilon^2-\bar n\bar{\epsilon^2})}{\bar n^2\bar{\epsilon^2}-\bar n\bar\epsilon^2}\frac{\bar\epsilon n-\bar n\epsilon}{\tau}
\end{equation}
Proceeding in the usual way, we linearize around the global equilibrium background defined by $n=\bar n$, $\epsilon=\bar\epsilon$ and read off the generalized relaxations $\tau_{\epsilon n}$ and $\tau_{\epsilon\epsilon}$ in terms of the microscopic time $\tau$. The effective relaxations depend either on $\alpha_1$ or $\alpha_2$, hence when we impose Onsager relations we can tune the parameters to ensure that time-reversal is preserved.

In conclusion, we showed that by considering a suitable modification of the BGK model we can obtain generalized relaxations from kinetic theory that identically obey Onsager relations and preserve charge conservation. These relaxations do not satisfy the second law of thermodynamics, in general, and depend on two phenomenological parameters.

\section{Summary, discussion and outlook}
In this chapter we studied the longitudinal thermoelectric magnetotransport of anomalous fluids, with a particular focus on the physics of Weyl semimetals. First, we pointed out a mistake present in previous models \cite{Landsteiner:NegativeMagnetoresistivityChiral,Lucas:HydrodynamicTheoryThermoelectric}, in which the anomalous conductivities were computed using order-one hydrodynamics with an order-one magnetic field. We showed that this approach leads to conductivities which are not anomalous: the magnetic-field dependent part appears only at order two in derivatives \eqref{eqn:ch6:conductivity_generic_frame}, thus it should be discarded in order-one hydrodynamics, since it is not physical. Indeed, this is confirmed by the fact that different hydrodynamic frames correspond to different results for the longitudinal magnetotransport, which should not happen. To properly study the longitudinal anomalous conductivities we suggest using a magnetic field that is order zero in derivatives $B\sim\mathcal{O}(1)$, such that it enters the thermodynamics.

Subsequently, we used ideal hydrodynamics with an order-zero magnetic field to study the DC transport of anomalous fluids. In agreement with previous works \cite{Lucas:HydrodynamicTheoryThermoelectric,Landsteiner:NegativeMagnetoresistivityChiral,Abbasi:MagnetotransportAnomalousFluid}, we found that in general energy, momentum and axial charge have to relax (all of which can be argued based on microscopic mechanisms) in order for the system to have finite DC conductivities. However, we showed that naive realizations of these relaxations lead to unphysical constraints between the relaxation rates when Onsager relations are taken into account. Thus, following phenomenological considerations, we set the goal to look for a system that obeys fundamental constraints: finite DC conductivity, microscopic time-reversal covariance (Onsager relations) and electric charge conservation. What we found is that by using the generalized relaxations introduced in Chapter~\ref{chapter:onsager} we could write a model with these properties, however entropy is not conserved in the system, implying that it is open. Moreover, our model gives new qualitative predictions for the thermoelectric transport of anomalous fluid, which could be used to probe the hydrodynamic regime of Weyl semimetals.

Finally, we discussed how to obtain generalized relaxations from kinetic theory. We found that this can be achieved by introducing a fixed reference distribution function $\bar f^{(0)}$ in an appropriate energy-dependent generalization of the RTA \eqref{eqn:ch6:generalised_relaxations_RTA}. This linearized collision integral can be argued from microscopic considerations regarding electron-phonon scatterings, and we can also modify it to identically conserve charge, as in Section~\ref{sec:ch6:BGK_model}. Remarkably, the generalized relaxations obtained from kinetic theory identically obey Onsager relations.

There are many interesting avenues that stem from our kinetic theory computation. In this chapter we focused our attention on relaxation rates with a simple power-law dependence \eqref{eqn:ch6:generalised_relaxations_RTA}, while also restricting ourselves to a linear dispersion relation $\epsilon_p=p$ (relevant for Weyl semimetals), however our result seems to be more general and could apply also to non-analytic relaxation rates $\tau(\epsilon_p)$ and different dispersion relation. Also, it would be interesting to check whether it is possible to modify the inter-valley RTA \eqref{eqn:ch6:inter_valley_RTA} in order to describe system with finite axial charge density in equilibrium $\mu_5\neq0\neq n_5$. To conclude, it would be fascinating to understand if the coefficients $\alpha_1$ and $\alpha_2$ introduced in Section~\ref{sec:ch6:BGK_model}, used to impose Onsager relations and charge conservation on the system via kinetic theory, can be constrained from microscopic fundamental principles or linear stability.
\chapter{Conclusions}
\epigraph{``No one should be discouraged, Theaetetus, who can make constant progress, even though it be slow.''}{Plato, \emph{Sophist}}

\section{Summary of main results}
In this thesis we analysed various aspects of relaxed hydrodynamics, broadly defined, with a specific focus on the transport properties in the linearized regime.

To begin, in Chapter~\ref{chapter:charge_density_waves} we studied a strongly-coupled electronic fluid in which spatial translations are broken by the presence of Charge Density Waves, a situation that often arises in cuprates. Hydrodynamics is then modified to account not only for the dynamics of conserved charges, but also of the (pseudo-)Goldstone bosons that appear due to the symmetry breaking. We considered two cases: one where translations are broken spontaneously, leading to the Goldstone fields remaining massless, and another where translations are broken pseudo-spontaneously. In the latter case, a small external source that explicitly breaks translations is turned on.

We constructed the hydrodynamic theory for such system, and analysed the transport properties focusing on a new regime in which both the lattice pressure $P_l$ and a strong external magnetic field $B\sim\mathcal{O}(1)$ are present. The former, in particular, is a new transport coefficient related to crystal fields, and must vanish in thermal equilibrium. Nonetheless, it is obtained naturally from holography in Q-lattice models, which describe metastable phases, and is therefore needed in the hydrodynamic description to correctly match the holographic correlators.

We studied the optical thermoelectric conductivities using hydrodynamics in linear response, obtaining analytic expressions which are functions of thermodynamic parameters, the DC values of certain non-universal conductivities, and, in the pseudo-spontaneous case, of the pinning frequency $\omega_0$. We obtained the values of the non-universal DC correlators from a specific Q-lattice holographic model in terms of horizon data, and for the same model we also computed the conductivities numerically.

We found a very good agreement between our analytic hydrodynamic correlators and the holographic ones over a large range of values of the parameters. Furthermore, we obtain a relation that expresses the Goldstone relaxation rate $\Omega^{IJ}$ in terms of the pinning frequency $\omega_0$, confirming a result previously obtained in other works.

Next, in Chapter~\ref{chapter:electrically_driven_fluids} we considered a hydrodynamic model that is analogue to the Drude model for a weakly-interacting gas. Specifically, we studied a charged fluid driven by an external strong electric field and that loses momentum and energy to impurities, which act as effective energy and momentum sinks for the system.

The presence of a momentum relaxation term in the (non-)conservation equations of hydrodynamics implies that boosts are broken in the theory, therefore we employ a boost-agnostic formalism in which the velocity is an integral part of the thermodynamics. In analogy to the Drude model, we focus on time-independent stationary states, and we use the hydrostatic generating functional to compute the constitutive relations up to order one in derivatives. Checking the consistency of the hydrodynamic equations order by order we infer that one of the hydrostatic constraints must be modified in the presence of relaxations. In particular, without momentum decay the difference between the electric field and the gradient of the chemical potential is zero on stationary flows, while the velocity is constant but unconstrained. However, we found that when momentum relaxation is present, the velocity of the fluid is fixed by the decay rate and the external driving force (the electric field), in agreement with phenomenological expectations.

Finally, we also found a kinematic constraint that momentum and energy relaxations must obey for the fluid not to produce entropy at order zero in derivatives. Furthermore, our theory gives new predictions for the thermoelectric transport of charged fluids in the presence of momentum relaxations.

Subsequently, in Chapter~\ref{chapter:onsager} we introduced the idea of generalized relaxations. Namely, we included in the linearized equations of hydrodynamics the most general relaxation terms proportional to the fluctuations of the conserved charges of the system. We obtained a set of very general constraints that the decay rates must obey in such systems, based on the requirement of microscopic time-reversal invariance (Onsager relations), positivity of entropy production, and linear stability.

However, because relaxation terms in general spoil the symmetries of the system, such as Lorentz and gauge invariance, it is not obvious how to couple quasihydrodynamic theories to curved spacetime and gauge fields to compute correlators using the variational approach. To amend the problem we included in the linearized equations of hydrodynamics all the possible source terms, each with its own free coefficient. What we found is that the simple requirement of time-reversal invariance of the microscopic theory uniquely fixes the values of all these extra parameters, giving access to the full set of Onsager-reciprocal correlators.

To conclude, in Chapter~\ref{chapter:anomalous_hydrodynamics} we studied models of anomalous hydrodynamics. First, we analysed the longitudinal transport in the presence of a background order-one magnetic field $B\sim\mathcal{O}(\partial)$, and we found that order-one hydrodynamics cannot capture the anomalous response of the fluid. This is because the magnetic-field corrections to the conductivities enter at order two in derivatives, while the hydrodynamic theory is order one. Indeed, this is confirmed by an explicit computation in a generic hydrodynamic frame, which shows that the result is frame dependent and therefore not physical. To solve this issue we argued that the best approach is to promote the magnetic field to be order zero in derivatives $B\sim\mathcal{O}(1)$, so that the background and conductivities are now frame independent.

Later, we analysed the DC transport of Weyl semimetals in the hydrodynamic regime. To obtain finite DC conductivities energy, momentum and axial charge must relax to equilibrium, however previous relaxation models failed to capture all the phenomenological and fundamental constraints that we expect in real systems. Our approach is based on generalized relaxations, and we showed that they are necessary if we want our system to have finite DC conductivities, obey Onsager relations, and conserve the electric charge. Furthermore, our model leads to new predictions for the thermoelectric transport, paving the road to experimental verifications of the hydrodynamic regime.

Finally, we also discussed how to derive generalized relaxations in hydrodynamics from kinetic theory. We found that we could introduce relaxations in the equations of hydrodynamics by considering RTA collision integrals that relax to fixed reference distribution functions. By allowing the RTA parameters to depend on the quasiparticle energy  $\tau=\tau(\epsilon_p)$ we obtained generalized relaxations from kinetic theory which identically obey the constraints imposed by Onsager relations. However, because the standard RTA does not conserve electric charge, we also suggested an improved version of the linearized collision integral which identically conserves charge and depends on two free parameters, which can be tuned to satisfy Onsager relations. Our results are very general, and apply to non-anomalous fluids too.

\section{Outlook and future avenues}
One interesting question that arises from this work is whether it is possible to recast the quasihydrodynamic models we analysed on a more formal ground. On this regard, there are many different approaches that we could try. First, it would be interesting to develop a framework along the lines of the Maxwell-Cattaneo model of \cite{Jain:SchwingerKeldyshEffectiveField}, in which relaxation terms could be introduced in the theory by including an explicit fixed vector $v^\mu$ that enters the thermodynamics
\begin{equation}
	\dif\epsilon=T\dif s+\mu\dif n+\frac{\chi_v}{2}\dif v^2
\end{equation}
and that we can align with the background clock-form. Another related approach, based on geometry, is to introduce extra degrees of freedom like the $v^\mu$ above, and to modify diffeomorphisms so that the equations of motion naturally include relaxation terms \cite{Abbasi:MagnetotransportAnomalousFluid}. Finally, we could also employ a Schwinger-Keldysh formalism to include extra non-dynamical fields that break the relevant symmetries \cite{Baggioli:QuasihydrodynamicsSchwingerKeldyshEffective,Jain:SchwingerKeldyshEffectiveField}.

On a parallel route, the topics covered in Chapter~\ref{chapter:electrically_driven_fluids} have potential applications to active matter systems. These have gained a lot of attention in recent years, due to their widespread presence in biophysics and ecology on one side, but also for their peculiar phenomenology and theoretical challenges. Active matter is characterized by the ability of its microscopic constituents to produce energy. Consequently, it describes intrinsically out-of-equilibrium phases with both microscopic and macroscopic dynamics significantly distinct from those observed in passive fluids. In \cite{Toner:FlocksHerdsSchools} the authors developed a macroscopic theory of transport for such systems, based on phenomenological considerations, which takes the form
\begin{subequations}
	\begin{align}
		&\partial_t\vect{v}+\lambda_1\left(\vect{v}\cdot\vect{\nabla}\right)\vect{v}+\lambda_2\left(\vect{\nabla}\cdot\vect{v}\right)\vect{v}+\lambda_3\vect{\nabla}\abs{\vect{v}}^2+\vect{\nabla}P\nonumber\\
		&\qquad-D_B\vect{\nabla}\left(\vect{\nabla}\cdot\vect{v}\right)-D_T\nabla^2\vect{v}-D_2\left(\vect{v}\cdot\vect{\nabla}\right)^2\vect{v}=\alpha\vect{v}-\beta\abs{\vect{v}}^2\vect{v}+\vect{f}\label{eqn:ch7:toner_model}\\
		&\partial_t\rho+\vect{\nabla}\cdot\left(\vect{v}\rho\right)=0\\
		&P=P(\rho)=\sum_{n=1}^\infty\sigma_n\left(\rho-\rho_0\right)^n
	\end{align}
\end{subequations}
These correspond respectively to the momentum equation of motion, particle number conservation and the equation of state. From a hydrodynamic perspective active matter systems do not have any boost symmetry, because they move with respect to some fixed background medium, and do not conserve momentum. Indeed, the LHS of \eqref{eqn:ch7:toner_model} generalizes Navier-Stokes for fluids without boost symmetry, while the RHS contains relaxation and noise terms which parametrize the non-conservation of momentum. Then, it would be interesting to apply the boost-agnostic hydrodynamic formalism of Chapter~\ref{chapter:electrically_driven_fluids} to derive the LHS of \eqref{eqn:ch7:toner_model} from a more rigorous approach, in order to check if terms are missing and if we can constrain the values of the free parameters $\lambda_i$ and $D_i$.

\pagestyle{plain}
\appendixpage

\appendix
\chapter{Integrals for kinetic theory}\label{appendix:useful_integrals}
In this Appendix we report some useful integrals \cite{Gorbar:ConsistentHydrodynamicTheory}, which we used to compute the thermodynamic quantities and the hydrodynamic equations that appear in Section~\ref{sec:ch6:kinetic_theory}. \texttt{Mathematica} correctly evaluates all these integrals.

Considering a linear dispersion relation $\epsilon_p=p$, relevant for Weyl semimetals, when integrating over momentum space we often encounter the integrals
\begin{subequations}
	\begin{align}
		\int\frac{\dif^3\vect{p}}{(2\pi)^3}p^{n-2}f^{(0)}&=-\frac{T^{n+1}\Gamma(n+1)}{2\pi^2}\text{Li}_{n+1}\left(-e^{\mu/T}\right)&\text{for }&n\geq0\\
		\int\frac{\dif^3\vect{p}}{(2\pi)^3}p^{n-2}\frac{\partial f^{(0)}}{\partial\epsilon_p}&=\frac{T^{n}\Gamma(n+1)}{2\pi^2}\text{Li}_{n}\left(-e^{\mu/T}\right)&\text{for }&n\geq0
	\end{align}
\end{subequations}
where $f^{(0)}$ is the Fermi-Dirac distribution function at zero velocity \eqref{eqn:ch6:small_velocity_expansion}, $\Gamma(n+1)$ is the Euler gamma function, while $\text{Li}_n(x)$ stands for the polylogarithm function.

When we compute thermodynamic quantities, we also need to sum over particles and holes contributions \eqref{eqn:ch6:two_currents_thermodynamics}. We can use the following expressions to simplify the results
\begin{subequations}
	\begin{align}
		\text{Li}_0\left(-e^x\right)+\text{Li}_0\left(-e^{-x}\right)&=-1\\
		\text{Li}_1\left(-e^x\right)-\text{Li}_1\left(-e^{-x}\right)&=-x\\
		\text{Li}_2\left(-e^x\right)+\text{Li}_2\left(-e^{-x}\right)&=-\frac{1}{2}\left(x^2+\frac{\pi^2}{3}\right)\\
		\text{Li}_3\left(-e^x\right)-\text{Li}_3\left(-e^{-x}\right)&=-\frac{x}{6}\left(x^2+\pi^2\right)\\
		\text{Li}_4\left(-e^x\right)+\text{Li}_4\left(-e^{-x}\right)&=-\frac{1}{4!}\left(x^4+2\pi^2x^2+\frac{7\pi^4}{15}\right)
	\end{align}
\end{subequations}
which can all be computed from the relation between the polylogarithm function for $n\geq0$ and Bernoulli polynomials $B_n(x)$
\begin{equation}
	\text{Li}_n\left(-e^{2\pi ix}\right)+(-1)^n\text{Li}_n\left(-e^{-2\pi ix}\right)=-\frac{\left(2\pi i\right)^n}{n!}B_n(x)
\end{equation}
Finally, we report here integrals over angular coordinates.
\begin{subequations}
	\begin{align}
		\int\frac{\dif^3\vect{p}}{(2\pi)^3}\vect{p}f(p)&=0\\
		\int\frac{\dif^3\vect{p}}{(2\pi)^3}\vect{p}\left(\vect{p}\cdot\vect{u}\right)f(p)&=\frac{u}{3}\int\frac{\dif^3\vect{p}}{(2\pi)^3}p^2f(p)\\
		\int\frac{\dif^3\vect{p}}{(2\pi)^3}\vect{p}\left(\vect{p}\cdot\vect{u}\right)\left(\vect{p}\cdot\vect{w}\right)f(p)&=0
	\end{align}
\end{subequations}
where $\vect{u}$ and $\vect{w}$ are generic vectors. These integrals are the reason why we can decouple the momentum equation from the energy and charge equations once we expand the distribution function at small velocity as in \eqref{eqn:ch6:small_velocity_expansion}.

\chapter{Aspects of Symmetry}\label{appendix:currents}
From the Nielsen-Ninomiya theorem \cite{Nielsen:NogoTheoremRegularizing,Nielsen:AdlerBellJackiwAnomalyWeyl,Nielsen:AbsenceNeutrinosLattice} we know that the net chirality in each Brillouin zone must be zero $\sum_i \lambda_i=0$. Thus, we focus on the simplest case with one left- and one right-handed Weyl cone, that behave as sources for the chirality (they are monopoles of Berry curvature in momentum space).

Consider a chiral system with independent $\mathrm{U(1)}$ symmetry for fermions of different chirality, which implies conserved currents at the classical level. We can trade the $\mathrm{U(1)}_L\times \mathrm{U(1)}_R$ currents in favour of a vector and axial description $\mathrm{U(1)}_V\times\mathrm{U(1)}_A$ as in \eqref{eqn:ch6:chiral_to_axial_currents}. Then, in the presence of external electromagnetic and pseudo-electromagnetic fields, the currents become anomalous and cannot be made both conserved (this is the true statement of the mixed 't~Hooft anomaly). Since we would like to interpret the vectorial current as an electric current in QED (i.e. $\partial_\mu F^{\mu\nu}=J^\nu$ for dynamical gauge fields), we add counterterms in the action such that the anomaly only sits in the axial current. This is the form of the so-called \emph{consistent} anomaly \cite{Landsteiner:NotesAnomalyInduced,Bardeen:ConsistentCovariantAnomalies}
\begin{equation}
	\partial_\mu J^\mu_\text{cons}=0\qquad\qquad\partial_\mu J^\mu_\text{5, cons}=-\frac{c}{24}\epsilon^{\mu\nu\rho\sigma}\left(3F_{\mu\nu}F_{\rho\sigma}+F_{\mu\nu}^5F_{\rho\sigma}^5\right)
\end{equation}
where the anomaly coefficient is $c=\frac{1}{2\pi^2}$ for a single fermion family. From the diffeomorphism invariance of the generating functional for the connected correlators $W=-i\ln Z$ (in the absence of gravitational anomalies), we obtain the conservation equation of the stress-energy tensor
\begin{equation}\label{eqn:appendix:consistent_stress_tensor_equation}
	\partial_\mu T^\mu_{\ \nu}=F_{\nu\lambda}J_\text{cons}^\lambda+F_{\nu\lambda}^5J_\text{5, cons}^\lambda-\partial_\mu J^\mu_\text{5, cons}A_\nu^5
\end{equation}

However, the currents so defined are not gauge covariant, meaning that the BRST operator does not commute with $\delta W/\delta A_\mu$. For this reason one usually define the \emph{covariant} currents by adding to the consistent current a Chern-Simons term (also known as Bardeen-Zumino polynomial)
\begin{subequations}
	\begin{align}
		J^\mu_\text{cov}&=J^\mu_\text{cons}+J^\mu_\text{BZ}=J^\mu_\text{cons}-\frac{c}{2}\epsilon^{\mu\nu\rho\sigma}A_\nu^5F_{\rho\sigma}\\
		J^\mu_\text{5, cov}&=J^\mu_\text{5, cons}+J^\mu_\text{5, BZ}=J^\mu_\text{5, cons}-\frac{c}{6}\epsilon^{\mu\nu\rho\sigma}A_\nu^5F_{\rho\sigma}^5
	\end{align}
\end{subequations}
The Chern-Simons currents cannot be obtained as variation of a 4 dimensional action, however they can appear via anomaly inflow as boundary variation of a 5 dimensional Chern-Simons action. These are topological currents that depend explicitly on $A_\nu^5$ in such a way to cancel the gauge dependence of the consistent currents\footnote{Remember that the Chern-Simons action does not depend on $g_{\mu\nu}$, hence the definition of the stress-energy tensor is the same in the covariant or consistent formalism.}. Then, for this choice of currents, the equations of motion become
\begin{subequations}
	\begin{align}
		\partial_\mu T^\mu_{\ \nu}&=F_{\nu\lambda}J^\lambda_\text{cov}+F_{\nu\lambda}^5J^\lambda_\text{5, cov}\\
		\partial_\mu J^\mu_\text{cov}&=-\frac{c}{4}\epsilon^{\mu\nu\rho\sigma}F_{\mu\nu}F_{\rho\sigma}^5\\
		\partial_\mu J^\mu_\text{5, cov}&=-\frac{c}{8}\epsilon^{\mu\nu\rho\sigma}\left(F_{\mu\nu}F_{\rho\sigma}+F_{\mu\nu}^5F_{\rho\sigma}^5\right)
	\end{align}
\end{subequations}

Notice that for the case of interest, namely the longitudinal magnetotransport in Weyl semimetals at $F_{\mu\nu}^5=0$, the Bardeen-Zumino currents are identically conserved ($J^\mu_\text{5, BZ}$ is actually zero) and thus the currents conservation equations are the same in both framework.

We remark that in Chapter~\ref{chapter:anomalous_hydrodynamics} we had to include an energy relaxation term to achieve finite DC conductivities. This was necessary because, from the covariant point of view, the RHS of the energy equations has a constant DC Maxwell term $F_{0\nu}J^\nu_\text{cov}\sim \vect{E}\cdot\vect{J}_\text{cov}\sim\vect{E}\cdot \vect{B}$, because $J^\mu_\text{cov}=\xi_B B^\mu$ in equilibrium, see \eqref{eqn:ch6:equilibrium_order_one}. From a consistent point of view, however, the equilibrium current vanishes, see Appendix~\ref{appendix:weyl_semimetals}, thus the Maxwell term is zero in the energy-conservation equation. Nonetheless, we still need energy relaxation because in the consistent description the RHS of the energy-conservation equation has an explicit anomaly term \eqref{eqn:appendix:consistent_stress_tensor_equation}.

\chapter{Weyl semimetals}\label{appendix:weyl_semimetals}
Consider a Weyl semimetal with two nodes in each Brillouin zone, one left- and one right-handed. We can describe its dynamics using Weyl equations, alternatively we can employ the Dirac equation in the presence of an axial gauge field $b_\mu$ \cite{Landsteiner:NotesAnomalyInduced}
\begin{equation}
	\left(i\slashed{\partial}-\gamma_5\slashed{b}\right)\psi=0
\end{equation}
The role of $b_\mu$ is to parametrize the distance in energy-momentum space between the two Weyl nodes. The dispersion relation is indeed
\begin{equation}
	\omega_{L,R}=\pm b_0\pm\sqrt{(\vect{p}-\vect{b})^2}
\end{equation}
and is shifted for the left- and right-handed components due to $b_\mu=A_\mu^5$. Notice that this axial gauge field is observable, contrary to the vector gauge field $A_\mu$ (it is not protected by gauge transformations). In materials without strain or boundaries $b_\mu$ is a constant, however in strained samples and near the boundaries $F_{\mu\nu}^5$ can be non-zero \cite{Cortijo:ElasticGaugeFields,Pikulin:ChiralAnomalyStraininduced}.

\begin{figure}
	\centering
	\includegraphics[width=0.8\textwidth]{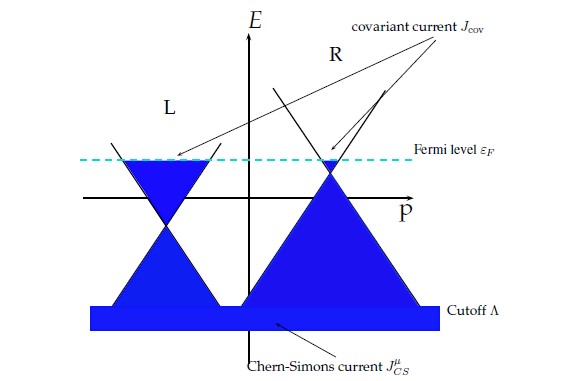}
	\caption{Figure taken from \cite{Landsteiner:NotesAnomalyInduced}. $2d$ projection of Weyl nodes shifted in energy-momentum space. The covariant current describes the dynamics near the Fermi surface, while the Bardeen-Zumino current close to the bottom of the Dirac sea.}
	\label{fig:appendix:weyl_nodes}
\end{figure}

Chiral Kinetic Theory naturally provides a description in terms of covariant currents \cite{Gorbar:ConsistentHydrodynamicTheory}, this is because kinetic theory by its nature studies the small fluctuations above the vacuum (the Fermi surface in this case), while the Chern-Simons currents are associated with the dynamics at the bottom of the Dirac sea, or with the cut-off from the point of view of the field theory computations \cite{Landsteiner:NotesAnomalyInduced}, see the Figure~\ref{fig:appendix:weyl_nodes}.

To shift from the covariant currents to the consistent ones, we need to add the Chern-Simons (Bardeen-Zumino) currents, see Appendix~\ref{appendix:currents}. The vector Bardeen-Zumino current takes the following form in terms of covariant electric and magnetic fields\footnote{Remember that $J^{\mu}_\text{5, BZ}=0$ when $F_{\mu\nu}^5=0$.}
\begin{equation}
	J^\mu_\text{BZ}=-c\varepsilon^{\mu\nu\rho\sigma}b_\nu u_\rho E_\sigma-c u^\mu(b\cdot B)+cB^\mu(b\cdot u)
\end{equation}
where we remind $b^\mu=A_\mu^5$ is the distance in momentum space between the two Weyl nodes. This current contains information about the anomalous Hall effect, and it is the current that appears in the Maxwell equations in a theory of magneto-hydrodynamics \cite{Gorbar:ConsistentHydrodynamicTheory}.

The consistent currents are the only one compatible with local charge conservation and Bloch theorem. Indeed, the consistent current in equilibrium is
\begin{equation}\label{eqn:appendix:consistent_current_equilibrium}
	\vect{J}_\text{cons}\bigr|_\text{eq}=c\left(\mu_5-b_0\right)\vect{B}=0
\end{equation}
because the chemical potential is fixed by the separation in energy between the nodes\footnote{Contrary to the vector chemical potential, the identification of $\mu_5$ with $A_0^5$ is more subtle \cite{Landsteiner:AnomalousTransportKubo}, so the two terms cancel only in equilibrium and not identically on any solution.} \cite{Landsteiner:AnomalousTransportKubo,Landsteiner:NotesAnomalyInduced}. However, this Chern-Simons current does not participate in the longitudinal magneto-transport, hence the results obtained in Chapter~\ref{chapter:anomalous_hydrodynamics} are correct and independent on the choice of framework (covariant or consistent currents).

\backmatter
\printbibliography[heading=bibintoc]

\end{document}